\documentclass[twocolumn,aps,prl,amsmath,amssymb,superscriptaddress]{revtex4-2}

\usepackage{newtxtext}
\usepackage{newtxmath}
\usepackage{microtype}
\usepackage{wasysym}
\usepackage{graphicx}
\usepackage{dcolumn}
\usepackage{bm}
\usepackage{multirow}
\usepackage{amssymb}
\usepackage{graphicx}
\usepackage{braket}
\usepackage{booktabs} 
\allowdisplaybreaks
\newcommand{\Tr}{\text{Tr} \ }
\newcommand{\beq}{\begin{eqnarray}}
\newcommand{\eeq}{\end{eqnarray}}
\renewcommand{\Re}{\mathop{\mathrm{Re}}}
\renewcommand{\Im}{\mathop{\mathrm{Im}}}

\renewcommand{\mod}{\,\mathrm{mod}\,}
\usepackage{xcolor}
\definecolor{TighnariBrown}{RGB}{141,87,41}
\definecolor{TighnariGreen}{RGB}{40,114,70}
\definecolor{TighnariYellow}{RGB}{247,191,99}
\definecolor{PRLBlue}{RGB}{46,48,146}
\usepackage[
    colorlinks,
    citecolor=PRLBlue,
    linkcolor=PRLBlue,
    urlcolor=PRLBlue,
]{hyperref}
\usepackage{indentfirst}
\usepackage{cases}

\begin{document}
\title{Transport and Temperature: Exact spectrum and Drude weight for the 
one-dimensional infinite-$U$ Hubbard model}

\author{Shuo Liu}
\affiliation{Department of Physics, Princeton University, Princeton, New Jersey 08544, USA}

\author{Yuhao Ma}
\affiliation{Department of Physics and Institute of Condensed Matter Theory, University of Illinois at Urbana-Champaign, Urbana, IL 61801, USA}

\author{Hitesh J. Changlani}
\affiliation{National High Magnetic Field Laboratory, Tallahassee, Florida 32310, USA}
\affiliation{Department of Physics, Florida State University, Tallahassee, Florida 32306, USA}

\author{Philip W. Phillips}
\affiliation{Department of Physics and Institute of Condensed Matter Theory, University of Illinois at Urbana-Champaign, Urbana, IL 61801, USA}

\author{B. Andrei Bernevig}
\email{bernevig@princeton.edu}
\affiliation{Department of Physics, Princeton University, Princeton, New Jersey 08544, USA}
\affiliation{Donostia International Physics Center (DIPC), Paseo Manuel de Lardizábal, 20018 San Sebastián, Spain}
\affiliation{IKERBASQUE, Basque Foundation for Science, 48013 Bilbao, Spain}

\date{\today}

\begin{abstract}
Understanding charge transport in strongly correlated systems remains a central challenge in condensed matter physics, while exact analytical results for finite-temperature transport remain rare. Here, we investigate charge transport in the one-dimensional Hubbard model in the infinite-interaction
limit.  We first construct the exact \emph{and explicit - i.e. beyond Bethe ansatz} energy spectrum and eigenstates. Building on these explicit eigenstates, we derive a closed-form analytical expression for the charge Drude weight at
arbitrary temperatures in the hole-dilute limit with a fixed number of doped holes, as well as analytical expressions in the low- and high-temperature limits at fixed hole density. We further show that, at low temperatures, the leading correction to the charge Drude weight is linear in $T$ when the number
of holes is fixed, whereas it is quadratic in $T$ when the hole density is fixed. This analytically controlled example highlights the importance of identifying the underlying thermodynamic trajectory when interpreting finite-size numerical studies of charge transport in more general models.
\end{abstract}

\maketitle

\textit{Introduction.---} Since the experimental discovery of high-temperature superconductivity (SC) in cuprates several decades ago, a consensus on the underlying microscopic mechanism has remained elusive~\cite{RevModPhys.66.763,RevModPhys.75.473,RevModPhys.82.1719, doi:10.1073/pnas.1316512110,RevModPhys.87.457, keimer_quantum_2015,PhysRevB.39.6880}. One point of agreement is that the answer lies in the solution to the doped Mott insulator problem~\cite{RevModPhys.82.1719}.  Central to this problem is the dynamical generation of a charge gap in half-filled band~\cite{RevModPhys.82.1719,PhysRevB.48.3916}.  In this regard, the two-dimensional (2D) Hubbard model~\cite{doi:10.1126/science.235.4793.1196, PhysRevB.37.3759, annurev:/content/journals/10.1146/annurev-conmatphys-090921-033948} is viewed as the gold standard for the copper-oxide layer.  Other popular models include the $t$-$J$ model which approximates the dynamical physics of the Hubbard model. However, these strongly correlated 2D systems remain analytically intractable. Numerical simulations also face severe limitations, such as the limited system widths accessible to density matrix renormalization group (DMRG)~\cite{annurev:/content/journals/10.1146/annurev-conmatphys-020911-125018} or the sign problem in quantum Monte Carlo (QMC) simulation~\cite{PhysRevB.41.9301,PhysRevLett.94.170201,PhysRevB.71.155115,PhysRevB.91.241117,PhysRevLett.117.267002,PhysRevLett.116.250601,annurev:/content/journals/10.1146/annurev-conmatphys-033117-054307, PhysRevB.106.L241109, PhysRevB.101.045108,PhysRevLett.134.016503}. Additionally, we note recent developments utilizing neural quantum states~\cite{roth2025superconductivitytwodimensionalhubbardmodel, chen2025neuralnetworkaugmentedpfaffianwavefunctions,gu2025solvinghubbardmodelneural}.

An extreme limit of the $t$-$J$ model is the $J=0$ case in which the high-energy sector is completely decoupled.  As $J$ is inversely related to the coupling constant in the Hubbard model, $J=0$ can be viewed as the infinite-interaction limit~\cite{PhysRev.147.392,PhysRevLett.108.126406}.  Even a single hole in this background is a non-trivial problem and the resultant transport properties remain unknown.  For certain classes of lattices (there are now generalizations to honeycomb and diamond lattices~\cite{PhysRevB.98.180101}), the Nagaoka theorem tells us that a fully polarized ferromagnetic state is the ground state~\cite{PhysRev.147.392}. 
Counterexamples to the Nagaoka theorem have also been investigated on frustrated lattices, including the $120^\circ$ antiferromagnetic ground state on the triangular lattice~\cite{PhysRevLett.95.087202,sherif2025haertershastrykineticmagnetismmetallicity,sharma2025instabilitynagaokastatequantum} and the exact resonating-valence-bond spin-liquid ground state on the checkerboard lattice~\cite{glittum_resonant_2025}.
Besides these analytical results, a number of intriguing phenomena have been reported in this infinite-interaction limit, based on numerical studies and experiments. These include kinetically induced hole--magnon bound states~\cite{PhysRevB.97.140507,PhysRevResearch.6.023196,10.21468/SciPostPhys.16.3.081,qiao2025kineticallyinducedboundstatesfrustrated} on the triangular lattice and, remarkably, signatures of strange-metal transport on the square lattice with a single doped hole~\cite{fratini2025strangemetaltransportcoupling}. However, even in this infinite-interaction limit, more general situations such as multiple doped holes remain highly nontrivial and analytically challenging.

A natural starting point for analytically understanding the physics of the 2D Hubbard model~\cite{doi:10.1126/science.235.4793.1196, PhysRevB.37.3759, annurev:/content/journals/10.1146/annurev-conmatphys-090921-033948, PhysRevX.13.011007, PhysRevB.94.235115, annurev:/content/journals/10.1146/annurev-conmatphys-031620-102024, PhysRevLett.110.216405,PhysRevB.75.085108, PhysRevB.82.155101, PhysRevLett.68.2402, 10.21468/SciPostPhys.6.6.067,lu2026thermalsignaturesslatermottcrossover,song2026computationthermalentropydoped, PhysRevX.11.031007} is its one-dimensional (1D) counterpart~\cite{essler2005one}. The 1D Hubbard model is integrable and thus admits an exact solution via the Bethe ansatz~\cite{PhysRevLett.20.1445, PhysRevLett.56.1529, 10.1143/PTP.47.69,10.1143/PTP.52.103,gohmann1998fermionic,IZERGIN1998537,maassarani1998n, PhysRevB.41.2326}. Despite this integrability, the Bethe ansatz solutions are not explicit, and a comprehensive analytical understanding of transport remains highly nontrivial~\cite{zotos2005issues,heidrich-meisner_transport_2007,RevModPhys.93.025003, 10.21468/SciPostPhys.15.2.073, 10.21468/SciPostPhys.12.1.028, PhysRevE.104.044106, PhysRevB.106.094303, PhysRevLett.131.037101,wl94-p2cd,PhysRevB.48.10595,fujimoto2026exactanomalouscurrentfluctuations}. At half filling, the system is a Mott insulator, and consequently the charge Drude weight vanishes~\cite{carmelo2018absence,carmelo2013zero,PhysRevB.70.205129,PhysRevB.90.155104}. Away from half filling, the charge Drude weight becomes finite~\cite{PhysRevLett.65.243,PhysRevB.43.13660,PhysRevB.96.081118,PhysRevLett.121.230602,PhysRevB.102.115121,zotos,garst2001transport}.  
Its exact expression at zero temperature has been obtained in Refs.~\cite{PhysRevLett.65.243,PhysRevB.43.13660,PhysRevB.61.5169}. The leading finite-temperature correction was later derived from the finite-size corrections to the thermodynamic Bethe ansatz equations, yielding a $T^{2}$ dependence~\cite{fujimoto1998exact}. However, the full temperature dependence of the Drude weight remains unknown, and no simple analytical expression has been established. Consequently, much of the existing understanding relies on numerical approaches~\cite{PhysRevB.42.8736,PhysRevB.44.6909,PhysRevB.92.205103,PhysRevLett.85.3910,PhysRevB.67.161103,PhysRevB.59.1825}.

In this work, we focus on the 1D Hubbard model in the infinite-interaction limit. We first construct the exact but more importantly \emph{explicit} eigenstates and energy spectrum of the Hamiltonian (see Sec. III
of the Supplemental Material (SM)~\cite{SupplementalMaterials}). Our systematic construction in the occupation basis captures the non-trivial spin-sector degeneracies and corresponding boundary twists. Building on these \emph{explicit} eigenstates, we then derive an analytical expression for the charge Drude weight at arbitrary temperatures with a single doped hole (see Sec. IV
of the SM~\cite{SupplementalMaterials}), whose form becomes particularly simple in the large-system-size limit (see Eq.~\eqref{eq:Drudeweight}). The result can be straightforwardly generalized to an arbitrary but finite number of doped holes (see 
Sec. IV of the SM~\cite{SupplementalMaterials}). In this hole-dilute limit, where the number of holes $n_h$ is kept fixed as the large-system-size limit is taken, the charge carriers propagate essentially independently. Consequently,
$n_h$ enters as an overall prefactor in the Drude weight. Furthermore, for a fixed density $\delta$ of doped holes, where $\delta =n_h/N$ is fixed in the thermodynamic limit $N\to \infty$, we derive analytical expressions for the charge Drude weight in both the low- and high-temperature limits (see Sec. V of the SM~\cite{SupplementalMaterials}). We note that these \emph{explicit} eigenstates may also provide insight into other properties of the present model, including spin transport~\cite{fujimoto2026exactanomalouscurrentfluctuations,10.21468/SciPostPhys.15.2.073}.

More importantly, based on the analytical expression, we further analyze the low-temperature scaling of the charge Drude weight for both a single doped hole and a fixed density of holes, as summarized in Table~\ref{tab:main_Drude}. We find that the leading temperature dependence of the Drude weight is linear in $T$ for a single doped hole, whereas it exhibits a $T^2$ correction at a fixed hole density. The latter scaling is consistent with the results in Ref.~\cite{fujimoto1998exact}; moreover, our calculation provides the exact analytical expression for the corresponding expansion coefficient.

The linear-in-$T$ resistivity behaviors associated with strange-metal phenomenology have been reported in numerical studies of the 2D Hubbard model~\cite{PhysRevLett.110.086401,PhysRevB.91.075124,PhysRevB.95.041110,PhysRevLett.123.036601,doi:10.1126/science.aau7063,fratini2025strangemetaltransportcoupling} and in experiments on cuprate superconductors at fixed doping~\cite{doi:10.1126/science.abh4273,marelQuantumCriticalBehaviour2003, PhysRevLett.81.4720, PhysRevLett.92.167001, jinLinkSpinFluctuations2011, PhysRevB.41.846, doi:10.1126/science.1165015, daouLinearTemperatureDependence2009, PhysRevB.95.224517, doiron-leyraudPseudogapPhaseCuprate2017,legrosUniversalTlinearResistivity2019, doi:10.1126/science.aat4134}. We also note that SYK-type models can produce linear-in-$T$ resistivity in the all-to-all interaction configuration~\cite{RevModPhys.94.035004,doi:10.1126/science.abq6011,doi:10.1073/pnas.2003179117,PhysRevLett.133.186502}, and similar behavior has also been observed in twisted bilayer graphene~\cite{polshynLargeLinearintemperatureResistivity2019, PhysRevB.99.165112,PhysRevLett.124.076801,jaouiQuantumCriticalBehaviour2022}. Although the present 1D infinite-$U$ Hubbard model is neither equivalent to the 2D Hubbard model nor a realistic microscopic description of cuprate materials, it provides an analytically controlled example demonstrating that the fixed-hole-number and fixed-hole-density thermodynamic limits can exhibit qualitatively distinct charge-transport behaviors. Therefore, extreme caution is required when extrapolating 2D finite-size numerical results—especially those from exact diagonalization (ED)—to finite-doping strange-metal phenomenology (see Sec.~VI of the SM~\cite{SupplementalMaterials} for further details, including a discussion of the respective limitations of ED, QMC, and DMRG). In such ED calculations, severely restricted cluster sizes often inherently constrain the system to a fixed-carrier-number trajectory rather than a true fixed-density limit. This limitation further underscores the need for analytical solutions.

\textit{Model and definition of optical conductivity.---} In this work, we consider the 1D Hubbard model in the infinite-interaction limit. 
The Hamiltonian is given by
\begin{eqnarray}
\label{eq:main_Hamiltonian_1D}
\hat{H}
= - t \sum_{\sigma=\{\uparrow,\downarrow\}} \sum_{j=0}^{N-1}
\left(
\tilde{c}^{\dagger}_{j+1,\sigma}\tilde{c}_{j,\sigma}
+ \tilde{c}^{\dagger}_{j,\sigma}\tilde{c}_{j+1,\sigma}
\right),
\end{eqnarray}
where
\(
\tilde{c}^{\dagger}_{j,\sigma}
= (1 - n_{j,\bar{\sigma}})\, c^{\dagger}_{j,\sigma}
\)
denotes the projected creation operator for a fermion with spin $\sigma$ at site $j$ and
\(
n_{j,\bar{\sigma}} = c^{\dagger}_{j,\bar{\sigma}} c_{j,\bar{\sigma}}
\)
is the density operator for fermions with the opposite spin $\bar{\sigma}$, thereby enforcing the no-double-occupancy constraint that encodes the
strong-correlation physics of the infinite-$U$ limit and distinguishes the
model from a non-interacting system. The hopping amplitude $t$ sets the overall energy scale, and we set $t=1$ throughout this work. We impose periodic boundary conditions (PBCs), such that
\(
\tilde{c}^{\dagger}_{N,\sigma} \equiv \tilde{c}^{\dagger}_{0,\sigma},
\)
where $N$ is the total number of lattice sites. Moreover, we denote by $n_{h}$ the number of doped holes in the system.

The real part of the optical conductivity 
(see Sec.~I of the SM~\cite{SupplementalMaterials})
can be expressed as
\begin{eqnarray}
\mathrm{Re}\,\sigma(\omega,T)
= 2\pi D(T)\,\delta(\omega)
+ \sigma_{\mathrm{reg}}(\omega,T),
\label{eq:main_optical}
\end{eqnarray}
where $T$ is the temperature, $\omega$ the frequency, $\delta(\cdot)$ the Dirac delta function, $D(T)$ the temperature-dependent charge Drude weight corresponding to the singular contribution at zero frequency, and $\sigma_{\mathrm{reg}}(\omega,T)$ the regular (finite-frequency) contribution. Here, ``regular'' indicates that this part does not contain a $\delta(\omega)$ singularity at zero frequency.
The DC conductivity at zero frequency can be formally decomposed as
\begin{align}
     \label{eq:main_dc}
    \sigma_{\mathrm{DC}}(T) = 2 \pi D(T) \delta(\omega) + \lim_{\omega \to 0} \sigma_{\mathrm{reg}}(\omega, T),
\end{align}
consisting of both the Drude weight contribution and the contribution obtained by extrapolating the regular part to $\omega=0$. Note that although $ \delta(\omega)$ is singular at $\omega=0$ and has no
ordinary pointwise value, we formally retain $\delta(\omega)$ to denote the
zero-frequency Drude contribution; the physically meaningful quantity
characterizing its spectral weight is the Drude weight $D(T)$. As a consequence, a non-zero Drude weight implies an ideal conductor. Whether the Drude weight is non-zero rests on the existence of conservation laws, as can be seen from the Mazur inequality~\cite{MAZUR1969533,zotos},
\beq
D\ge \frac{\beta}{2N} \sum_n\frac{\langle \hat J \hat{Q}_n\rangle^2}{\langle \hat{Q}_n^2\rangle},
\label{drude}
\eeq
where $\hat{Q}_n$ is conserved. In typical notation~\cite{zotos}, $\hat{Q}_2$ is the Hamiltonian.  We find that $\hat J$ commutes with the Hamiltonian, implying a non-zero overlap on the RHS of Eq. (\ref{drude}) and hence a finite Drude weight (see Sec.~I of the SM~\cite{SupplementalMaterials}).

$D(T)$ and  $\sigma_{\mathrm{reg}}(\omega,T)$
(see Refs.~\cite{PhysRevLett.65.243, PhysRevB.16.2437, PhysRevB.35.8391, PhysRevLett.68.2830,PhysRevB.59.1825, PhysRevLett.134.176501} or 
Sec. I of the SM~\cite{SupplementalMaterials}
for a detailed derivation) are given by 
\begin{align}
\label{eq:main_Drudeweight}
D(T)
&= \frac{1}{N}
\left[
\frac{1}{2}\,\langle -\hat{K} \rangle
- \frac{1}{Z}
\sum_{\substack{m,n\\ E_m \neq E_n}}
e^{-\beta E_n}
\frac{|J_{mn}|^2}{E_m - E_n}
\right],
\end{align}
\begin{align}
\label{eq:main_regular}
\sigma_{\mathrm{reg}}(\omega,T)
&= \pi \frac{1 - e^{-\beta \omega}}{\omega Z N}
\sum_{n,m}
e^{-\beta E_n}
\left| J_{mn} \right|^2
\delta(E_n + \omega - E_m),
\end{align}
respectively. In Eq.~\eqref{eq:main_Drudeweight} the first term is the diamagnetic response whereas the latter term is the paramagnetic contribution, both of which enter the DC response. Since the paramagnetic contribution is nonnegative, the diamagnetic response provides an upper bound on the Drude weight (see Sec.~I of the SM~\cite{SupplementalMaterials}). Note that the expression for the regular part in Eq.~\eqref{eq:main_regular} does not apply strictly at zero frequency, as is clear from the derivation presented in 
Sec. I~\cite{SupplementalMaterials}.
Here $\beta = 1/T$ is the inverse temperature, $\hat{K}$ denotes the kinetic-energy operator, which coincides with the Hamiltonian $\hat{H}$ in the present model (see Sec.~I of the SM~\cite{SupplementalMaterials})
and \(
\langle -\hat{K} \rangle
=\mathrm{Tr}( -\hat{K} e^{-\beta \hat{H}} )/Z
\)
denotes its thermal expectation value, where 
\( Z = \mathrm{Tr}\!\left( e^{-\beta \hat{H}} \right) \) is the partition function,  $E_n$ is the eigenenergy of the $n$-th eigenstate $\ket{n}$ of the Hamiltonian $\hat{H}$. 
The charge current operator $\hat{J}$ (see Sec.~II of the SM~\cite{SupplementalMaterials})
is 
\begin{eqnarray}
\label{eq:main_currentoperator_1D}
\hat{J}
= i \sum_{\sigma=\{\uparrow,\downarrow\}} \sum_{j=0}^{N-1}
\left(
\tilde{c}^{\dagger}_{j+1,\sigma}\tilde{c}_{j,\sigma}
- \tilde{c}^{\dagger}_{j,\sigma}\tilde{c}_{j+1,\sigma}
\right),
\end{eqnarray}
and its matrix elements in the eigenbasis of $\hat{H}$ are defined as
\(
J_{mn} = \bra{m} \hat{J} \ket{n}.
\)

\textit{Exact solutions of the energy spectrum.---} 
Inspired by the frozen-spin structure, namely, the fact that the action of $\hat{H}$ preserves the relative cyclic ordering of the spin-up and spin-down fermions, we have obtained the complete set of eigenstates and the energy spectrum explicitly. The detailed derivations including the explicit expressions for the eigenstates for the cases with one hole, two holes, and an arbitrary number of holes are presented in Sec.~III~\cite{SupplementalMaterials}.
Below, we present only the energy spectrum with a single doped hole.

In the fully polarized sector with 
$S_{z} = \frac{1}{2}\sum_{j}(n_{j, \uparrow} - n_{j,\downarrow}) = \pm \frac{N-1}{2}$, corresponding to $N-1$ spin-up fermions (or $N-1$ spin-down fermions),
the energy spectrum (see Sec.~III~\cite{SupplementalMaterials})
is
\begin{eqnarray}
\label{eq:main_EWithoutDown}
E(m) = 2 \cos\!\left( \frac{2\pi m}{N} \right),
\end{eqnarray}
where $m = 0,\dots,N-1$ labels the distinct eigenstates.

More generally, in the sector with 
$S_{z} = \frac{N-1}{2} - n_{\downarrow}$ and $0<n_{\downarrow}<N-1$,
where $n_{\downarrow}$ denotes the number of spin-down fermions, 
the energy levels take the form (see Sec.~III of the SM~\cite{SupplementalMaterials})
\begin{eqnarray}
\label{eq:main_spectrum}
E(m,m_{\downarrow}) 
= 2 \cos\!\left(
\frac{2\pi m}{N}
+ \frac{2\pi m_{\downarrow}}{N(N-1)}
\right),
\end{eqnarray}
with $m=0,\dots,N-1$ and $m_{\downarrow}=0,\dots,N-2$. The degeneracy of each such energy level is $\frac{1}{N-1}\binom{N-1}{n_{\downarrow}}$.
For finite-size systems, additional eigenvalues not captured by Eq.~\eqref{eq:main_spectrum} may appear. However, as discussed in Sec.~III~\cite{SupplementalMaterials},
the number of such states is negligible in the thermodynamic limit and therefore does not affect the analytical results for charge transport presented below.

We note that Eqs.~\eqref{eq:main_EWithoutDown} and~\eqref{eq:main_spectrum} are consistent with the spin--charge separation exhibited in the present model~\cite{PhysRevB.41.2326}: its charge degrees of freedom can be mapped onto non-interacting spinless fermions. However, the spin degrees of freedom affect the spectrum by effectively introducing twisted boundary conditions, as shown in Eq.~\eqref{eq:main_spectrum} and Sec.~III of the SM~\cite{SupplementalMaterials}; determining these boundary twists explicitly is nontrivial. Moreover, because the explicit many-body eigenstates retain the full
information about both the charge and spin degrees of freedom beyond that
captured by the noninteracting spinless-fermion model, they may also provide
insight into other properties, such as spin transport~\cite{fujimoto2026exactanomalouscurrentfluctuations,10.21468/SciPostPhys.15.2.073}.

\begin{table*}[t]
\caption{Analytical expressions for the charge Drude weight in different limits. }
\label{tab:main_Drude}
\begin{ruledtabular}
\begin{tabular}{lcc}
Case & High-temperature limit & Low-temperature limit \\
\hline
Single hole 
& $D_{1h}(T\to\infty)=\dfrac{1}{NT}$ 
& $D_{1h}(T\to 0)=\dfrac{1}{N}\!\left(1-\dfrac{T}{4}\right)$ \\[6pt]

Fixed number of holes $n_h$ 
& $D_{n_h}(T\to\infty)=\dfrac{n_h}{NT}$ 
& $D_{n_h}(T\to 0)=\dfrac{n_h}{N}\!\left(1-\dfrac{T}{4}\right)$ \\[6pt]

Fixed hole density $\delta=n_h/N$ 
& $D_{\delta}(T\to\infty)=\dfrac{\delta(1-\delta)}{T}$ 
& $D_{\delta}(T\to 0)=\dfrac{\sin(\pi\delta)}{\pi}
-\dfrac{\pi T^2}{24\,\sin(\pi\delta)}$
\end{tabular}
\end{ruledtabular}
\end{table*}

\textit{Analytical expression for the Drude weight.---} In this section, we present the analytical expression for the Drude weight building upon the explicit energy spectrum and eigenstates. We focus on the case of a single doped hole, and also discuss the cases of an arbitrary fixed number of holes and of a fixed density of holes.

First, we note that in the present model the current operator $\hat{J}$ commutes with the Hamiltonian $\hat{H}$, i.e., $[\hat{J},\hat{H}] = 0$ (see Sec.~II of the SM~\cite{SupplementalMaterials}). By Eq. (\ref{drude}), the Drude weight must be finite.
This property is special to the infinite-interaction limit and does not hold for the 1D Hubbard model with finite interaction strength, where generally $[\hat{J},\hat{H}] \neq 0$.  As a consequence, $\hat{H}$ and $\hat{J}$ share a common set of eigenstates $\ket{n}$, and thus we have
\begin{eqnarray}
\ \ \vert J_{mn} \vert =0 \ \text{when} \ m \neq n,
\end{eqnarray} 
which is also true for degenerate states (see Sec.~II of the SM~\cite{SupplementalMaterials}).
Together with $\hat{K}=\hat{H}$ for the present model,
the Drude weight in Eq.~\eqref{eq:main_Drudeweight} simplifies to 
\begin{align}
\label{eq:main_Drudeweight2}
D(T)
&= \frac{1}{2N}\,\langle -\hat{H} \rangle,
\end{align}
and the regular part in Eq.~\eqref{eq:main_regular} becomes
\begin{align}
\label{eq:main_regular2}
\sigma_{\mathrm{reg}}(\omega,T)
&= \pi \frac{1 - e^{-\beta \omega}}{\omega Z N}
\sum_{n}
e^{-\beta E_n}
\left| J_{nn} \right|^2
\delta(\omega).
\end{align}
Therefore, $\sigma_{\mathrm{reg}}(\omega,T)$ vanishes at any nonzero frequency and thus $\lim_{\omega\to 0}\sigma_{\mathrm{reg}}(\omega,T)=0$. Consequently, according to Eq.~\eqref{eq:main_dc}, only the Drude weight shown in Eq.~\eqref{eq:main_Drudeweight2} contributes to the DC conductivity for the present model.

Moreover, although Eq.~\eqref{eq:main_regular2} is not directly applicable at strictly zero frequency, taking the $\omega\to0$ limit of its prefactor, i.e., $\lim_{\omega \to 0} \frac{1-e^{-\beta \omega}}{\omega} = \beta$, and dividing by $2\pi$ yields, 
\begin{align}
\label{eq:main_sigmadiff_definition}
D_{\mathrm{diff}}(T)
&= \frac{\beta}{2ZN}
\sum_{n}
e^{-\beta E_n}
\left| J_{nn} \right|^2.
\end{align}
It is related to the Drude weight through
\begin{align}
\label{eq:main_sigmadiff_Drude_Meissner}
D_{\mathrm{diff}}(T)
=  D(T) - D_m(T),
\end{align}
that is, it measures the difference between the Drude weight and the Meissner stiffness $D_m(T)$~\cite{PhysRevB.73.085117,PhysRevB.77.245131}. 
In Sec.~IV~\cite{SupplementalMaterials},
we have proven that $D_{\mathrm{diff}}(T) =  D(T)$,
i.e., $D_m(T)=0$, consistent with the absence of SC in one dimension.
Furthermore, because $\hat{J}$ in
Eq.~\eqref{eq:main_currentoperator_1D} is itself conserved, it can be chosen
as a conserved charge in the Mazur inequality (see Eq.~\eqref{drude}), which provides a lower bound
on the Drude weight~\cite{MAZUR1969533}. With this choice, the Mazur bound
reproduces Eq.~\eqref{eq:main_sigmadiff_definition}. Therefore, for the present model,
the Mazur lower bound coincides with the exact charge Drude weight.

In the sector with a single doped hole, the analytical expression for the Drude weight in the large-system-size limit reads (see Sec.~IV of the SM~\cite{SupplementalMaterials})
\begin{eqnarray}
\label{eq:Drudeweight}
D_{1h}(T)
= \frac{\beta}{N}
\left(
1 - \frac{I_2(2\beta)}{I_0(2\beta)}
\right),
\end{eqnarray} 
where $I_l(\cdot)$ denotes the $l$-th modified Bessel function of the first kind, arising from the $\cos(\cdot)$ energy spectrum shown in Eqs.~\eqref{eq:main_EWithoutDown} and \eqref{eq:main_spectrum}. We note that Eq.~\eqref{eq:Drudeweight} is the same as that of the non-interacting spinless fermion model due to the spin-charge separation~\cite{PhysRevB.41.2326}. For a fixed number of doped holes $n_h$, in the thermodynamic limit, the Drude weight generalizes straightforwardly by acquiring an overall multiplicative factor of $n_h$ (see Sec.~IV of the SM~\cite{SupplementalMaterials}), reflecting the fact that, in the hole-dilute limit, the holes propagate essentially independently. Beyond this limit, this does not apply. The analytical expressions obtained for the specific heat in Sec.~IV~\cite{SupplementalMaterials}
also support this point.
In the thermodynamic limit, the hole density $n_h/N$ vanishes, and consequently the Drude weight also vanishes, consistent with the Mott-insulating nature of the system at half filling~\cite{carmelo2018absence,carmelo2013zero,PhysRevB.70.205129, PhysRevB.90.155104}.

At finite doping with fixed $\delta$, although we do not obtain an analytical expression for the Drude weight $D_{\delta}(T)$ at arbitrary temperatures, we derive analytical expressions for the Drude weight in both the low- and high-temperature limits (see Sec.~V of the SM~\cite{SupplementalMaterials}).
The analytical expressions for the Drude weight for a fixed number of holes $n_h$ and for a fixed hole density $\delta$ in the high- and low-temperature limits are summarized in Table~\ref{tab:main_Drude}.

As shown in Table~\ref{tab:main_Drude}, at zero temperature, $D_{1h}(T=0)=\frac{1}{N}$ and $D_{\delta}(T=0)=\frac{\sin(\pi\delta)}{\pi}$, both of which are consistent with the Luttinger-liquid prediction $\frac{K v_s}{\pi}$, where $K=1/2$ is the Luttinger parameter for the present model~\cite{PhysRevLett.64.2831,PhysRevB.83.205113}, and $v_s=\left.\frac{d\varepsilon(k)}{dk}\right|_{k=k_F}=2\sin(\pi\delta)$ is the sound velocity. Note that with fixed $n_h$, the Fermi momentum scales as $k_F\sim n_h/N$, and the corresponding Fermi energy behaves as $E_F\sim k_F^2\sim n_h^2/N^2$. Therefore, the Luttinger-liquid description, which is valid as a low-energy effective theory only for $T\ll E_F$, applies strictly only at zero temperature.

\textit{Temperature scaling of the Drude weight.---} As summarized in Table~\ref{tab:main_Drude}, in the high-temperature limit,
the Drude weight scales as $D(T)\propto 1/T$, irrespective of whether the
number or density of holes is held fixed. Previous studies of the Hubbard
model have reported robust high-temperature scaling of
charge-transport quantities~\cite{PhysRevB.111.035123,y7fg-lhsm}. More interestingly, at low temperatures as shown in Table~\ref{tab:main_Drude}, the single-hole Drude weight $D_{1h}(T)$ exhibits a linear-in-$T$ dependence, whereas $D_{\delta}(T)$ exhibits a quadratic-in-$T$ dependence~\cite{fujimoto1998exact}. This qualitative distinction calls for caution when extrapolating conclusions from finite-size numerical calculations. This issue is particularly relevant to small-cluster ED studies in 2D, for which severe restrictions on accessible system sizes make the finite-density thermodynamic limit difficult to reach.

\begin{figure}[t]
    \centering
\includegraphics[width=0.95\linewidth]{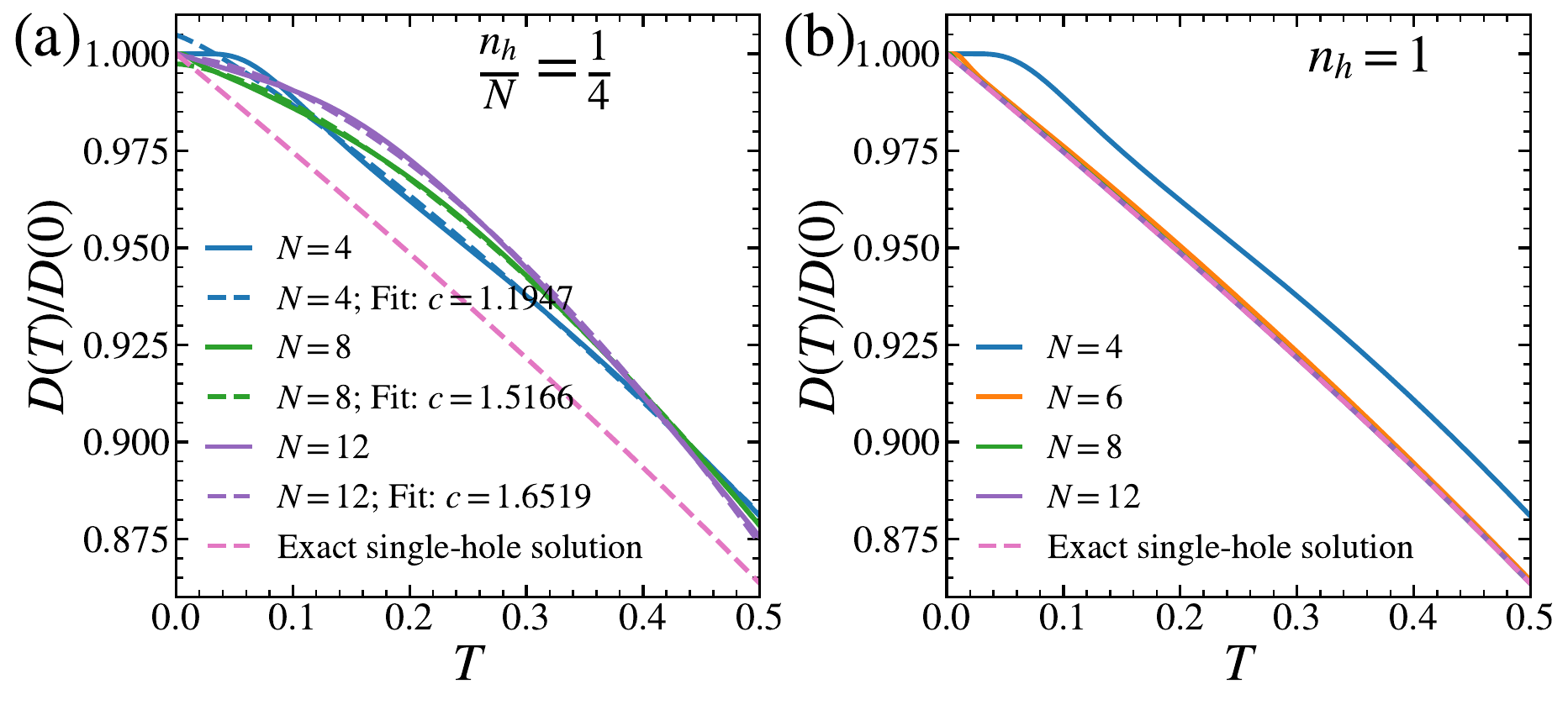}
    \caption{Normalized Drude weight $D(T)/D(0)$ as a function of temperature $T$ for (a) fixed density of holes $\delta = n_h/N = 1/4$ and (b) a single doped hole. In panel (b), the numerical Drude weight rapidly converges to the analytical prediction as the system size increases. 
In panel (a), a clearly distinct scaling behavior is observed. 
By fitting the data to the form $a + b T^{c}$, we find that the extracted exponent $c$ is larger than $1$ and approaches $2$ as the system size increases.
}
    \label{fig:DrudeWeightComparison}
\end{figure}

We have numerically verified these two distinct scaling behaviors using ED. To compare different system sizes, we normalize the Drude weight by its zero-temperature value, i.e., $D(T)/D(0)$. The results are shown in Fig.~\ref{fig:DrudeWeightComparison}. In Fig.~\ref{fig:DrudeWeightComparison}(b), corresponding to a single doped hole, the Drude weight obtained from ED rapidly approaches the analytical prediction (Eq.~\eqref{eq:Drudeweight}) as the system size increases. In Fig.~\ref{fig:DrudeWeightComparison}(a), where the density of holes is fixed at $\delta=n_h/N = 1/4$, a direct observation of the quadratic correction is limited by the exponential growth of the Hilbert space.
Instead, we fit the numerical data using the form $a + b T^{c}$ and find that the extracted exponent $c$ approaches the predicted value of $2$ as the system size increases. 
Therefore, the two ways of taking the limit do not commute.
As explained in Sec.~IV~\cite{SupplementalMaterials}, for a fixed number of holes, the system remains in the hole-dilute, nondegenerate regime, where the low-energy excitations are controlled by the quadratic dispersion near the band minimum. As a result, thermal averaging produces a linear-in-$T$ correction to the Drude weight. By contrast, at fixed hole density, the system enters a degenerate Fermi-gas regime, in which the low-energy excitations are governed by the linearized dispersion near the Fermi points. In this case, the Sommerfeld expansion applies and yields the standard $T^{2}$ correction to the Drude weight.

\textit{Discussion and outlook.---} In conclusion, we have obtained the explicit energy spectrum and eigenstates (see Sec.~III of the SM~\cite{SupplementalMaterials})
of the one-dimensional Hubbard model in the infinite-interaction limit and derived a closed-form analytical expression for the charge Drude weight in the hole-dilute regime at arbitrary temperatures, as well as analytical expressions for the Drude weight at finite doping in both the low- and high-temperature limits (see Sec.~IV and Sec.~V~\cite{SupplementalMaterials}). The Drude weight we find is consistent with the Mazur bound (Eq. (\ref{drude})). At low temperatures, the leading correction to the Drude weight is linear in $T$ at infinitesimal hole concentration, whereas it is quadratic in $T$ at a fixed finite hole density. Nevertheless, both limits exhibit the same high-temperature scaling.

The distinct temperature scalings of the Drude weight revealed by our
analytically controlled example highlight the importance of identifying the
underlying thermodynamic trajectory when interpreting finite-size numerical
studies of charge transport in more general models. In a finite system, the Dirac delta functions associated with both the Drude and regular contributions in Eq.~\eqref{eq:main_optical} should be broadened to obtain a smooth conductivity. Consequently, linear-in-$T$ and
quadratic-in-$T$ corrections to the Drude weight give rise to corresponding
leading temperature corrections to the regularized effective resistivity, as
discussed in Sec.~VI of the SM~\cite{SupplementalMaterials}. This observation
underscores the need to distinguish between fixed-carrier-number and
fixed-carrier-density thermodynamic trajectories before relating transport
scaling obtained from finite-size calculations to finite-doping phenomena,
such as the linear-in-$T$ resistivity associated with strange-metal
behavior~\cite{doi:10.1126/science.abh4273,marelQuantumCriticalBehaviour2003,
PhysRevLett.81.4720,PhysRevLett.92.167001,jinLinkSpinFluctuations2011,
PhysRevB.41.846,doi:10.1126/science.1165015,
daouLinearTemperatureDependence2009,PhysRevB.95.224517,
doiron-leyraudPseudogapPhaseCuprate2017,legrosUniversalTlinearResistivity2019,
doi:10.1126/science.aat4134}.

Although the present 1D model is neither equivalent to the
2D Hubbard model, in which signatures of strange-metal
transport have been reported numerically~\cite{PhysRevLett.110.086401,PhysRevB.91.075124,PhysRevB.95.041110,PhysRevLett.123.036601,doi:10.1126/science.aau7063,fratini2025strangemetaltransportcoupling}, nor a realistic microscopic
description of cuprate materials, where such behavior has been observed
experimentally~\cite{doi:10.1126/science.abh4273,marelQuantumCriticalBehaviour2003, PhysRevLett.81.4720, PhysRevLett.92.167001, jinLinkSpinFluctuations2011, PhysRevB.41.846, doi:10.1126/science.1165015, daouLinearTemperatureDependence2009, PhysRevB.95.224517, doiron-leyraudPseudogapPhaseCuprate2017,legrosUniversalTlinearResistivity2019, doi:10.1126/science.aat4134}, an interesting direction for future work is to develop
perturbative frameworks built on the exact 1D solutions, with
the goal of gaining analytical insight into strange-metal transport.
For example, a vanishing Drude weight requires the
overlap on the RHS of Eq. (\ref{drude}) to vanish.
Consequently, it is the lack of conservation of the current that must be at the heart of $T$-linear resistivity in strongly interacting models.  As a result, expansions around the conservation limit are a promising route to strange metallicity.  Another promising direction is to seek exact solutions directly in the 2D Hubbard model~\cite{2DHubbardExact} or in the recently developed momentum-mixing variant of the Hatsugai–Kohmoto model~\cite{maiTwistingHubbardModel2026, ma2025charge}, for which combined numerical and analytical approaches remain feasible, and then to extract analytical insight from such solutions.

\textit{Acknowledgement.---} We thank Aman Kumar for discussions on a different collaboration.
S.L. and B.A.B.
were supported by the Gordon and Betty Moore Foundation through Grant No. GBMF8685 towards the Princeton theory program, the Gordon and Betty Moore Foundation’s EPiQS Initiative (Grant No. GBMF11070), the Global Collaborative
Network Grant at Princeton University, the Simons Investigator Grant No. 404513, the Princeton Global
Network, the NSF-MERSEC (Grant No. MERSEC DMR 2011750), the Simons Collaboration on New Frontiers in Superconductivity (Grant No. SFI-MPS-NFS-00006741-01 and No. SFI-MPS-NFS-00006741-06), the Princeton Catalysis
Initiative, the Schmidt Foundation at the Princeton University,
European Research Council (ERC) under the European Union’s Horizon 2020 research and innovation program (Grant Agreement No. 101020833), the National Science Foundation through the AI Research Institutes program
Award No. DMR-2433348. H.J.C. acknowledges funding from National Science Foundation Grant No. DMR 2046570 and the National High Magnetic Field Laboratory (NHMFL). NHMFL is supported by the National Science Foundation through DMR-2128556 and the state of Florida.

%

\clearpage
\newpage
\widetext

\begin{center}
\textbf{\large Supplemental Material for ``Transport and Temperature: Exact spectrum and Drude weight for the 
one-dimensional infinite-$U$ Hubbard model''}
\end{center}

\renewcommand{\thefigure}{S\arabic{figure}}
\setcounter{figure}{0}
\renewcommand{\theequation}{S\arabic{equation}}
\setcounter{equation}{0}
\renewcommand{\thesection}{\Roman{section}}
\setcounter{section}{0}
\setcounter{secnumdepth}{4}

\addtocontents{toc}{\protect\setcounter{tocdepth}{0}}
{
\tableofcontents
}

\section{Linear-response derivation of the optical conductivity}
\label{sec:linear_response_optical_conductivity}

In this section, we first review linear response theory in Sec.~\ref{sec:Reviewoflinearresponsetheory}, then use it to derive the optical conductivity in Sec.~\ref{sec:optical_conductivity_from_LRT}. We next present the expressions for the Drude weight and the regular part of the optical conductivity in Sec.~\ref{sec:ExpressionForDrudeWeightAndRegularPart}, and give their Lehmann representations in Sec.~\ref{DrudeweightandregularpartintheLehmannrepresentation}. In Sec.~\ref{sec:bounds_on_Drude_weight}, we discuss the lower and upper bounds on the Drude weight. For further details, see Refs.~\cite{PhysRevLett.65.243,PhysRevB.16.2437,PhysRevB.35.8391,PhysRevLett.68.2830,PhysRevB.59.1825}. 

\subsection{Review of linear response theory}
\label{sec:Reviewoflinearresponsetheory}

We briefly review linear response theory~\cite{kubo1957statistical}.

We first consider a system in thermal equilibrium described by a time-independent Hamiltonian
\(\hat H\).
The corresponding equilibrium density matrix is
\begin{align}
    \hat \rho
    =
    \frac{1}{Z} e^{-\beta \hat H},
\end{align}
where \(\beta = 1/T\), \(T\) is the temperature and $ Z = \Tr\!\left(e^{-\beta \hat H}\right)$ is the partition function.
The thermal expectation value of an operator \(\hat O\) is defined as
\begin{align}
    \langle \hat O \rangle
    =
    \Tr\!\left(\hat \rho \hat O\right).
\end{align}

We now turn on a time-dependent perturbation \(\hat V(t)\), so that the full Hamiltonian becomes
\begin{align}
    \hat H(t)
    =
    \hat H + \hat V(t).
\end{align}
As a result, the density matrix becomes time dependent, which we denote by \(\hat \rho(t)\).
Its time evolution is given by
\begin{align}
    \frac{d\hat \rho(t)}{dt}
    =
    \frac{1}{i\hbar}
    \left[
        \hat H(t), \hat \rho(t)
    \right]
    =
    \frac{1}{i\hbar}
    \left(
        \hat H(t)\hat \rho(t) - \hat \rho(t)\hat H(t)
    \right).
    \label{eq:liouville_equation}
\end{align}
The expectation value of operator \(\hat O\) also becomes time dependent
\begin{align}
    \langle \hat O(t)\rangle
    =
    \Tr\!\left(\hat \rho(t)\hat O\right).
\end{align}

To treat the perturbation more conveniently, we introduce the interaction picture.
For an operator \(\hat O\), its interaction-picture form is defined as
\begin{align}
    \hat O_I(t)
    =
    e^{i\hat H t/\hbar}\,
    \hat O\,
    e^{-i\hat H t/\hbar} = \hat S^{-1}(t)\,\hat O\,\hat S(t),
\end{align}
where $\hat S(t)=e^{-i\hat H t/\hbar}$. Similarly, the interaction-picture density matrix is defined as
\begin{align}
    \hat \rho_I(t)
    =
    e^{i\hat H t/\hbar}\,
    \hat \rho(t)\,
    e^{-i\hat H t/\hbar}
    =
    \hat S^{-1}(t)\,\hat \rho(t)\,\hat S(t),
\end{align}
and the interaction-picture perturbation is
\begin{align}
    \hat V_I(t)
    =
    e^{i\hat H t/\hbar}\,
    \hat V(t)\,
    e^{-i\hat H t/\hbar} = \hat S^{-1}(t)\,\hat V(t)\,\hat S(t).
\end{align}

We now derive the equation of motion for \(\hat \rho_I(t)\).
Differentiating
\(\hat \rho_I(t)=\hat S^{-1}(t)\hat \rho(t)\hat S(t)\),
we obtain
\begin{align}
    \frac{d\hat \rho_I(t)}{dt}
    =
    \frac{d\hat S^{-1}(t)}{dt}\hat \rho(t)\hat S(t)
    +
    \hat S^{-1}(t)\frac{d\hat \rho(t)}{dt}\hat S(t)
    +
    \hat S^{-1}(t)\hat \rho(t)\frac{d\hat S(t)}{dt}.
\end{align}
Using
\begin{align}
    \frac{d\hat S(t)}{dt}
    =
    -\frac{i}{\hbar}\hat H \hat S(t),
    \qquad
    \frac{d\hat S^{-1}(t)}{dt}
    =
    \frac{i}{\hbar}\hat S^{-1}(t)\hat H,
\end{align}
together with Eq.~\eqref{eq:liouville_equation}, we have
\begin{align}
    \frac{d\hat \rho_I(t)}{dt}
    &=
    \frac{i}{\hbar}\hat S^{-1}(t)\hat H\hat \rho(t)\hat S(t)
    +
    \hat S^{-1}(t)
    \frac{1}{i\hbar}
    \left[
        \hat H + \hat V(t), \hat \rho(t)
    \right]
    \hat S(t)
    -
    \frac{i}{\hbar}\hat S^{-1}(t)\hat \rho(t)\hat H\hat S(t)
    \nonumber\\
    &=
    \frac{1}{i\hbar}
    \hat S^{-1}(t)
    \left[
        \hat V(t), \hat \rho(t)
    \right]
    \hat S(t)
    \nonumber\\
    &=
    \frac{1}{i\hbar}
    \left[
        \hat V_I(t), \hat \rho_I(t)
    \right].
    \label{eq:rhoI_equation_of_motion}
\end{align}
Integrating the above equation from an initial time \(t_0\) to \(t\), we obtain
\begin{align}
    \hat \rho_I(t)
    =
    \hat \rho_I(t_0)
    +
    \frac{1}{i\hbar}
    \int_{t_0}^{t} dt'\,
    \left[
        \hat V_I(t'), \hat \rho_I(t')
    \right].
    \label{eq:rhoI_integral_equation}
\end{align}
This is an exact integral equation.
To make further progress, we iteratively substitute \(\hat \rho_I(t')\) on the right-hand side.
To first order in the iteration,
\begin{align}
    \hat \rho_I(t')
    =
    \hat \rho_I(t_0)
    +
    \frac{1}{i\hbar}
    \int_{t_0}^{t'} dt''\,
    \left[
        \hat V_I(t''), \hat \rho_I(t'')
    \right].
\end{align}
Substituting this back into Eq.~\eqref{eq:rhoI_integral_equation}, we have
\begin{align}
    \hat \rho_I(t)
    =
    \hat \rho_I(t_0)
    +
    \frac{1}{i\hbar}
    \int_{t_0}^{t} dt'\,
    \left[
        \hat V_I(t'), \hat \rho_I(t_0)
    \right]
    +
    \left(\frac{1}{i\hbar}\right)^2
    \int_{t_0}^{t} dt'
    \int_{t_0}^{t'} dt''\,
    \left[
        \hat V_I(t'),
        \left[
            \hat V_I(t''), \hat \rho_I(t'')
        \right]
    \right].
\end{align}
When the perturbation is weak, we keep only the term linear in \(\hat V_I\), namely
\begin{align}
    \hat \rho_I(t)
    \simeq
    \hat \rho_I(t_0)
    +
    \frac{1}{i\hbar}
    \int_{t_0}^{t} dt'\,
    \left[
        \hat V_I(t'), \hat \rho_I(t_0)
    \right].
    \label{eq:rhoI_linear_order}
\end{align}

We now assume that the perturbation is adiabatically switched on from the remote past,
so that \(t_0\to -\infty\), and the system is in equilibrium at \(t=-\infty\).
Since the equilibrium density matrix commutes with \(\hat H\),
it is time independent in the interaction picture, and hence
\begin{align}
    \hat \rho_I(-\infty)=\hat \rho.
\end{align}
Therefore Eq.~\eqref{eq:rhoI_linear_order} becomes
\begin{align}
    \hat \rho_I(t)
    \simeq
    \hat \rho
    +
    \frac{1}{i\hbar}
    \int_{-\infty}^{t} dt'\,
    \left[
        \hat V_I(t'), \hat \rho
    \right].
    \label{eq:rhoI_linear_response_final}
\end{align}

We now compute the time-dependent expectation value of operator \(\hat O\). Using the cyclic property of the trace, the expectation value of operator $\hat{O}$ can be rewritten as
\begin{align}
    \langle \hat O(t)\rangle
    &=
    \Tr\!\left(\hat \rho(t)\hat O\right)
    \nonumber\\
    &=
    \Tr\!\left(
        \hat S^{-1}(t)\hat \rho(t)\hat S(t)\,
        \hat S^{-1}(t)\hat O\hat S(t)
    \right)
    \nonumber\\
    &=
    \Tr\!\left(
        \hat \rho_I(t)\hat O_I(t)
    \right).
    \label{eq:expectation_interaction_picture}
\end{align}
Using Eq.~\eqref{eq:rhoI_linear_response_final},
we obtain
\begin{align}
    \langle \hat O(t)\rangle
    &=
    \Tr\!\left(
        \hat \rho_I(t)\hat O_I(t)
    \right)
    \nonumber\\
    &\simeq
    \Tr\!\left(
        \hat \rho\,\hat O_I(t)
    \right)
    +
    \frac{1}{i\hbar}
    \int_{-\infty}^{t} dt'\,
    \Tr\!\left(
        \left[
            \hat V_I(t'), \hat \rho
        \right]
        \hat O_I(t)
    \right) \nonumber \\
   &= \Tr\!\left(
        \hat \rho\,\hat O_I(t)
    \right)
    +
    \frac{1}{i\hbar}
    \int_{-\infty}^{t} dt'\,
    \Tr\!\left(
        \hat \rho
        \left[
            \hat O_I(t), \hat V_I(t')
        \right]
    \right),
\end{align}
where in the last equality we have used the cyclic property of the trace.

Since \(\hat O_I(t)\) is just the Heisenberg-picture operator evolved by the unperturbed Hamiltonian $\hat{H}$,
we may write it as \(\hat O_H(t)\), and similarly \(\hat V_I(t')=\hat V_H(t')\).
Moreover, for the equilibrium term,
\begin{align}
    \Tr\!\left(
        \hat \rho\,\hat O_H(t)
    \right)
    =
    \Tr\!\left(
        \hat \rho\,\hat O
    \right)
    =
    \langle \hat O\rangle,
\end{align}
because the equilibrium expectation value is time independent.
Therefore,
\begin{align}
    \langle \hat O(t)\rangle
    =
    \langle \hat O\rangle
    +
    \frac{1}{i\hbar}
    \int_{-\infty}^{t} dt'\,
    \left\langle
        \left[
            \hat O_H(t), \hat V_H(t')
        \right]
    \right\rangle.
    \label{eq:general_linear_response_observable}
\end{align}

It is convenient to define the deviation from equilibrium by
\begin{align}
    \delta \langle \hat O(t)\rangle
    \equiv
    \langle \hat O(t)\rangle - \langle \hat O\rangle.
\end{align}
Then Eq.~\eqref{eq:general_linear_response_observable} becomes
\begin{align}
    \delta \langle \hat O(t)\rangle
    =
    \frac{1}{i\hbar}
    \int_{-\infty}^{t} dt'\,
    \left\langle
        \left[
            \hat O_H(t), \hat V_H(t')
        \right]
    \right\rangle,
    \label{eq:deltaO_general}
\end{align}
which describes how the expectation value of an operator deviates from its thermal-equilibrium value under a time-dependent perturbation within linear response theory.

We now specialize to the case in which the time-dependent perturbation can be expressed as  
\begin{align}
    \hat V(t)
    =
    \hat A\,f(t),
\end{align}
where $\hat{A}$ is a time-independent operator and \(f(t)\) is a time-dependent scalar function.
Substituting this into Eq.~\eqref{eq:deltaO_general}, we obtain
\begin{align}
    \delta \langle \hat O(t)\rangle
    &=
    \frac{1}{i\hbar}
    \int_{-\infty}^{t} dt'\,
    f(t')
    \left\langle
        \left[
            \hat O_H(t), \hat A_H(t')
        \right]
    \right\rangle
    \nonumber\\
    &=
    \int_{-\infty}^{\infty} dt'\,
    \chi(t,t')\,f(t'),
    \label{eq:kubo_time_domain}
\end{align}
where we introduced the generalized susceptibility
\begin{align}
    \chi(t,t')
    =
    \frac{1}{i\hbar}\,
    \Theta(t-t')
    \left\langle
        \left[
            \hat O_H(t), \hat A_H(t')
        \right]
    \right\rangle.
    \label{eq:generalized_susceptibility_def}
\end{align}
The step function \(\Theta(t-t')\) explicitly enforces causality:
the response at time \(t\) depends only on the perturbation applied at earlier times \(t'<t\). Since the susceptibility depends only on the time difference \(t-t'\),
we write
\begin{align}
    \chi(t,t')
    =
    \chi^R(t-t'),
\end{align}
where superscript $R$ denotes the retarded response. Thus the linear response takes the convolution form of Kubo formula
\begin{align}
    \delta \langle \hat O(t)\rangle
    =
    \int_{-\infty}^{\infty} dt'\,
    \chi^R(t-t')\,f(t').
    \label{eq:linear_response_convolution}
\end{align}

We now Fourier transform Eq.~\eqref{eq:linear_response_convolution}.
Using the convention
\begin{align}
    \chi^R(t-t')
    &=
    \int \frac{d\omega}{2\pi}\,
    e^{-i\omega (t-t')}\,
    \chi^R(\omega),
    \\
    f(\omega)
    &=
    \int_{-\infty}^{\infty} dt'\,
    e^{i\omega t'} f(t'),
\end{align}
we have
\begin{align}
    \delta \langle \hat O(t)\rangle
    &=
    \int \frac{d\omega}{2\pi}\,
    e^{-i\omega t}\,
    \chi^R(\omega)
    \int_{-\infty}^{\infty} dt'\,
    e^{i\omega t'} f(t') \nonumber \\ 
    & = \int \frac{d\omega}{2\pi}\,
    e^{-i\omega t}\,
    \chi^R(\omega)\,f(\omega),
\end{align}
and thus
\begin{align}
    \delta \langle \hat O(\omega)\rangle
    =
    \chi^R(\omega)\,f(\omega).
    \label{eq:linear_response_frequency}
\end{align}
Hence the Fourier transform of the retarded susceptibility is
\begin{align}
    \chi^R(\omega)
    =
    \frac{\delta \langle \hat O(\omega)\rangle}{f(\omega)}.
\end{align}

Finally, the frequency-domain retarded susceptibility can be written as
\begin{align}
    \chi^R(\omega)
    =
    \int_{-\infty}^{\infty} dt\,
    e^{i\omega t}\,
    \chi^R(t)
    =
    \frac{1}{i\hbar}
    \int_0^\infty dt\,
    e^{i(\omega+i0^+)t}
    \left\langle
        \left[
            \hat O_H(t), \hat A_H(0)
        \right]
    \right\rangle,
    \label{eq:retarded_susceptibility_frequency}
\end{align}
where the infinitesimal \(i0^+\) ensures convergence and selects the retarded boundary condition, i.e., the response  depends only on the perturbation applied at earlier times.

\subsection{Optical conductivity from linear response theory}
\label{sec:optical_conductivity_from_LRT}

We now apply the general formalism of linear response theory shown in Sec.~\ref{sec:Reviewoflinearresponsetheory} to the optical conductivity.
In the following, we set \(\hbar=1\) for simplicity.
We also assume that the electron charge and the lattice spacing are set to unity unless otherwise stated.

We present the derivation for a one-dimensional lattice model.
The generalization to higher dimensions is straightforward. We start from the Hamiltonian
\begin{align}
    \label{eq:GenericH}
    \hat H
    =
    \hat K + \hat H_{\rm int},
\end{align}
where \(\hat H_{\rm int}\) denotes the density--density interaction term, and the kinetic-energy term is
\begin{align}
    \hat K
    =
    -t_h \sum_{\sigma =\{\uparrow, \downarrow \} }\sum_{l=0}^{N-1}
    \left(
         c^{\dagger}_{l+1,\sigma} c_{l,\sigma}
        +
     c^{\dagger}_{l,\sigma}  c_{l+1,\sigma}
    \right),
\label{eq:kinetic_term_for_conductivity}
\end{align}
where \(c^{\dagger}_{l,\sigma}\) and \(c_{l,\sigma}\) are, respectively, the creation and annihilation operators for a fermion with spin \(\sigma\) at site \(l\). Here, $t_h$ is the hopping amplitude (we avoid the notation $t$ to distinguish it from the time variable), and $N$ is the number of lattice sites. We impose periodic boundary conditions (PBCs) throughout.

To derive the optical conductivity, we consider an external time-dependent vector potential projected onto the hopping direction \(A(l,t)\) ($l$ is the site index and $t$ denotes time)
coupled to the system through the Peierls substitution~\cite{peierlsZurTheorieDiamagnetismus1933a}. The hopping term is modified according to
\begin{align}
     c^{\dagger}_{l+1,\sigma}  c_{l,\sigma}
    \;\to\;
    e^{+iA(l,t)}
     c^{\dagger}_{l+1,\sigma}  c_{l,\sigma}, \\ 
     c^{\dagger}_{l,\sigma}  c_{l+1,\sigma}
    \;\to\;
    e^{-iA(l,t)}
     c^{\dagger}_{l,\sigma} c_{l+1,\sigma}.
\end{align}
Thus the Hamiltonian becomes
\begin{align}
    \hat H[A]
    =
    -t_h \sum_{\sigma = \{\uparrow,\downarrow \}}\sum_{l=0}^{N-1}
    \left(
        e^{+iA(l,t)}
         c^{\dagger}_{l+1,\sigma} c_{l,\sigma}
        +
        e^{-iA(l,t)}
         c^{\dagger}_{l,\sigma} c_{l+1,\sigma}
    \right)
    +
    \hat H_{\rm int},
    \label{eq:H_with_A}
\end{align}
where the density-density interaction part $\hat H_{\rm int}$ remains unchanged.

We are interested in the linear response to a weak external field and thus expand Eq.~\eqref{eq:H_with_A} to the second order in \(A(l,t)\) (to get the total current operator in Eq.~\eqref{eq:DefinitionOfTotalJ}).
Using
\begin{align}
    e^{\pm iA(l,t)}
    =
    1 \pm iA(l,t) - \frac{1}{2}A^2(l,t) + \mathcal O(A^3),
\end{align}
we obtain
\begin{align}
    \hat H[A]
    =
    \hat H
    -
    \sum_{l=0}^{N-1} \hat J_l\,A(l,t)
    -
    \frac{1}{2}\sum_{l=0}^{N-1} \hat K_l\,A^2(l,t)
    + \mathcal O(A^3),
    \label{eq:H_expand_A}
\end{align}
where we have introduced the local paramagnetic current operator
\begin{align}
    \hat J_l
    =
    it_h\sum_{\sigma = \{\uparrow, \downarrow \}}
    \left(
        \hat c^{\dagger}_{l+1,\sigma}\hat c_{l,\sigma}
        -
        \hat c^{\dagger}_{l,\sigma}\hat c_{l+1,\sigma}
    \right),
    \label{eq:local_current_operator}
\end{align}
and the local kinetic operator
\begin{align}
    \hat K_l
    =
    -t_h\sum_{\sigma= \{\uparrow, \downarrow \}}
    \left(
        \hat c^{\dagger}_{l+1,\sigma}\hat c_{l,\sigma}
        +
        \hat c^{\dagger}_{l,\sigma}\hat c_{l+1,\sigma}
    \right).
    \label{eq:local_K_operator}
\end{align}
We note that \(\sum_l \hat K_l = \hat K\), namely the sum of the local kinetic operators
is precisely the kinetic-energy operator appearing in Eq.~\eqref{eq:kinetic_term_for_conductivity}.

The total current operator at site $l$ is given by
\begin{align}
    \hat J_{\rm tot}(l,t)
    \equiv
    -\frac{\delta \hat H[A]}{\delta A(l,t)}=   \hat J_l
    +
    \hat K_l\,A(l,t)
    + \mathcal O(A^2),
\end{align}
and thus the total current operator is given by 
\begin{align}
    \label{eq:DefinitionOfTotalJ}
    \hat J_{\rm tot}(t) = \sum_{l=0}^{N-1} \hat J_l
    + \sum_{l=0}^{N-1}
    \hat K_l\,A(l,t) = \hat{J}
    + \hat{J}_{\mathrm{dia}}(t),
\end{align}
where we have neglected higher-order terms in the external field. We note that 
\begin{align}
    \label{eq:DefinitionOfParaJ}
    \hat{J} = \sum_{l=0}^{N-1} \hat J_l
\end{align}
is the paramagnetic current operator and 
\begin{align}
    \label{eq:DefinitionOfDiaJ}
    \hat{J}_{\mathrm{dia}}(t) = \sum_{l=0}^{N-1}
    \hat K_l\,A(l,t)
\end{align}
is the diamagnetic part. Therefore, to linear order in the external field, the total current operator consists of two contributions: the usual paramagnetic term $\hat{J}$, which is already present at \(A=0\), and the diamagnetic term $\hat{J}_{\mathrm{dia}}(t)$, which arises from the explicit dependence of the Hamiltonian on the vector potential.

According to Eq.~\eqref{eq:H_expand_A}, the time-dependent perturbation to the leading order in the vector potential is
\begin{align}
    \label{eq:DefinitionOfPerturbationV}
    \hat V(t)
    =
    -\sum_{l=0}^{N-1} \hat J_l\,A(l,t).
\end{align}

Next, within linear response theory, we derive the expectation value of the total current operator $\langle \hat J_{\rm tot}(t) \rangle$ in Eq.~\eqref{eq:DefinitionOfTotalJ} in the presence of the time-dependent perturbation $\hat V(t)$ in Eq.~\eqref{eq:DefinitionOfPerturbationV}. We then obtain the corresponding expression for the conductivity. For simplicity, we use $\sum_l$ as shorthand for $\sum_{l=0}^{N-1}$ below.

\subsubsection{Paramagnetic part} 
We first consider the paramagnetic part. We set the paramagnetic current operator $\hat{J}$ in Eq.~\eqref{eq:DefinitionOfParaJ} as the operator \(\hat O\) in Eq.~\eqref{eq:deltaO_general}. We have
\begin{align}
    \delta\langle \hat J(t)\rangle
    &=
    \frac{1}{i}
    \int_{-\infty}^{t}dt'\,
    \left\langle
        \left[
            \hat J_H(t),\hat V_H(t')
        \right]
    \right\rangle
    \nonumber \\ 
    &=
    \frac{1}{i}
    \int_{-\infty}^{t}dt'\,
    \left\langle
        \left[
            \sum_l \hat J_{l,H}(t),
            -\sum_{l'} \hat J_{l',H}(t')A(l',t')
        \right]
    \right\rangle
    \nonumber \\ 
    &=
    -\frac{1}{i}
    \sum_{l,l'}
    \int_{-\infty}^{t}dt'\,
    \left\langle
        \left[
            \hat J_{l,H}(t),
            \hat J_{l',H}(t')
        \right]
    \right\rangle
    A(l',t')  \nonumber \\
    &=
   i
    \sum_{l,l'}
    \int_{-\infty}^{t}dt'\,
    \left\langle
        \left[
            \hat J_{l,H}(t),
            \hat J_{l',H}(t')
        \right]
    \right\rangle
    A(l',t') \nonumber \\ 
    &=
   i
    \sum_{l,l'}
    \int_{-\infty}^{t}dt'\,
    \left\langle
        \left[
            \hat J_{l}(t),
            \hat J_{l'}(t')
        \right]
    \right\rangle
    A(l',t'),
    \label{eq:deltaJ_general_again}
\end{align}
where $\hat{J}_{H}(t) = e^{i\hat{H} t} \hat{J} e^{-i\hat{H}t}$ and $\hat{J}_{l, H}(t) = e^{i \hat{H} t} \hat{J}_l e^{-i\hat{H}t}$. Note that $A(l',t')$ is a time-dependent function rather than an operator. In the last equality, we have omitted the subscript $H$ since $\hat{J}_{l}(t) = \hat{J}_{l,H}(t)$.

We now Fourier transform the above expression to momentum space.
Starting from the response of the local current operator
\begin{align}
    \delta\langle \hat J_l(t)\rangle
    =
    i\sum_{l'}
    \int_{-\infty}^{t}dt'\,
    \left\langle
        \left[
            \hat J_{l}(t),\hat J_{l'}(t')
        \right]
    \right\rangle
    A(l',t'),
\end{align}
we define the finite-momentum current operator as
\begin{align}
    \hat J(q,t)
    =
    \sum_l e^{-iql}\hat J_l(t),
\end{align}
and thus
\begin{align}
    \delta\langle \hat J(q,t)\rangle
    &=
    \sum_l e^{-iql}\delta\langle \hat J_l(t)\rangle
    \nonumber\\
    &=
    i
    \sum_{l,l'}
    \int_{-\infty}^{t}dt'\,
    e^{-iql}
    \left\langle
        \left[
            \hat J_{l}(t),\hat J_{l'}(t')
        \right]
    \right\rangle
    A(l',t').
\end{align}
Substituting
\begin{align}
    A(l',t')
    =
    \frac{1}{N}\sum_{q'} e^{iq'l'}A(q',t'),
\end{align}
we have
\begin{align}
    \delta\langle \hat J(q,t)\rangle
    &=
    \frac{i}{N}
    \sum_{q'}
    \int_{-\infty}^{t}dt'\,
    \sum_{l,l'}
    e^{-iql}e^{iq'l'}
    \left\langle
        \left[
            \hat J_{l}(t),\hat J_{l'}(t')
        \right]
    \right\rangle
    A(q',t').
    \label{eq:deltaJ_qt_before_translation}
\end{align}
Using the definition
\begin{align}
    \hat J(q,t)
    &=
    \sum_l e^{-iql}\hat J_{l}(t), \\ 
    \hat J(-q',t')
    &=
    \sum_{l'} e^{iq'l'}\hat J_{l'}(t'),
\end{align}
Eq.~\eqref{eq:deltaJ_qt_before_translation} can be rewritten as
\begin{align}
    \label{eq:deltaJ_qt_after_translation}
    \delta\langle \hat J(q,t)\rangle
    &=
    \frac{i}{N}
    \sum_{q'}
    \int_{-\infty}^{t}dt'\,
    \left\langle
        \left[
            \hat J(q,t),\hat J(-q',t')
        \right]
    \right\rangle
    A(q',t'),
\end{align}
in which by translation invariance, only the contribution with \(q'=q\) survives. 
To see why only the contribution with \(q'=q\) survives, we define
\begin{align}
    C_{q,q'}(t,t')
    \equiv
    \left\langle
        \left[
            \hat J(q,t),\hat J(-q',t')
        \right]
    \right\rangle
    =
    \Tr\!\left(
        \hat \rho
        \left[
            \hat J(q,t),\hat J(-q',t')
        \right]
    \right).
\end{align}
Since the system is translationally invariant, the translation operator \(\hat T_x\) satisfies
\begin{align}
    [\hat T_x,\hat H]=0.
\end{align}
Therefore, the thermal density matrix $\hat{\rho}$
also commutes with \(\hat T_x\). On the other hand, the current operator in momentum space transforms under translation as
\begin{align}
    \hat T_x\,\hat J(q,t)\,\hat T_x^{-1}
    &=
    e^{iq}\hat J(q,t), \\
    \hat T_x\,\hat J(-q',t')\,\hat T_x^{-1}
    &=
    e^{-iq'}\hat J(-q',t').
\end{align}
Using these relations, together with \([\hat T_x,\hat \rho]=0\) and the cyclic property of the trace,
we obtain
\begin{align}
    C_{q,q'}(t,t')
    &=
    \Tr\!\left(
        \hat \rho
        \left[
            \hat T_x\hat J(q,t)\hat T_x^{-1},
            \hat T_x\hat J(-q',t')\hat T_x^{-1}
        \right]
    \right)
    \nonumber\\
    &=
    e^{i(q-q')}
    \Tr\!\left(
        \hat \rho
        \left[
            \hat J(q,t),\hat J(-q',t')
        \right]
    \right)
    \nonumber\\
    &=
    e^{i(q-q')} C_{q,q'}(t,t').
\end{align}
Hence
\begin{align}
    \bigl(1-e^{i(q-q')}\bigr)C_{q,q'}(t,t')=0.
\end{align}
Therefore, \(C_{q,q'}(t,t')\) can be nonzero only if
\begin{align}
    e^{i(q-q')}=1,
\end{align}
namely \(q=q'\) modulo \(2\pi\). As a result, in the linear-response expression only the contribution with \(q'=q\) survives.

Therefore, Eq.~\eqref{eq:deltaJ_qt_after_translation} becomes
\begin{align}
    \delta\langle \hat J(q,t)\rangle
    =
    \int_{-\infty}^{\infty}dt'\,
    \chi^R_{JJ}(q,t-t')\,A(q,t'),
    \label{eq:deltaJ_qt_final}
\end{align}
where the retarded current--current correlator, which depends only on the time difference $t-t'$, is defined as
\begin{align}
    \chi^R_{JJ}(q,t-t')
    =
    \frac{i}{N}\Theta(t-t')
    \left\langle
        \left[
            \hat J(q,t),\hat J(-q,t')
        \right]
    \right\rangle.
    \label{eq:chiJJ_q_time}
\end{align}
Using the Fourier-transform convention
\begin{align}
    \chi^R_{JJ}(q,t-t')
    &=
    \int \frac{d\omega}{2\pi}\,
    e^{-i\omega(t-t')}\,
    \chi^R_{JJ}(q,\omega),
    \\
    A(q,\omega)
    &=
    \int_{-\infty}^{\infty}dt\,e^{i\omega t}A(q,t),
\end{align}
with inverse transform
\begin{align}
    A(q,t)
    =
    \int \frac{d\omega}{2\pi}\,
    e^{-i\omega t}A(q,\omega),
\end{align}
we obtain
\begin{align}
    \delta\langle \hat J(q,t)\rangle
    &=
    \int_{-\infty}^{\infty}dt'\,
    \chi^R_{JJ}(q,t-t')A(q,t')
    \nonumber\\
    &=
    \int \frac{d\omega}{2\pi}\,
    e^{-i\omega t}\,
    \chi^R_{JJ}(q,\omega)
    \int_{-\infty}^{\infty}dt'\,
    e^{i\omega t'}A(q,t')
    \nonumber\\
    &=
    \int \frac{d\omega}{2\pi}\,
    e^{-i\omega t}\,
    \chi^R_{JJ}(q,\omega)A(q,\omega).
\end{align}
Therefore, in frequency space we arrive at
\begin{align}
    \delta\langle \hat J(q,\omega)\rangle
    =
    \chi^R_{JJ}(q,\omega)A(q,\omega),
    \label{eq:deltaJ_qomega}
\end{align}
with
\begin{align}
    \chi^R_{JJ}(q,\omega)
    &=
    \int_{-\infty}^{\infty}dt\,e^{i\omega t}\chi^R_{JJ}(q,t)
    \nonumber\\
    &=
    \frac{i}{N}
    \int_0^\infty dt\,
    e^{i(\omega+i0^+)t}
    \left\langle
        \left[
            \hat J(q,t),\hat J(-q,0)
        \right]
    \right\rangle.
    \label{eq:chiJJ_qomega}
\end{align}
where the infinitesimal \(i0^{+}\) has been introduced to implement the retarded prescription; equivalently, it contributes a damping factor \(e^{i(i0^{+})t}=e^{-0^{+}t}\), which ensures the convergence of the time integral for \(t>0\).

\subsubsection{Diamagnetic part} 
We now add back the diamagnetic contribution.
Since the diamagnetic part shown in Eq.~\eqref{eq:DefinitionOfDiaJ}
is already linear in the external field, within linear response theory
it is sufficient to evaluate its thermal-equilibrium expectation value
\begin{align}
    \langle \hat J_{\rm dia}(t)\rangle
    =
    \sum_{l=0}^{N-1}\langle \hat K_l\rangle A(l,t).
\end{align}

Since
\begin{align}
    \sum_{l=0}^{N-1}\hat K_l A(l,t)
    &=
    \sum_{l=0}^{N-1}
    \left(
        \frac{1}{N}\sum_q e^{iql}\hat K(q)
    \right)
    \left(
        \frac{1}{N}\sum_{q'} e^{iq'l}A(q',t)
    \right)
    \nonumber \\ 
    &=
    \frac{1}{N^2}\sum_{q,q'}\hat K(q)A(q',t)
    \sum_{l=0}^{N-1} e^{i(q+q')l} \nonumber \\ 
    & = \frac{1}{N}\sum_q \hat K(-q)A(q,t),
\end{align}
the diamagnetic contribution to the current expectation value is
\begin{align}
    \langle \hat J_{\rm dia}(t)\rangle
    =
    \frac{1}{N}\sum_q \langle \hat K(-q)\rangle A(q,t).
    \label{eq:Jdia_time}
\end{align}

We now Fourier transform with respect to time. Using
\begin{align}
    A(q,\omega)
    =
    \int_{-\infty}^{\infty}dt\,e^{i\omega t}A(q,t),
\end{align}
Eq.~\eqref{eq:Jdia_time} becomes
\begin{align}
    \langle \hat J_{\rm dia}(\omega)\rangle
    =
    \frac{1}{N}\sum_q \langle \hat K(-q)\rangle A(q,\omega).
    \label{eq:Jdia_omega}
\end{align}

Therefore, combining Eq.~\eqref{eq:deltaJ_qomega} and Eq.~\eqref{eq:Jdia_omega}, the expectation value of the total current operator is 
\begin{align}
    \langle \hat{J}_{\mathrm{tot}}(q, \omega) \rangle & = \langle \hat{J}(q,\omega) \rangle + \delta \langle \hat{J}(q,\omega) \rangle + \langle \hat{J}_{\mathrm{dia}}(q, \omega) \rangle \nonumber \\ 
    & =  \left( \chi^{R}_{JJ}(q, \omega) +  \frac{1}{N} \langle \hat{K}(-q) \rangle \right) A(q, \omega),
\end{align}
where in the second equality we have used $\langle \hat{J}(q,\omega) \rangle=0$, i.e., the expectation value of paramagnetic current operator in thermal equilibrium is zero.

The frequency-dependent uniform electrical conductivity $\sigma(\omega)$ is given by the current response to an electric field with $q=0$
\begin{align}
    E(q=0, \omega) = i \omega A(q=0, \omega).
\end{align}
Moreover, since 
\begin{align}
    \langle \hat J_{\rm tot}(q=0, \omega)\rangle
    =
    \sigma(\omega)\,E(q=0, \omega),
    \label{eq:sigma_definition}
\end{align}
we obtain the expression for conductivity
\begin{align}
    \label{eq:sigma_q_omega_final}
    \sigma(\omega)
   & =    \frac{
        \frac{1}{N}\langle \hat K \rangle
        +
        \chi^R_{JJ}(\omega)
    }{
        i (\omega+i0^+)
    } \nonumber \\ 
    & = i\frac{
        \frac{1}{N} \langle -\hat K \rangle
        -
        \chi^R_{JJ}(\omega)
    }{
        \omega+i0^+
    }, 
\end{align}
where
\begin{align}
    \label{eq:DefinitionOfRChi}
    \chi^R_{JJ}(\omega)
    &=
    \frac{i}{N}
    \int_0^\infty dt\,
    e^{i(\omega+i0^+)t}
    \left\langle
        \left[
            \hat J(t),\hat J(0)
        \right]
    \right\rangle.
\end{align}

\subsection{Drude weight and regular part of the optical conductivity}
\label{sec:ExpressionForDrudeWeightAndRegularPart}

In Sec.~\ref{sec:optical_conductivity_from_LRT}, as shown in Eq.~\eqref{eq:sigma_q_omega_final}, we have derived the expression for conductivity $\sigma(\omega)$ which is complex. Its real part characterizes the dissipative response of the system, namely the absorption of energy from the external electric field. For this reason, its real part is directly associated with charge transport and is the quantity measured in standard optical conductivity experiments~\cite{RevModPhys.83.471, kuzmenko2005kramers}. By contrast, the imaginary part describes the reactive or nondissipative response. It encodes the phase shift between the applied electric field and the induced current, and is related to the temporary storage rather than irreversible dissipation of electromagnetic energy~\cite{RevModPhys.83.471, kuzmenko2005kramers}. In the following, we focus on the real part of optical conductivity.

The real part of optical conductivity is usually decomposed as 
\begin{equation}
\Re\,\sigma(\omega,T)=2\pi D(T)\,\delta(\omega)+\sigma_{\mathrm{reg}}(\omega,T),
\label{eq:Drude_decomp}
\end{equation}
where \(2\pi D(T)\,\delta(\omega)\) is the singular contribution at zero frequency, with \(D(T)\) the Drude weight, and \(\sigma_{\mathrm{reg}}(\omega,T)\) denotes the regular part at finite frequency \(\omega>0\). Here ``regular'' means the part that does not contain a $\delta(\omega)$ singularity at zero frequency. In the following, we present the expressions for $D(T)$ and $\sigma_{\mathrm{reg}}(\omega,T)$.

Using the distribution identity
\begin{equation}
\frac{1}{\omega+i0^+}
=
\mathcal{P}\frac{1}{\omega}-i\pi\delta(\omega),
\label{eq:dist_id}
\end{equation}
we rewrite Eq.~\eqref{eq:sigma_q_omega_final}
as
\begin{equation}
\sigma(\omega,T)
=
i\Big( \frac{\langle -\hat K\rangle}{N}-\chi^{R}_{JJ}(\omega)\Big)\,
\mathcal{P}\frac{1}{\omega}
+\pi \Big(\frac{\langle -\hat K\rangle}{N}-\chi^{R}_{JJ}(\omega)\Big)\,\delta(\omega).
\label{eq:sigma_split}
\end{equation}
Since $f(\omega)\delta(\omega)=f(0)\delta(\omega)$ in the sense of distributions, only the value at $\omega=0$
enters the $\delta$-peak. Taking the real part 
of Eq.~\eqref{eq:sigma_split} yields
\begin{align}
\Re\,\sigma(\omega,T)
&=
\pi \Big[\frac{\langle -\hat K\rangle}{N}-\Re\,\chi^{R}_{JJ}(0)\Big]\delta(\omega)
+\Re\!\left[
i\Big( \frac{\langle -\hat K\rangle}{N} - \chi^{R}_{JJ}(\omega)\Big)\,
\mathcal{P}\frac{1}{\omega}
\right]
\nonumber\\
&=
\pi\Big[\frac{\langle -\hat K\rangle}{N} - \Re\,\chi^{R}_{JJ}(0)\Big]\delta(\omega)
-\Re\!\left[
i\chi^{R}_{JJ}(\omega)\,
\mathcal{P}\frac{1}{\omega}
\right],
\label{eq:Re_sigma_general_step}
\end{align}
where we used $\langle -\hat K\rangle\in\mathbb{R}$ and hence
$\Re\!\left[\frac{i}{N}\langle -\hat K\rangle\,\mathcal{P}\frac{1}{\omega}\right]=0$.
Writing $\chi^{R}_{JJ}(\omega)=\Re\chi^{R}_{JJ}(\omega)+i\,\Im\chi^{R}_{JJ}(\omega)$, we obtain
\begin{equation}
\Re\,\sigma(\omega,T)
=
\pi\Big[ \frac{\langle -\hat K\rangle}{N} - \Re\,\chi^{R}_{JJ}(0)\Big]\delta(\omega)
+\,\Im \chi^{R}_{JJ}(\omega)\,
\mathcal{P}\frac{1}{\omega}.
\end{equation}
For $\omega\neq 0$, the $\delta(\omega)$ term vanishes and $\mathcal{P}(1/\omega)=1/\omega$, so that
\begin{equation}
\Re\,\sigma(\omega,T)
=
\pi\Big[ \frac{\langle -\hat K\rangle}{N} -\Re\,\chi^{R}_{JJ}(0)\Big]\delta(\omega)
+\frac{1}{\omega}\,\Im \chi^{R}_{JJ}(\omega)\,.
\label{eq:Re_sigma_general}
\end{equation}

Comparing Eqs.~\eqref{eq:Re_sigma_general} and \eqref{eq:Drude_decomp}
gives
\begin{equation}
D(T)
=
\frac{1}{2}\left( \frac{\langle -\hat K\rangle}{N} - \Re\,\chi^{R}_{JJ}(0)\right),
\label{eq:D_from_chi}
\end{equation}
and 
\begin{equation}
\sigma_{\mathrm{reg}}(\omega,T)
= \frac{\Im\,\chi^{R}_{JJ}(\omega)}{\omega},
\qquad (\omega\neq 0).
\label{eq:sigma_reg_Imchi}
\end{equation}
Note that the expression for the regular part does not apply at strictly zero frequency.
The DC conductivity at $\omega=0$ is given by
\begin{align}
    \label{eq:appendix_dcconductivity}
    \sigma_{\mathrm{DC}}(T) = 2\pi D(T) \delta(\omega) + \lim_{\omega \to 0} \sigma_{\mathrm{reg}}(\omega, T).
\end{align}

\subsection{Drude weight and regular part in the Lehmann representation}
\label{DrudeweightandregularpartintheLehmannrepresentation}

Next, we provide the expression for Drude weight and regular part in the Lehmann representation.

Let $\ket{n}$ denote the eigenstates of $\hat H$ with eigenenergies $E_n$, i.e.,
$\hat H\ket{n}=E_n\ket{n}$, and define $J_{mn}\equiv\bra{m}\hat J\ket{n}$.
Inserting a complete set of eigenstates yields
\begin{align}
\langle [ \hat J(t), \hat J(0) ] \rangle
&=
\frac{1}{Z}
\sum_{n} e^{-\beta E_n}
\bra{n}
[ \hat J(t), \hat J(0) ]
\ket{n}
\nonumber\\
&=
\frac{1}{Z}
\sum_{n,m}
e^{-\beta E_n}
\Big(
\bra{n} \hat{J}(t) \ket{m}
\bra{m} \hat{J}(0) \ket{n}
-
\bra{n} \hat{J}(0) \ket{m}
\bra{m} \hat{J}(t) \ket{n}
\Big)
\nonumber\\
&=
\frac{1}{Z}
\sum_{n,m}
e^{-\beta E_n}
\Big(
e^{i(E_n-E_m)t}
-
e^{-i(E_n-E_m)t}
\Big)
|J_{mn}|^2.
\label{eq:comm_time}
\end{align}
Using
\begin{equation}
\int_0^{\infty} dt\,
e^{i(\omega-\Delta+i0^+)t}
=
\frac{i}{\omega-\Delta+i0^+},
\label{eq:time_integral_identity}
\end{equation}
we obtain the Lehmann representation of the retarded current--current correlator defined in Eq.~\eqref{eq:DefinitionOfRChi}  
\begin{equation}
\chi^{R}_{JJ}(\omega)
=
-\frac{1}{NZ}\sum_{n,m}
\frac{e^{-\beta E_n}-e^{-\beta E_m}}{\omega+i0^+-(E_m-E_n)}\,|J_{mn}|^2.
\label{eq:chi_lehmann}
\end{equation}

Using $\Im\,(x+i0^+)^{-1}=-\pi\delta(x)$, Eq.~\eqref{eq:chi_lehmann} gives
\begin{align}
\Im\,\chi^{R}_{JJ}(\omega)
&=
\frac{\pi}{NZ}\sum_{n,m}
\big(e^{-\beta E_n}-e^{-\beta E_m}\big)\,
|J_{mn}|^2\,
\delta\!\big(\omega-(E_m-E_n)\big)
\nonumber\\
&=
\pi\,\frac{1-e^{-\beta\omega}}{NZ}\sum_{n,m}
e^{-\beta E_n}\,
|J_{mn}|^2\,
\delta(E_n+\omega-E_m),
\label{eq:Imchi_final}
\end{align}
where in the second line we used the delta-function constraint $E_m=E_n+\omega$.

Substituting Eq.~\eqref{eq:Imchi_final} into Eq.~\eqref{eq:sigma_reg_Imchi} yields
\begin{equation}
\sigma_{\mathrm{reg}}(\omega,T)
=
\pi\,\frac{1-e^{-\beta\omega}}{\omega Z N}
\sum_{n,m}
e^{-\beta E_n}\,
|J_{mn}|^2\,
\delta(E_n+\omega-E_m),
\qquad (\omega\neq 0).
\label{eq:sigma_reg_final}
\end{equation}

From Eq.~\eqref{eq:chi_lehmann}, the real part at $\omega=0$ is understood as a principal value which makes it explicit that only the terms with $E_m \neq E_n$ contribute 
\begin{align}
\Re\,\chi^{R}_{JJ}(0)
&=
-\frac{1}{NZ}\sum_{n,m}
\big(e^{-\beta E_n}-e^{-\beta E_m}\big)\,
\mathcal P\frac{1}{-(E_m-E_n)}\,|J_{mn}|^2
\nonumber\\
&=
-\frac{2}{NZ}\sum_{\substack{m,n\\ E_m \neq E_n}}e^{-\beta E_n}\frac{|J_{mn}|^2}{E_n-E_m} \nonumber \\
&=
\frac{2}{NZ}\sum_{\substack{m,n\\ E_m \neq E_n}}e^{-\beta E_n}\frac{|J_{mn}|^2}{E_m-E_n}.
\label{eq:Rechi0_final}
\end{align}

Substituting Eq.~\eqref{eq:Rechi0_final} into Eq.~\eqref{eq:D_from_chi}, we obtain
\begin{equation}
D(T)
=
\frac{1}{N}
\left[
\frac{1}{2}\,\langle -\hat{K} \rangle
-\frac{1}{Z}
\sum_{\substack{m,n\\ E_m \neq E_n}}e^{-\beta E_n}\frac{|J_{mn}|^2}{E_m - E_n}
\right].
\label{eq:D_final}
\end{equation}
Therefore, we have proved Eq.~(5) and Eq.~(6) in the main text.

\subsection{Lower and upper bounds on the Drude weight}
\label{sec:bounds_on_Drude_weight}

The Lehmann representation derived above also provides useful lower and
upper bounds on the Drude weight. The lower bound follows from the overlap
of the current operator with conserved quantities, whereas the upper bound
follows from the optical sum rule and the positivity of the regular
optical conductivity.

\subsubsection{Mazur lower bound}

We first discuss the Mazur inequality~\cite{MAZUR1969533,zotos}.

Let $\{\hat Q_n\}$ be a set of Hermitian conserved quantities satisfying
$[\hat Q_n,\hat H]=0$. Following the standard convention, we choose the
conserved quantities to have vanishing thermal expectation values,
$\langle\hat Q_n\rangle=0$, and assume first that they are mutually
orthogonal with respect to the thermal inner product:
\begin{align}
    \langle \hat Q_n\hat Q_m \rangle
    =
    \delta_{nm}
    \langle \hat Q_n^2 \rangle.
    \label{eq:orthogonal_conserved_quantities}
\end{align}
The equilibrium expectation value of the current also vanishes,
$\langle\hat J\rangle=0$.

The nondecaying part of the current autocorrelation function is characterized
by its infinite-time average,
\begin{align}
    \overline{C}_{JJ}
    =
    \lim_{\mathcal T\to\infty}
    \frac{1}{\mathcal T}
    \int_0^{\mathcal T}dt\,
    \Re\langle
        \hat J(t)\hat J(0)
    \rangle.
    \label{eq:long_time_current_correlation}
\end{align}
The nondecaying contribution defines the persistent-current weight
$D_{\mathrm{diff}}(T)$ according to
\begin{align}
    D_{\mathrm{diff}}(T)
    =
    \frac{\beta}{2N}\,
    \overline{C}_{JJ}.
    \label{eq:Ddiff_long_time_correlation}
\end{align}
It is related to the Drude weight defined in Eq.~\eqref{eq:Drude_decomp}
through
\begin{align}
    D_{\mathrm{diff}}(T)
    =
    D(T)-D_m(T),
    \label{eq:Ddiff_Drude_Meissner_bounds}
\end{align}
where $D_m(T)$ is the Meissner stiffness
~\cite{PhysRevB.73.085117,PhysRevB.77.245131}. For a nonsuperconducting system, $D_m(T)=0$, such that
$D_{\mathrm{diff}}(T)=D(T)$. We focus on this case throughout.
The Mazur inequality provides a lower bound on the persistent part of the
current correlation:
\begin{align}
    \overline{C}_{JJ}
    \geq
    \sum_n
    \frac{
        \langle \hat J\hat Q_n \rangle^2
    }{
        \langle \hat Q_n^2 \rangle
    }.
    \label{eq:Mazur_current_bound}
\end{align}
Consequently, the persistent-current weight satisfies
\begin{align}
    D(T)
    \geq
    D_{\mathrm{Mazur}}(T)
    \equiv
    \frac{\beta}{2N}
    \sum_n
    \frac{
        \langle \hat J\hat Q_n \rangle^2
    }{
        \langle \hat Q_n^2 \rangle
    }.
    \label{eq:Mazur_Drude_bound}
\end{align}
Thus, a nonzero overlap between the current and at least one conserved
quantity guarantees a nonzero Drude weight and hence a
ballistic contribution to charge transport. 

For completeness, if the conserved quantities are not mutually orthogonal,
we define their correlation matrix as
\begin{align}
    C_{nm}
    =
    \langle \hat Q_n\hat Q_m \rangle.
\end{align}
The corresponding basis-independent form of the Mazur bound is
\begin{align}
    D(T)
    \geq
    \frac{\beta}{2N}
    \sum_{n,m}
    \langle \hat J\hat Q_n \rangle
    \left(C^{-1}\right)_{nm}
    \langle \hat Q_m\hat J \rangle.
    \label{eq:Mazur_Drude_bound_nonorthogonal}
\end{align}
Equation~\eqref{eq:Mazur_Drude_bound} follows after orthogonalizing the
conserved quantities. If the chosen conserved quantities form a complete
basis for the conserved component of the current, the Mazur inequality is
saturated. Conversely, vanishing overlaps with an incomplete set of
conserved quantities only make the corresponding lower bound vanish and
do not, by themselves, imply a vanishing 
Drude weight.

A particularly important case occurs when the current is itself conserved,
$[\hat J,\hat H]=0$. One may then choose $\hat Q_1=\hat J$. Since
$\hat J(t)=\hat J$, the current autocorrelation function does not decay,
and Eqs.~\eqref{eq:long_time_current_correlation} and
\eqref{eq:Ddiff_long_time_correlation} give
\begin{align}
    D(T)
    =
    D_{\mathrm{Mazur}}(T)
    =
    \frac{\beta}{2N}
    \langle \hat J^2 \rangle.
    \label{eq:Mazur_bound_conserved_current}
\end{align}
Because $\hat J$ already lies entirely within the conserved subspace, all
additional conserved quantities can be orthogonalized with respect to
$\hat J$. Their overlaps with $\hat J$ then vanish, so including them does
not change the bound. Equivalently, the projection of the current onto the
conserved subspace is already the current itself.

\subsubsection{Upper bound from the optical sum rule}

An upper bound follows from the optical $f$-sum rule
~\cite{PhysRevB.16.2437,RevModPhys.93.025003},
which, with the normalization adopted in
Eq.~\eqref{eq:Drude_decomp}, reads
\begin{align}
    \int_0^{\infty}d\omega\,
    \Re\,\sigma(\omega,T)
    =
    \frac{\pi}{2N}
    \langle
        -\hat K
   \rangle.
    \label{eq:optical_sum_rule_positive_frequency}
\end{align}
Here the zero-frequency delta function contributes one-half of its total
weight because the integration is restricted to nonnegative frequencies.
Substituting Eq.~\eqref{eq:Drude_decomp} gives
\begin{align}
    \pi D(T)
    +
    \int_0^{\infty}d\omega\,
    \sigma_{\mathrm{reg}}(\omega,T)
    =
    \frac{\pi}{2N}
    \langle
        -\hat K
    \rangle.
    \label{eq:Drude_optical_sum_rule}
\end{align}
Since the regular part is nonnegative,
$\sigma_{\mathrm{reg}}(\omega,T)\geq0$ for $\omega>0$, we obtain
\begin{align}
    0
    \leq
    D(T)
    \leq
    \frac{1}{2N}
    \langle
        -\hat K
    \rangle.
    \label{eq:Drude_upper_bound}
\end{align}

The same upper bound follows directly from the Lehmann representation in
Eq.~\eqref{eq:D_final}. Indeed, the paramagnetic contribution can be
symmetrized as
\begin{align}
&\frac{1}{Z}
\sum_{\substack{m,n\\E_m\neq E_n}}
e^{-\beta E_n}
\frac{|J_{mn}|^2}{E_m-E_n}=
\frac{1}{2Z}
\sum_{\substack{m,n\\E_m\neq E_n}}
\frac{
    e^{-\beta E_n}-e^{-\beta E_m}
}{
    E_m-E_n
}
|J_{mn}|^2
\geq0,
\label{eq:positive_paramagnetic_contribution}
\end{align}
where the last inequality follows because $e^{-\beta E}$ is a monotonically
decreasing function of $E$. Equation~\eqref{eq:D_final} therefore immediately
implies Eq.~\eqref{eq:Drude_upper_bound}. The upper bound is saturated when
there is no finite-frequency optical spectral weight, or equivalently when
$J_{mn}=0$ for all pairs with $E_m\neq E_n$.

For a nonsuperconducting system, $D_{\mathrm{diff}}(T)=D(T)$, and combining
Eqs.~\eqref{eq:Mazur_Drude_bound} and
\eqref{eq:Drude_upper_bound} gives
\begin{align}
    \frac{\beta}{2N}
    \sum_n
    \frac{
        \langle
            \hat J\hat Q_n
        \rangle^2
    }{
        \langle
            \hat Q_n^2
        \rangle
    }
    \leq
    D(T)
    \leq
    \frac{1}{2N}
    \langle
        -\hat K
    \rangle.
    \label{eq:combined_Drude_bounds}
\end{align}

For the infinite-$U$ Hubbard model considered in this work, we show in
Sec.~\ref{sec:Currentoperatorandproof} that
$[\hat J,\hat H]=0$. It follows that $J_{mn}=0$ whenever
$E_m\neq E_n$, so the upper bound is saturated. Since
$\hat K=\hat H$ for the present model,
\begin{align}
    D(T)
    =
    \frac{1}{2N}
    \left\langle
        -\hat H
    \right\rangle.
    \label{eq:upper_bound_saturated_present_model}
\end{align}
At the same time, choosing $\hat Q_1=\hat J$ in the Mazur inequality gives
Eq.~\eqref{eq:Mazur_bound_conserved_current}. Therefore, both bounds
are saturated.

\section{Model and observables}
\label{sec:Model_and_observables}

In this section, we introduce the one-dimensional Hubbard model in the infinite-interaction limit and present the definitions of the current operator, the optical conductivity, and the specific heat. Finally, we provide a summary of the analytical results obtained in the subsequent sections.

\subsection{1D Hubbard model}
The Hamiltonian of the one-dimensional (1D) Hubbard model is
\begin{eqnarray}
    \label{eq:Hamiltonian_1D_withU}
    \hat{H} &=& - t \sum_{\sigma=\{ \uparrow,\downarrow \}}\sum_{j=0}^{N-1} (c^{\dagger}_{j+1, \sigma} c_{j, \sigma} + c^{\dagger}_{j, \sigma} c_{j+1, \sigma}) + U \sum_{j} n_{j,\uparrow} n_{j, \downarrow},
\end{eqnarray}
where $c^{\dagger}_{j,\sigma}$ and $c_{j,\sigma}$ denote, respectively, the creation and annihilation operators for a fermion with spin $\sigma$ at site $j$. The corresponding density operator is $n_{j,\sigma}=c^{\dagger}_{j,\sigma}c_{j,\sigma}$. Here, $t$ and $U$ denote the hopping amplitude and the on-site interaction strength, respectively.

The model considered in the main text is the 1D Hubbard model in the infinite-interaction limit, which is equivalent to the $t$--$J$ model with $J=0$. The Hamiltonian reads
\begin{eqnarray}
    \label{eq:Hamiltonian_1D}
    \hat{H} &=& - t \sum_{\sigma=\{ \uparrow,\downarrow \}}\sum_{j=0}^{N-1} (\tilde{c}^{\dagger}_{j+1, \sigma} \tilde{c}_{j, \sigma} + \tilde{c}^{\dagger}_{j, \sigma} \tilde{c}_{j+1, \sigma}).
\end{eqnarray}
Here $\tilde{c}^{\dagger}_{j,\sigma} = (1 - n_{j,\bar{\sigma}})\, c^{\dagger}_{j,\sigma}$ denotes the projected creation operator of a fermion with spin $\sigma$ at site $j$ in the subspace without double occupancy, and $n_{j,\bar{\sigma}} = c^{\dagger}_{j,\bar{\sigma}} c_{j,\bar{\sigma}}$ is the corresponding density operator of fermion with opposite spin $\bar{\sigma}$. The hopping amplitude $t$ sets the energy scale and is fixed to $t=1$ throughout. Periodic boundary conditions (PBCs) are imposed, such that $\tilde{c}^{\dagger}_{N,\sigma} \equiv \tilde{c}^{\dagger}_{0,\sigma}$, where $N$ denotes the system size, i.e., the total number of lattice sites.

The Hamiltonian in Eq.~\eqref{eq:Hamiltonian_1D} preserves translational symmetry as well as a $U(1)\times U(1)$ symmetry. The former corresponds to lattice translation invariance, while the latter reflects the conservation of total charge ($\sum_{j=0}^{N-1} (n_{j,\uparrow} + n_{j, \downarrow})$) and total magnetization ($\sum_{j=0}^{N-1} (n_{j,\uparrow} - n_{j, \downarrow})/2$). 

\subsection{Current operator and proof of $[\hat{J}, \hat{H}]=0$}
\label{sec:Currentoperatorandproof}
For the 1D Hubbard model in the infinite-interaction limit defined in Eq.~\eqref{eq:Hamiltonian_1D}, the current operator $\hat{J}$ is 
\begin{eqnarray}
    \label{eq:currentoperator_1D}
    \hat{J} = i \sum_{\sigma=\{ \uparrow,\downarrow \}}\sum_{j=0}^{N-1} (\tilde{c}^{\dagger}_{j+1, \sigma} \tilde{c}_{j, \sigma} - \tilde{c}^{\dagger}_{j, \sigma} \tilde{c}_{j+1, \sigma}).
\end{eqnarray} 
Note that we have set $t=1$. We find that the current operator $\hat{J}$ commutes with the Hamiltonian $\hat{H}$, i.e., $[\hat{J}, \hat{H}] = 0$, whereas in the conventional 1D Hubbard model one generally has $[\hat{J}, \hat{H}] \neq 0$. We now present an explicit proof that $[\hat{J}, \hat{H}] = 0$ in the infinite-interaction limit.
In our case, with doped holes, double occupancy is still forbidden, and we work directly in the no-double-occupancy subspace when deriving the eigenstates. Therefore, $(1-n_{j,\bar{\sigma}})$ in $\tilde{c}^{\dagger}_{j,\sigma}$ becomes redundant. Here, however, to verify $[\hat{J}, \hat{H}] = 0$, we keep it for completeness.
We define the local current operator
\begin{equation}
\hat{J}_{j,j+1,\sigma}
=
i\bigl(
\tilde{c}^{\dagger}_{j+1,\sigma}\tilde{c}_{j,\sigma}
-
\tilde{c}^{\dagger}_{j,\sigma}\tilde{c}_{j+1,\sigma}
\bigr)
=
i(1-n_{j+1,\bar{\sigma}})
\bigl(
c^{\dagger}_{j+1,\sigma}c_{j,\sigma}
-
c^{\dagger}_{j,\sigma}c_{j+1,\sigma}
\bigr)
(1-n_{j,\bar{\sigma}}),
\end{equation}
such that $\hat{J} = \sum_{j,\sigma} \hat{J}_{j,j+1,\sigma}$. We first evaluate the commutator $[\hat{J}_{j,j+1,\sigma},\hat{H}]$, which can be decomposed as
\begin{equation}
[\hat{J}_{j,j+1,\sigma},\hat{H}]
=
[\hat{J}_{j,j+1,\sigma},\hat{H}_{\sigma}]
+
[\hat{J}_{j,j+1,\sigma},\hat{H}_{\bar{\sigma}}],
\end{equation}
where $\hat{H}_{\sigma}$ and $\hat{H}_{\bar{\sigma}}$ denote the hopping terms for spin $\sigma$ and opposite spin $\bar{\sigma}$, respectively.

For the same-spin contribution $[\hat{J}_{j,j+1,\sigma},\hat{H}_{\sigma}]$, only the hopping terms in $\hat{H}_{\sigma}$ involving sites $j-1$, $j$, $j+1$, and $j+2$ contribute. A straightforward calculation yields
\begin{align}
[\hat{J}_{j,j+1,\sigma},\hat{H}_{\sigma}]
&=
\Bigl(
- i (1-n_{j+2,\bar{\sigma}})(1-n_{j+1,\bar{\sigma}})(1-n_{j,\bar{\sigma}})
(-c^{\dagger}_{j+2,\sigma}c_{j,\sigma})
\nonumber \\
&\quad
- i (1-n_{j+1,\bar{\sigma}})(1-n_{j,\bar{\sigma}})
(n_{j+1,\sigma}-n_{j,\sigma})
\nonumber \\
&\quad
- i (1-n_{j+1,\bar{\sigma}})(1-n_{j,\bar{\sigma}})(1-n_{j-1,\bar{\sigma}})
c^{\dagger}_{j+1,\sigma}c_{j-1,\sigma}
\Bigr)
+ \text{H.c.}
\end{align}
Summing over $j$ and $\sigma$, these terms cancel pairwise, leading to
\begin{equation}
\sum_{j,\sigma}
[\hat{J}_{j,j+1,\sigma},\hat{H}_{\sigma}]
=
0.
\end{equation}

For the opposite-spin contribution $[\hat{J}_{j,j+1,\sigma},\hat{H}_{\bar{\sigma}}]$, one finds
\begin{align}
[\hat{J}_{j,j+1,\sigma},\hat{H}_{\bar{\sigma}}]
&=
\Bigl(
- i (1-n_{j,\bar{\sigma}})(1-n_{j+2,\sigma})
c^{\dagger}_{j+1,\sigma}c_{j,\sigma}
c^{\dagger}_{j+2,\bar{\sigma}}c_{j+1,\bar{\sigma}}
\nonumber \\
&\quad
+ i (1-n_{j+1,\bar{\sigma}})(1-n_{j-1,\sigma})
c^{\dagger}_{j+1,\sigma}c_{j,\sigma}
c^{\dagger}_{j,\bar{\sigma}}c_{j-1,\bar{\sigma}}
\Bigr)
+ \text{H.c.}
\end{align}
Upon summation over $j$ and $\sigma$, 
\begin{eqnarray}
    \sum_{j,\sigma} [\hat{J}_{j,j+1, \sigma}, \hat{H}_{\bar{\sigma}}] &=& \sum_{j,\sigma} \left( - i(1-n_{j, \bar{\sigma}})(1-n_{j+2, \sigma}) c^{\dagger}_{j+1, \sigma} c_{j, \sigma} c^{\dagger}_{j+2, \bar{\sigma}} c_{j+1, \bar{\sigma}} \right. \nonumber \\  
    && \left. + i(1-n_{j+1, \bar{\sigma}})(1-n_{j-1, \sigma}) c^{\dagger}_{j+1, \sigma} c_{j, \sigma} c^{\dagger}_{j, \bar{\sigma}} c_{j-1, \bar{\sigma}} \right)  + \text{H.c.} \nonumber\\  
    &=& \left( \sum_{j,\sigma} - i(1-n_{j, \bar{\sigma}})(1-n_{j+2, \sigma}) c^{\dagger}_{j+1, \sigma} c_{j, \sigma} c^{\dagger}_{j+2, \bar{\sigma}} c_{j+1, \bar{\sigma}} \right. \nonumber \\ 
    && \left. + \sum_{j, \sigma} i(1-n_{j+1, \sigma})(1-n_{j-1, \bar{\sigma}}) c^{\dagger}_{j+1, \bar{\sigma}} c_{j, \bar{\sigma}}c^{\dagger}_{j, \sigma} c_{j-1, \sigma}) \right) + \text{H.c.} \nonumber\\ 
    &=& \left( \sum_{j,\sigma} - i(1-n_{j, \bar{\sigma}})(1-n_{j+2, \sigma}) c^{\dagger}_{j+1, \sigma} c_{j, \sigma} c^{\dagger}_{j+2, \bar{\sigma}} c_{j+1, \bar{\sigma}} \right. \nonumber \\ 
    && \left. + \sum_{j, \sigma} i(1-n_{j+2, \sigma})(1-n_{j, \bar{\sigma}}) c^{\dagger}_{j+2, \bar{\sigma}} c_{j+1, \bar{\sigma}}c^{\dagger}_{j+1, \sigma} c_{j, \sigma} \right) + \text{H.c.},
\end{eqnarray}
the contributions again cancel pairwise, yielding
\begin{equation}
\sum_{j,\sigma}
[\hat{J}_{j,j+1,\sigma},\hat{H}_{\bar{\sigma}}]
=
0.
\end{equation}

Combining the two parts, we obtain
\begin{equation}
[\hat{J},\hat{H}] = 0.
\end{equation} The Hamiltonian and the current operator hence share common eigenstates.

\subsection{Optical conductivity}
\label{sec:Optical_conductivity_model}

We have obtained the expressions for Drude weight (see Eq.~\eqref{eq:D_final}) and regular part (see Eq.~\eqref{eq:sigma_reg_final}). Now we apply these results to the present model.

Let \(\ket{n}\) denote an eigenstate of the Hamiltonian \(\hat H\) with eigenenergy \(E_n\), i.e.,
\begin{align}
    \hat H\ket{n}=E_n\ket{n}.
\end{align}
If \([\hat J,\hat H]=0\), then
\begin{align}
    \bra{n}[\hat J,\hat H]\ket{m} &= 0 \nonumber\\
    \Rightarrow\quad
    \bra{n}(\hat J\hat H-\hat H\hat J)\ket{m} &= 0 \nonumber\\
    \Rightarrow\quad
    (E_m-E_n)\bra{n}\hat J\ket{m} &= 0.
\end{align}
Therefore, when \(E_m\neq E_n\), we have
\begin{align}
    \bra{n}\hat J\ket{m}=0.
\end{align}

Now consider the case in which \(\hat H\) has a degenerate eigenspace. Let
\(\ket{E,1},\cdots,\ket{E,g_E}\) be an orthonormal basis of the \(g_E\)-fold degenerate eigenspace with eigenenergy \(E\). Since \([\hat J,\hat H]=0\), for each \(i=1,\cdots,g_E\) we have
\begin{align}
    \hat H\hat J\ket{E,i}
    =
    \hat J\hat H\ket{E,i}
    =
    E\,\hat J\ket{E,i}.
\end{align}
Therefore, \(\hat J\ket{E,i}\) also belongs to the same eigenspace with energy \(E\). We may thus restrict \(\hat J\) to this \(g_E\)-dimensional subspace and define the corresponding matrix
\begin{align}
    J^{(E)}_{ij}
    \equiv
    \bra{E,i}\hat J\ket{E,j},
    \qquad
    i,j=1,\cdots,g_E.
\end{align}
Since \(\hat J\) is Hermitian, the matrix \(J^{(E)}\) is also Hermitian. Hence there exists a unitary matrix \(U^{(E)}\) such that
\begin{align}
    \bigl(U^{(E)}\bigr)^\dagger
    J^{(E)}
    U^{(E)}
    =
    \mathrm{diag}\!\bigl(j_{E,1},j_{E,2},\cdots,j_{E,g_E}\bigr).
\end{align}

We then define a new basis within the same degenerate eigenspace by
\begin{align}
    \ket{E,\gamma}'
    \equiv
    \sum_{i=1}^{g_E}
    \ket{E,i}\,U^{(E)}_{i\gamma},
    \qquad
    \gamma=1,\cdots,g_E.
\end{align}
Because this transformation is performed entirely within the degenerate subspace, the new basis states are still eigenstates of \(\hat H\) with the same energy \(E\):
\begin{align}
    \hat H\ket{E,\gamma}'
    =
    E\ket{E,\gamma}'.
\end{align}
Moreover, in this new basis, the matrix elements of \(\hat J\) are
\begin{align}
    \bra{E,\gamma_1}' \hat J \ket{E,\gamma_2}'
    &=
    \sum_{i,j}
    \bigl(U^{(E)}_{i\gamma_1}\bigr)^*
    \bra{E,i}\hat J\ket{E,j}
    U^{(E)}_{j\gamma_2}
    \nonumber\\
    &=
    \left[
        \bigl(U^{(E)}\bigr)^\dagger
        J^{(E)}
        U^{(E)}
    \right]_{\gamma_1 \gamma_2}
    \nonumber\\
    &=
    j_{E,\gamma_1}\,\delta_{\gamma_1 \gamma_2}.
\end{align}
Therefore, within each degenerate eigenspace of \(\hat H\), one can always choose a basis in which \(\hat J\) is diagonal.

Combining all energy eigenspaces, we conclude that if \([\hat H,\hat J]=0\), then one can always choose a complete orthonormal basis of simultaneous eigenstates of \(\hat H\) and \(\hat J\). In such a basis,
\begin{align}
    \bra{m}\hat J\ket{n}=0,
    \qquad
    m\neq n.
\end{align}
Equivalently, all off-diagonal matrix elements of \(\hat J\) vanish in the common eigenbasis.

Therefore, for the present model, the Drude weight (see Eq.~\eqref{eq:D_final}) simplifies to 
\begin{align}
D(T)
&= \frac{1}{2N}\,\langle -\hat{K}  \rangle = \frac{1}{2N} \langle-\hat{H} \rangle,
\end{align}
where we have used the fact that in the present model the kinetic-energy operator $\hat{K}$ coincides with the Hamiltonian $\hat{H}$, and the regular part (see Eq.~\eqref{eq:sigma_reg_final}) is
\begin{align}
\sigma_{\mathrm{reg}}(\omega,T)
&= \pi \frac{1 - e^{-\beta \omega}}{\omega Z N}
\sum_{n}
e^{-\beta E_n}
\left| J_{nn} \right|^2
\delta(\omega).
\end{align}
The finite-frequency regular part vanishes and thus
\begin{align}
    \lim_{\omega \to 0} \sigma_{\mathrm{reg}}(\omega, T) = 0.
\end{align}
Consequently, the dc conductivity consists solely of the Drude peak.

Moreover, although the expression for the regular part is not strictly applicable at zero frequency, it is related to the Drude weight~\cite{PhysRevB.73.085117,PhysRevB.77.245131}. 
By formally setting $\omega=0$, we obtain
\begin{align}
\label{eq:DefinitionOfSigmadiff}
\sigma_{\mathrm{diff}}(\omega,T)
&= \pi \frac{\beta}{Z N}
\sum_{n}
e^{-\beta E_n}
\left| J_{nn} \right|^2
\delta(\omega) = 2 \pi D_{\mathrm{diff}}(T) \delta(\omega),
\end{align}
where we have used $\lim_{\omega \to 0} \frac{1-e^{-\beta \omega}}{\omega} = \beta$.
Here we denote it by $D_{\mathrm{diff}}(T)$ ($\sigma_{\mathrm{diff}}(\omega,T)$) because it corresponds to the difference between the Drude weight and the Meissner stiffness, and to distinguish it from the regular part, which only applies at $\omega>0$.
As shown in Refs.~\cite{PhysRevB.73.085117,PhysRevB.77.245131}, we have the following equality
\begin{align}
D(T)- D_m(T)
= D_{\mathrm{diff}}(T),
\end{align}
where $D_m(T)$ is the Meissner stiffness~\cite{PhysRevB.73.085117,PhysRevB.77.245131}. $D_m(T)=0$ indicates that the system is not a superconductor. In Ref.~\cite{PhysRevB.77.245131}, the authors have numerically demonstrated that $D_m(T)=0$ in a one dimensional spinless fermion model.
In Secs.~\ref{sec:Exact_solutions_with_single_hole} and \ref{sec:Exact_solutions_with_two_holes}, we have proven that for the present model $ D(T) = D_{\mathrm{diff}}(T)$, i.e., $D_m(T)=0$, consistent with the fact that there is no superconductivity in one dimension.

\subsection{Specific heat}
In addition to the optical conductivity and resistivity, we also consider the specific heat, which is defined as
\begin{eqnarray}
\label{eq:specificheat}
C_{V}(T)
=
\beta^2
\left(
\langle \hat{H}^2 \rangle
-
\langle \hat{H} \rangle^2
\right),
\end{eqnarray}
where the thermal expectation value of an operator $\hat{O}$ is given by
\begin{equation}
\langle \hat{O} \rangle
=
\frac{\mathrm{Tr}\left(e^{-\beta \hat{H}} \hat{O}\right)}{Z} = \frac{\sum_{n} e^{-\beta E_n} \bra{n} \hat{O} \ket{n}}{Z}.
\end{equation}
The specific heat is therefore determined by the energy fluctuations.

\subsection{Summary of analytical results}

For a fixed number of holes, we summarize the energy spectrum $E$ and the eigenvalues $J$ of the current operator $\hat{J}$ in Table~\ref{tab:fixed-holes-results-E-J}. The corresponding analytical expressions for the Drude weight and specific heat are summarized in Table~\ref{tab:fixed-holes-results-Drude-Specificheat}. Here, $n_h$ and $n_{\downarrow}$ denote the numbers of holes and spin-down fermions, respectively.
$I_l(\cdot)$ denotes the $l$-th modified Bessel function of the first kind. Detailed derivations are provided in the following sections.

In Sec.~\ref{sec:finitedoping}, we have also obtained the analytical expressions for the Drude weight at fixed hole density in both the high-temperature and low-temperature limits (see Eq.~\eqref{eq:appendix_Drude_highT} and Eq.~\eqref{eq:final_lowT_energy_density_fixed_delta}).

\begin{table}[t]
\centering
\caption{Eigenenergies $E$ and eigenvalues $J$ of the current operator $\hat{J}$.}
\label{tab:fixed-holes-results-E-J}
\begin{tabular}{c @{\hspace{1.2cm}} c @{\hspace{1.2cm}} c}
\toprule
 & $n_{\downarrow}=0$ or $n_{\downarrow}=N-n_h$ 
 & $1 \le n_{\downarrow} \le N-n_h-1$ \\
\midrule
$E$ &
$\displaystyle \sum_{a=1}^{n_h} 2\cos\!\left(\frac{2\pi m_a}{N}\right)$ &
$\displaystyle \sum_{a=1}^{n_h} 
2\cos\!\left(\frac{2\pi m_a}{N}+\frac{2\pi m_{\downarrow}}{N(N-n_h)}\right)$ \\

$J$ &
$\displaystyle \sum_{a=1}^{n_h} 2\sin\!\left(\frac{2\pi m_a}{N}\right)$ &
$\displaystyle \sum_{a=1}^{n_h} 
2\sin\!\left(\frac{2\pi m_a}{N}+\frac{2\pi m_{\downarrow}}{N(N-n_h)}\right)$ \\

Choice of $m_a$ and $m_{\downarrow}$ &
$0 \le m_1 < \cdots < m_{n_h} \le N-1$ &
$\begin{aligned}
0 &\le m_1 < \cdots < m_{n_h} \le N-1,\\
&m_{\downarrow}=0,\ldots,N-n_h-1
\end{aligned}$ \\

Degeneracy for given $m_a,m_{\downarrow}$ &
$1$ &
$\displaystyle \binom{N-n_h}{n_{\downarrow}}/(N-n_h)$ \\

\bottomrule
\end{tabular}
\end{table}

\begin{table}[t]
\centering
\caption{Analytical expressions for specific heat and Drude weight with a fixed number of holes.}
\label{tab:fixed-holes-results-Drude-Specificheat}
\begin{tabular}{c @{\hspace{1.2cm}} c @{\hspace{1.2cm}} c}
\toprule
\textbf{Number of holes} & Specific heat $C_V(T)$ & Drude weight $D(T)$  \\
\midrule
1 & $C_V^{1h}(T) =  \beta^2 \left(2 + 2\frac{I_2(2\beta)}{I_0(2\beta)} - 4 \left(\frac{I_1(2\beta)}{I_0(2\beta)}\right)^2\right)$ & $D_{1h}(T) = \frac{\beta}{N} \left(1 - \frac{I_2(2\beta)}{I_0(2\beta)}\right)$ \\
2 & $2C_V^{1h}(T)$ & $2D_{1h}(T)$  \\
$n_h$ & $n_h C_V^{1h}(T)$  & $n_h D_{1h}(T)$  \\
\bottomrule
\end{tabular}
\end{table}

\section{Explicit eigenstates and energy spectrum}
In this section, we derive the explicit eigenstates and energy spectrum of the present model. We begin with the single-hole case (Sec.~\ref{sec:Exact_solutions_with_single_hole}), then proceed to the two-hole case (Sec.~\ref{sec:Exact_solutions_with_two_holes}), and finally generalize to the case with an arbitrary number of holes (Sec.~\ref{sec:Exact_solutions_with_any_holes}).

Since we work directly in the subspace without double occupancy, the projection is implicitly enforced, and for simplicity we omit the tilde notation in the following discussion, i.e., $\tilde{c}\rightarrow c$.

\subsection{Single-hole case}
\label{sec:Exact_solutions_with_single_hole}
We first specify the convention for the Fock states. We use $n_{\downarrow}$ to denote the number of spin-down fermions and $\vert \mathrm{vac} \rangle$ to denote the vacuum state. We use the symbol $0$ to denote a hole, and $\sigma=1,-1$ to denote a spin-up or spin-down fermion, respectively.

A state with the hole at site $j$ and $N-1$ fermions occupying the remaining sites $j'$ with spin $\sigma_{j'}$ is written as
\begin{eqnarray}
\label{eq:OneHoleConvention}
\vert \sigma_{0}, \cdots, \sigma_{j-1}, 0, \sigma_{j+1}, \cdots, \sigma_{N-1} \rangle
=
c^{\dagger}_{0,\sigma_{0}}
\cdots
c^{\dagger}_{j-1,\sigma_{j-1}}
c^{\dagger}_{j+1,\sigma_{j+1}}
\cdots
c^{\dagger}_{N-1,\sigma_{N-1}}
\vert \mathrm{vac} \rangle,
\end{eqnarray}
where the fermion creation operators are arranged in ascending order of the site index when acting on the vacuum state. When $n_{\downarrow}=0$ or $N-1$, the state can be denoted simply by $\ket{j}$ where $j$ is the position of the hole. When $0<n_{\downarrow}<N-1$, there are $N \binom{N-1}{n_{\downarrow}}$ possible Fock states, and the state can be denoted by $\ket{j,\alpha}$: in the reduced $(N-1)$-site chain obtained by removing the site occupied by the hole, we label by site $0$ the site immediately to the right of the hole in the original $N$-site chain. Let the sites occupied by the $n_{\downarrow}$ spin-down fermions in the reduced chain be indexed by $j_1<j_2<\cdots<j_{n_{\downarrow}}$.
Then $\alpha$ is given by
\begin{align}
\alpha(j_1,\ldots,j_{n_\downarrow})
=
\binom{N-1}{n_\downarrow}
-1
-\sum_{a=1}^{n_\downarrow}
\binom{N-2-j_a}{\,n_\downarrow-a+1\,}.
\label{eq:ExpressionForAlpha}
\end{align}
Here we adopt the convention that $\binom{m}{r}=0
\ \text{if } m<r \text{ or } r<0$. When $n_{\downarrow}=1$, Eq.~\eqref{eq:ExpressionForAlpha} reduces to
\begin{align}
\alpha(j_1)
&=
\binom{N-1}{1}
-1
-\binom{N-2-j_1}{1}
= j_1,
\end{align}
i.e., $\alpha$ is simply the index of the site occupied by the spin-down fermion in the reduced $(N-1)$-site chain. For example, the following state can be denoted by $\ket{j,\alpha}$ with $j=0$ and $\alpha=1$:
\begin{equation}
\vert 0,1,-1,1,1 \rangle
=
c^{\dagger}_{1,\uparrow}
c^{\dagger}_{2,\downarrow}
c^{\dagger}_{3,\uparrow}
c^{\dagger}_{4,\uparrow}
\vert \mathrm{vac} \rangle.
\end{equation}

\subsubsection{No spin-down fermion}
\label{sec:Single_hole_and_no_spin-down_fermion}
We first consider the simplest case of a single hole and $N-1$ spin-up fermions, i.e., no spin-down fermions, corresponding to $S_z = \frac{N-1}{2}$. 
The Hilbert space dimension of this sector is $N$. 
We choose the following basis states
\begin{eqnarray}
\vert j \rangle
= (\prod_{j^{\prime} \neq j}  c^{\dagger}_{j^{\prime}, \uparrow} )\ket{\text{vac}} =
c^{\dagger}_{0,\uparrow}
\cdots
c^{\dagger}_{j-1,\uparrow}
c^{\dagger}_{j+1,\uparrow}
\cdots
c^{\dagger}_{N-1,\uparrow}
\vert \mathrm{vac} \rangle,
\end{eqnarray}
where the hole is located at site $j$ with $j=0,\ldots,N-1$.

The Hamiltonian (with hopping amplitude $t=1$) acts on these basis states as
\begin{eqnarray}
\hat{H}\vert 0 \rangle
&=&
-\bigl(
\vert 1 \rangle
+
(-1)^N \vert N-1 \rangle
\bigr),
\nonumber \\
\hat{H}\vert j \rangle
&=&
-\bigl(
\vert j-1 \rangle
+
\vert j+1 \rangle
\bigr),
\quad j=1,\ldots,N-2,
\nonumber \\
\hat{H}\vert N-1 \rangle
&=&
-\bigl(
(-1)^N \vert 0 \rangle
+
\vert N-2 \rangle
\bigr),
\end{eqnarray}
where the phase factor $(-1)^N$ arises from the fermionic sign accumulated when the hole hops across the boundary.

Since $\hat{H}$ preserves translational symmetry, (unnormalized) eigenstates can be constructed as momentum states
\begin{eqnarray}
\vert k \rangle
=
\sum_{j=0}^{N-1}
e^{ikj}
(T_x)^j
\vert 0 \rangle,
\end{eqnarray}
where $T_x$ denotes the translation operator that shifts all fermions to the right by one lattice site. The state $\vert k \rangle$ is an eigenstate of the translation operator $T_x$ with eigenvalue $e^{-ik}$. Under the action of the translation operator $T_x$, one has
\begin{align}
T_x \vert 0 \rangle
&=
T_x \left(
c^{\dagger}_{1,\uparrow}
\cdots
c^{\dagger}_{N-1,\uparrow}
\vert \mathrm{vac} \rangle
\right)
\nonumber \\
&=
c^{\dagger}_{2,\uparrow}
\cdots
c^{\dagger}_{N,\uparrow}
\vert \mathrm{vac} \rangle
\nonumber \\
&=
(-1)^N
c^{\dagger}_{0,\uparrow}
c^{\dagger}_{2,\uparrow}
\cdots
c^{\dagger}_{N-1,\uparrow}
\vert \mathrm{vac} \rangle
\nonumber \\
&=
(-1)^N
\vert 1 \rangle,
\end{align}
where the factor $(-1)^N$ arises from reordering the fermionic creation operators based on the convention in Eq.~\eqref{eq:OneHoleConvention}. The momentum eigenstate $\vert k \rangle$ can thus be written as
\begin{eqnarray}
    \vert k \rangle = \sum_{j=0}^{N-1} e^{ikj} (-1)^{Nj} \vert j \rangle.
\end{eqnarray}
It is straightforward to verify that $\vert k \rangle$ is an eigenstate of $\hat{H}$ provided $e^{ikN}=1$. Acting with $\hat{H}$ yields
\begin{eqnarray}
\hat{H} \vert k \rangle
&=&
\hat{H}\vert 0 \rangle
+
\sum_{j=1}^{N-2}
e^{ikj}(-1)^{Nj}
\hat{H}\vert j \rangle
+
e^{ik(N-1)}(-1)^{N(N-1)}
\hat{H}\vert N-1 \rangle
\nonumber \\
&=&
-\left(
\vert 1 \rangle
+
(-1)^N \vert N-1 \rangle
\right)
\nonumber \\
&&
-
\sum_{j=1}^{N-2}
e^{ikj}(-1)^{Nj}
\left(
\vert j-1 \rangle
+
\vert j+1 \rangle
\right)
\nonumber \\
&&
-
e^{ik(N-1)}(-1)^{N(N-1)}
\left(
(-1)^N \vert 0 \rangle
+
\vert N-2 \rangle
\right).
\end{eqnarray}
Rearranging the terms, one finds that all coefficients become proportional to $(e^{ik}+e^{-ik})$ when $e^{ikN}=1$. 
Consequently,
\begin{eqnarray}
\hat{H}\vert k \rangle
&=&
(-1)^{N+1}
\left(
e^{ik}
+
e^{-ik}
\right)
\vert k \rangle
\nonumber \\
&=&
2(-1)^{N+1}\cos(k)\,\vert k \rangle,
\end{eqnarray}
which gives the expected $2\cos(k)$ dispersion of a hole in the ferromagnetic background (up to an overall sign determined by $N$).

Since the current operator $\hat{J}$ commutes with $\hat{H}$ as discussed in Sec.~\ref{sec:Model_and_observables}, they share common eigenstates. It can be verified directly that $\vert k \rangle$ is also an eigenstate of $\hat{J}$ (see Eq.~\eqref{eq:ProofOfEigenstateJWithoutSpinDown} for the proof), and 
\begin{eqnarray}
\label{eq:1HNoDownJexpression}
\frac{
\bigl|
\langle k' \vert \hat{J} \vert k \rangle
\bigr|^2
}{
\bigl|
\langle k' \vert k \rangle
\bigr|^2
}
=
\left(
2\sin k
\right)^2
\delta_{k,k'},
\end{eqnarray}
where $\langle k' \ket{k} = N \delta_{k,k^{\prime}}$.

Since the calculation of the Drude weight and the specific heat depends only on the energy spectrum, we label the eigenenergies by $m=0,\cdots, N-1$ for simplicity.
For odd $N$, 
$k=\frac{2\pi m}{N}$. 
For even $N$, $k=\frac{2\pi (m-N/2)}{N} = \frac{2\pi m}{N} - \pi$. Therefore, the energy spectrum can be expressed as $2\cos(\frac{2\pi m}{N})$ with $m=0,\cdots,N-1$.

\subsubsection{One spin-down fermion}
\label{sec:Single_hole_and_one_spin-down_fermion}
Next, we consider the sector with one hole and one spin-down fermion, i.e., $N-2$ spin-up fermions, corresponding to $S_z = \frac{N-3}{2}$. The Hilbert space dimension of this sector is $N(N-1)$.

The $N(N-1)$ Fock states can be constructed systematically as $\vert j,\alpha \rangle$ (see Eq.~\eqref{eq:ExpressionForAlpha}). As an illustrative example, for $N=4$, the $12$ Fock states are given below:
\begin{align}
\vert j=0, \alpha=0 \rangle & \equiv \vert 0, -1, 1,1 \rangle, \nonumber\\
\vert j=1, \alpha=0 \rangle & \equiv \vert 1,0,-1, 1 \rangle, \nonumber\\
\vert j=2, \alpha=0 \rangle & \equiv \vert 1, 1,0,-1 \rangle, \nonumber\\
\vert j=3, \alpha=0 \rangle & \equiv \vert -1, 1, 1,0 \rangle, \nonumber\\
\vert j=0, \alpha=1 \rangle & \equiv \vert 0, 1, -1,1 \rangle, \nonumber\\
\vert j=1, \alpha=1 \rangle & \equiv \vert 1,0,1, -1 \rangle, \nonumber\\
\vert j=2, \alpha=1 \rangle & \equiv \vert -1, 1,0,1 \rangle, \nonumber\\
\vert j=3, \alpha=1 \rangle & \equiv \vert 1, -1, 1,0 \rangle, \nonumber\\
\vert j=0, \alpha=2 \rangle & \equiv \vert 0, 1, 1,-1 \rangle, \nonumber\\
\vert j=1, \alpha=2 \rangle & \equiv \vert -1,0,1, 1 \rangle, \nonumber\\
\vert j=2, \alpha=2 \rangle & \equiv \vert 1, -1,0,1 \rangle, \nonumber\\
\vert j=3, \alpha=2 \rangle & \equiv \vert 1, 1, -1,0 \rangle.
\end{align}
Note that the Fock states $\ket{j,\alpha}$ with the same $\alpha$ share the same configuration on the $(N-1)$-site chain (i.e. the chain without the hole).

The Hamiltonian acts on the Fock states $\ket{j,\alpha}$ as
\begin{eqnarray}
\hat{H}\ket{0,\alpha}
&=&
-\ket{1,\alpha-1}
-
(-1)^N \ket{N-1,\alpha+1},
\nonumber \\
\hat{H}\ket{j,\alpha}
&=&
-\ket{j+1,\alpha-1}
-
\ket{j-1,\alpha+1},
\qquad j=1,\ldots,N-2,
\nonumber \\
\hat{H}\ket{N-1,\alpha}
&=&
-(-1)^{N}\ket{0,\alpha-1}
-
\ket{N-2,\alpha+1},
\end{eqnarray}
where $\alpha$ is defined modulo $N-1$.

We then construct the momentum eigenstates
\begin{eqnarray}
    \ket{k, \alpha}
    =
    \sum_{j=0}^{N-1} e^{ikj} T_x^{\,j} \ket{0, \alpha}
    =
    \sum_{j=0}^{N-1} e^{ikj} (-1)^{Nj} \ket{j,\alpha},
\end{eqnarray}
where $T_{x}$ is the translation operator that shifts all fermions one site to the right, and the sign factor $(-1)^{Nj}$ originates from reordering the fermion creation operators according to the convention in Eq.~\eqref{eq:OneHoleConvention}. Therefore, the Hamiltonian acts on the momentum states as
\begin{eqnarray}
\hat{H} \ket{k, \alpha}
&=&
- e^{-ik} (-1)^{N} \ket{k, \alpha-1}
-
(-1)^{N} e^{-ik(N-1)} (-1)^{N(N-1)} \ket{k, \alpha+1}
\nonumber\\
&=&
(-1)^{N+1}
\left(
e^{-ik}\ket{k, \alpha-1}
+
e^{ik}\ket{k, \alpha+1}
\right),
\end{eqnarray}
where we used $e^{ikN}=1$ and the fact that $N(N-1)$ is even.

The following state is an eigenstate of $\hat{H}$
\begin{eqnarray}
    \label{eq:1H1DownDefinitionOfEigenstate}
    \ket{k,k_{\downarrow}} = \sum_{\alpha=0}^{N-2} e^{-ik_{\downarrow} \alpha} \ket{k, \alpha},
\end{eqnarray}
provided $e^{ik_{\downarrow}(N-1)}=1$. To verify this, we consider the action of the Hamiltonian $\hat{H}$ on $\ket{k,k_{\downarrow}}$:
\begin{eqnarray}
\hat{H} \ket{k,k_{\downarrow}}
&=&
\hat{H} \sum_{\alpha=0}^{N-2} e^{-ik_{\downarrow} \alpha} \ket{k, \alpha}
\nonumber\\
&=&
\hat{H} \ket{k, 0}
+
\hat{H} \sum_{\alpha=1}^{N-3} e^{-ik_{\downarrow} \alpha} \ket{k, \alpha}
+
\hat{H} \ket{k, N-2}
\nonumber\\
&=&
(-1)^{N+1} \Big(
e^{-ik} \ket{k, N-2}
+
e^{ik} \ket{k, 1}
\Big)
\nonumber\\
&&
+\, (-1)^{N+1}
\sum_{\alpha=1}^{N-3} e^{-ik_{\downarrow} \alpha}
\Big(
e^{-ik} \ket{k, \alpha-1}
+
e^{ik} \ket{k, \alpha+1}
\Big)
\nonumber\\
&&
+\, (-1)^{N+1} e^{-i k_{\downarrow} (N-2)}
\Big(
e^{-ik} \ket{k, N-3}
+
e^{ik} \ket{k, 0}
\Big)
\nonumber\\
&=&
(-1)^{N+1} \Bigg[
\Big(
e^{-ik_{\downarrow}} e^{-ik}
+
e^{-i k_{\downarrow} (N-2)} e^{ik}
\Big) \ket{k, 0}
\nonumber\\
&&
+\,
\sum_{\alpha=1}^{N-3}
\Big(
e^{-ik_{\downarrow}} e^{-ik}
+
e^{ik_{\downarrow}} e^{ik}
\Big) e^{-ik_{\downarrow} \alpha}\ket{k, \alpha}
\nonumber\\
&&
+\,
\Big(
e^{i(N-2)k_{\downarrow}} e^{-ik}
+
e^{i k_{\downarrow}} e^{ik}
\Big) e^{-i(N-2)k_{\downarrow}}\ket{k, N-2}
\Bigg].
\end{eqnarray}
Therefore, $\ket{k, k_{\downarrow}}$ is an eigenstate with eigenenergy
\begin{align}
E = (-1)^{N+1} 2\cos\!\left(k+k_{\downarrow}\right),
\end{align}
when $e^{i(N-1)k_{\downarrow}}=1$. 

Moreover, $k$ and $k_{\downarrow}$ can be expressed as 
\begin{align}
    k &= \frac{2 \pi m}{N} - \frac{2\pi m_{\downarrow}}{N}, \label{eq:1H1DownKOdd} \\ 
    k_{\downarrow} &= \frac{2\pi m_{\downarrow}}{N-1}, \label{eq:1H1DownKDownOdd}
\end{align}
when $N$ is odd and 
\begin{align}
        k &= \frac{2 \pi m}{N} - \frac{2\pi m_{\downarrow}}{N} + \pi, \label{eq:1H1DownKEven} \\ 
    k_{\downarrow} &= \frac{2\pi m_{\downarrow}}{N-1}, \label{eq:1H1DownKDownEven}
\end{align}
when $N$ is even. Here, $m = 0, \ldots, N-1$ and $m_{\downarrow} = 0, \ldots, N-2$.

Consequently, the eigenstate $\ket{k, k_{\downarrow}}$ in Eq.~\eqref{eq:1H1DownDefinitionOfEigenstate} 
can be rewritten as $\ket{m, m_{\downarrow}}$.  
We then have
\begin{eqnarray}
\hat{H}\ket{m, m_{\downarrow}}
&=&
2\cos\!\left(
\frac{2\pi m}{N}
+
\frac{2\pi m_{\downarrow}}{N(N-1)}
\right)
\ket{m, m_{\downarrow}},
\end{eqnarray}
\begin{eqnarray}
\label{eq:1H1DJexpression}
\frac{
\bigl|
\bra{m, m_{\downarrow}}
\hat{J}
\ket{m, m_{\downarrow}}
\bigr|^2
}{
\langle m, m_{\downarrow} \vert m, m_{\downarrow} \rangle^2
}
&=&
\left[
2\sin\!\left(
\frac{2\pi m}{N}
+
\frac{2\pi m_{\downarrow}}{N(N-1)}
\right)
\right]^2.
\end{eqnarray}
See Eq.~\eqref{eq:ProofOfEigenstateJWithSpinDown} for the proof that $\ket{m, m_{\downarrow}}$ is also an eigenstate of $\hat{J}$.
The presence of a spin-down fermion effectively imposes a twisted boundary condition with twist angle $2\pi m_{\downarrow}/(N-1)$.

\subsubsection{Two spin-down fermions}
\label{sec:Single_hole_and_two_spin-down_fermions}
In the sector with one hole and two spin-down fermions, the Hilbert space dimension is $N\binom{N-1}{2}$. 

In this sector, the eigenenergies are still given by
\begin{eqnarray}
E(m,m_{\downarrow})
=
2\cos\!\left(
\frac{2\pi m}{N}
+
\frac{2\pi m_{\downarrow}}{N(N-1)}
\right),
\end{eqnarray}
with $m=0,\ldots,N-1$ and $m_{\downarrow}=0,\ldots,N-2$. 
However, compared to the single spin-down case discussed in Sec.~\ref{sec:Single_hole_and_one_spin-down_fermion}, the energy levels now exhibit additional degeneracies which we now show.

Similar to the sector with one hole and one spin-down fermion, we again use
$\ket{j,\alpha}$ (see Eq.~\eqref{eq:ExpressionForAlpha}) to label the $N\binom{N-1}{2}$ Fock states, where the hole is located at site $j$.
As an illustrative example, when $N=5$, the six Fock states with the hole located at
site $0$ are
\begin{eqnarray}
    \ket{j=0, \alpha = 0} &\equiv& \ket{0,-1,-1,1,1}, \\
    \ket{j=0, \alpha = 1} &\equiv& \ket{0,-1,1,-1,1}, \nonumber\\
    \ket{j=0, \alpha = 2} &\equiv& \ket{0,-1,1,1,-1}, \nonumber\\
    \ket{j=0, \alpha = 3} &\equiv& \ket{0,1,-1,-1,1}, \nonumber\\
    \ket{j=0, \alpha = 4} &\equiv& \ket{0,1,-1,1,-1}, \nonumber \\
    \ket{j=0, \alpha = 5} &\equiv& \ket{0,1,1,-1,-1}. \nonumber 
\end{eqnarray}

Owing to translational invariance, we first construct momentum eigenstates by applying the translation operator $T_x$ acting on all the fermions,
\begin{eqnarray}
\vert k,\alpha \rangle
=
\sum_{j=0}^{N-1}
e^{ikj}
T_x^{j}
\vert 0,\alpha \rangle,
\end{eqnarray}
where the operator $T_x$ translates all fermions to the right by one lattice site. Here $k$ is the discrete crystal momentum satisfying $e^{iNk}=1$, so that there are $N$ allowed momentum values. In a fixed momentum-$k$ sector, the Hilbert space dimension is thus $\binom{N-1}{2}$ since the number of possible values of $\alpha$ is $\binom{N-1}{2}$. As an example, for $N=5$, the state $\vert k,\alpha=0 \rangle$ reads
\begin{align}
\vert k,0 \rangle
=
\vert 0,-1,-1,1,1 \rangle
-
e^{ik}
\vert 1,0,-1,-1,1 \rangle
+
e^{2ik}
\vert 1,1,0,-1,-1 \rangle
-
e^{3ik}
\vert -1,1,1,0,-1 \rangle
+
e^{4ik}
\vert -1,-1,1,1,0 \rangle.
\end{align}
Other basis states $\ket{k,\alpha}$ can also be constructed in a similar way. For simplicity, we use the notation $\{0,-1,-1,1,1\}_{k}$ to denote the translation-invariant state $\ket{k,0}$. 
That is, we represent the state $\ket{k,\alpha}$ by the configuration of the corresponding product state with the hole fixed at site $0$ within the translation-symmetry sector labeled by momentum $k$. For example, when $N=5$, the states $\ket{k, \alpha}$ can be represented as  
\begin{eqnarray}
     \{0,-1,-1,1,1\}_k &\equiv \ket{k, 0}, \\ \nonumber
     \{0,-1,1,-1,1\}_k &\equiv \ket{k, 1}, \\ \nonumber
     \{0,-1,1,1,-1\}_k &\equiv \ket{k, 2}, \\ \nonumber
     \{0,1,-1,-1,1\}_k &\equiv \ket{k, 3}, \\ \nonumber 
    \{0,1,-1,1,-1\}_k &\equiv \ket{k, 4}, \\ \nonumber
    \{0,1,1,-1,-1\}_k &\equiv \ket{k, 5}.
\end{eqnarray}
After removing the site occupied by the hole, we obtain a four-site chain with PBCs. 
For the states $\{0,-1,-1,1,1\}_k$, $\{0,1,-1,-1,1\}_k$, $\{0,1,1,-1,-1\}_k$, and $\{0,-1,1,1,-1\}_k$, 
the corresponding spin configurations in the four-site chain are 
$\{-1,-1,1,1\}$, $\{1,-1,-1,1\}$, $\{1,1,-1,-1\}$, and $\{-1,1,1,-1\}$, respectively. 
Therefore, the distance between the two spin-down fermions in this four-site chain with PBCs is the same in all these four states, namely $d=1$. For the states $\{0,-1,1,-1,1\}_k$ and $\{0,1,-1,1,-1\}_k$, 
the corresponding spin configurations in the four-site chain are 
$\{-1,1,-1,1\}$ and $\{1,-1,1,-1\}$, respectively. 
In these two states, the distance between the two spin-down fermions is the same, namely $d=2$.
Since the action of the Hamiltonian on a basis state does not change the 
distance
between the two spin-down fermions on the $(N-1)$-site chain~\cite{PhysRev.147.392}, the states $\{0,-1,-1,1,1\}_k$, 
$\{0,1,-1,-1,1\}_k$, 
$\{0,1,1,-1,-1\}_k$, 
and $\{0,-1,1,1,-1\}_k$ form an orbit of length 4, while the states $\{0,-1,1,-1,1\}_k$ and 
$\{0,1,-1,1,-1\}_k$ form an orbit of length 2. 
The meaning of forming orbits is that, for the Hamiltonian matrix elements $\bra{k,\alpha}\hat{H}\ket{k,\alpha'}$, it is nonzero only when the two states $\ket{k,\alpha}$ and $\ket{k,\alpha'}$ belong to the same orbit. Otherwise, the matrix element vanishes if the two states are from different orbits.
Note that in the case of one spin-down fermion, there is a single orbit of length $N-1$.

In the above, we used the example with system size $N=5$ to introduce the concept of the distance $d$ between two spin-down fermions on the $(N-1)$-site chain obtained after removing the site occupied by the hole, as well as the orbits formed by the states $\ket{k,\alpha}$ that share the same distance $d$. We now turn to the general case.

For a given $N$, there are $\binom{N-1}{2}$ basis states with fixed $k$. 
On the $(N-1)$-site chain, the distance $d$ can take values 
$1, \cdots, \frac{N-1}{2}$ when $N$ is odd, and 
$1, \cdots, \frac{N-2}{2}$ when $N$ is even. 

When $N$ is odd, the number of states with $d=\frac{N-1}{2}$ is $\frac{N-1}{2}$. 
For other values $d=1, \cdots, \frac{N-1}{2}-1$, the number of states for each given $d$ is $N-1$. 
Therefore, the $\binom{N-1}{2}$ basis states $\ket{k,\alpha}$ with fixed $k$ split into one orbit of length $(N-1)/2$ and $(N-3)/2$ orbits of length $N-1$.
The total dimension is correctly reproduced,
\begin{equation}
\frac{N-1}{2}
+
\frac{N-3}{2}(N-1)
=
\binom{N-1}{2}.
\end{equation}

The orbit of length $(N-1)/2$ contains basis states of the form
\begin{eqnarray}
\label{eq:1H2DownSpecialOrbits}
\{
0,\;
-1,\;
\underbrace{1,\ldots,1}_{\frac{N-3}{2}},\;
-1,\;
\underbrace{1,\ldots,1}_{\frac{N-3}{2}}
\}_k,
\end{eqnarray}
together with the other $(N-3)/2$ states obtained by applying the translation operator within $(N-1)$-site chain to it.
The corresponding eigenenergies are
\begin{eqnarray}
E(m,m_\downarrow^{\prime})
=
2\cos(\frac{2\pi m}{N}+\frac{4\pi m_\downarrow^{\prime}}{N(N-1)}),
\end{eqnarray}
with $m = 0,\cdots, N-1$ and $m_\downarrow^{\prime}=0,\ldots,\frac{N-1}{2}-1$. 

The orbits of length $N-1$ contain basis states of the form
\begin{eqnarray}
\label{eq:1H1DownNOddOrbitsLengthN-1}
\{
0,\;
-1,\;
\underbrace{1,\ldots,1}_{d-1},\;
-1,\;
\underbrace{1,\ldots,1}_{N-d-2}
\}_k,
\end{eqnarray}
together with other $N-2$ states obtained by applying translation operator within $(N-1)$-site chain to it. Here $d=1, \cdots, \frac{N-1}{2}-1$.
The $N-1$ eigenenergies in such an orbit are
\begin{eqnarray}
E(m,m_\downarrow)
=
2\cos(\frac{2\pi m}{N}+ \frac{2\pi m_{\downarrow}}{N(N-1)}),
\end{eqnarray}
with $m_\downarrow=0,\ldots,N-2$.

When $N$ is even, the special orbit of length $(N-1)/2$ is absent,
and the states decompose into $(N-2)/2$ orbits of length $N-1$. Each orbit with $d$ contains basis states of the form
\begin{eqnarray}
\{
0,\;
-1,\;
\underbrace{1,\ldots,1}_{d-1},\;
-1,\;
\underbrace{1,\ldots,1}_{N-d-2}
\}_k,
\end{eqnarray}
and the other $N-2$ states obtained by applying the translation operator within $(N-1)$-site chain to it. Here, $d=1,\cdots,\frac{N-2}{2}$. In this case, all eigenenergies take the uniform form
\begin{eqnarray}
E(m,m_{\downarrow})
=
2\cos\!\left(
\frac{2\pi m}{N}
+
\frac{2\pi m_{\downarrow}}{N(N-1)}
\right),
\end{eqnarray}
with $m=0,\ldots,N-1$ and
$m_{\downarrow}=0,\ldots,N-2$,
and each level has degeneracy $(N-2)/2$.

Note that even for odd $N$, the contribution from the single orbit of length $(N-1)/2$ is subleading compared to the $(N-3)/2$ orbits of length $N-1$ since the ratio between their numbers is $\frac{1}{(N-3)/2}$ which vanishes in the thermodynamic limit. 
For the purpose of deriving exact expressions for the Drude weight and the specific heat, it is therefore convenient to restrict to even $N$, where the special orbit of length $(N-1)/2$ is absent. Consequently, for the eigenstates $\ket{m, m_{\downarrow}}$, we obtain
\begin{eqnarray}
\hat{H}\ket{m, m_{\downarrow}}
&=&
2\cos\!\left(
\frac{2\pi m}{N}
+
\frac{2\pi m_{\downarrow}}{N(N-1)}
\right)
\ket{m, m_{\downarrow}},
\\
\frac{
\left|
\bra{m, m_{\downarrow}}
\hat{J}
\ket{m, m_{\downarrow}}
\right|^2
}{
\langle m, m_{\downarrow} \vert m, m_{\downarrow} \rangle^2
}
&=&
\left[
2\sin\!\left(
\frac{2\pi m}{N}
+
\frac{2\pi m_{\downarrow}}{N(N-1)}
\right)
\right]^2,
\end{eqnarray}
with $m = 0, \ldots, N-1$ and $m_{\downarrow} = 0, \ldots, N-2$. Each energy level has degeneracy $(N-2)/2$. See Sec.~\ref{sec:Anyholesandanyspin-downfermions} for the explicit expressions for eigenstates $\ket{m, m_{\downarrow}}$ and see Eq.~\eqref{eq:ProofOfEigenstateJWithSpinDown} for the proof that $\ket{m, m_{\downarrow}}$ is also an eigenstate of $\hat{J}$.

\subsubsection{Any number of spin-down fermions}
\label{sec:Single_hole_and_any_spin-down_fermions}
The above analysis in Sec.~\ref{sec:Single_hole_and_two_spin-down_fermions} can be generalized 
to the sector with one hole and an arbitrary number of spin-down fermions. In this section, we mainly outline the key ideas about orbits in this generalization. 
For further details, including the explicit form of the eigenstates, see Sec.~\ref{sec:Anyholesandanyspin-downfermions}. We denote by $n_{\downarrow}$ the number of spin-down fermions. The total Hilbert-space dimension is $N \binom{N-1}{n_{\downarrow}}$.

We denote by $\ket{j,\alpha}$ (see Eq.~\eqref{eq:ExpressionForAlpha}) the Fock states where the hole is located at site $j$, with $j = 0, \cdots, N-1$, and the $n_{\downarrow}$ spin-down fermions occupy the $\alpha$-th configuration among the $\binom{N-1}{n_{\downarrow}}$ possible arrangements, where $\alpha = 0, \cdots, \binom{N-1}{n_{\downarrow}} - 1$. As discussed in Sec.~\ref{sec:Single_hole_and_two_spin-down_fermions}, we then construct the corresponding momentum eigenstates $\ket{k,\alpha}$
\begin{align}
    \label{eq:1HAnyDownDefinitionOfK}
    \ket{k, \alpha} = \sum_{j=0}^{N-1} e^{ikj} T_x^j \ket{0,\alpha}.
\end{align}
Within each momentum sector labeled by $k$, the Hilbert-space dimension is $\binom{N-1}{n_{\downarrow}}$.

We prove that when $N-1$ is a prime number, the $\binom{N-1}{n_{\downarrow}}$ momentum states $\ket{k,\alpha}$ 
with fixed $k$ form $\binom{N-1}{n_{\downarrow}}/(N-1)$ distinct orbits, each of length $N-1$. The momentum state $\ket{k,\alpha}$ is characterized by a configuration of $n_{\downarrow}$ spin-down fermions on the $(N-1)$-site chain obtained after removing the site occupied by the hole, i.e., $\alpha(j_1, \cdots, j_{n_{\downarrow}})$ (see Eq.~\eqref{eq:ExpressionForAlpha}).

For a given momentum state $\ket{k,\alpha(j_1,\cdots, j_{n_\downarrow})}$, under the action of the Hamiltonian there are two connected states, 
$\ket{k,\alpha^{(-1)}}$ 
and 
$\ket{k,\alpha^{(+1)}}$,
where
$\alpha^{(-1)}=\alpha(j_1-1, \cdots, j_{n_{\downarrow}}-1)$ and $\alpha^{(+1)}=\alpha(j_1+1, \cdots, j_{n_{\downarrow}}+1)$ 
correspond to the configurations obtained by shifting all fermions to the left or to the right by one site on the $(N-1)$-site chain, respectively. Similarly, $\alpha^{(\pm i)} = \alpha(j_1\pm i, \cdots, j_{n_{\downarrow}} \pm i)$.
Under the action of $\hat{H}$, the relative positions of the $n_{\downarrow}$ spin-down fermions remain unchanged. Consequently, the momentum states $\ket{k,\alpha}$ that share the same relative configuration of these $n_{\downarrow}$ spin-down fermions on the $(N-1)$-site chain form an orbit under $\hat{H}$. See the example with $N=5$ and $n_{\downarrow}=2$ in Sec.~\ref{sec:Single_hole_and_two_spin-down_fermions}.

For an orbit of length $L$, it contains the states
\begin{eqnarray}
    \ket{k,\alpha},\,
\ket{k,\alpha^{(\pm 1)}},\,
\ket{k,\alpha^{(\pm 2)}},\,
\cdots,\,
\ket{k,\alpha^{(\pm (L-1)})},
\end{eqnarray}
with the identification $\ket{k,\alpha^{(\pm L)}} = \ket{k,\alpha}$. It may appear that the number of states in the orbit exceeds $L$. However, since these states form a closed orbit, $\ket{k,\alpha^{(-i)}}$ and $\ket{k,\alpha^{(+ (L - i))}}$ represent the same state. Therefore, the total number of distinct states contained in the orbit is $L$. Clearly, the maximal possible orbit length is $N-1$, i.e., $L \leq N-1$.

Now suppose that $N-1$ is a prime number and assume that there exists an orbit of length $l$ with $0 < l < N-1$. Then $\ket{k,\alpha^{(\pm l)}} = \ket{k,\alpha}$, which further implies
\begin{align}
\ket{k,\alpha}
= \ket{k,\alpha^{(\pm l)}}
= \ket{k,\alpha^{(\pm 2l)}}
= \cdots.
\end{align}
On the other hand, we always have $\ket{k,\alpha} = \ket{k,\alpha^{(+(N-1))}}$. Therefore, $l$ must be a divisor of $N-1$. Since $N-1$ is prime, the only positive divisors are $1$ and $N-1$. Excluding the trivial case $l=1$, we conclude that $l = N-1$. 

Consequently, when $N-1$ is a prime number, there are $\binom{N-1}{n_{\downarrow}}/(N-1)$ distinct orbits in the given momentum sector, each of length $N-1$. Even when $N-1$ is not a prime number, although additional orbits may exist, their number is negligible. See Eq.~\eqref{eq:1H2DownSpecialOrbits} for an example.

We note that the above proof can be straightforwardly generalized to the case with $n_h$ holes by replacing $1$ with $n_h$. When $N-n_h$ is a prime number, the $\binom{N-n_h}{n_\downarrow}$ states with fixed $k$ form $\binom{N-n_h}{n_{\downarrow}}/(N-n_h)$ orbits, each of length $N-n_h$. This will be used in the derivation of explicit eigenstates for arbitrary numbers of holes and spin-down fermions in Sec.~\ref{sec:Anyholesandanyspin-downfermions}.

Here we only present the energy spectrum and the corresponding eigenvalues of the current operator. 
For further details, including the explicit form of the eigenstates and the relation between $m$, $m_{\downarrow}$, and the momentum $k$, see Sec.~\ref{sec:Anyholesandanyspin-downfermions}.
\begin{eqnarray}
E(m,m_{\downarrow})
=
2\cos\!\left(
\frac{2\pi m}{N}
+
\frac{2\pi m_{\downarrow}}{N(N-1)}
\right),
\end{eqnarray}
with $m=0,\ldots,N-1$ and $m_{\downarrow}=0,\ldots,N-2$. The degeneracy of each energy level is $\frac{\binom{N-1}{n_{\downarrow}}}{N-1}$.
The corresponding squared diagonal matrix element of the current operator is
\begin{equation}
\bigl|
\hat{J}_{mm_{\downarrow},\,mm_{\downarrow}}
\bigr|^2
=
\left[
2\sin\!\left(
\frac{2\pi m}{N}
+
\frac{2\pi m_{\downarrow}}{N(N-1)}
\right)
\right]^2.
\end{equation}

\subsection{Two-hole case}
\label{sec:Exact_solutions_with_two_holes}

\subsubsection{No spin-down fermion}
We first consider the simplest case of two holes and $N-2$ spin-up
fermions. The Hilbert space dimension of this sector is $\binom{N}{2}$.

We now prove that in this sector, the eigenenergy is simply the sum of the
single-hole energies corresponding to two distinct momentum modes.
The eigenenergies are
\begin{eqnarray}
E(m_1,m_2)
=
2\cos\!\left(\frac{2\pi m_1}{N}\right)
+
2\cos\!\left(\frac{2\pi m_2}{N}\right),
\end{eqnarray}
with $0\le m_1 < m_2 \le N-1$.
There are $\binom{N}{2}$ eigenstates, in agreement with the Hilbert
space dimension of this sector.

To verify this result explicitly, we construct the (unnormalized)
two-hole eigenstate
\begin{eqnarray}
\vert k_1,k_2\rangle
=
\sum_{j_1=0}^{N-1}
\sum_{j_2=0}^{N-1}
e^{ij_1k_1}
e^{ij_2k_2}
c_{j_1,\uparrow}
c_{j_2,\uparrow}
\ket{\mathrm{FM}},
\end{eqnarray}
where
\begin{equation}
\ket{\mathrm{FM}}
=
c^{\dagger}_{0,\uparrow}
\cdots
c^{\dagger}_{N-1,\uparrow}
\vert \mathrm{vac}\rangle
\end{equation}
is the fully polarized ferromagnetic state with $N$ spin-up fermions.
The annihilation operators
$c_{j_1,\uparrow}c_{j_2,\uparrow}$
create two holes at sites $j_1$ and $j_2$.
When $j_1=j_2$, the state vanishes identically due to fermionic
statistics.

Since we work in the subspace without double occupancy, the projection
operators in $\hat{H}$ can be omitted. For simplicity, we use $\sum_{j_1,j_2}$ as a shorthand for
$\sum_{j_1=0}^{N-1}\sum_{j_2=0}^{N-1}$. Acting with the Hamiltonian (with $t=1$) yields
\begin{eqnarray}
\hat{H}\vert k_1,k_2\rangle
&=& -\sum_{j_1,j_2} e^{ij_1k_1} e^{ij_2k_2} (c^{\dagger}_{j_{1},\uparrow}c_{j_1+1,\uparrow} + c^{\dagger}_{j_{1},\uparrow}c_{j_1-1,\uparrow} + c^{\dagger}_{j_{2},\uparrow}c_{j_2+1,\uparrow} + c^{\dagger}_{j_{2},\uparrow}c_{j_2-1,\uparrow}) c_{j_{1}, \uparrow} c_{j_{2}, \uparrow} 
\ket{\mathrm{FM}}
\nonumber  \\ 
&=& \sum_{j_1,j_2}
e^{ij_1k_1}
e^{ij_2k_2}
\left(
c_{j_1+1,\uparrow}c_{j_2,\uparrow}
+
c_{j_1-1,\uparrow}c_{j_2,\uparrow}
+
c_{j_1,\uparrow}c_{j_2+1,\uparrow}
+
c_{j_1,\uparrow}c_{j_2-1,\uparrow}
\right)
\ket{\mathrm{FM}}
\nonumber\\
&=&
\left(
e^{ik_1}
+
e^{-ik_1}
+
e^{ik_2}
+
e^{-ik_2}
\right)
\vert k_1,k_2\rangle
\nonumber\\
&=&
\left(
2\cos k_1
+
2\cos k_2
\right)
\vert k_1,k_2\rangle,
\end{eqnarray}
where we have used $e^{iNk_1}=e^{iNk_2}=1$.
Hence, the eigenenergy is indeed the sum of two distinct single-hole
mode energies. Moreover, $k_1$ and $k_2$ can be expressed as 
\begin{align}
    k_{1} &= \frac{2 \pi m_1}{N}, \label{eq:2H0DownK1} \\ 
    k_{2} &= \frac{2 \pi m_2}{N}, \label{eq:2H0DownK2}
\end{align}
with $0 \leq m_{1} < m_{2} \leq N-1$. Thus the eigenstate $\ket{k_1, k_2}$ can also be labeled by $\ket{m_1, m_2}$.

Since $[\hat{J},\hat{H}]=0$, the Hamiltonian and the current operator
share common eigenstates 
$\ket{m_1, m_2}$.
For the eigenstate $\vert m_1,m_2\rangle$, we obtain
\begin{eqnarray}
    \hat{H} \ket{m_1, m_2} = \left[ 2 \cos(\frac{2 \pi m_1}{N}) + 2\cos(\frac{2 \pi m_2}{N}) \right] \ket{m_1, m_2},
\end{eqnarray}
\begin{eqnarray}
\left|
\langle m_1,m_2\vert
\hat{J}
\vert m_1,m_2\rangle
\right|^2
=
\left[
2\sin\!\left(\frac{2\pi m_1}{N}\right)
+
2\sin\!\left(\frac{2\pi m_2}{N}\right)
\right]^2.
\end{eqnarray}
See Eq.~\eqref{eq:ProofOfEigenstateJWithoutSpinDown} for the proof that $\ket{m_1, m_2}$ is also an eigenstate of $\hat{J}$.

\subsubsection{One spin-down fermion}
\label{sec:Two_holes_and_one_spin-down_fermion}

In the sector with two holes and one spin-down fermion, the eigenenergy
is still given by the sum of the single-hole mode energies associated
with two distinct modes. However, in contrast to
Sec.~\ref{sec:Single_hole_and_one_spin-down_fermion}, after removing the
sites occupied by the two holes, the length of the chain becomes $N-2$,
and the effective twist angles are consequently modified, as detailed
below. The Hilbert space dimension of this sector is
$(N-2)\binom{N}{2}$.

In this sector, the eigenstates can be written as
\begin{eqnarray}
\vert k_1,k_2,k_{\alpha} \rangle
&=&
\sum_{j_1,j_2=0}^{N-1}
e^{i j_1 k_1}
e^{i j_2 k_2}
\sum_{\substack{j_3=0 \\ j_3\neq j_1,j_2}}^{N-1}
e^{i j_{\alpha}(j_1,j_2; j_3) k_{\alpha}}
c^{\dagger}_{j_3,\downarrow}
c_{j_3,\uparrow}
c_{j_1,\uparrow}
c_{j_2,\uparrow}
\ket{\mathrm{FM}},
\end{eqnarray}
where
$c^{\dagger}_{j_3,\downarrow}
c_{j_3,\uparrow}
c_{j_1,\uparrow}
c_{j_2,\uparrow}
\ket{\mathrm{FM}}
$
denotes a Fock state in which the two holes are located at sites
$j_1$ and $j_2$, and the spin-down fermion is located at site $j_3$.
Here $j_{\alpha}(j_1,j_2; j_3)$ represents the position of the
spin-down fermion at site $j_3$ within the $(N-2)$-site chain
obtained after removing the hole sites $j_1$ and $j_2$. For example, for $N=5$ with $j_{1}=0$ and $j_{2}=3$, we have
$j_{\alpha}(j_1,j_2; 1) = 0$, 
$j_{\alpha}(j_1,j_2; 2) = 1$, and 
$j_{\alpha}(j_1,j_2; 4) = 2$.
The analytical expression for $j_{\alpha}(j_1, j_2; j_3)$ is 
\begin{align}
    \label{eq:DefinitionOfJalpha}
    j_{\alpha}(j_1, j_2; j_3) = j_3 - \Theta(j_3-j_{1}) - \Theta(j_3-j_{2}),
\end{align}
where $\Theta(x)$ is the Heaviside step function
\begin{align}
    \Theta(x)=
\begin{cases}
0, & x<0,\\
1, & x>0.
\end{cases}
\end{align}

For state $\ket{k_1, k_2, k_{\alpha}}$, when $k_1$, $k_2$, $k_{\alpha}$ satisfy
\begin{equation}
e^{i(N-2)k_{\alpha}}
=
e^{iNk_1} e^{ik_{\alpha}}
=
e^{iNk_2} e^{ik_{\alpha}}
=
1,
\end{equation}
it is an eigenstate of
$\hat{H}$. To demonstrate this, we now consider the action of $\hat{H}$ on
$\vert k_1,k_2,k_{\alpha} \rangle$.
\begin{eqnarray}
\label{eq:TwoHoleOneSpinH}
\hat{H} \vert k_1,k_2,k_{\alpha} \rangle
&=&
-\sum_{j_1,j_2=0}^{N-1}
\sum_{\substack{j_3=0 \\ j_3\neq j_1,j_2}}^{N-1}
e^{i j_{\alpha}(j_1,j_2;j_3) k_{\alpha}}
e^{i j_1 k_1}
e^{i j_2 k_2}
\nonumber\\
&&\times
\Bigl(
c^{\dagger}_{j_1,\uparrow}c_{j_1+1,\uparrow}
+
c^{\dagger}_{j_1,\uparrow}c_{j_1-1,\uparrow}
+
c^{\dagger}_{j_2,\uparrow}c_{j_2+1,\uparrow}
+
c^{\dagger}_{j_2,\uparrow}c_{j_2-1,\uparrow}
\nonumber\\
&&\qquad
+
c^{\dagger}_{j_3,\uparrow}c_{j_3+1,\uparrow}
+
c^{\dagger}_{j_3,\uparrow}c_{j_3-1,\uparrow}
+
c^{\dagger}_{j_3+1,\downarrow}c_{j_3,\downarrow}
+
c^{\dagger}_{j_3-1,\downarrow}c_{j_3,\downarrow}
\Bigr)
c^{\dagger}_{j_3,\downarrow}
c_{j_3,\uparrow}
c_{j_1,\uparrow}
c_{j_2,\uparrow}
\ket{\mathrm{FM}}
\nonumber\\[6pt]
&=&
-\sum_{j_1,j_2=0}^{N-1}
\sum_{\substack{j_3=0 \\ j_3\neq j_1,j_2}}^{N-1}
e^{i j_{\alpha}(j_1,j_2;j_3) k_{\alpha}}
e^{i j_1 k_1}
e^{i j_2 k_2}
\nonumber\\
&&\times
c^{\dagger}_{j_3,\downarrow}
c_{j_3,\uparrow}
\Bigl(
- c_{j_1+1,\uparrow} c_{j_2,\uparrow}
- c_{j_1-1,\uparrow} c_{j_2,\uparrow}
- c_{j_1,\uparrow} c_{j_2+1,\uparrow}
- c_{j_1,\uparrow} c_{j_2-1,\uparrow}
\Bigr)
\ket{\mathrm{FM}}
\nonumber\\
&&
-\sum_{j_1,j_2=0}^{N-1}
\sum_{\substack{j_3=0 \\ j_3\neq j_1,j_2}}^{N-1}
e^{i j_{\alpha}(j_1,j_2;j_3) k_{\alpha}}
e^{i j_1 k_1}
e^{i j_2 k_2}
\nonumber\\
&&\times
\Bigl(
- c^{\dagger}_{j_3,\downarrow} c_{j_3+1,\uparrow}
- c^{\dagger}_{j_3,\downarrow} c_{j_3-1,\uparrow}
+ c^{\dagger}_{j_3+1,\downarrow} c_{j_3,\uparrow}
+ c^{\dagger}_{j_3-1,\downarrow} c_{j_3,\uparrow}
\Bigr)
c_{j_1,\uparrow}
c_{j_2,\uparrow}
\ket{\mathrm{FM}}
\nonumber\\[6pt]
&=&
\sum_{j_1,j_2=0}^{N-1}
\sum_{\substack{j_3=0 \\ j_3\neq j_1,j_2}}^{N-1}
e^{i j_{\alpha}(j_1,j_2;j_3) k_{\alpha}}
e^{i j_1 k_1}
e^{i j_2 k_2}
\nonumber\\
&&\times
c^{\dagger}_{j_3,\downarrow}
c_{j_3,\uparrow}
\Bigl(
c_{j_1+1,\uparrow} c_{j_2,\uparrow}
+ c_{j_1-1,\uparrow} c_{j_2,\uparrow}
+ c_{j_1,\uparrow} c_{j_2+1,\uparrow}
+ c_{j_1,\uparrow} c_{j_2-1,\uparrow}
\Bigr)
\ket{\mathrm{FM}}
\nonumber\\
&&
-\sum_{j_1,j_2=0}^{N-1}
\sum_{\substack{j_3=0 \\ j_3\neq j_1,j_2}}^{N-1}
e^{i j_{\alpha}(j_1,j_2;j_3) k_{\alpha}}
e^{i j_1 k_1}
e^{i j_2 k_2}
\nonumber\\
&&\times
\Bigl(
c^{\dagger}_{j_3+1,\downarrow} c_{j_3,\uparrow}
+
c^{\dagger}_{j_3-1,\downarrow} c_{j_3,\uparrow}
\Bigr)
c_{j_1,\uparrow}
c_{j_2,\uparrow}
\ket{\mathrm{FM}},
\end{eqnarray}
where the last equality follows because the operator $c^{\dagger}_{j_3,\downarrow}$ would otherwise create a doubly occupied site at $j_3$, which is forbidden in the projected Hilbert space.

To simplify the subsequent derivation, we decompose $\hat{H} \vert k_1,k_2,k_{\alpha} \rangle$ in Eq.~\eqref{eq:TwoHoleOneSpinH} into two parts $A$ and $B$, where $A$ and $B$ correspond to the first and second sets of sums in Eq.~\eqref{eq:TwoHoleOneSpinH}, respectively, such that $\hat{H} \vert k_1,k_2,k_{\alpha} \rangle = A + B$. We then analyze each part separately. The first part $A$ is given by 
\begin{align}
\text{Part A} &=    \sum_{j_1,j_2=0}^{N-1}
\sum_{\substack{j_3=0 \\ j_3\neq j_1,j_2}}^{N-1}
e^{i j_{\alpha}(j_1,j_2;j_3) k_{\alpha}}
e^{i j_1 k_1}
e^{i j_2 k_2}
c^{\dagger}_{j_3,\downarrow}
c_{j_3,\uparrow}
\Bigl(
c_{j_1+1,\uparrow} c_{j_2,\uparrow}
+ c_{j_1-1,\uparrow} c_{j_2,\uparrow}
+ c_{j_1,\uparrow} c_{j_2+1,\uparrow}
+ c_{j_1,\uparrow} c_{j_2-1,\uparrow}
\Bigr)
\ket{\mathrm{FM}}
\nonumber \\ 
&= A_{1,-} + A_{1,+} + A_{2,-} + A_{2,+},
\end{align}
where 
\begin{align}
\label{eq:1H2DA1-}
    A_{1,-} = e^{-i k_1} \sum_{j_1,j_2=0}^{N-1}
\sum_{\substack{j_3=0 \\ j_3\neq j_1, j_1+1, j_2}}^{N-1}
e^{i j_{\alpha}(j_1,j_2;j_3) k_{\alpha}}
e^{i (j_1+1) k_1}
e^{i j_2 k_2}
c^{\dagger}_{j_3,\downarrow}
c_{j_3,\uparrow}
\Bigl(
c_{j_1+1,\uparrow} c_{j_2,\uparrow}
\Bigr)
\ket{\mathrm{FM}},
\end{align}
\begin{align}
    \label{eq:1H2DA1+}
    A_{1,+} = e^{i k_1} \sum_{j_1,j_2=0}^{N-1}
\sum_{\substack{j_3=0 \\ j_3\neq j_1, j_1-1, j_2}}^{N-1}
e^{i j_{\alpha}(j_1,j_2;j_3) k_{\alpha}}
e^{i (j_1-1) k_1}
e^{i j_2 k_2}
c^{\dagger}_{j_3,\downarrow}
c_{j_3,\uparrow}
\Bigl(
c_{j_1-1,\uparrow} c_{j_2,\uparrow}
\Bigr)
\ket{\mathrm{FM}},
\end{align}
\begin{align}
\label{eq:1H2DA2-}
    A_{2,-} = e^{-i k_2} \sum_{j_1,j_2=0}^{N-1}
\sum_{\substack{j_3=0 \\ j_3\neq j_1,j_2, j_2+1}}^{N-1}
e^{i j_{\alpha}(j_1,j_2;j_3) k_{\alpha}}
e^{i j_1 k_1}
e^{i (j_2 +1 ) k_2}
c^{\dagger}_{j_3,\downarrow}
c_{j_3,\uparrow}
\Bigl(
c_{j_1,\uparrow} c_{j_2+1,\uparrow}
\Bigr)
\ket{\mathrm{FM}},
\end{align}
\begin{align}
\label{eq:1H2DA2+}
    A_{2,+} = e^{i k_2} \sum_{j_1,j_2=0}^{N-1}
\sum_{\substack{j_3=0 \\ j_3\neq j_1,j_2, j_2-1}}^{N-1}
e^{i j_{\alpha}(j_1,j_2;j_3) k_{\alpha}}
e^{i j_1 k_1}
e^{i (j_2-1) k_2}
c^{\dagger}_{j_3,\downarrow}
c_{j_3,\uparrow}
\Bigl(
c_{j_1,\uparrow} c_{j_2-1,\uparrow}
\Bigr)
\ket{\mathrm{FM}}.
\end{align}

The second part $B$ is given by 
\begin{align}
   \text{Part B} &=  -\sum_{j_1,j_2=0}^{N-1}
\sum_{\substack{j_3=0 \\ j_3\neq j_1,j_2}}^{N-1}
e^{i j_{\alpha}(j_1,j_2;j_3) k_{\alpha}}
e^{i j_1 k_1}
e^{i j_2 k_2}
\Bigl(
c^{\dagger}_{j_3+1,\downarrow} c_{j_3,\uparrow}
+
c^{\dagger}_{j_3-1,\downarrow} c_{j_3,\uparrow}
\Bigr)
c_{j_1,\uparrow}
c_{j_2,\uparrow}
\ket{\mathrm{FM}}
\nonumber \\ 
&=
-\sum_{j_1,j_2=0}^{N-1}
\sum_{\substack{j_3=0 \\ j_3\neq j_1,j_2}}^{N-1}
e^{i j_{\alpha}(j_1,j_2;j_3) k_{\alpha}}
e^{i j_1 k_1}
e^{i j_2 k_2}
\Bigl(
c^{\dagger}_{j_1,\downarrow} c_{j_1-1,\uparrow}\,\delta_{j_1,j_3+1}
+
c^{\dagger}_{j_2,\downarrow} c_{j_2-1,\uparrow}\,\delta_{j_2,j_3+1}
\nonumber\\
&\qquad\qquad
+
c^{\dagger}_{j_1,\downarrow} c_{j_1+1,\uparrow}\,\delta_{j_1,j_3-1}
+
c^{\dagger}_{j_2,\downarrow} c_{j_2+1,\uparrow}\,\delta_{j_2,j_3-1}
\Bigr)
c_{j_1,\uparrow} c_{j_2,\uparrow}
\ket{\mathrm{FM}}
\nonumber\\
&=
\sum_{j_1,j_2=0}^{N-1}
e^{i j_{\alpha}(j_1,j_2;j_1+1) k_{\alpha}}
e^{i j_1 k_1}
e^{i j_2 k_2}
\;
c^{\dagger}_{j_1,\downarrow} c_{j_1,\uparrow}
c_{j_1+1,\uparrow} c_{j_2,\uparrow}
\ket{\mathrm{FM}}
\nonumber\\
& + \sum_{j_1,j_2=0}^{N-1}
e^{i j_{\alpha}(j_1,j_2;j_1-1) k_{\alpha}}
e^{i j_1 k_1}
e^{i j_2 k_2}
\;
c^{\dagger}_{j_1,\downarrow} c_{j_1,\uparrow}
c_{j_1-1,\uparrow} c_{j_2,\uparrow}
\ket{\mathrm{FM}}
\nonumber\\
&
+
\sum_{j_1,j_2=0}^{N-1}
e^{i j_{\alpha}(j_1,j_2;j_2+1) k_{\alpha}}
e^{i j_1 k_1}
e^{i j_2 k_2}
\;
c^{\dagger}_{j_2,\downarrow} c_{j_2,\uparrow}
c_{j_1,\uparrow} c_{j_2+1,\uparrow}
\ket{\mathrm{FM}}
\nonumber \\ 
&
+
\sum_{j_1,j_2=0}^{N-1}
e^{i j_{\alpha}(j_1,j_2;j_2-1) k_{\alpha}}
e^{i j_1 k_1}
e^{i j_2 k_2}
\;
c^{\dagger}_{j_2,\downarrow} c_{j_2,\uparrow}
c_{j_1,\uparrow} c_{j_2-1,\uparrow}
\ket{\mathrm{FM}}
\nonumber \\ 
& = B_{1,-} + B_{1,+} + B_{2,-} + B_{2,+},
\label{eq:part2}
\end{align}
where 
\begin{align}
\label{eq:1H2DB1-}
     B_{1,-} = e^{-ik_1} \sum_{j_1,j_2=0}^{N-1}
e^{i j_{\alpha}(j_1,j_2;j_1+1) k_{\alpha}}
e^{i (j_1+1) k_1}
e^{i j_2 k_2}
\;
c^{\dagger}_{j_1,\downarrow} c_{j_1,\uparrow}
c_{j_1+1,\uparrow} c_{j_2,\uparrow}
\ket{\mathrm{FM}},
\end{align}
\begin{align}
\label{eq:1H2DB1+}
    B_{1,+} = e^{ik_1} \sum_{j_1,j_2=0}^{N-1}
e^{i j_{\alpha}(j_1,j_2;j_1-1) k_{\alpha}}
e^{i (j_1-1) k_1}
e^{i j_2 k_2}
\;
c^{\dagger}_{j_1,\downarrow} c_{j_1,\uparrow}
c_{j_1-1,\uparrow} c_{j_2,\uparrow}
\ket{\mathrm{FM}},
\end{align}
\begin{align}
\label{eq:1H2DB2-}
   B_{2,-} = e^{-ik_2} \sum_{j_1,j_2=0}^{N-1}
e^{i j_{\alpha}(j_1,j_2;j_2+1) k_{\alpha}}
e^{i j_1 k_1}
e^{i (j_2+1) k_2}
\;
c^{\dagger}_{j_2,\downarrow} c_{j_2,\uparrow}
c_{j_1,\uparrow} c_{j_2+1,\uparrow}
\ket{\mathrm{FM}},
\end{align}
\begin{align}
\label{eq:1H2DB2+}
    B_{2,+} = e^{ik_2} \sum_{j_1,j_2=0}^{N-1}
e^{i j_{\alpha}(j_1,j_2;j_2-1) k_{\alpha}}
e^{i j_1 k_1}
e^{i (j_2-1) k_2}
\;
c^{\dagger}_{j_2,\downarrow} c_{j_2,\uparrow}
c_{j_1,\uparrow} c_{j_2-1,\uparrow}
\ket{\mathrm{FM}}.
\end{align}

Next, we first combine $A_{1,-}$ in Eq.~\eqref{eq:1H2DA1-} and $B_{1,-}$ in Eq.~\eqref{eq:1H2DB1-},
\begin{eqnarray} 
\label{eq:combination1H2DA1-B1-}
A_{1,-} + B_{1,-}&=& e^{-ik_1} \sum_{j_1,j_2=0}^{N-1} \sum_{\substack{j_3=0 \\ j_3\neq j_1,j_2,j_1+1}}^{N-1} e^{i j_{\alpha}(j_1,j_2;j_3) k_{\alpha}} e^{i (j_1+1) k_1} e^{i j_2 k_2} \; c^{\dagger}_{j_3,\downarrow} c_{j_3,\uparrow} c_{j_1+1,\uparrow} c_{j_2,\uparrow} 
\ket{\mathrm{FM}}
\nonumber\\ &+& \sum_{j_1,j_2=0}^{N-1} e^{i j_{\alpha}(j_1,j_2;j_1+1) k_{\alpha}} e^{i j_1 k_1} e^{i j_2 k_2} \; c^{\dagger}_{j_1,\downarrow} c_{j_1,\uparrow} c_{j_1+1,\uparrow} c_{j_2,\uparrow} 
\ket{\mathrm{FM}}
\nonumber\\ &=& e^{-ik_1} \Biggl[ \sum_{j_1,j_2=0}^{N-1} \sum_{\substack{j_3=0 \\ j_3\neq j_1,j_2,j_1+1}}^{N-1} e^{i j_{\alpha}(j_1,j_2;j_3) k_{\alpha}} e^{i (j_1+1) k_1} e^{i j_2 k_2} \; c^{\dagger}_{j_3,\downarrow} c_{j_3,\uparrow} c_{j_1+1,\uparrow} c_{j_2,\uparrow} \nonumber\\ &&\qquad\qquad + \sum_{j_1,j_2=0}^{N-1} e^{i j_{\alpha}(j_1,j_2;j_1+1) k_{\alpha}} e^{i (j_1+1) k_1} e^{i j_2 k_2} \; c^{\dagger}_{j_1,\downarrow} c_{j_1,\uparrow} c_{j_1+1,\uparrow} c_{j_2,\uparrow} \Biggr] 
\ket{\mathrm{FM}}.
\end{eqnarray}

If $\vert k_1,k_2,k_{\alpha}\rangle$ is an eigenstate with eigenenergy
$e^{-ik_1}+e^{ik_1}+e^{-ik_2}+e^{ik_2}$, i.e.,
\begin{align}
    \hat{H }\vert k_1,k_2,k_{\alpha}\rangle = (e^{-ik_1} + e^{ik_1} + e^{-ik_2} + e^{ik_2} ) \ket{k_1,k_2, k_{\alpha}},
\end{align}
we have $A_{1,-} + B_{1,-} = e^{-ik_1} \ket{k_1, k_2, k_{\alpha}}$. From Eq.~\eqref{eq:combination1H2DA1-B1-}, we have the following equality
\begin{eqnarray}
\label{eq:Twoholes_samecoefficient}
&&
\Biggl[
\sum_{j_1,j_2=0}^{N-1}
\sum_{\substack{j_3=0 \\ j_3\neq j_1,j_2,j_1+1}}^{N-1}
e^{i j_{\alpha}(j_1,j_2;j_3) k_{\alpha}}
e^{i (j_1+1) k_1}
e^{i j_2 k_2}
\;
c^{\dagger}_{j_3,\downarrow} c_{j_3,\uparrow}
c_{j_1+1,\uparrow} c_{j_2,\uparrow}
\nonumber\\
&&\qquad\qquad
+
\sum_{j_1,j_2=0}^{N-1}
e^{i j_{\alpha}(j_1,j_2;j_1+1) k_{\alpha}}
e^{i (j_1+1) k_1}
e^{i j_2 k_2}
\;
c^{\dagger}_{j_1,\downarrow} c_{j_1,\uparrow}
c_{j_1+1,\uparrow} c_{j_2,\uparrow}
\Biggr]
\ket{\mathrm{FM}}
\nonumber\\
&=&
\sum_{j_1,j_2=0}^{N-1}
\sum_{\substack{j_3=0 \\ j_3\neq j_1,j_2}}^{N-1}
e^{i j_{\alpha}(j_1,j_2;j_3) k_{\alpha}}
e^{i j_1 k_1}
e^{i j_2 k_2}
\;
c^{\dagger}_{j_3,\downarrow} c_{j_3,\uparrow}
c_{j_1,\uparrow} c_{j_2,\uparrow}
\ket{\mathrm{FM}}.
\end{eqnarray}
The coefficients of each basis state
$c^{\dagger}_{j_3,\downarrow} c_{j_3,\uparrow}
c_{j_1+1,\uparrow} c_{j_2,\uparrow}
\ket{\mathrm{FM}}
$
must match on both sides of Eq.~\eqref{eq:Twoholes_samecoefficient}.

Consider the basis state
$c^{\dagger}_{j_3,\downarrow} c_{j_3,\uparrow}
c_{j_1+1,\uparrow} c_{j_2,\uparrow}
\ket{\mathrm{FM}}
$
with $j_1+1 \neq 0 \ (\mod \  N)$. The coefficient on the left-hand side of
Eq.~\eqref{eq:Twoholes_samecoefficient} depends on whether
$j_3=j_1$ or $j_3\neq j_1$. 

\medskip
\noindent
\textit{Case 1: $j_3 \neq j_1$.}
In this case, only the first term on the left-hand side contributes, and the coefficient is
\begin{eqnarray}
&&
e^{i j_{\alpha}(j_1,j_2;j_3) k_{\alpha}}
e^{i (j_1+1) k_1}
e^{i j_2 k_2}
\nonumber\\
&=&
e^{i j_{\alpha}(j_1+1,j_2;j_3) k_{\alpha}}
e^{i (j_1+1) k_1}
e^{i j_2 k_2}.
\end{eqnarray}
Since we also have $j_3\neq j_1+1$, the ordering of site $j_3$ within the
$(N-2)$-site chain is unchanged, i.e.,
$j_{\alpha}(j_1,j_2;j_3) = j_{\alpha}(j_1+1,j_2;j_3)$,
and the coefficients on both sides are identical.

\medskip
\noindent
\textit{Case 2: $j_3 = j_1$.}
The coefficient on the left-hand side becomes
\begin{eqnarray}
&&
e^{i j_{\alpha}(j_1,j_2;j_1+1) k_{\alpha}}
e^{i (j_1+1) k_1}
e^{i j_2 k_2}
\nonumber\\
&=&
e^{i j_{\alpha}(j_1+1,j_2;j_1) k_{\alpha}}
e^{i (j_1+1) k_1}
e^{i j_2 k_2},
\end{eqnarray}
which again matches the coefficient on the right-hand side of
Eq.~\eqref{eq:Twoholes_samecoefficient}.

\medskip
\noindent
Next consider the boundary case $j_1+1=N$, i.e.,
$j_1+1 =0 \pmod N$.

\medskip
\noindent
\textit{Case 3: $j_3 \neq N-1$.}
The coefficients on the left- and right-hand sides are
\begin{eqnarray}
\text{LHS:}\quad
e^{i j_{\alpha}(N-1,j_2;j_3) k_{\alpha}}
e^{i N k_1}
e^{i j_2 k_2},
\qquad
\text{RHS:}\quad
e^{i j_{\alpha}(0,j_2;j_3) k_{\alpha}}
e^{i j_2 k_2}.
\end{eqnarray}
Using
$j_{\alpha}(N-1,j_2;j_3)
=
j_{\alpha}(0,j_2;j_3)+1$,
we obtain
\begin{eqnarray}
\label{eq:minusk1_1}
e^{i k_{\alpha}} e^{i N k_1} = 1 .
\end{eqnarray}

\medskip
\noindent
\textit{Case 4: $j_3 = N-1$.}
The coefficients are
\begin{eqnarray}
\text{LHS:}\quad
e^{i j_{\alpha}(N-1,j_2;0) k_{\alpha}}
e^{i N k_1}
e^{i j_2 k_2},
\qquad
\text{RHS:}\quad
e^{i j_{\alpha}(0,j_2;N-1) k_{\alpha}}
e^{i j_2 k_2}.
\end{eqnarray}
Since
$j_{\alpha}(N-1,j_2;0)=0$
and
$j_{\alpha}(0,j_2;N-1)=N-3$,
we obtain
\begin{eqnarray}
\label{eq:minusk1_2}
e^{i N k_1} = e^{i (N-3) k_{\alpha}}.
\end{eqnarray}
Combining Eqs.~\eqref{eq:minusk1_1} and \eqref{eq:minusk1_2}, we have 
\begin{align}
        \label{eq:1H2DCondition1}
        e^{i(N-2)k_{\alpha}} = 1, \\ 
        e^{iNk_{1}} e^{ik_{\alpha}} = 1. \label{eq:1H2DCondition2}
\end{align}
Therefore, when the conditions in Eqs.~\eqref{eq:1H2DCondition1} and \eqref{eq:1H2DCondition2} are satisfied, we obtain $A_{1,-} + B_{1,-} = e^{-ik_1} \ket{k_1,k_2,k_{\alpha}}$ and $A_{1,+} + B_{1,+} = e^{ik_1} \ket{k_1,k_2,k_{\alpha}}$. Moreover, imposing $e^{iNk_2}e^{ik_{\alpha}}=1$ gives $A_{2,-} + B_{2,-} = e^{-ik_2} \ket{k_1,k_2,k_{\alpha}}$ and $A_{2,+} + B_{2,+} = e^{ik_2} \ket{k_1,k_2,k_{\alpha}}$.

In conclusion, the state $\ket{k_1,k_2,k_{\alpha}}$ is an eigenstate of $\hat{H}$
with eigenenergy $2\cos(k_1)+2\cos(k_2)$ when
\begin{align}
e^{i(N-2)k_{\alpha}} &= 1, \\
e^{iNk_{1}} e^{ik_{\alpha}} &= 1, \\
e^{iNk_{2}} e^{ik_{\alpha}} &= 1 .
\end{align}
It follows that
\begin{align}
    k_{\alpha} = -\frac{2\pi m_{\downarrow}}{N-2},
\qquad
m_{\downarrow} = 0, \cdots, N-3, \label{eq:2H1DownKalpha}
\end{align}
and
\begin{align}
k_{1}
&=
\frac{2\pi m_1}{N} - \frac{k_{\alpha}}{N}
=
\frac{2\pi m_1}{N}
+
\frac{2\pi m_{\downarrow}}{N(N-2)},
\label{eq:2H1DownK1} \\
k_{2}
&=
\frac{2\pi m_2}{N} - \frac{k_{\alpha}}{N}
=
\frac{2\pi m_2}{N}
+
\frac{2\pi m_{\downarrow}}{N(N-2)}, \label{eq:2H1DownK2}
\end{align}
with $0 \le m_1 < m_2 \le N-1$.

In conclusion, the corresponding eigenenergies are
\begin{eqnarray}
E(m_1,m_2,m_\downarrow)
=
2\cos\!\left(\frac{2\pi m_1}{N} + \frac{2\pi m_\downarrow}{N(N-2)}\right)
+
2\cos\!\left(\frac{2\pi m_2}{N} + \frac{2\pi m_\downarrow}{N(N-2)}\right),
\end{eqnarray}
and 
\begin{eqnarray}
\left|
\langle m_1, m_2,m_\downarrow \vert \hat{J} \vert m_1,m_2,m_\downarrow \rangle
\right|^2
=
\left[
2\sin\!\left(\frac{2\pi m_1}{N} + \frac{2\pi m_\downarrow}{N(N-2)}\right)
+
2\sin\!\left(\frac{2\pi m_2}{N} + \frac{2\pi m_\downarrow}{N(N-2)}\right)
\right]^2,
\end{eqnarray}
with $0 \le m_1 < m_2 \le N-1$ and $m_{\downarrow}=0, \cdots, N-3$. See Eq.~\eqref{eq:ProofOfEigenstateJWithSpinDown} for the proof that $\ket{m_1, m_2, m_{\downarrow}}$ is also an eigenstate of $\hat{J}$.

\subsubsection{Two spin-down fermions}
In the sector with two holes and two spin-down fermions, the Hilbert space dimension is $\binom{N}{2}\binom{N-2}{2}$.

In the $(N-2)$-site chain after removing the sites occupied by two holes, there are $\binom{N-2}{2}$ spin configurations. For example, when $N=7$, the $10$ spin configurations in the $5$-site chain are shown below:
\begin{align}
   s_{1} = \{ -1, -1, +1, +1, +1 \} \\ \nonumber
   s_{2} = \{ +1, -1, -1, +1, +1 \} \\ \nonumber
   s_{3} = \{ +1, +1, -1, -1, +1 \} \\ \nonumber
   s_{4} = \{ +1, +1, +1, -1, -1 \} \\ \nonumber 
   s_{5} = \{ -1, +1, +1, +1, -1 \} \\ \nonumber
   s_{6} = \{ -1, +1, -1, +1, +1 \} \\ \nonumber
   s_{7} = \{ +1, -1, +1, -1, +1 \} \\ \nonumber
   s_{8} = \{ +1, +1, -1, +1, -1 \} \\ \nonumber
   s_{9} = \{ -1, +1, +1, -1, +1 \} \\ \nonumber
   s_{10} = \{ +1, -1, +1, +1, -1 \} \\ \nonumber
\end{align}
These $10$ spin configurations form two orbits of length $5$ under the action of translation operator $\hat{T}$ in the $(N-2)$-site chain
\begin{align}
    \label{eq:2H2DOrbitsExample}
    s_{1} \xrightarrow{\hat{T}} s_{2}  \xrightarrow{\hat{T}} s_{3}  \xrightarrow{\hat{T}} s_{4}  \xrightarrow{\hat{T}} s_{5}  \xrightarrow{\hat{T}} s_{1} \nonumber \\ 
    s_{6} \xrightarrow{\hat{T}} s_{7}  \xrightarrow{\hat{T}} s_{8}  \xrightarrow{\hat{T}} s_{9}  \xrightarrow{\hat{T}} s_{10}  \xrightarrow{\hat{T}} s_{6}
\end{align}
We note that for spin configurations within a given orbit, the distance $d$ under PBCs between the two spin-down fermions is the same. For the first orbit in Eq.~\eqref{eq:2H2DOrbitsExample}, the distance between the two spin-down fermions is $d=1$ (under PBCs), while for the second orbit in Eq.~\eqref{eq:2H2DOrbitsExample}, the distance between the two spin-down fermions is $d=2$.

When $N-2$ is a prime number, the $\binom{N-2}{2}$ allowed spin configurations in the $(N-2)$-site chain form $\frac{\binom{N-2}{2}}{N-2} = \frac{N-3}{2}$ orbits, each of length $N-2$. The proof follows directly from that in Sec.~\ref{sec:Single_hole_and_any_spin-down_fermions} upon replacing $N-1$ by $N-2$ throughout. For spin configurations within the same orbit, the distance between the two spin-down fermions is fixed. This distance can take values $1, \dots, \frac{N-3}{2}$, consistent with the total number of orbits.

Moreover, for any given Fock state, under the action of the Hamiltonian, the distance $d$ between the two spin-down fermions in the $(N-2)$-site chain remains unchanged. 
Therefore, for each orbit in which the distance between the two spin-down fermions is $d$, we can define the following states, which can be explicitly shown to be eigenstates.
\begin{align}
    \ket{k_1, k_2, k_{\alpha}, d} = \sum_{j_1,j_2 =0}^{N-1} \sum_{j'=0}^{N-1} e^{i j_{\alpha}(j_{1}, j_{2}; j') k_{\alpha}} e^{i j_1 k_1} e^{i j_2 k_2} c^{\dagger}_{j', \downarrow} c_{j',\uparrow} c^{\dagger}_{r(j_1, j_2; j',d), \downarrow} c_{r(j_1, j_2; j',d), \uparrow} c_{j_1, \uparrow} c_{j_2, \uparrow} \ket{\text{FM}}.
\end{align}
Here $c^{\dagger}_{j', \downarrow} c_{j',\uparrow}$ creates a spin-down fermion at site $j'$ and $r(j_1,j_2; j',d)$ denotes the position of the other spin-down fermion in the $N$-site chain, such that in the $(N-2)$-site chain obtained after removing the sites $j_1$ and $j_2$ occupied by the two holes, this spin-down fermion is located $d$ sites to the right of the first spin-down fermion (at site $j'$ in the $N$-site chain).
Similarly, we can also use $l(j_1, j_2; j'', d)$ to denote the position of a spin-down fermion in the $N$-site chain, such that in the $(N-2)$-site chain, this spin-down fermion is located at $d$ sites to the left of another spin-down fermion located at site $j''$ in the original $N$-site chain.

$j_{\alpha}(j_1, j_2; j')$ defined in Eq.~\eqref{eq:DefinitionOfJalpha} denotes the position of the spin-down fermion located at site $j'$ in the $N$-site chain, expressed in the $(N-2)$-site chain obtained after removing the sites $j_1$ and $j_2$. For example, with $N=7$, $j_{1}=1$, $j_{2}=3$, $d=1$, 
\begin{align}
    \ket{\downarrow,0,\downarrow,0, \uparrow, \uparrow, \uparrow}: j_{\alpha} = 0, \\ \nonumber
    \ket{\uparrow,0,\downarrow,0, \downarrow, \uparrow, \uparrow}: j_{\alpha} = 1, \\ \nonumber
    \ket{\uparrow,0,\uparrow,0, \downarrow, \downarrow, \uparrow}: j_{\alpha} = 2, \\ \nonumber
    \ket{\uparrow,0,\uparrow,0, \uparrow, \downarrow, \downarrow}: j_{\alpha} = 3, \\ \nonumber
    \ket{\downarrow,0,\uparrow,0, \uparrow, \uparrow, \downarrow}: j_{\alpha} = 4. \nonumber
\end{align}

We use $j_{\alpha}^{-1}(j_1,j_2; j)$ to denote the position of the spin-down fermion in the original $N$-site chain when its position in the reduced $(N-2)$-site chain is $j$. It is defined as 
\begin{align}
    j_{\alpha}^{-1}(j_1,j_2; j) = j + \Theta(j - \mathrm{min}(j_{1}, j_{2}) + \frac{1}{2}) + \Theta(j- \mathrm{max}(j_{1}, j_{2}) + \frac{3}{2}).
\end{align}
Therefore,
\begin{align}
    r(j_1,j_2;j',d) = j_{\alpha}^{-1} (j_1,j_2; j_{\alpha}(j_1,j_2; j') + d \pmod{N-2}),
\end{align}
and 
\begin{align}
    l(j_1,j_2; j',d) = j_{\alpha}^{-1} (j_1,j_2; j_{\alpha}(j_1,j_2; j') - d \pmod{N-2}).
\end{align}

After introducing these notations, we explicitly prove that $\ket{k_1, k_2, k_{\alpha}, d}$ is an eigenstate when $e^{iNk_1} e^{ik_{\alpha}} = e^{iNk_2} e^{ik_{\alpha}} = e^{i(N-2) k_{\alpha}}=1$.

Under the action of Hamiltonian $\hat{H}$, 
\begin{align}
     \label{eq:2H2DActionH}
    \hat{H} \ket{k_1, k_2, k_{\alpha}, d} &= - \sum_{j_1,j_2=0}^{N-1} e^{ij_1 k_1} e^{ij_2 k_2} \sum_{j'=0}^{N-1} e^{i j_{\alpha}(j_1, j_2; j') k_{\alpha}} \nonumber \\ 
    &\quad \quad \left( \left( c^{\dagger}_{j_1, \uparrow} c_{j_1+1,\uparrow} + c^{\dagger}_{j_1, \uparrow} c_{j_1-1,\uparrow} + c^{\dagger}_{j_2, \uparrow} c_{j_2+1,\uparrow} + c^{\dagger}_{j_2, \uparrow} c_{j_2-1,\uparrow}
    \right) \right. \nonumber \\ 
    & \quad \quad \left.  \left( c^{\dagger}_{j'+1, \downarrow} c_{j',\downarrow} +  c^{\dagger}_{j'-1, \downarrow} c_{j',\downarrow}+ c^{\dagger}_{r(j_1,j_2;j',d)+1, \downarrow} c_{r(j_1,j_2;j',d), \downarrow} + c^{\dagger}_{r(j_1,j_2;j',d)-1, \downarrow} c_{r(j_1,j_2;j',d), \downarrow} \right)  \right) \nonumber \\ 
    &\quad \quad \quad  c^{\dagger}_{j', \downarrow} c_{j',\uparrow} c^{\dagger}_{r(j_1, j_2; j',d), \downarrow} c_{r(j_1, j_2; j',d), \uparrow} c_{j_1, \uparrow} c_{j_2, \uparrow} \ket{\text{FM}} \nonumber \\ 
    & =  \sum_{j_1,j_2=0}^{N-1} e^{ij_1 k_1} e^{ij_2 k_2} \sum_{j'=0}^{N-1} e^{i j_{\alpha}(j_1, j_2; j') k_{\alpha}} \nonumber \\ 
    &\quad \quad c^{\dagger}_{j', \downarrow} c_{j',\uparrow} c^{\dagger}_{r(j_1, j_2; j',d), \downarrow} c_{r(j_1, j_2; j',d), \uparrow} \left( c_{j_1+1,\uparrow} c_{j_2, \uparrow} +  c_{j_1-1,\uparrow} c_{j_2, \uparrow} + c_{j_1, \uparrow}  c_{j_2+1,\uparrow} + c_{j_1, \uparrow}  c_{j_2-1,\uparrow}
    \right)  \ket{\text{FM}} \nonumber \\ 
    & - \sum_{j_1,j_2=0}^{N-1} e^{ij_1 k_1} e^{ij_2 k_2} \sum_{j'=0}^{N-1} e^{i j_{\alpha}(j_1, j_2; j') k_{\alpha}} \left(  \left( c^{\dagger}_{j'+1, \downarrow} c_{j',\uparrow} +  c^{\dagger}_{j'-1, \downarrow} c_{j',\uparrow} \right)c^{\dagger}_{r(j_1, j_2; j',d), \downarrow} c_{r(j_1, j_2; j',d), \uparrow} \right. \nonumber \\ 
    & \quad \quad \left. +  c^{\dagger}_{j', \downarrow} c_{j',\uparrow} \left( c^{\dagger}_{r(j_1,j_2;j',d)+1, \downarrow} c_{r(j_1,j_2;j',d), \uparrow} + c^{\dagger}_{r(j_1,j_2;j',d)-1, \downarrow} c_{r(j_1,j_2;j',d), \uparrow} \right) \right) c_{j_1, \uparrow} c_{j_2, \uparrow} \ket{\text{FM}}.  
\end{align}
Next, we denote by Parts $A$ and $B$ the first and second sets of sums in Eq.~\eqref{eq:2H2DActionH},  such that $ \hat{H} \ket{k_1, k_2, k_{\alpha}, d} = A+B$. The first part $A$ is given by
\begin{align}
    \text{Part A} = A_{1,-} + A_{1, +} + A_{2,-} + A_{2,+},
\end{align}
where
\begin{align}
    \label{eq:2H2DA1-}
    A_{1,-} &= e^{-ik_1} \sum_{j_1,j_2=0}^{N-1} e^{i(j_1+1) k_1} e^{ij_2 k_2} \sum_{j'=0}^{N-1} e^{i j_{\alpha}(j_1, j_2; j') k_{\alpha}} c^{\dagger}_{j', \downarrow} c_{j',\uparrow} c^{\dagger}_{r(j_1, j_2; j',d), \downarrow} c_{r(j_1, j_2; j',d), \uparrow} \left( c_{j_1+1,\uparrow} c_{j_2, \uparrow}
    \right)  \ket{\text{FM}},
\end{align}
\begin{align}
    \label{eq:2H2DA1+}
    A_{1,+} &= e^{ik_1} \sum_{j_1,j_2=0}^{N-1} e^{i(j_1-1) k_1} e^{ij_2 k_2} \sum_{j'=0}^{N-1} e^{i j_{\alpha}(j_1, j_2; j') k_{\alpha}} c^{\dagger}_{j', \downarrow} c_{j',\uparrow} c^{\dagger}_{r(j_1, j_2; j',d), \downarrow} c_{r(j_1, j_2; j',d), \uparrow} \left( c_{j_1-1,\uparrow} c_{j_2, \uparrow}
    \right)  \ket{\text{FM}}, 
\end{align}
\begin{align}
   \label{eq:2H2DA2-}
A_{2,-} &= e^{-ik_2} \sum_{j_1,j_2=0}^{N-1} e^{i j_1 k_1} e^{i (j_2+1) k_2} \sum_{j'=0}^{N-1} e^{i j_{\alpha}(j_1, j_2; j') k_{\alpha}} c^{\dagger}_{j', \downarrow} c_{j',\uparrow} c^{\dagger}_{r(j_1, j_2; j',d), \downarrow} c_{r(j_1, j_2; j',d), \uparrow} \left( c_{j_1,\uparrow} c_{j_2+1, \uparrow}
    \right)  \ket{\text{FM}},
\end{align}
\begin{align}
   \label{eq:2H2DA2+}
    A_{2,+} &= e^{ik_2} \sum_{j_1,j_2=0}^{N-1} e^{i j_1 k_1} e^{i(j_2-1) k_2} \sum_{j'=0}^{N-1} e^{i j_{\alpha}(j_1, j_2; j') k_{\alpha}} c^{\dagger}_{j', \downarrow} c_{j',\uparrow} c^{\dagger}_{r(j_1, j_2; j',d), \downarrow} c_{r(j_1, j_2; j',d), \uparrow} \left( c_{j_1,\uparrow} c_{j_2-1, \uparrow}
    \right)  \ket{\text{FM}}. 
\end{align}
Note that in $A_{1,-}$ and $A_{1,+}$, we require $j' \neq j_1$ and $r(j_1,j_2; j',d) \neq j_1$, while in $A_{2,-}$ and $A_{2,+}$, we require $j' \neq j_2$ and $r(j_1,j_2; j',d) \neq j_2$. All other constraints are automatically enforced by the creation operators.

The second part $B$ is given by
\begin{align}
    \text{Part B} = B_{1,-} + B_{1,+} + B_{2,-} + B_{2,+},
\end{align}
where
\begin{align}
   \label{eq:2H2DB1-}
    B_{1,-} &= - \sum_{j_1,j_2=0}^{N-1} e^{ij_1 k_1} e^{ij_2 k_2} \sum_{j'=0}^{N-1} e^{i j_{\alpha}(j_1, j_2; j') k_{\alpha}} \left(  \left(  c^{\dagger}_{j'-1, \downarrow} c_{j',\uparrow} \right) \delta_{j'-1, j_1} c^{\dagger}_{r(j_1, j_2; j',d), \downarrow} c_{r(j_1, j_2; j',d), \uparrow} \right. \nonumber \\ 
    & \quad \quad \left. +  c^{\dagger}_{j', \downarrow} c_{j',\uparrow} \left( c^{\dagger}_{r(j_1,j_2;j',d)-1, \downarrow} c_{r(j_1,j_2;j',d), \uparrow} \right) \delta_{r(j_1,j_2;j',d)-1, j_1} \right) c_{j_1, \uparrow} c_{j_2, \uparrow} \ket{\text{FM}} \nonumber \\
    &= -e^{-ik_1} \sum_{j_1,j_2=0}^{N-1} e^{i (j_1+1) k_1} e^{ij_2 k_2}  e^{i j_{\alpha}(j_1, j_2; j_1+1) k_{\alpha}}  c^{\dagger}_{j_1, \downarrow} c_{j_1+1, \uparrow} c^{\dagger}_{r(j_1, j_2; j_1+1,d), \downarrow} c_{r(j_1, j_2; j_1+1,d), \uparrow} c_{j_1, \uparrow} c_{j_2, \uparrow} \ket{\text{FM}} \nonumber \\ 
    & \quad  - e^{-ik_1} \sum_{j_1,j_2=0}^{N-1} e^{i (j_1+1) k_1} e^{ij_2 k_2}  e^{i j_{\alpha}(j_1, j_2; l(j_1, j_2; j_1+1, d)) k_{\alpha}} c^{\dagger}_{l(j_1, j_2; j_1+1, d), \downarrow} c_{l(j_1, j_2; j_1+1, d), \uparrow} c^{\dagger}_{j_{1}, \downarrow} c_{j_{1}+1, \uparrow} c_{j_1, \uparrow} c_{j_2, \uparrow} \ket{\text{FM}} \nonumber \\ 
    & =  e^{-ik_1} \sum_{j_1,j_2=0}^{N-1} e^{i (j_1+1) k_1} e^{ij_2 k_2}  e^{i j_{\alpha}(j_1, j_2; j_1+1) k_{\alpha}}  c^{\dagger}_{j_1, \downarrow} c_{j_1, \uparrow} c^{\dagger}_{r(j_1, j_2; j_1+1,d), \downarrow} c_{r(j_1, j_2; j_1+1,d), \uparrow} c_{j_1+1, \uparrow} c_{j_2, \uparrow} \ket{\text{FM}}  \nonumber \\ 
    & + e^{-ik_1} \sum_{j_1,j_2=0}^{N-1} e^{i (j_1+1) k_1} e^{ij_2 k_2}  e^{i j_{\alpha}(j_1, j_2; l(j_1, j_2; j_1+1, d)) k_{\alpha}} c^{\dagger}_{l(j_1, j_2; j_1+1, d), \downarrow} c_{l(j_1, j_2; j_1+1, d), \uparrow} c^{\dagger}_{j_{1}, \downarrow} c_{j_{1}, \uparrow} c_{j_1+1, \uparrow} c_{j_2, \uparrow} \ket{\text{FM}}, 
\end{align}

\begin{align}
\label{eq:2H2DB1+}
    B_{1,+} & =  e^{ik_1} \sum_{j_1,j_2=0}^{N-1} e^{i (j_1-1) k_1} e^{ij_2 k_2}  e^{i j_{\alpha}(j_1, j_2; j_1-1) k_{\alpha}}  c^{\dagger}_{j_1, \downarrow} c_{j_1, \uparrow} c^{\dagger}_{r(j_1, j_2; j_1-1,d), \downarrow} c_{r(j_1, j_2; j_1-1,d), \uparrow} c_{j_1-1, \uparrow} c_{j_2, \uparrow} \ket{\text{FM}} \\ \nonumber
    & + e^{ik_1} \sum_{j_1,j_2=0}^{N-1} e^{i (j_1-1) k_1} e^{ij_2 k_2}  e^{i j_{\alpha}(j_1, j_2; l(j_1, j_2; j_1-1, d)) k_{\alpha}} c^{\dagger}_{l(j_1, j_2; j_1-1, d), \downarrow} c_{l(j_1, j_2; j_1-1, d), \uparrow} c^{\dagger}_{j_{1}, \downarrow} c_{j_{1}, \uparrow} c_{j_1-1, \uparrow} c_{j_2, \uparrow} \ket{\text{FM}}, 
\end{align}

\begin{align}
\label{eq:2H2DB2-}
    B_{2,-} 
    & =  e^{-ik_2} \sum_{j_1,j_2=0}^{N-1} e^{i j_1 k_1} e^{i (j_2+1) k_2}  e^{i j_{\alpha}(j_1, j_2; j_2+1) k_{\alpha}}  c^{\dagger}_{j_2, \downarrow} c_{j_2, \uparrow} c^{\dagger}_{r(j_1, j_2; j_2+1,d), \downarrow} c_{r(j_1, j_2; j_2+1,d), \uparrow} c_{j_1, \uparrow} c_{j_2+1, \uparrow} \ket{\text{FM}} \\ \nonumber
    & + e^{-ik_2} \sum_{j_1,j_2=0}^{N-1} e^{i j_1 k_1} e^{i (j_2+1) k_2}  e^{i j_{\alpha}(j_1, j_2; l(j_1, j_2; j_2+1, d)) k_{\alpha}} c^{\dagger}_{l(j_1, j_2; j_2+1, d), \downarrow} c_{l(j_1, j_2; j_2+1, d), \uparrow} c^{\dagger}_{j_{2}, \downarrow} c_{j_{2}, \uparrow} c_{j_1, \uparrow} c_{j_2+1, \uparrow} \ket{\text{FM}}, 
\end{align}

\begin{align}
\label{eq:2H2DB2+}
    B_{2,+} & =  e^{ik_2} \sum_{j_1,j_2=0}^{N-1} e^{i j_1 k_1} e^{i(j_2-1) k_2}  e^{i j_{\alpha}(j_1, j_2; j_2-1) k_{\alpha}}  c^{\dagger}_{j_2, \downarrow} c_{j_2, \uparrow} c^{\dagger}_{r(j_1, j_2; j_2-1,d), \downarrow} c_{r(j_1, j_2; j_2-1,d), \uparrow} c_{j_1, \uparrow} c_{j_2-1, \uparrow} \ket{\text{FM}} \\ \nonumber
    & + e^{ik_2} \sum_{j_1,j_2=0}^{N-1} e^{i j_1 k_1} e^{i(j_2-1) k_2}  e^{i j_{\alpha}(j_1, j_2; l(j_1, j_2; j_2-1, d)) k_{\alpha}} c^{\dagger}_{l(j_1, j_2; j_2-1, d), \downarrow} c_{l(j_1, j_2; j_2-1, d), \uparrow} c^{\dagger}_{j_{2}, \downarrow} c_{j_{2}, \uparrow} c_{j_1, \uparrow} c_{j_2-1, \uparrow} \ket{\text{FM}}. 
\end{align}

We first combine $A_{1,-}$ in Eq.~\eqref{eq:2H2DA1-} and $B_{1,-}$ in Eq.~\eqref{eq:2H2DB1-}. $A_{1,-} + B_{1,-}$ is given by
\begin{align}
    \label{eq:2H2DA1-B1-}
    A_{1,-} + B_{1,-} &= e^{-ik_1} \sum_{j_1,j_2=0}^{N-1} e^{i(j_1+1) k_1} e^{ij_2 k_2} \sum_{j'=0}^{N-1} e^{i j_{\alpha}(j_1, j_2; j') k_{\alpha}} c^{\dagger}_{j', \downarrow} c_{j',\uparrow} c^{\dagger}_{r(j_1, j_2; j',d), \downarrow} c_{r(j_1, j_2; j',d), \uparrow} \left( c_{j_1+1,\uparrow} c_{j_2, \uparrow}
    \right)  \ket{\text{FM}}   \nonumber \\
    &+  e^{-ik_1} \sum_{j_1,j_2=0}^{N-1} e^{i (j_1+1) k_1} e^{ij_2 k_2}  e^{i j_{\alpha}(j_1, j_2; j_1+1) k_{\alpha}}  c^{\dagger}_{j_1, \downarrow} c_{j_1, \uparrow} c^{\dagger}_{r(j_1, j_2; j_1+1,d), \downarrow} c_{r(j_1, j_2; j_1+1,d), \uparrow} c_{j_1+1, \uparrow} c_{j_2, \uparrow} \ket{\text{FM}}  \nonumber \\ 
    & + e^{-ik_1} \sum_{j_1,j_2=0}^{N-1} e^{i (j_1+1) k_1} e^{ij_2 k_2}  e^{i j_{\alpha}(j_1, j_2; l(j_1, j_2; j_1+1, d)) k_{\alpha}} c^{\dagger}_{l(j_1, j_2; j_1+1, d), \downarrow} c_{l(j_1, j_2; j_1+1, d), \uparrow} c^{\dagger}_{j_{1}, \downarrow} c_{j_{1}, \uparrow} c_{j_1+1, \uparrow} c_{j_2, \uparrow} \ket{\text{FM}}  \nonumber \\ 
    & = e^{-ik_1} \sum_{j_2=0}^{N-1} e^{ij_2 k_2} \sum_{j_1=0}^{N-2}e^{i(j_1+1) k_1}  
    \left( \sum_{j'=0}^{N-1} e^{i j_{\alpha}(j_1, j_2; j') k_{\alpha}} c^{\dagger}_{j', \downarrow} c_{j',\uparrow} c^{\dagger}_{r(j_1, j_2; j',d), \downarrow} c_{r(j_1, j_2; j',d), \uparrow}  \right. \nonumber \\ 
    &  + e^{i j_{\alpha}(j_1, j_2; j_1+1) k_{\alpha}}  c^{\dagger}_{j_1, \downarrow} c_{j_1, \uparrow} c^{\dagger}_{r(j_1, j_2; j_1+1,d), \downarrow} c_{r(j_1, j_2; j_1+1,d), \uparrow}  \nonumber \\ 
    & \left. + e^{i j_{\alpha}(j_1, j_2; l(j_1, j_2; j_1+1, d)) k_{\alpha}} c^{\dagger}_{l(j_1, j_2; j_1+1, d), \downarrow} c_{l(j_1, j_2; j_1+1, d), \uparrow} c^{\dagger}_{j_1, \downarrow} c_{j_1, \uparrow} \right) c_{j_1+1, \uparrow} c_{j_2, \uparrow} \ket{\text{FM}} \nonumber \\ 
    &+ e^{-ik_1} \sum_{j_2=0}^{N-1} e^{ij_2 k_2} e^{iNk_1 } \left( \sum_{j'=0}^{N-1} e^{i j_{\alpha}(N-1, j_2; j') k_{\alpha}} c^{\dagger}_{j', \downarrow} c_{j',\uparrow} c^{\dagger}_{r(N-1, j_2; j',d), \downarrow} c_{r(N-1, j_2; j',d), \uparrow}  \right. \nonumber \\ 
    &  + e^{i j_{\alpha}(N-1, j_2; 0) k_{\alpha}}  c^{\dagger}_{N-1, \downarrow} c_{N-1, \uparrow} c^{\dagger}_{r(N-1, j_2; 0,d), \downarrow} c_{r(N-1, j_2; 0,d), \uparrow}  \nonumber \\ 
    & \left. + e^{i j_{\alpha}(N-1, j_2; l(N-1, j_2; 0, d)) k_{\alpha}} c^{\dagger}_{l(N-1, j_2; 0, d), \downarrow} c_{l(N-1, j_2; 0, d), \uparrow} c^{\dagger}_{N-1, \downarrow} c_{N-1, \uparrow} \right) c_{0, \uparrow} c_{j_2, \uparrow} \ket{\text{FM}}. 
\end{align}
We define $A_{1,-}^{j_1 < N-1}$, $B_{1,-}^{j_1 < N-1}$, $A_{1,-}^{j_1 = N-1}$, $B_{1,-}^{j_1 = N-1}$ as follows such that 
\begin{align}
    \label{eq:2H2DA1-B1-rewrite}
    A_{1,-}+ B_{1,-} = A_{1,-}^{j_1 <N-1} + B_{1,-}^{j_1< N-1} + A_{1,-}^{j_1=N-1} + B_{1,-}^{j_1=N-1}.
\end{align}
$A_{1,-}^{j_1 < N-1}$ in Eq.~\eqref{eq:2H2DA1-B1-rewrite} is given by
\begin{align}
    A_{1,-}^{j_1<N-1} &= e^{-ik_1} \sum_{j_2=0}^{N-1} e^{ij_2 k_2} \sum_{j_1=0}^{N-2}e^{i(j_1+1) k_1}  
    \left( \sum_{j'=0}^{N-1} e^{i j_{\alpha}(j_1, j_2; j') k_{\alpha}} c^{\dagger}_{j', \downarrow} c_{j',\uparrow} c^{\dagger}_{r(j_1, j_2; j',d), \downarrow} c_{r(j_1, j_2; j',d), \uparrow}  \right)
c_{j_1+1, \uparrow} c_{j_2, \uparrow} \ket{\text{FM}}  \nonumber \\ 
&= e^{-ik_1} \sum_{j_2=0}^{N-1} e^{ij_2 k_2} \sum_{j_1=0}^{N-2}e^{i(j_1+1) k_1}  
    \left( \sum_{\substack{j' =0 \\ j' \neq j_1, r(j_1, j_2; j',d) \neq j_1}}^{N-1} e^{i j_{\alpha}(j_1, j_2; j') k_{\alpha}} c^{\dagger}_{j', \downarrow} c_{j',\uparrow} c^{\dagger}_{r(j_1, j_2; j',d), \downarrow} c_{r(j_1, j_2; j',d), \uparrow}  \right)
c_{j_1+1, \uparrow} c_{j_2, \uparrow} \ket{\text{FM}} \nonumber \\ 
& = e^{-ik_1} \sum_{j_2=0}^{N-1} e^{ij_2 k_2} \sum_{j_1=0}^{N-2}e^{i(j_1+1) k_1}  
    \left( \sum_{\substack{j' =0 \\ j' \neq j_1, r(j_1, j_2; j',d) \neq j_1}}^{N-1} e^{i j_{\alpha}(j_1+1, j_2; j') k_{\alpha}} c^{\dagger}_{j', \downarrow} c_{j',\uparrow} c^{\dagger}_{r(j_1+1, j_2; j',d), \downarrow} c_{r(j_1+1, j_2; j',d), \uparrow}  \right) \nonumber \\
 & \quad \quad \quad c_{j_1+1, \uparrow} c_{j_2, \uparrow} \ket{\text{FM}}.
\end{align}
In the second equality, we have explicitly written down the constraints in the sum $\sum_{j'=0}^{N-1}$: $j' \neq j_1$ and $r(j_1, j_2; j' ,d) \neq j_1$, while other constraints, such as $j' \neq j_1 + 1$, $r(j_1,j_2; j',d) \neq j_1 + 1$ are forced by the fermion creation operators. In the third equality, we have used 
\begin{align}
    j_{\alpha}(j_1,j_2;j') = j_{\alpha}(j_1+1,j_2;j'), \\ 
    r(j_1,j_2;j',d) = r(j_1+1,j_2; j',d),
\end{align}
when $j' \neq j_1, j_1+1$ and $r(j_1,j_2; j', d) \neq j_1, j_1+1$. See Sec.~\ref{sec:Proofofequalitiesusedintheeigenstateconstruction} for the proofs of these equalities.

$B_{1,-}^{j_1 < N-1}$ in Eq.~\eqref{eq:2H2DA1-B1-rewrite} is given by
\begin{align}
    B_{1,-}^{j_1<N-1} &= e^{-ik_1} \sum_{j_2=0}^{N-1} e^{ij_2 k_2} \sum_{j_1=0}^{N-2}e^{i(j_1+1) k_1}  
 \left( e^{i j_{\alpha}(j_1, j_2; j_1+1) k_{\alpha}}  c^{\dagger}_{j_1, \downarrow} c_{j_1, \uparrow} c^{\dagger}_{r(j_1, j_2; j_1+1,d), \downarrow} c_{r(j_1, j_2; j_1+1,d), \uparrow} \right. \nonumber \\ 
    &  \left. + e^{i j_{\alpha}(j_1, j_2; l(j_1, j_2; j_1+1, d)) k_{\alpha}} c^{\dagger}_{l(j_1, j_2; j_1+1, d), \downarrow} c_{l(j_1, j_2; j_1+1, d), \uparrow} c^{\dagger}_{j_1, \downarrow} c_{j_1, \uparrow} \right) c_{j_1+1, \uparrow} c_{j_2, \uparrow} \ket{\text{FM}} \nonumber \\ 
    & =e^{-ik_1} \sum_{j_2=0}^{N-1} e^{ij_2 k_2} \sum_{j_1=0}^{N-2}e^{i(j_1+1) k_1}  
 \left( e^{i j_{\alpha}(j_1+1, j_2; j_1) k_{\alpha}}  c^{\dagger}_{j_1, \downarrow} c_{j_1, \uparrow} c^{\dagger}_{r(j_1+1, j_2; j_1,d), \downarrow} c_{r(j_1+1, j_2; j_1,d), \uparrow} \right. \nonumber \\ 
    &  \left. + e^{i j_{\alpha}(j_1+1, j_2; l(j_1+1, j_2; j_1, d)) k_{\alpha}} c^{\dagger}_{l(j_1+1, j_2; j_1, d), \downarrow} c_{l(j_1+1, j_2; j_1, d), \uparrow} c^{\dagger}_{j_1, \downarrow} c_{j_1, \uparrow} \right) c_{j_1+1, \uparrow} c_{j_2, \uparrow} \ket{\text{FM}}, 
\end{align}
where we have used
\begin{align}
    j_{\alpha}(j_1,j_2; j_1+1) &= j_{\alpha}(j_1+1, j_2; j_1),  \\ 
    j_{\alpha}(j_1,j_2; l(j_1,j_2;j_1+1,d)) &= j_{\alpha}(j_1+1,j_2; l(j_1+1,j_2;j_1,d)), \\ \nonumber
\end{align}
and 
\begin{align}
        r(j_1,j_2; j_1+1, d) &= r(j_1+1, j_2; j_1,d) \quad \mathrm{when} \quad r(j_1,j_2; j_1+1, d) \neq j_1+1,  \\
        l(j_1,j_2;j_1+1,d) &= l(j_1+1,j_2;j_1,d) \quad \mathrm{when} \quad l(j_1,j_2;j_1+1,d) \neq j_1+1.
\end{align}
See Sec.~\ref{sec:Proofofequalitiesusedintheeigenstateconstruction} for the proofs of these equalities.

$ A_{1,-}^{j_1 = N-1}$ in Eq.~\eqref{eq:2H2DA1-B1-rewrite} is given by
\begin{align}
    A_{1,-}^{j_1 = N-1} &=e^{-ik_1} \sum_{j_2=0}^{N-1} e^{ij_2 k_2} e^{iNk_1 } \left( \sum_{j'=0}^{N-1} e^{i j_{\alpha}(N-1, j_2; j') k_{\alpha}} c^{\dagger}_{j', \downarrow} c_{j',\uparrow} c^{\dagger}_{r(N-1, j_2; j',d), \downarrow} c_{r(N-1, j_2; j',d), \uparrow}  \right) c_{0, \uparrow} c_{j_2, \uparrow} \ket{\text{FM}} \nonumber \\ 
    &=e^{-ik_1} \sum_{j_2=0}^{N-1} e^{ij_2 k_2} e^{iNk_1 } \left( \sum_{ \substack{j'=0 \\ j' \neq N-1, r(N-1,j_2;j',d) \neq N-1}}^{N-1} e^{i j_{\alpha}(N-1, j_2; j') k_{\alpha}} c^{\dagger}_{j', \downarrow} c_{j',\uparrow} c^{\dagger}_{r(N-1, j_2; j',d), \downarrow} c_{r(N-1, j_2; j',d), \uparrow}  \right) \nonumber \\
    & \quad \quad \quad c_{0, \uparrow} c_{j_2, \uparrow} \ket{\text{FM}}  \nonumber \\
    & =e^{-ik_1} \sum_{j_2=0}^{N-1} e^{ij_2 k_2} e^{iNk_1 } \left( \sum_{ \substack{j'=0 \\ j' \neq N-1, r(N-1,j_2;j',d) \neq N-1}}^{N-1} e^{ik_{\alpha}}e^{i j_{\alpha}(0, j_2; j') k_{\alpha}} c^{\dagger}_{j', \downarrow} c_{j',\uparrow} c^{\dagger}_{r(0, j_2; j',d), \downarrow} c_{r(0, j_2; j',d), \uparrow}  \right) \nonumber \\
    & \quad \quad \quad c_{0, \uparrow} c_{j_2, \uparrow} \ket{\text{FM}}. 
\end{align}
In the second equality, we have explicitly written down the constraints in sum $\sum_{j'=0}^{N-1}$. In the third equality, we have used 
\begin{align}
    j_{\alpha}(N-1,j_2;j')& = j_{\alpha}(0,j_2;j') +1 \quad \mathrm{when} \quad j' \neq 0, j_2, N-1,\\ 
    r(N-1,j_2; j',d) &= r(0,j_2; j',d) \quad \mathrm{when} \quad j' \neq 0, j_2, N-1; r(N-1,j_2; j',d) \neq 0.
\end{align}
See Sec.~\ref{sec:Proofofequalitiesusedintheeigenstateconstruction} for the proofs of these equalities.

$B_{1,-}^{j_1 = N-1}$ in Eq.~\eqref{eq:2H2DA1-B1-rewrite} is given by
\begin{align}
    B_{1,-}^{j_1 = N-1} &=e^{-ik_1} \sum_{j_2=0}^{N-1} e^{ij_2 k_2} e^{iNk_1 } \left( e^{i j_{\alpha}(N-1, j_2; 0) k_{\alpha}}  c^{\dagger}_{N-1, \downarrow} c_{N-1, \uparrow} c^{\dagger}_{r(N-1, j_2; 0,d), \downarrow} c_{r(N-1, j_2; 0,d), \uparrow} \right.  \nonumber \\ 
    & \left. + e^{i j_{\alpha}(N-1, j_2; l(N-1, j_2; 0, d)) k_{\alpha}} c^{\dagger}_{l(N-1, j_2; 0, d), \downarrow} c_{l(N-1, j_2; 0, d), \uparrow} c^{\dagger}_{N-1, \downarrow} c_{N-1, \uparrow} \right) c_{0, \uparrow} c_{j_2, \uparrow} \ket{\text{FM}} \nonumber \\ 
    &=e^{-ik_1} \sum_{j_2=0}^{N-1} e^{ij_2 k_2} e^{iNk_1 } \left( e^{-i(N-3) k_{\alpha}}e^{i j_{\alpha}(0, j_2; N-1) k_{\alpha}}  c^{\dagger}_{N-1, \downarrow} c_{N-1, \uparrow} c^{\dagger}_{r(0, j_2; N-1,d), \downarrow} c_{r(0, j_2; N-1,d), \uparrow} \right.  \nonumber \\ 
        & \left. +  e^{ik_{\alpha}}e^{i j_{\alpha}(0, j_2; l(0, j_2; N-1, d)) k_{\alpha}} c^{\dagger}_{l(0, j_2; N-1, d), \downarrow} c_{l(0, j_2; N-1, d), \uparrow} c^{\dagger}_{N-1, \downarrow} c_{N-1, \uparrow} \right) c_{0, \uparrow} c_{j_2, \uparrow} \ket{\text{FM}}, 
\end{align}
where we have used
\begin{align}
    j_{\alpha}(N-1,j_2;0) &= 0, \\
    j_{\alpha}(0,j_2; N-1) &= N-3, \\ 
    r(N-1,j_2; 0, d) & = r(0,j_2;N-1,d ), \\ 
    j_{\alpha}(N-1, j_2; l(N-1, j_2; 0, d)) &= j_{\alpha}(0, j_2; l(0, j_2; N-1, d)) + 1, \\
    l(N-1, j_2; 0, d) & = l(0, j_2; N-1, d).
\end{align}
See Sec.~\ref{sec:Proofofequalitiesusedintheeigenstateconstruction} for the proofs of these equalities.

Therefore, when $e^{i(N-2)k_{\alpha}} = 1$ and  $e^{iN k_1} e^{ik_{\alpha}} = 1$,
we have 
\begin{align}
        A_{1,-}+ B_{1,-} = A_{1,-}^{j_1 <N-1} + B_{1,-}^{j_1< N-1} + A_{1,-}^{j_1=N-1} + B_{1,-}^{j_1=N-1} = e^{-ik_1} \ket{k_1, k_2, k_{\alpha}, d}.
\end{align}
We can also verify under the conditions $e^{i(N-2)k_{\alpha}} = 1$ and  $e^{iN k_1} e^{ik_{\alpha}} = 1$
\begin{align}
    A_{1,+}+ B_{1,+} =  e^{ik_1} \ket{k_1, k_2, k_{\alpha}, d}.
\end{align}
Moreover, when $e^{iNk_2}e^{ik_{\alpha}}=1$, we have 
\begin{align}
    A_{2,-}+ B_{2,-} =  e^{-ik_2} \ket{k_1, k_2, k_{\alpha}, d}, \\ 
    A_{2,+}+ B_{2,+} =  e^{ik_2} \ket{k_1, k_2, k_{\alpha}, d}.
\end{align}
In conclusion, we have verified that when
\begin{align}
    e^{i(N-2)k_{\alpha}} &=1, \\ 
    e^{iNk_1} e^{ik_{\alpha}}   &= 1, \\ 
     e^{iNk_2} e^{ik_{\alpha}}  &= 1,
\end{align}
$\ket{k_1,k_2,k_{\alpha},d}$ is an eigenstate with eigenenergy $2\cos(k_1) + 2\cos(k_2)$. Note that the above derivation does not depend on the distance $d$ between two spin-down fermions. Therefore, the degeneracy of each energy level is $\frac{N-3}{2}$, i.e., the number of distinct orbits.

Moreover, $k_1$, $k_2$ and $k_{\alpha}$ can be expressed as 
\begin{align}
    k_{\alpha} &= -\frac{2 \pi m_{\downarrow}}{N-2}, \quad \mathrm{with} \quad m_{\downarrow}=0,\cdots, N-3, \\ 
    k_{1} & = \frac{2\pi m_{1}}{N} - \frac{k_{\alpha}}{N} = \frac{2\pi m_{1}}{N} + \frac{2\pi m_{\downarrow}}{N(N-2)}, \label{eq:TwoHolesTwoDownk1}\\
    k_{2} & = \frac{2\pi m_{2}}{N} - \frac{k_{\alpha}}{N} = \frac{2\pi m_{2}}{N} + \frac{2\pi m_{\downarrow}}{N(N-2)}, \label{eq:TwoHolesTwoDownk2}
\end{align}
where $0 \leq m_1 < m_2 \leq N-1$ since we require $k_{1} \neq k_{2}$ otherwise the state $\ket{k_1, k_2, k_{\alpha}}$ vanishes.

\subsubsection{Any number of spin-down fermions}
For a sector with $n_{\downarrow}$ spin-down fermions and
$N-n_{\downarrow}-2$ spin-up fermions, the eigenenergies still take the form $2\cos(k_1) + 2\cos(k_2)$ where $k_1$ and $k_2$ are given by Eqs.~\eqref{eq:TwoHolesTwoDownk1} and ~\eqref{eq:TwoHolesTwoDownk2}. 
Each energy level has degeneracy
$\frac{\binom{N-2}{n_{\downarrow}}}{N-2}$.
See Sec.~\ref{sec:Anyholesandanyspin-downfermions} for the derivation in the case of an arbitrary number of holes and spin-down fermions.

\subsection{Any number of holes}
\label{sec:Exact_solutions_with_any_holes}

In this section, we present the derivation of the eigenstates and the energy spectrum for arbitrary numbers of holes and spin-down fermions. We begin with the sector containing an arbitrary number of holes and no spin-down fermions in Sec.~\ref{sec:Anyholesandnospin-downfermion}. We then discuss the sector with an arbitrary number of holes and one spin-down fermion in Sec.~\ref{sec:Anyholesandonespin-downfermion}. Finally, we present the derivation for arbitrary numbers of holes and spin-down fermions in Sec.~\ref{sec:Anyholesandanyspin-downfermions}. In Sec.~\ref{sec:Proofofequalitiesusedintheeigenstateconstruction}, we provide the proof of the equalities used in the eigenstate construction in Sec.~\ref{sec:Anyholesandonespin-downfermion} and Sec.~\ref{sec:Anyholesandanyspin-downfermions}. 

\subsubsection{No spin-down fermion}
\label{sec:Anyholesandnospin-downfermion}
In the sector with $n_{h}$ holes and $N-n_{h}$ spin-up fermions, i.e., no spin-down fermion, the Hilbert-space dimension is $\binom{N}{n_{h}}$.

The following state is an eigenstate \begin{align} \ket{k_1, k_2, \cdots, k_{n_h}} = \sum_{j_1, j_2, \cdots, j_{n_h}=0}^{N-1} e^{ij_1 k_1} e^{ij_2 k_2} \cdots e^{ij_{n_h}k_{n_{h}}} c_{j_{1}, \uparrow} c_{j_{2}, \uparrow} \cdots c_{j_{n_h}, \uparrow} \ket{\mathrm{FM}}, \end{align}
when $e^{iN k_1} = e^{iN k_2} = \cdots = e^{iN k_{n_h}} = 1$. Note that any two $k_i$ and $k_j$ are distinct, otherwise the state vanishes.
In the following, we explicitly verify that this is an eigenstate with eigenenergy $2\sum_{i=1}^{n_h} \cos(k_{i})$.

Acting with the Hamiltonian $\hat{H}$ (with $t=1$) yields 
\begin{align} 
\label{eq:AnyholeNoDownActionOfH}
\hat{H} \ket{k_1, k_2, \cdots, k_{n_h}} &= -\sum_{j_1, j_2, \cdots, j_{n_h}=0}^{N-1} e^{ij_1 k_1} e^{ij_2 k_2} \cdots e^{ij_{n_h}k_{n_{h}}} \left( \sum_{a=1}^{n_h} (c^{\dagger}_{j_a, \uparrow} c_{j_a+1, \uparrow} + c^{\dagger}_{j_a, \uparrow} c_{j_a-1, \uparrow}) \right) c_{j_{1}, \uparrow} c_{j_{2}, \uparrow} \cdots c_{j_{n_h}, \uparrow} \ket{\mathrm{FM}}  \nonumber  \\
&= \sum_{j_1, j_2, \cdots, j_{n_h}=0}^{N-1} e^{ij_1 k_1} e^{ij_2 k_2} \cdots e^{ij_{n_h}k_{n_{h}}} (c_{j_{1}+1, \uparrow} + c_{j_{1}-1, \uparrow}) c_{j_{2}, \uparrow} \cdots c_{j_{n_h}, \uparrow} \ket{\mathrm{FM}} \nonumber \\ 
&+ \sum_{j_1, j_2, \cdots, j_{n_h}=0}^{N-1} e^{ij_1 k_1} e^{ij_2 k_2} \cdots e^{ij_{n_h}k_{n_{h}}} c_{j_{1}, \uparrow} (c_{j_{2}+1, \uparrow} + c_{j_{2}-1, \uparrow}) \cdots c_{j_{n_h}, \uparrow} \ket{\mathrm{FM}} \nonumber \\  
&+\,\cdots\, + \sum_{j_1, j_2, \cdots, j_{n_h}=0}^{N-1} e^{ij_1 k_1} e^{ij_2 k_2} \cdots e^{ij_{n_h}k_{n_{h}}} c_{j_{1}, \uparrow} c_{j_{2}, \uparrow} \cdots (c_{j_{n_h}+1, \uparrow} + c_{j_{n_h}-1, \uparrow}) \ket{\mathrm{FM}} \nonumber \\  
&= e^{-ik_1 } \left( \sum_{j_1 = 0}^{N-2} \sum_{j_2, \cdots, j_{n_h}=0}^{N-1} e^{i(j_1+1) k_1} e^{ij_2 k_2} \cdots e^{ij_{n_h}k_{n_{h}}} c_{j_{1}+1, \uparrow} c_{j_{2}, \uparrow} \cdots c_{j_{n_h}, \uparrow} \ket{\mathrm{FM}} \right. \nonumber \\  
& \quad \quad \left. + \sum_{j_2, \cdots, j_{n_h}=0}^{N-1} e^{i N k_1} e^{ij_2 k_2} \cdots e^{ij_{n_h}k_{n_{h}}} c_{0, \uparrow} c_{j_{2}, \uparrow} \cdots c_{j_{n_h}, \uparrow} \ket{\mathrm{FM}} \right) \nonumber \\ 
&+ e^{ik_1 } \left( \sum_{j_1 = 1}^{N-1} \sum_{j_2, \cdots, j_{n_h}=0}^{N-1} e^{i(j_1-1) k_1} e^{ij_2 k_2} \cdots e^{ij_{n_h}k_{n_{h}}} c_{j_{1}-1, \uparrow} c_{j_{2}, \uparrow} \cdots c_{j_{n_h}, \uparrow} \ket{\mathrm{FM}} \right. \nonumber \\  
& \quad \quad \left. + \sum_{j_2, \cdots, j_{n_h}=0}^{N-1} e^{-i k_1} e^{ij_2 k_2} \cdots e^{ij_{n_h}k_{n_{h}}} c_{N-1, \uparrow} c_{j_{2}, \uparrow} \cdots c_{j_{n_h}, \uparrow} \ket{\mathrm{FM}} \right) \nonumber \\  
&+ e^{-ik_2 } \left( \sum_{j_2 = 0}^{N-2} \sum_{j_1, j_3, \cdots, j_{n_h}=0}^{N-1} e^{i j_1 k_1} e^{i (j_2+1) k_2} \cdots e^{ij_{n_h}k_{n_{h}}} c_{j_{1}, \uparrow} c_{j_{2}+1, \uparrow} \cdots c_{j_{n_h}, \uparrow} \ket{\mathrm{FM}} \right. \nonumber \\ 
& \quad \quad \left. + \sum_{j_1, j_3, \cdots, j_{n_h}=0}^{N-1} e^{i j_1 k_1} e^{i N k_2} \cdots e^{ij_{n_h}k_{n_{h}}} c_{j_1, \uparrow} c_{0, \uparrow} \cdots c_{j_{n_h}, \uparrow} \ket{\mathrm{FM}} \right) \nonumber \\ 
&+ e^{ik_2 } \left( \sum_{j_2 = 1}^{N-1} \sum_{j_1, j_3, \cdots, j_{n_h}=0}^{N-1} e^{i j_1 k_1} e^{i(j_2-1) k_2} \cdots e^{ij_{n_h}k_{n_{h}}} c_{j_{1}, \uparrow} c_{j_{2}-1, \uparrow} \cdots c_{j_{n_h}, \uparrow} \ket{\mathrm{FM}} \right. \nonumber \\ 
& \quad \quad \left. + \sum_{j_1, j_3, \cdots, j_{n_h}=0}^{N-1} e^{i j_1 k_1} e^{-i k_2} \cdots e^{ij_{n_h}k_{n_{h}}} c_{j_1, \uparrow} c_{N-1, \uparrow} \cdots c_{j_{n_h}, \uparrow} \ket{\mathrm{FM}} \right) \nonumber\\ 
&+\,\cdots\, +  e^{-ik_{n_h} } \left( \sum_{j_{n_h} = 0}^{N-2} \sum_{j_1, \cdots, j_{n_h-1}=0}^{N-1} e^{i j_1 k_1} e^{ij_2 k_2} \cdots e^{i(j_{n_h}+1) k_{n_{h}}} c_{j_{1}, \uparrow} c_{j_{2}, \uparrow} \cdots c_{j_{n_h}+1, \uparrow} \ket{\mathrm{FM}} \right. \nonumber \\  
& \quad \quad \left. + \sum_{j_1, \cdots, j_{n_h-1}=0}^{N-1} e^{i j_1 k_1} e^{ij_2 k_2} \cdots e^{i N k_{n_{h}}} c_{j_{1}, \uparrow} c_{j_{2}, \uparrow} \cdots c_{0, \uparrow} \ket{\mathrm{FM}} \right) \nonumber \\ 
&+ e^{ik_{n_h} } \left( \sum_{j_{n_h} = 1}^{N-1} \sum_{j_1, \cdots, j_{n_h-1}=0}^{N-1} e^{i j_1 k_1} e^{ij_2 k_2} \cdots e^{i (j_{n_h} - 1) k_{n_{h}}} c_{j_{1}, \uparrow} c_{j_{2}, \uparrow} \cdots c_{j_{n_h}-1, \uparrow} \ket{\mathrm{FM}} \right. \nonumber\\  
& \quad \quad \left. + \sum_{j_1, \cdots, j_{n_h-1}=0}^{N-1} e^{i j_1 k_1} e^{ij_2 k_2} \cdots e^{-i k_{n_{h}}} c_{j_1, \uparrow} c_{j_{2}, \uparrow} \cdots c_{N-1, \uparrow} \ket{\mathrm{FM}} \right).  
\end{align}
Therefore, when $e^{iNk_1} = e^{iNk_2} = \cdots =e^{iNk_{n_h}}=1$, 
\begin{align} \hat{H} \ket{k_1, k_2, \cdots, k_{n_h}} = 2 \sum_{i=1}^{n_h} \cos(k_{i}) \ket{k_1,k_2, \cdots, k_{n_{h}}}.
\end{align} 

Since $e^{iNk_1} = e^{iNk_2} = \cdots =e^{iNk_{n_h}}=1$ and any two $k_{i}$ and $k_{j}$ are distinct, $k_i$ can be expressed as 
\begin{align}
    k_{i} = \frac{2 \pi m_{i}}{N},  \quad i = 1, \cdots, n_{h},
\end{align}
with $0 \leq m_{1} < m_{2} < \cdots < m_{n_h} \leq N-1$.

Moreover, under the action of translation operator $\hat{T}_{x}$, we have
\begin{align}
    \hat{T}_{x} \ket{k_1, k_2, \cdots, k_{n_{h}}} &= \sum_{j_1, j_2, \cdots, j_{n_h}=0}^{N-1} e^{ij_1 k_1} e^{ij_2 k_2} \cdots e^{ij_{n_h} k_{n_h}} c_{j_1+1, \uparrow} c_{j_2+1, \uparrow} \cdots c_{j_{n_h}+1, \uparrow} \ket{1\cdots1} \nonumber \\ 
    & = e^{-i \sum_{a=1}^{n_h} k_{a}}  \sum_{j_1, j_2, \cdots, j_{n_h}=0}^{N-1} e^{i(j_1+1) k_1} e^{i(j_2+1) k_2} \cdots e^{i(j_{n_h}+1) k_{n_h}} c_{j_1+1, \uparrow} c_{j_2+1, \uparrow} \cdots c_{j_{n_h}+1, \uparrow} \ket{1\cdots1}  \nonumber \\ 
    & = e^{-i \sum_{a=1}^{n_h} k_{a}} \ket{k_1, k_2, \cdots, k_{n_h}}.
\end{align}
Therefore, the total momentum is $k = \sum_{a=1}^{n_h} k_{a}$.

Next, we prove that $\ket{k_1, k_2, \cdots, k_{n_h}}$ is an eigenstate of the current operator $\hat{J}$ with eigenvalue $2 \sum_{i=1}^{n_h} \sin(k_i)$. Under the action of the current operator $\hat{J}$, we have
\begin{align}
\hat{J} \ket{k_1, k_2, \cdots, k_{n_h}}
&= i \sum_{j_1, j_2, \cdots, j_{n_h}=0}^{N-1}
e^{i j_1 k_1} e^{i j_2 k_2} \cdots e^{i j_{n_h} k_{n_h}}
\left(
\sum_{a=1}^{n_h}
(-c^{\dagger}_{j_a,\uparrow} c_{j_a+1,\uparrow}
+ c^{\dagger}_{j_a,\uparrow} c_{j_a-1,\uparrow})
\right)
c_{j_1,\uparrow} c_{j_2,\uparrow} \cdots c_{j_{n_h},\uparrow}
\ket{\mathrm{FM}} .
\end{align}

Comparing with Eq.~\eqref{eq:AnyholeNoDownActionOfH}, the only difference is that
$-\sum_{a=1}^{n_h} c^{\dagger}_{j_a,\uparrow} c_{j_a+1,\uparrow}$ in
Eq.~\eqref{eq:AnyholeNoDownActionOfH} becomes
$-i \sum_{a=1}^{n_h} c^{\dagger}_{j_a,\uparrow} c_{j_a+1,\uparrow}$,
and
$-\sum_{a=1}^{n_h} c^{\dagger}_{j_a,\uparrow} c_{j_a-1,\uparrow}$ becomes
$i \sum_{a=1}^{n_h} c^{\dagger}_{j_a,\uparrow} c_{j_a-1,\uparrow}$.
From Eq.~\eqref{eq:AnyholeNoDownActionOfH}, we know that the former contributes
$\sum_{a=1}^{n_h} e^{-i k_a} \ket{k_1, k_2, \cdots, k_{n_h}}$
and the latter contributes
$\sum_{a=1}^{n_h} e^{i k_a} \ket{k_1, k_2, \cdots, k_{n_h}}$.
Therefore,
\begin{align}
\label{eq:ProofOfEigenstateJWithoutSpinDown}
\hat{J} \ket{k_1, k_2, \cdots, k_{n_h}}
&= \left(i \sum_{a=1}^{n_h} e^{-i k_a}
      - i \sum_{a=1}^{n_h} e^{i k_a}\right)
   \ket{k_1, k_2, \cdots, k_{n_h}} \nonumber\\
&= \sum_{a=1}^{n_h} 2\sin(k_a)\,
   \ket{k_1, k_2, \cdots, k_{n_h}} .
\end{align}

\subsubsection{One spin-down fermion}
\label{sec:Anyholesandonespin-downfermion}
In the sector with $n_h$ holes, one spin-down fermion, and $N-n_{h}-1$ spin-up fermions, the Hilbert space dimension of this sector is $\binom{N}{n_h} (N-n_h)$.

We define the following state
\begin{align}
     \label{eq:DefinitionOfEigenstateAnyholesOnedown}
    \ket{k_1, \cdots, k_{n_h}, k_{\alpha}}  = \sum_{j_{1}, \cdots, j_{n_h} = 0}^{N-1} e^{ij_1 k_1} \cdots e^{ij_{n_h} k_{n_h}} \sum_{j'=0}^{N-1} e^{ij_{\alpha}(j_1, \cdots, j_{n_h}; j') k_{\alpha}} c^{\dagger}_{j', \downarrow} c_{j', \uparrow} c_{j_1, \uparrow} \cdots c_{j_{n_h}, \uparrow}  \ket{\mathrm{FM}}
\end{align}
where $c^{\dagger}_{j', \downarrow} c_{j', \uparrow} c_{j_1, \uparrow} \cdots c_{j_{n_h}, \uparrow}  \ket{\mathrm{FM}}$ denotes a Fock state in which the $n_{h}$ holes are located at sites $j_{1}, \cdots, j_{n_h}$ and the spin-down fermion is located at site $j'$. Therefore, $j' \neq j_a$ with $a = 1, \cdots, n_{h}$. Here, we do not explicitly include these constraints in the summation 
$\sum_{j'=0}^{N-1}$, since if there exists $j_{a}$ such that $j_a = j'$,  
$c_{j',\uparrow} c_{j_a,\uparrow} = 0$ and the corresponding Fock state vanishes.

Here, $j_{\alpha}(j_1, \cdots, j_{n_{h}}; j')$ is the generalization of $j_{\alpha}(j_1, j_{2}; j')$ used in Sec.~\ref{sec:Two_holes_and_one_spin-down_fermion} and 
denotes the position of the spin-down fermion at site $j'$ within the $(N-n_{h})$-site chain obtained after removing the hole sites $j_{1}, \cdots, j_{n_{h}}$. $j_{\alpha}(j_1, \cdots, j_{n_{h}}; j')$ is given by
\begin{align}
    \label{eq:DefinitionOfGeneralJalpha}
    j_{\alpha}(j_1, \cdots, j_{n_{h}}; j') = j' - \sum_{i=1}^{n_h} \Theta(j'-j_i).
\end{align}
For example, with $N=6$, $n_{h}=3$, $j_{1}=1$, $j_{2}=3$, and $j_{3}=4$, we have
$j_{\alpha}(j_1, j_{2}, j_{3}; 0) = 0$,
$j_{\alpha}(j_1, j_{2}, j_{3}; 2) = 1$, and
$j_{\alpha}(j_1, j_{2}, j_{3}; 5) = 2$.

More importantly, we have the following equalities, which will be used in the proof below.
\begin{align}
    \label{eq:equalities_for_jalpha}
    & j_{\alpha}(j_1-1, j_{2}, \cdots, j_{n_{h}}; j') = j_{\alpha}(j_1, j_{2}, \cdots, j_{n_{h}}; j'),  \ \text{when} \ j' \neq j_1, j_1-1,     \\ \nonumber
    & j_{\alpha}(j_1-1, j_{2}, \cdots, j_{n_{h}}; j_1) = j_{\alpha}(j_1, j_{2}, \cdots, j_{n_{h}}; j_1-1),     \\ \nonumber
    & j_{\alpha}(N-1, j_{2}, \cdots, j_{n_{h}}; j') = j_{\alpha}(0, j_{2}, \cdots, j_{n_{h}}; j') + 1, \\ \nonumber
    &j_{\alpha}(N-1, j_{2}, \cdots, j_{n_{h}}; 0) = j_{\alpha}(0, j_{2}, \cdots, j_{n_{h}}; N-1) - ( N-n_{h}-1).
\end{align}
The proof of these equalities is provided in Sec.~\ref{sec:Proofofequalitiesusedintheeigenstateconstruction}.

Next, we explicitly verify that $\ket{k_1,\cdots, k_{n_h}, k_{\alpha}}$ in Eq.~\eqref{eq:DefinitionOfEigenstateAnyholesOnedown} is an eigenstate with eigenenergy $\sum_{i=1}^{n_h} 2 \cos(k_{i})$ when $e^{i(N-n_h)k_{\alpha}} = 1$ and $e^{ik_{\alpha}}e^{iNk_i}=1$ for $i=1,\cdots, n_{h}$.

Under the action of the Hamiltonian $\hat{H}$,
\begin{align}
    \label{eq:AnyholesAnyDownHaction}
    \hat{H} \ket{k_1, \cdots, k_{n_{h}}, k_{\alpha}} &= - \sum_{j_{1}, \cdots, j_{n_h} = 0}^{N-1} e^{ij_1 k_1} \cdots e^{ij_{n_h} k_{n_h}} \sum_{j'=0}^{N-1} e^{ij_{\alpha}(j_1, \cdots, j_{n_h}; j') k_{\alpha}} \nonumber \\ 
    &  \left( \sum_{a=1}^{n_{h}} (c^{\dagger}_{j_a, \uparrow} c_{j_{a}+1, \uparrow} +  c^{\dagger}_{j_a, \uparrow} c_{j_{a}-1, \uparrow}) + (c^{\dagger}_{j',\uparrow} c_{j'+1,\uparrow} + c^{\dagger}_{j',\uparrow} c_{j'-1,\uparrow} + c^{\dagger}_{j'+1,\downarrow} c_{j',\downarrow} +c^{\dagger}_{j'-1,\downarrow} c_{j',\downarrow}) \right) \nonumber \\ 
    & c^{\dagger}_{j', \downarrow} c_{j', \uparrow} c_{j_1, \uparrow} \cdots c_{j_{n_h}, \uparrow}  \ket{\mathrm{FM}}  \nonumber \\ 
    &= \sum_{j_{1}, \cdots, j_{n_h} = 0}^{N-1} e^{ij_1 k_1} \cdots e^{ij_{n_h} k_{n_h}} \sum_{j'=0}^{N-1} e^{ij_{\alpha}(j_1, \cdots, j_{n_h}; j') k_{\alpha}} \nonumber \\ 
    &  c^{\dagger}_{j', \downarrow}c_{j', \uparrow} \left( (c_{j_1+1, \uparrow} + c_{j_1-1, \uparrow}) c_{j_2,\uparrow} \cdots c_{j_{n_h}, \uparrow} + \cdots + c_{j_1, \uparrow} \cdots c_{j_{n_h-1}, \uparrow} (c_{j_{n_h}+1, \uparrow} + c_{j_{n_h}-1, \uparrow}) \right) \ket{\mathrm{FM} } \nonumber \\ 
    & - \sum_{j_{1}, \cdots, j_{n_h} = 0}^{N-1} e^{ij_1 k_1} \cdots e^{ij_{n_h} k_{n_h}} \sum_{j'=0}^{N-1} e^{ij_{\alpha}(j_1, \cdots, j_{n_h}; j') k_{\alpha}}  \nonumber \\
    & \left( c_{j'+1, \uparrow} c_{j', \downarrow} +  c_{j'-1, \uparrow} c_{j', \downarrow} + c^{\dagger}_{j'+1, \downarrow} c_{j', \uparrow} + c^{\dagger}_{j'-1, \downarrow} c_{j', \uparrow}   \right)c_{j_1, \uparrow} \cdots c_{j_{n_h}, \uparrow} \ket{\mathrm{FM}} \nonumber \\ 
    &= \sum_{j_{1}, \cdots, j_{n_h} = 0}^{N-1} e^{ij_1 k_1} \cdots e^{ij_{n_h} k_{n_h}} \sum_{j'=0}^{N-1} e^{ij_{\alpha}(j_1, \cdots, j_{n_h}; j') k_{\alpha}}  \nonumber \\
    &  c^{\dagger}_{j', \downarrow}c_{j', \uparrow} \left( (c_{j_1+1, \uparrow} + c_{j_1-1, \uparrow}) c_{j_2,\uparrow} \cdots c_{j_{n_h}, \uparrow} + \cdots + c_{j_1, \uparrow} \cdots c_{j_{n_h-1}, \uparrow} (c_{j_{n_h}+1, \uparrow} + c_{j_{n_h}-1, \uparrow}) \right) \ket{\mathrm{FM} } \nonumber \\ 
    & - \sum_{j_{1}, \cdots, j_{n_h} = 0}^{N-1} e^{ij_1 k_1} \cdots e^{ij_{n_h} k_{n_h}} \sum_{j'=0}^{N-1} e^{ij_{\alpha}(j_1, \cdots, j_{n_h}; j') k_{\alpha}} \nonumber \\ 
    & \left( c^{\dagger}_{j'+1, \downarrow} c_{j', \uparrow} + c^{\dagger}_{j'-1, \downarrow} c_{j', \uparrow}   \right) c_{j_1, \uparrow} \cdots c_{j_{n_h}, \uparrow} \ket{\mathrm{FM}},
\end{align}
where the last equality follows because the operator $c^{\dagger}_{j', \downarrow}$ would otherwise create a doubly occupied site at $j'$ which is forbidden in the projected Hilbert space.

To simplify the subsequent derivation, we decompose $\hat{H} \ket{k_1, \cdots, k_{n_{h}}, k_{\alpha}}$ in the above into two parts $A$ and $B$ such that $\hat{H} \ket{k_1, \cdots, k_{n_h}, k_{\alpha}} = A + B$ and analyze each part separately. The first part $A$ of $\hat{H} \ket{k_1, \cdots, k_{n_{h}}, k_{\alpha}}$ is given by
\begin{align}
    \label{eq:AnyholesOnedown_PartA}
    \text{Part A} &= \sum_{j_{1}, \cdots, j_{n_h} = 0}^{N-1} e^{ij_1 k_1} \cdots e^{ij_{n_h} k_{n_h}} \sum_{j'=0}^{N-1} e^{ij_{\alpha}(j_1, \cdots, j_{n_h}; j') k_{\alpha}} \nonumber \\ 
    &  c^{\dagger}_{j', \downarrow}c_{j', \uparrow} \left( (c_{j_1+1, \uparrow} + c_{j_1-1, \uparrow}) c_{j_2,\uparrow} \cdots c_{j_{n_h}, \uparrow} + \cdots + c_{j_1, \uparrow} \cdots c_{j_{n_h-1}, \uparrow} (c_{j_{n_h}+1, \uparrow} + c_{j_{n_h}-1, \uparrow}) \right) \ket{\mathrm{FM}}  \nonumber  \\
    &= e^{-ik_1} \sum_{j_{1}, \cdots, j_{n_h} = 0}^{N-1} e^{i(j_1+1) k_1} \cdots e^{ij_{n_h} k_{n_h}} \sum_{\substack{j'=0 \\ j' \ne j_1}}^{N-1} e^{ij_{\alpha}(j_1, \cdots, j_{n_h}; j') k_{\alpha}}   c^{\dagger}_{j', \downarrow}c_{j', \uparrow} \left( c_{j_1+1, \uparrow}  c_{j_2,\uparrow} \cdots c_{j_{n_h}, \uparrow}   \right) \ket{\mathrm{FM}} \nonumber \\ 
    &+ e^{ik_1} \sum_{j_{1}, \cdots, j_{n_h} = 0}^{N-1} e^{i(j_1-1) k_1} \cdots e^{ij_{n_h} k_{n_h}} \sum_{\substack{j'=0 \\ j' \ne j_1}}^{N-1} e^{ij_{\alpha}(j_1, \cdots, j_{n_h}; j') k_{\alpha}}   c^{\dagger}_{j', \downarrow}c_{j', \uparrow} \left( c_{j_1-1, \uparrow}  c_{j_2,\uparrow} \cdots c_{j_{n_h}, \uparrow}   \right) \ket{\mathrm{FM}} \nonumber \\ 
    &+\,\cdots\, + e^{-ik_{n_{h}}} \sum_{j_{1}, \cdots, j_{n_h} = 0}^{N-1} e^{i j_1 k_1} \cdots e^{i(j_{n_h}+1) k_{n_h}} \sum_{\substack{j'=0 \\ j' \ne j_{n_h}}}^{N-1} e^{ij_{\alpha}(j_1, \cdots, j_{n_h}; j') k_{\alpha}}   c^{\dagger}_{j', \downarrow}c_{j', \uparrow} \left( c_{j_1, \uparrow}  c_{j_2,\uparrow} \cdots c_{j_{n_h}+1, \uparrow}  \right) \ket{\mathrm{FM} } \nonumber \\ 
    &+ e^{ik_{n_{h}}} \sum_{j_{1}, \cdots, j_{n_h} = 0}^{N-1} e^{i j_1 k_1} \cdots e^{i (j_{n_h}-1) k_{n_h}} \sum_{\substack{j'=0 \\ j' \ne j_{n_h}}}^{N-1} e^{ij_{\alpha}(j_1, \cdots, j_{n_h}; j') k_{\alpha}}   c^{\dagger}_{j', \downarrow}c_{j', \uparrow} \left( c_{j_1, \uparrow}  c_{j_2,\uparrow} \cdots c_{j_{n_h}-1, \uparrow}   \right) \ket{\mathrm{FM} }. 
\end{align}
Note that here we explicitly impose the constraint in the summation 
$\sum_{j'=0}^{N-1}$, since it cannot be automatically guaranteed by the fermionic creation operator. 
Moreover, we define
\begin{subequations}
\begin{align}
    A_{1,-} &= e^{-ik_1} \sum_{j_{1}, \cdots, j_{n_h} = 0}^{N-1} e^{i(j_1+1) k_1} \cdots e^{ij_{n_h} k_{n_h}} \sum_{\substack{j'=0 \\ j' \ne j_1}}^{N-1} e^{ij_{\alpha}(j_1, \cdots, j_{n_h}; j') k_{\alpha}}   c^{\dagger}_{j', \downarrow}c_{j', \uparrow} \left( c_{j_1+1, \uparrow}  c_{j_2,\uparrow} \cdots c_{j_{n_h}, \uparrow}   \right) \ket{\mathrm{FM} }, \label{eq:A1-} \\ 
    A_{1,+} &= e^{ik_1} \sum_{j_{1}, \cdots, j_{n_h} = 0}^{N-1} e^{i(j_1-1) k_1} \cdots e^{ij_{n_h} k_{n_h}} \sum_{\substack{j'=0 \\ j' \ne j_1}}^{N-1} e^{ij_{\alpha}(j_1, \cdots, j_{n_h}; j') k_{\alpha}}   c^{\dagger}_{j', \downarrow}c_{j', \uparrow} \left( c_{j_1-1, \uparrow}  c_{j_2,\uparrow} \cdots c_{j_{n_h}, \uparrow}   \right) \ket{\mathrm{FM}},  \label{eq:A1+} \\ \nonumber
    & \cdots  \\ 
     A_{n_{h}, -} & = e^{-ik_{n_{h}}} \sum_{j_{1}, \cdots, j_{n_h} = 0}^{N-1} e^{i j_1 k_1} \cdots e^{i(j_{n_h}+1) k_{n_h}} \sum_{\substack{j'=0 \\ j' \ne j_{n_h}}}^{N-1} e^{ij_{\alpha}(j_1, \cdots, j_{n_h}; j') k_{\alpha}}   c^{\dagger}_{j', \downarrow}c_{j', \uparrow} \left( c_{j_1, \uparrow}  c_{j_2,\uparrow} \cdots c_{j_{n_h}+1, \uparrow}  \right) \ket{\mathrm{FM}}, \label{eq:Anh-} \\ 
    A_{n_{h}, +} &= e^{ik_{n_{h}}} \sum_{j_{1}, \cdots, j_{n_h} = 0}^{N-1} e^{i j_1 k_1} \cdots e^{i (j_{n_h}-1) k_{n_h}} \sum_{\substack{j'=0 \\ j' \ne j_{n_h}}}^{N-1} e^{ij_{\alpha}(j_1, \cdots, j_{n_h}; j') k_{\alpha}}   c^{\dagger}_{j', \downarrow}c_{j', \uparrow} \left( c_{j_1, \uparrow}  c_{j_2,\uparrow} \cdots c_{j_{n_h}-1, \uparrow}   \right) \ket{\mathrm{FM}}, \label{eq:Anh+}
\end{align}
\end{subequations}
and such that $A$ defined in Eq.~\eqref{eq:AnyholesOnedown_PartA} is given by $A = \sum_{i=1}^{n_{h}} (A_{i,-} + A_{i, +})$.

The second part $B$ of $\hat{H} \ket{k_1, \cdots, k_{n_{h}}, k_{\alpha}}$ is given by 
\begin{align}
\label{eq:AnyholesOnedown_PartB}
 \text{Part B} &= - \sum_{j_{1}, \cdots, j_{n_h} = 0}^{N-1} e^{ij_1 k_1} \cdots e^{ij_{n_h} k_{n_h}} \sum_{j'=0}^{N-1} e^{ij_{\alpha}(j_1, \cdots, j_{n_h}; j') k_{\alpha}} \left( c^{\dagger}_{j'+1, \downarrow} c_{j', \uparrow} + c^{\dagger}_{j'-1, \downarrow} c_{j', \uparrow}   \right) c_{j_1, \uparrow} \cdots c_{j_{n_h}, \uparrow} \ket{\mathrm{FM}}  \nonumber \\
    &= - \sum_{j_{1}, \cdots, j_{n_h} = 0}^{N-1} e^{ij_1 k_1} \cdots e^{ij_{n_h} k_{n_h}} \sum_{j'=0}^{N-1} e^{ij_{\alpha}(j_1, \cdots, j_{n_h}; j') k_{\alpha}}  \nonumber \\
    & \quad \quad \quad \quad  \sum_{a=1}^{n_h}\left( c^{\dagger}_{j'+1, \downarrow} c_{j', \uparrow} \delta_{j'+1, j_{a}} + c^{\dagger}_{j'-1, \downarrow} c_{j', \uparrow} \delta_{j'-1, j_{a}}  \right) c_{j_1, \uparrow} \cdots c_{j_{n_h}, \uparrow} \ket{\mathrm{FM}}  \nonumber \\
&= - \sum_{j_{1}, \cdots, j_{n_h} = 0}^{N-1} e^{ij_1 k_1} \cdots e^{ij_{n_h} k_{n_h}} \sum_{a=1}^{n_h} e^{ij_{\alpha}(j_1, \cdots, j_{n_h}; j_a-1) k_{\alpha}}  \left( c^{\dagger}_{j_a, \downarrow} c_{j_a-1, \uparrow} \right) c_{j_1, \uparrow} \cdots c_{j_{n_h}, \uparrow} \ket{\mathrm{FM}}  \nonumber \\ 
& \quad  - \sum_{j_{1}, \cdots, j_{n_h} = 0}^{N-1} e^{ij_1 k_1} \cdots e^{ij_{n_h} k_{n_h}} \sum_{a=1}^{n_h} e^{ij_{\alpha}(j_1, \cdots, j_{n_h}; j_a+1) k_{\alpha}}  \left( c^{\dagger}_{j_a, \downarrow} c_{j_a+1, \uparrow}  \right) c_{j_1, \uparrow} \cdots c_{j_{n_h}, \uparrow} \ket{\mathrm{FM}}  \nonumber \\ 
&= e^{ik_1} \sum_{j_{1}, \cdots, j_{n_h} = 0}^{N-1} e^{i(j_1-1) k_1} \cdots e^{ij_{n_h} k_{n_h}}  e^{ij_{\alpha}(j_1, \cdots, j_{n_h}; j_1-1) k_{\alpha}}  \left( c^{\dagger}_{j_1, \downarrow} c_{j_1, \uparrow} \right) c_{j_1-1, \uparrow} \cdots c_{j_{n_h}, \uparrow} \ket{\mathrm{FM}}  \nonumber \\ 
&   + e^{-ik_1} \sum_{j_{1}, \cdots, j_{n_h} = 0}^{N-1} e^{i(j_1+1) k_1} \cdots e^{ij_{n_h} k_{n_h}}  e^{ij_{\alpha}(j_1, \cdots, j_{n_h}; j_1+1) k_{\alpha}}  \left( c^{\dagger}_{j_1, \downarrow} c_{j_1, \uparrow}  \right) c_{j_1+1, \uparrow} \cdots c_{j_{n_h}, \uparrow} \ket{\mathrm{FM}}  \nonumber \\ 
& +\,\cdots\, + e^{ik_{n_{h}}} \sum_{j_{1}, \cdots, j_{n_h} = 0}^{N-1} e^{i j_1 k_1} \cdots e^{i(j_{n_h}-1) k_{n_h}}  e^{ij_{\alpha}(j_1, \cdots, j_{n_h}; j_{n_{h}}-1) k_{\alpha}}  \left( c^{\dagger}_{j_{n_{h}}, \downarrow} c_{j_{n_{h}}, \uparrow} \right) c_{j_1, \uparrow} \cdots c_{j_{n_h}-1, \uparrow} \ket{\mathrm{FM}}  \nonumber \\
& + e^{-ik_{n_{h}}} \sum_{j_{1}, \cdots, j_{n_h} = 0}^{N-1} e^{i j_1 k_1} \cdots e^{i (j_{n_h}+1) k_{n_h}}  e^{ij_{\alpha}(j_1, \cdots, j_{n_h}; j_{n_{h}}+1) k_{\alpha}}  \left( c^{\dagger}_{j_{n_{h}}, \downarrow} c_{j_{n_{h}}, \uparrow}  \right) c_{j_1, \uparrow} \cdots c_{j_{n_h}+1, \uparrow} \ket{\mathrm{FM}}. 
\end{align}

Moreover, we define
\begin{subequations}
    \begin{align}
        B_{1,-} & =  e^{-ik_1} \sum_{j_{1}, \cdots, j_{n_h} = 0}^{N-1} e^{i(j_1+1) k_1} \cdots e^{ij_{n_h} k_{n_h}}  e^{ij_{\alpha}(j_1, \cdots, j_{n_h}; j_1+1) k_{\alpha}}  \left( c^{\dagger}_{j_1, \downarrow} c_{j_1, \uparrow}  \right) c_{j_1+1, \uparrow} \cdots c_{j_{n_h}, \uparrow} \ket{1 \cdots 1},  \label{eq:B1-} \\ 
        B_{1,+} &= e^{ik_1} \sum_{j_{1}, \cdots, j_{n_h} = 0}^{N-1} e^{i(j_1-1) k_1} \cdots e^{ij_{n_h} k_{n_h}}  e^{ij_{\alpha}(j_1, \cdots, j_{n_h}; j_1-1) k_{\alpha}}  \left( c^{\dagger}_{j_1, \downarrow} c_{j_1, \uparrow} \right) c_{j_1-1, \uparrow} \cdots c_{j_{n_h}, \uparrow} \ket{1 \cdots 1},  \label{eq:B1+} \\ \nonumber
        & \cdots \\ 
        B_{n_{h},-} & = e^{-ik_{n_{h}}} \sum_{j_{1}, \cdots, j_{n_h} = 0}^{N-1} e^{i j_1 k_1} \cdots e^{i (j_{n_h}+1) k_{n_h}}  e^{ij_{\alpha}(j_1, \cdots, j_{n_h}; j_{n_{h}}+1) k_{\alpha}}  \left( c^{\dagger}_{j_{n_{h}}, \downarrow} c_{j_{n_{h}}, \uparrow}  \right) c_{j_1, \uparrow} \cdots c_{j_{n_h}+1, \uparrow} \ket{1 \cdots 1},  \label{eq:Bnh-} \\ 
        B_{n_{h},+} &= e^{ik_{n_{h}}} \sum_{j_{1}, \cdots, j_{n_h} = 0}^{N-1} e^{i j_1 k_1} \cdots e^{i(j_{n_h}-1) k_{n_h}}  e^{ij_{\alpha}(j_1, \cdots, j_{n_h}; j_{n_{h}}-1) k_{\alpha}}  \left( c^{\dagger}_{j_{n_{h}}, \downarrow} c_{j_{n_{h}}, \uparrow} \right) c_{j_1, \uparrow} \cdots c_{j_{n_h}-1, \uparrow} \ket{1 \cdots 1}, \label{eq:Bnh+} 
    \end{align}
\end{subequations}
such that part $B$ defined in Eq.~\eqref{eq:AnyholesOnedown_PartB} is given by $B = \sum_{i=1}^{n_{h}} (B_{i,-} + B_{i,+})$.

Therefore, 
\begin{align}
    \hat{H} \ket{k_1, \cdots, k_{n_{h}}, k_{\alpha}}  &=\sum_{i=1}^{n_{h}} (A_{i,-} + A_{i,+}) +  \sum_{i=1}^{n_{h}} (B_{i,-} + B_{i,+})=\sum_{i=1}^{n_{h}} ( (A_{i,-} + B_{i,-}) + (A_{i,+} + B_{i,+})). 
\end{align}

We first consider $A_{1,-} + B_{1,-}$. We have
\begin{align}
   A_{1,-} + B_{1,-} & = e^{-ik_1} \sum_{j_{1}, \cdots, j_{n_h} = 0}^{N-1} e^{i(j_1+1) k_1} \cdots e^{ij_{n_h} k_{n_h}}   \left( (\sum_{\substack{j'=0 \\ j' \ne j_1}}^{N-1} e^{ij_{\alpha}(j_1, \cdots, j_{n_h}; j') k_{\alpha}}   c^{\dagger}_{j', \downarrow}c_{j', \uparrow} ) +  e^{ij_{\alpha}(j_1, \cdots, j_{n_h}; j_1+1) k_{\alpha}}    c^{\dagger}_{j_1, \downarrow} c_{j_1, \uparrow} \right) \nonumber \\
   & \quad \quad \quad \quad \quad \left( c_{j_1+1, \uparrow}  c_{j_2,\uparrow} \cdots c_{j_{n_h}, \uparrow}    \right) \ket{\mathrm{FM} } \nonumber \\ 
    & =e^{-ik_1} \sum_{j_2, \cdots, j_{n_{h}}=0}^{N-1} e^{ij_2 k_{2}} \cdots e^{ij_{n_h} k_{n_h}}  \sum_{j_{1} = 0}^{N-2} e^{i(j_1+1) k_1}  \nonumber \\ 
    & \quad \quad \quad \left( (\sum_{\substack{j'=0 \\ j' \ne j_1}}^{N-1} e^{ij_{\alpha}(j_1, \cdots, j_{n_h}; j') k_{\alpha}}   c^{\dagger}_{j', \downarrow}c_{j', \uparrow} ) +  e^{ij_{\alpha}(j_1, \cdots, j_{n_h}; j_1+1) k_{\alpha}} c^{\dagger}_{j_1, \downarrow} c_{j_1, \uparrow} \right)  \left( c_{j_1+1, \uparrow}  c_{j_2,\uparrow} \cdots c_{j_{n_h}, \uparrow}    \right) \ket{\mathrm{FM}} \nonumber  \\ 
    & + e^{-ik_1} \sum_{j_2, \cdots, j_{n_{h}}=0}^{N-1} e^{ij_2 k_{2}} \cdots e^{ij_{n_h} k_{n_h}}    e^{i N k_1} \nonumber\\ 
    & \quad \quad \quad  \left( (\sum_{\substack{j'=0 \\ j' \ne N-1}}^{N-1} e^{ij_{\alpha}(N-1, \cdots, j_{n_h}; j') k_{\alpha}}   c^{\dagger}_{j', \downarrow}c_{j', \uparrow} ) +  e^{ij_{\alpha}(N-1, \cdots, j_{n_h}; N) k_{\alpha}} c^{\dagger}_{N-1, \downarrow} c_{N-1, \uparrow} \right)  \left( c_{0, \uparrow}  c_{j_2,\uparrow} \cdots c_{j_{n_h}, \uparrow}    \right) \ket{\mathrm{FM}}  \nonumber \\ 
    & =e^{-ik_1} \sum_{j_2, \cdots, j_{n_{h}}=0}^{N-1} e^{ij_2 k_{2}} \cdots e^{ij_{n_h} k_{n_h}}  \sum_{j_{1} = 1}^{N-1} e^{i j_1 k_1}  \nonumber \\ 
    & \quad \quad \quad \left( (\sum_{\substack{j'=0 \\ j' \ne j_1-1}}^{N-1} e^{ij_{\alpha}(j_1-1, \cdots, j_{n_h}; j') k_{\alpha}}   c^{\dagger}_{j', \downarrow}c_{j', \uparrow} ) +  e^{ij_{\alpha}(j_1-1, \cdots, j_{n_h}; j_1) k_{\alpha}} c^{\dagger}_{j_1-1, \downarrow} c_{j_1-1, \uparrow} \right)  \left( c_{j_1, \uparrow}  c_{j_2,\uparrow} \cdots c_{j_{n_h}, \uparrow}    \right) \ket{\mathrm{FM}}  \nonumber \\ 
    & + e^{-ik_1} \sum_{j_2, \cdots, j_{n_{h}}=0}^{N-1} e^{ij_2 k_{2}} \cdots e^{ij_{n_h} k_{n_h}}    e^{i N k_1} \nonumber \\ 
    & \quad \quad \quad  \left( (\sum_{\substack{j'=0 \\ j' \ne N-1}}^{N-1} e^{ij_{\alpha}(N-1, \cdots, j_{n_h}; j') k_{\alpha}}   c^{\dagger}_{j', \downarrow}c_{j', \uparrow} ) +  e^{ij_{\alpha}(N-1, \cdots, j_{n_h}; 0) k_{\alpha}} c^{\dagger}_{N-1, \downarrow} c_{N-1, \uparrow} \right)  \left( c_{0, \uparrow}  c_{j_2,\uparrow} \cdots c_{j_{n_h}, \uparrow}    \right) \ket{\mathrm{FM}}  \nonumber \\ 
    &=e^{-ik_1} \sum_{j_2, \cdots, j_{n_{h}}=0}^{N-1} e^{ij_2 k_{2}} \cdots e^{ij_{n_h} k_{n_h}}  \sum_{j_{1} = 1}^{N-1} e^{i j_1 k_1}  \nonumber \\ 
    & \quad \quad \quad \left( (\sum_{\substack{j'=0 \\ j' \ne j_1-1}}^{N-1} e^{ij_{\alpha}(j_1, \cdots, j_{n_h}; j') k_{\alpha}}   c^{\dagger}_{j', \downarrow}c_{j', \uparrow} ) +  e^{ij_{\alpha}(j_1, \cdots, j_{n_h}; j_1-1) k_{\alpha}} c^{\dagger}_{j_1-1, \downarrow} c_{j_1-1, \uparrow} \right)  \left( c_{j_1, \uparrow}  c_{j_2,\uparrow} \cdots c_{j_{n_h}, \uparrow}    \right) \ket{\mathrm{FM}}  \nonumber \\ 
    & + e^{-ik_1} \sum_{j_2, \cdots, j_{n_{h}}=0}^{N-1} e^{ij_2 k_{2}} \cdots e^{ij_{n_h} k_{n_h}}    e^{i N k_1} \nonumber \\ 
    & \left( (\sum_{\substack{j'=0 \\ j' \ne N-1}}^{N-1} e^{ik_{\alpha}} e^{ij_{\alpha}(0, \cdots, j_{n_h}; j') k_{\alpha}}   c^{\dagger}_{j', \downarrow}c_{j', \uparrow} ) +  e^{-i (N-n_{h}-1) k_{\alpha}}e^{ij_{\alpha}(0, \cdots, j_{n_h}; N-1) k_{\alpha}} c^{\dagger}_{N-1, \downarrow} c_{N-1, \uparrow} \right) \nonumber \\
   &  \quad \quad \left( c_{0, \uparrow}  c_{j_2,\uparrow} \cdots c_{j_{n_h}, \uparrow}    \right) \ket{\mathrm{FM}}, 
\end{align}
where in the last equality we have used equalities in Eq.~\eqref{eq:equalities_for_jalpha}. Therefore, when $e^{iNk_1} e^{ik_{\alpha}} = 1$ and $e^{i(N-n_{h}) k_{\alpha}} = 1$,
\begin{align}
    A_{1,-} + B_{1,-} = e^{-ik_1} \ket{k_1, \cdots, k_{n_{h}}, k_{\alpha}}.
\end{align}
Similarly, we can also prove that when $e^{i(N-n_{h})k_{\alpha}}=1$ and $e^{iNk_{a}} e^{ik_{\alpha}}=1$ for $a=1,\cdots, n_{h}$, we have 
\begin{align}
        A_{a,-} + B_{a,-} = e^{-ik_a} \ket{k_1, \cdots, k_{n_{h}}, k_{\alpha}}.
\end{align}
See Sec.~\ref{sec:Anyholesandanyspin-downfermions} for the proof and further details.

We then consider $A_{1,+} + B_{1,+}$. We have 
\begin{align}
A_{1,+} + B_{1,+} &=
  e^{ik_1} \sum_{j_{1}, \cdots, j_{n_h} = 0}^{N-1} e^{i(j_1-1) k_1} \cdots e^{ij_{n_h} k_{n_h}}   \left( (\sum_{\substack{j'=0 \\ j' \ne j_1}}^{N-1} e^{ij_{\alpha}(j_1, \cdots, j_{n_h}; j') k_{\alpha}}   c^{\dagger}_{j', \downarrow}c_{j', \uparrow} ) +  e^{ij_{\alpha}(j_1, \cdots, j_{n_h}; j_1-1) k_{\alpha}}    c^{\dagger}_{j_1, \downarrow} c_{j_1, \uparrow} \right) \nonumber \\ 
   & \quad \quad \quad \quad \quad \left( c_{j_1-1, \uparrow}  c_{j_2,\uparrow} \cdots c_{j_{n_h}, \uparrow}    \right) \ket{\mathrm{FM}} \nonumber \\
    & =e^{ik_1} \sum_{j_2, \cdots, j_{n_{h}}=0}^{N-1} e^{ij_2 k_{2}} \cdots e^{ij_{n_h} k_{n_h}}  \sum_{j_{1} = 1}^{N-1} e^{i(j_1-1) k_1}  \nonumber \\ 
    & \quad \quad \quad \left( (\sum_{\substack{j'=0 \\ j' \ne j_1}}^{N-1} e^{ij_{\alpha}(j_1, \cdots, j_{n_h}; j') k_{\alpha}}   c^{\dagger}_{j', \downarrow}c_{j', \uparrow} ) +  e^{ij_{\alpha}(j_1, \cdots, j_{n_h}; j_1-1) k_{\alpha}} c^{\dagger}_{j_1, \downarrow} c_{j_1, \uparrow} \right)  \left( c_{j_1-1, \uparrow}  c_{j_2,\uparrow} \cdots c_{j_{n_h}, \uparrow}    \right) \ket{\mathrm{FM}}  \nonumber \\ 
    & + e^{ik_1} \sum_{j_2, \cdots, j_{n_{h}}=0}^{N-1} e^{ij_2 k_{2}} \cdots e^{ij_{n_h} k_{n_h}}    e^{-i k_1} \nonumber \\ 
    & \quad \quad \quad  \left( (\sum_{\substack{j'=0 \\ j' \ne 0}}^{N-1} e^{ij_{\alpha}(0, \cdots, j_{n_h}; j') k_{\alpha}}   c^{\dagger}_{j', \downarrow}c_{j', \uparrow} ) +  e^{ij_{\alpha}(0, \cdots, j_{n_h}; N-1) k_{\alpha}} c^{\dagger}_{0, \downarrow} c_{0, \uparrow} \right)  \left( c_{N-1, \uparrow}  c_{j_2,\uparrow} \cdots c_{j_{n_h}, \uparrow}    \right) \ket{\mathrm{FM}}  \nonumber \\ 
    & =e^{ik_1} \sum_{j_2, \cdots, j_{n_{h}}=0}^{N-1} e^{ij_2 k_{2}} \cdots e^{ij_{n_h} k_{n_h}}  \sum_{j_{1} = 0}^{N-2} e^{i j_1 k_1}  \nonumber \\ 
    & \quad \quad \quad \left( (\sum_{\substack{j'=0 \\ j' \ne j_1+1}}^{N-1} e^{ij_{\alpha}(j_1+1, \cdots, j_{n_h}; j') k_{\alpha}}   c^{\dagger}_{j', \downarrow}c_{j', \uparrow} ) +  e^{ij_{\alpha}(j_1+1, \cdots, j_{n_h}; j_1 ) k_{\alpha}} c^{\dagger}_{j_1+1, \downarrow} c_{j_1+1, \uparrow} \right)  \left( c_{j_1, \uparrow}  c_{j_2,\uparrow} \cdots c_{j_{n_h}, \uparrow}    \right) \ket{\mathrm{FM}}  \nonumber \\ 
    & + e^{ik_1} \sum_{j_2, \cdots, j_{n_{h}}=0}^{N-1} e^{ij_2 k_{2}} \cdots e^{ij_{n_h} k_{n_h}}    e^{-i k_1} \nonumber \\ 
    & \quad \quad \quad  \left( (\sum_{\substack{j'=0 \\ j' \ne 0}}^{N-1} e^{ij_{\alpha}(0, \cdots, j_{n_h}; j') k_{\alpha}}   c^{\dagger}_{j', \downarrow}c_{j', \uparrow} ) +  e^{ij_{\alpha}(0, \cdots, j_{n_h}; N-1) k_{\alpha}} c^{\dagger}_{0, \downarrow} c_{0, \uparrow} \right)  \left( c_{N-1, \uparrow}  c_{j_2,\uparrow} \cdots c_{j_{n_h}, \uparrow}    \right) \ket{\mathrm{FM}}  \nonumber \\ 
& =e^{ik_1} \sum_{j_2, \cdots, j_{n_{h}}=0}^{N-1} e^{ij_2 k_{2}} \cdots e^{ij_{n_h} k_{n_h}}  \sum_{j_{1} = 0}^{N-2} e^{i j_1 k_1}   \nonumber \\
    & \quad \quad \quad \left( (\sum_{\substack{j'=0 \\ j' \ne j_1+1}}^{N-1} e^{ij_{\alpha}(j_1, \cdots, j_{n_h}; j') k_{\alpha}}   c^{\dagger}_{j', \downarrow}c_{j', \uparrow} ) +  e^{ij_{\alpha}(j_1, \cdots, j_{n_h}; j_1+1 ) k_{\alpha}} c^{\dagger}_{j_1+1, \downarrow} c_{j_1+1, \uparrow} \right)  \left( c_{j_1, \uparrow}  c_{j_2,\uparrow} \cdots c_{j_{n_h}, \uparrow}    \right) \ket{\mathrm{FM}} \nonumber \\ 
    & + e^{ik_1} \sum_{j_2, \cdots, j_{n_{h}}=0}^{N-1} e^{ij_2 k_{2}} \cdots e^{ij_{n_h} k_{n_h}}    e^{-i k_1} \nonumber \\ 
    &   \left( (\sum_{\substack{j'=0 \\ j' \ne 0}}^{N-1} e^{-ik_\alpha}e^{ij_{\alpha}(N-1, \cdots, j_{n_h}; j') k_{\alpha}}   c^{\dagger}_{j', \downarrow}c_{j', \uparrow} ) +  e^{i (N-n_h-1) k_{\alpha}}  e^{ij_{\alpha}(N-1, \cdots, j_{n_h}; 0) k_{\alpha}} c^{\dagger}_{0, \downarrow} c_{0, \uparrow} \right)  \left( c_{N-1, \uparrow}  c_{j_2,\uparrow} \cdots c_{j_{n_h}, \uparrow}    \right) \ket{\mathrm{FM}}, 
\end{align}
where in the last equality we have used equalities in Eq.~\eqref{eq:equalities_for_jalpha}. Therefore, when $e^{i(N-n_h) k_{\alpha}} = 1$ and $e^{iNk_1} e^{ik_{\alpha}} = 1$,
\begin{align}
A_{1,+} + B_{1,+} = e^{ik_1} \ket{k_1, \cdots, k_{n_h}, k_{\alpha}}.
\end{align}
Similarly, we can also prove that when $e^{i(N-n_{h})k_{\alpha}}=1$ and $e^{iNk_{a}} e^{ik_{\alpha}}=1$ for $a=1,\cdots, n_{h}$, we have 
\begin{align}
        A_{a,+} + B_{a,+} = e^{ik_a} \ket{k_1, \cdots, k_{n_{h}}, k_{\alpha}}.
\end{align}
See Sec.~\ref{sec:Anyholesandanyspin-downfermions} for the proof and further details.

In conclusion, when
\begin{align}
    e^{i(N-n_{h})k_{\alpha}}&=1, \\ 
    e^{ik_{\alpha}} e^{iNk_{a}}&=1, \quad a = 1, \cdots, n_{h},
\end{align}
we have
\begin{align}
    \hat{H} \ket{k_1,\cdots, k_{n_h}, k_{\alpha}} = \sum_{a=1}^{n_h} 2 \cos(k_{a}) \ket{k_1,\cdots, k_{n_h}, k_{\alpha}}.
\end{align}
Moreover, $k_{\alpha}$ and $k_{a}$ with $a=1, \cdots, n_{h}$ can be expressed as 
\begin{align}
    k_{\alpha} &= -\frac{2\pi m_{\downarrow}}{N-n_{h}}, \quad \mathrm{with} \quad m_{\downarrow} = 0, \cdots, N-n_h-1, \\ 
    k_{a} &= \frac{2\pi m_{a}}{N} - \frac{k_{\alpha}}{N} = \frac{2\pi m_a}{N} + \frac{2 \pi m_{\downarrow}}{N(N-n_h)}, \quad a = 1, \cdots, n_{h}.
\end{align}
Since any two $k_{i}$ and $k_{j}$ are distinct, $0 \leq m_1 < m_2 < \cdots < m_{n_h} \leq N-1$.

\subsubsection{Any number of spin-down fermions}
\label{sec:Anyholesandanyspin-downfermions}
In this section, we derive the eigenstates and the energy spectrum for arbitrary numbers of holes and spin-down fermions.

In the sector with $n_{h}$ holes and $n_{\downarrow}$ spin-down fermions, the Hilbert-space dimension is $\binom{N}{n_h}\binom{N-n_h}{n_{\downarrow}}$.
We assume that $N-n_{h}$ is a prime number. After removing the $n_{h}$ sites occupied by holes, the remaining system forms a $(N-n_{h})$-site chain. In this chain there are $\binom{N-n_h}{n_{\downarrow}}$
possible spin configurations and these $\binom{N-n_h}{n_{\downarrow}}$ spin configurations form $\frac{\binom{N-n_{h}}{n_{\downarrow}}}{N-n_{h}}$
orbits under translation, and each orbit has length $N-n_{h}$. See the proof in Sec.~\ref{sec:Single_hole_and_any_spin-down_fermions}.

For a given spin configuration, we denote by $x_{i}$ with $i=1, \cdots, n_{\downarrow}$ the position of $i$-th spin-down fermion in the $(N-n_{h})$-chain and denote by $d_{i}=x_{i+1}-x_{i}$ the distance between $(i+1)$-th and $i$-th spin-down fermions in the $(N-n_{h})$-site chain. Note that $d_{i}$ is defined modulo $N-n_{h}$. Here, we show an example. When $N=9$, $n_{h}=2$, $n_{\downarrow}=3$, in the remaining $7$-site chain, there are $35$ possible spin configurations:
\begin{align}
    \label{eq:ExampleOfOrbit1}
    s^{(1)}_{1} &= \{ -1, -1, -1, +1, +1, +1, +1 \}; \quad (d_1, d_2, d_3) = (1,1,5) \\ \nonumber
    s^{(1)}_{2} &= \{ +1, -1, -1, -1, +1, +1, +1 \}; \quad (d_1, d_2, d_3) = (1,1,5) \\ \nonumber
    s^{(1)}_{3} &= \{ +1, +1, -1, -1, -1, +1, +1 \}; \quad (d_1, d_2, d_3) = (1,1,5) \\ \nonumber
    s^{(1)}_{4} &= \{ +1, +1, +1, -1, -1, -1, +1 \}; \quad (d_1, d_2, d_3) = (1,1,5) \\ \nonumber
    s^{(1)}_{5} &= \{ +1, +1, +1, +1, -1, -1, -1 \}; \quad (d_1, d_2, d_3) = (1,1,5) \\ \nonumber
    s^{(1)}_{6} &= \{ -1, +1, +1, +1, +1, -1, -1 \}; \quad (d_1, d_2, d_3) = (5,1,1) \\ \nonumber
    s^{(1)}_{7} &= \{ -1, -1, +1, +1, +1, +1, -1 \}; \quad (d_1, d_2, d_3) = (1,5,1)
\end{align}

\begin{align}
\label{eq:ExampleOfOrbit2}
    s^{(2)}_{1} &= \{ -1, -1, +1, -1, +1, +1, +1 \}; \quad (d_1, d_2, d_3) = (1,2,4) \\ \nonumber
    s^{(2)}_{2} &= \{ +1, -1, -1, +1, -1, +1, +1 \}; \quad (d_1, d_2, d_3) = (1,2,4) \\ \nonumber
    s^{(2)}_{3} &= \{ +1, +1, -1, -1, +1, -1, +1 \}; \quad (d_1, d_2, d_3) = (1,2,4) \\ \nonumber
    s^{(2)}_{4} &= \{ +1, +1, +1, -1, -1, +1, -1 \}; \quad (d_1, d_2, d_3) = (1,2,4) \\ \nonumber
    s^{(2)}_{5} &= \{ -1, +1, +1, +1, -1, -1, +1 \}; \quad (d_1, d_2, d_3) = (4,1,2) \\ \nonumber
    s^{(2)}_{6} &= \{ +1, -1, +1, +1, +1, -1, -1 \}; \quad (d_1, d_2, d_3) = (4,1,2) \\ \nonumber
    s^{(2)}_{7} &= \{ -1, +1, -1, +1, +1, +1, -1 \}; \quad (d_1, d_2, d_3) = (2,4,1)
\end{align}

\begin{align}
\label{eq:ExampleOfOrbit3}
    s^{(3)}_{1} &= \{ -1, -1, +1, +1, -1, +1, +1 \}; \quad (d_1, d_2, d_3) = (1,3,3) \\ \nonumber
    s^{(3)}_{2} &= \{ +1, -1, -1, +1, +1, -1, +1 \}; \quad (d_1, d_2, d_3) = (1,3,3) \\ \nonumber
    s^{(3)}_{3} &= \{ +1, +1, -1, -1, +1, +1, -1 \}; \quad (d_1, d_2, d_3) = (1,3,3) \\ \nonumber
    s^{(3)}_{4} &= \{ -1, +1, +1, -1, -1, +1, +1 \}; \quad (d_1, d_2, d_3) = (3,1,3) \\ \nonumber
    s^{(3)}_{5} &= \{ +1, -1, +1, +1, -1, -1, +1 \}; \quad (d_1, d_2, d_3) = (3,1,3) \\ \nonumber
    s^{(3)}_{6} &= \{ +1, +1, -1, +1, +1, -1, -1 \}; \quad (d_1, d_2, d_3) = (3,1,3) \\ \nonumber
    s^{(3)}_{7} &= \{ -1, +1, +1, -1, +1, +1, -1 \}; \quad (d_1, d_2, d_3) = (3,3,1)
\end{align}

\begin{align}
\label{eq:ExampleOfOrbit4}
    s^{(4)}_{1} &= \{ -1, -1, +1, +1, +1, -1, +1 \}; \quad (d_1, d_2, d_3) = (1,4,2) \\ \nonumber
    s^{(4)}_{2} &= \{ +1, -1, -1, +1, +1, +1, -1 \}; \quad (d_1, d_2, d_3) = (1,4,2) \\ \nonumber
    s^{(4)}_{3} &= \{ -1, +1, -1, -1, +1, +1, +1 \}; \quad (d_1, d_2, d_3) = (2,1,4) \\ \nonumber
    s^{(4)}_{4} &= \{ +1, -1, +1, -1, -1, +1, +1 \}; \quad (d_1, d_2, d_3) = (2,1,4) \\ \nonumber
    s^{(4)}_{5} &= \{ +1, +1, -1, +1, -1, -1, +1 \}; \quad (d_1, d_2, d_3) = (2,1,4) \\ \nonumber
    s^{(4)}_{6} &= \{ +1, +1, +1, -1, +1, -1, -1 \}; \quad (d_1, d_2, d_3) = (2,1,4) \\ \nonumber
    s^{(4)}_{7} &= \{ -1, +1, +1, +1, -1, +1, -1 \}; \quad (d_1, d_2, d_3) = (4,2,1)
\end{align}

\begin{align}
\label{eq:ExampleOfOrbit5}
    s^{(5)}_{1} &= \{ -1, +1, -1, +1, -1, +1, +1 \}; \quad (d_1, d_2, d_3) = (2,2,3) \\ \nonumber
    s^{(5)}_{2} &= \{ +1, -1, +1, -1, +1, -1, +1 \}; \quad (d_1, d_2, d_3) = (2,2,3) \\ \nonumber
    s^{(5)}_{3} &= \{ +1, +1, -1, +1, -1, +1, -1 \}; \quad (d_1, d_2, d_3) = (2,2,3) \\ \nonumber
    s^{(5)}_{4} &= \{ -1, +1, +1, -1, +1, -1, +1 \}; \quad (d_1, d_2, d_3) = (3,2,2) \\ \nonumber
    s^{(5)}_{5} &= \{ +1, -1, +1, +1, -1, +1, -1 \}; \quad (d_1, d_2, d_3) = (3,2,2) \\ \nonumber
    s^{(5)}_{6} &= \{ -1, +1, -1, +1, +1, -1, +1 \}; \quad (d_1, d_2, d_3) = (2,3,2) \\ \nonumber
    s^{(5)}_{7} &= \{ +1, -1, +1, -1, +1, +1, -1 \}; \quad (d_1, d_2, d_3) = (2,3,2)
\end{align}

If $(d_1, d_2, \cdots, d_{n_{\downarrow}})$ and $(d'_1, d'_2, \cdots, d'_{n_{\downarrow}})$ are related by a cyclic permutation, the corresponding two spin configurations belong to the same orbit. For example, the $7$ spin configurations shown in Eq.~\eqref{eq:ExampleOfOrbit1} belong to the same orbit
\begin{align}
   s^{(1)}_1 \xrightarrow{\hat{T}} s^{(1)}_2  \xrightarrow{\hat{T}} s^{(1)}_3 \xrightarrow{\hat{T}} s^{(1)}_4 \xrightarrow{\hat{T}} s^{(1)}_5 \xrightarrow{\hat{T}} s^{(1)}_6 \xrightarrow{\hat{T}} s^{(1)}_7
   \xrightarrow{\hat{T}} s^{(1)}_1.
 \end{align}
The spin configurations shown in Eqs.~\eqref{eq:ExampleOfOrbit2}–\eqref{eq:ExampleOfOrbit5} form four additional orbits.

For each orbit, we choose a representative spin configuration
with $(d_1, \cdots, d_{n_{\downarrow}})$ and designate the left-most
spin-down fermion as the first spin-down fermion.
Once the representative configuration is fixed, the other spin
configurations in the orbit are obtained by applying the translation
operator to it. In these translated configurations, the first
spin-down fermion is defined as the translation of the first
spin-down fermion in the representative configuration.

Then we define the following state
\begin{align}
    \ket{k_1, \cdots, k_{n_h}, k_{\alpha}, (d_1,\cdots, d_{n_{\downarrow}})} &= \sum_{j_1, \cdots, j_{n_{h}}=0}^{N-1} \sum_{j'=0}^{N-1} e^{ij_{\alpha}(j_1, \cdots, j_{n_h};j')k_{\alpha}} e^{ij_1 k_1} \cdots e^{ij_{n_h} k_{n_h}}  c^{\dagger}_{j', \downarrow} c_{j',\uparrow} \nonumber \\ 
    & c^{\dagger}_{r(j_1, \cdots, j_{n_h}; j', (d_1, \cdots, d_{n_{\downarrow}}))_2, \downarrow} c_{r(j_1, \cdots, j_{n_h}; j', (d_1, \cdots, d_{n_{\downarrow}}))_2, \uparrow} \cdots \nonumber\\  & c^{\dagger}_{r(j_1, \cdots, j_{n_h}; j', (d_1, \cdots, d_{n_{\downarrow}}))_{n_{\downarrow}}, \downarrow} c_{r(j_1, \cdots, j_{n_h}; j', (d_1, \cdots, d_{n_{\downarrow}}))_{n_{\downarrow}}, \uparrow} \nonumber \\ 
    & c_{j_{1}, \uparrow} \cdots c_{j_{n_{h}},\uparrow} \ket{\mathrm{FM}}  \nonumber \\
    & = \sum_{j_1, \cdots, j_{n_{h}}=0}^{N-1} \sum_{j'=0}^{N-1} e^{ij_{\alpha}(j_1, \cdots, j_{n_h};j')k_{\alpha}} e^{ij_1 k_1} \cdots e^{ij_{n_h} k_{n_h}}   \nonumber \\ 
    & \left( \prod_{i=1}^{n_\downarrow} c^{\dagger}_{r(j_1, \cdots, j_{n_h}; j', (d_1, \cdots, d_{n_{\downarrow}}))_i, \downarrow} c_{r(j_1, \cdots, j_{n_h}; j', (d_1, \cdots, d_{n_{\downarrow}}))_i, \uparrow} \right)  \left(  \prod_{i=1}^{n_{h}} c_{j_{i}, \uparrow} \right) \ket{\mathrm{FM}}.
\end{align}
For simplicity, we first introduce the shorthands used in the following derivation
\begin{subequations}\label{eq:definitions_vectors_and_sums}
\begin{align}
    \mathbf{k} &\equiv (k_1, \cdots, k_{n_h}),
    \label{eq:def_k} \\ 
    \mathbf{d} &\equiv (d_1, \cdots, d_{n_{\downarrow}}),
    \label{eq:def_d} \\ 
    \mathbf{j} &\equiv (j_1, \cdots, j_{n_h}),
    \label{eq:def_j} \\ 
    \mathbf{j}^{(a+1)} &\equiv (j_1, \cdots, j_{a-1}, j_a+1, j_{a+1}, \cdots, j_{n_h}),
    \label{eq:def_jap1} \\ 
    \mathbf{j}^{(a-1)} &\equiv (j_1, \cdots, j_{a-1}, j_a-1, j_{a+1}, \cdots, j_{n_h}),
    \label{eq:def_jam1} \\ 
    \mathbf{j}_{a} &\equiv (j_1, \cdots, j_{a-1}, j_{a+1}, \cdots, j_{n_h}),
    \label{eq:def_ja} \\ 
    \sum_{\mathbf{j}} &\equiv \sum_{j_1,\cdots,j_{n_h}=0}^{N-1},
    \label{eq:def_sumj} \\ 
    \sum_{\mathbf{j}_a} &\equiv
    \sum_{j_1,\cdots,j_{a-1},\,j_{a+1},\cdots,j_{n_h}=0}^{N-1}.
    \label{eq:def_sumja}
\end{align}
\end{subequations}
Since $\mathbf{j}_a$ excludes the component $j_a$, we may also write $\mathbf{j}=(\mathbf{j}_a,j_a)$.

For $i=1,\ldots,n_{\downarrow}$, the function $r(j_1,\ldots,j_{n_h};j',(d_1,\ldots,d_{n_{\downarrow}}))_i\equiv r(\mathbf{j};j',\mathbf{d})_i$ gives the position of the $i$-th spin-down fermion on the $N$-site chain. Here, the holes occupy the sites specified by $\mathbf{j}=(j_1,\ldots,j_{n_h})$, the first spin-down fermion is located at site $j'$, and the spin configuration is specified by $\mathbf{d}=(d_1,\ldots,d_{n_\downarrow})$. The function $r(\mathbf{j};j',\mathbf{d})_i$ is given by
\begin{align}
\label{eq:DefinitionOfRfunction}
r(\mathbf{j};j', \mathbf{d})_i
=
j_{\alpha}^{-1}\!\left(
\mathbf{j};
\,
j_{\alpha}(\mathbf{j};j')
+\sum_{\ell=1}^{i-1} d_\ell
\;\; (\mathrm{mod}\; N-n_h)
\right).
\end{align}
See Eq.~\eqref{eq:DefinitionOfGeneralJalpha} for the definition of $j_{\alpha}(\mathbf{j}; j')$. The function $j_{\alpha}^{-1}$ denotes the inverse of $j_{\alpha}$, i.e., $j_{\alpha}^{-1}(\mathbf{j}; j')$ gives the position of the spin-down fermion in the $N$-site chain when its position in the $(N-n_h)$-site chain is $j'$, with the hole positions specified by $\mathbf{j}$. It is defined as 
\begin{align}
    j_\alpha^{-1}(\mathbf j; j')
    =
     j'+\sum_{m=1}^{n_h}\Theta\!\bigl(j'-(\tilde{j}_m-m +\frac{1}{2})\bigr),
\end{align}
where $
    \tilde j_1<\tilde j_2<\cdots<\tilde j_{n_h} $
    are the hole positions obtained by sorting the set $\{j_1,\cdots,j_{n_h}\}$ in ascending order.

We further define
$l(\mathbf{j}; j'', \mathbf{d})_i$
to denote the position of the first spin-down fermion in the original
$N$-site chain when the $i$-th spin-down fermion is located at site $j''$.
Since the position of the $i$-th spin-down fermion in the reduced
$(N-n_h)$-site chain differs from that of the first spin-down fermion by
$\sum_{\ell=1}^{i-1} d_\ell$, we have
\begin{align}
\label{eq:DefinitionOfLfunction}
l(\mathbf{j}; j'', \mathbf{d})_i
=
j_{\alpha}^{-1}\!\left(
\mathbf{j};
\, j_{\alpha}(\mathbf{j}; j'')
-\sum_{\ell=1}^{i-1} d_\ell
\pmod{N-n_h}
\right).
\end{align}
Consequently, we have
\begin{align}
r(\mathbf{j};
\, l(\mathbf{j}; j'', \mathbf{d})_i,
\, \mathbf{d})_i
= j''.
\end{align}

With the shorthand notation defined above (Eqs.~\eqref{eq:def_k}-\eqref{eq:def_sumja}), the state $ \ket{k_1, \cdots, k_{n_h}, k_{\alpha}, (d_1,\cdots, d_{n_{\downarrow}})}$ can be rewritten as
\begin{align}
    \label{eq:anyholesanydown_eigenstate}
    \ket{\mathbf{k},k_{\alpha}, \mathbf{d}} = \sum_{\mathbf{j}} \sum_{j'=0}^{N-1} e^{ij_{\alpha}(\mathbf{j};j')k_{\alpha}} e^{i \mathbf{j} \cdot \mathbf{k}} \left( \prod_{i=1}^{n_{\downarrow}} c^{\dagger}_{r(\mathbf{j}; j', \mathbf{d})_i, \downarrow} c_{r(\mathbf{j}; j', \mathbf{d})_i, \uparrow} \right) \left( \prod_{i=1}^{n_{h}} c_{j_{i}, \uparrow} \right) \ket{\mathrm{FM}}.
\end{align}

Under the action of Hamiltonian $\hat{H}$,
\begin{align}
\label{eq:anyholesanydown_eigenstate_withH}
    \hat{H} \ket{\mathbf{k}, k_{\alpha}, \mathbf{d}} &= - \sum_{\mathbf{j}}\sum_{j'=0}^{N-1} e^{ij_{\alpha}(\mathbf{j}; j') k_{\alpha}} e^{i \mathbf{j} \cdot \mathbf{k}}  \nonumber \\ 
    & \left(\sum_{a=1}^{n_h} c^{\dagger}_{j_a, \uparrow}( c_{j_a+1, \uparrow}  + c_{j_a-1,\uparrow}) +  \sum_{b=1}^{n_{\downarrow}} (c^{\dagger}_{r(\mathbf{j};j',\mathbf{d})_b +1, \downarrow} + c^{\dagger}_{r(\mathbf{j};j',\mathbf{d})_b -1, \downarrow})c_{r(\mathbf{j};j', \mathbf{d})_b, \downarrow}  \right) \nonumber \\ 
&  \left( \prod_{i=1}^{n_\downarrow} c^{\dagger}_{r(\mathbf{j}; j', \mathbf{d})_i, \downarrow} c_{r(\mathbf{j}; j', \mathbf{d})_i, \uparrow} \right)  \left(  \prod_{i=1}^{n_{h}} c_{j_{i}, \uparrow} \right) \ket{\mathrm{FM}} \nonumber \\ 
& = \sum_{a=1}^{n_{h}} (A_{a,-} + A_{a,+} + B_{a,-} + B_{a,+}).
\end{align}

The $A_{a,-}$ with $a=1,\cdots,n_{h}$ in Eq.~\eqref{eq:anyholesanydown_eigenstate_withH} is given by
\begin{align}
    \label{eq:anyholeanydownAa-}
    A_{a,-} & = - \sum_{\mathbf{j}} \sum_{j'=0}^{N-1} e^{ij_{\alpha}(\mathbf{j}; j') k_{\alpha}} e^{i \mathbf{j} \cdot \mathbf{k}}  c^{\dagger}_{j_a, \uparrow} c_{j_a+1, \uparrow}  \left( \prod_{i=1}^{n_\downarrow} c^{\dagger}_{r(\mathbf{j}; j', \mathbf{d})_i, \downarrow} c_{r(\mathbf{j}; j', \mathbf{d})_i, \uparrow} \right)  \left(  \prod_{i=1}^{n_{h}} c_{j_{i}, \uparrow} \right) \ket{\mathrm{FM}} \nonumber \\ 
    & = e^{-ik_a} \sum_{\mathbf{j}}\sum_{j'=0}^{N-1} e^{ij_{\alpha}(\mathbf{j}; j')k_{\alpha}} e^{i \mathbf{j}^{(a+1)} \cdot \mathbf{k}}   \left( \prod_{i=1}^{n_\downarrow} c^{\dagger}_{r(\mathbf{j}; j', \mathbf{d})_i, \downarrow} c_{r(\mathbf{j}; j', \mathbf{d})_i, \uparrow} \right)  c_{j_1, \uparrow} \cdots c_{j_{a-1}, \uparrow} c_{j_{a}+1, \uparrow} c_{j_{a+1}, \uparrow} \cdots c_{j_{n_h}, \uparrow} \ket{\mathrm{FM}},
\end{align}
where we have used $e^{i\mathbf{j} \cdot \mathbf{k}} = e^{-ik_{a}} e^{i \mathbf{j}^{(a+1)} \cdot \mathbf{k}}$ since $\mathbf{j}^{(a+1)} = (j_{1}, \cdots, j_{a-1}, j_{a}+1, j_{a+1}, \cdots, j_{n_h})$. Note that in the summation over $j'$ in $A_{a,-}$, we impose the constraints for any $i=1,\cdots,n_{\downarrow}$
\begin{align}
    \label{eq:anyholeanydown_Aa-_constraint}
    r(\mathbf{j};j',\mathbf{d})_i &\neq j_{b}, \quad b=1,\cdots, n_{h}, \\ \nonumber 
    r(\mathbf{j};j',\mathbf{d})_i &\neq j_{a} + 1.  
\end{align}
Apart from the constraint $r(\mathbf{j}; j', \mathbf{d})_i \neq j_{a}$, the remaining constraints are automatically satisfied by the fermionic creation operators.

The $A_{a,+}$ with $a=1,\cdots,n_{h}$ in Eq.~\eqref{eq:anyholesanydown_eigenstate_withH} is given by
\begin{align}
    \label{eq:anyholeanydownAa+}
    A_{a,+} & = - \sum_{\mathbf{j}} \sum_{j'=0}^{N-1} e^{ij_{\alpha}(\mathbf{j}; j') k_{\alpha}} e^{i \mathbf{j} \cdot \mathbf{k}}  c^{\dagger}_{j_a, \uparrow} c_{j_a-1, \uparrow}  \left( \prod_{i=1}^{n_\downarrow} c^{\dagger}_{r(\mathbf{j}; j', \mathbf{d})_i, \downarrow} c_{r(\mathbf{j}; j', \mathbf{d})_i, \uparrow} \right)  \left(  \prod_{i=1}^{n_{h}} c_{j_{i}, \uparrow} \right) \ket{\mathrm{FM}} \nonumber \\ 
    & = e^{ik_a} \sum_{\mathbf{j}}\sum_{j'=0}^{N-1} e^{ij_{\alpha}(\mathbf{j}; j')k_{\alpha}} e^{i \mathbf{j}^{(a-1)} \cdot \mathbf{k}}   \left( \prod_{i=1}^{n_\downarrow} c^{\dagger}_{r(\mathbf{j}; j', \mathbf{d})_i, \downarrow} c_{r(\mathbf{j}; j', \mathbf{d})_i, \uparrow} \right)  c_{j_1, \uparrow} \cdots c_{j_{a-1}, \uparrow} c_{j_{a}-1, \uparrow} c_{j_{a+1}, \uparrow} \cdots c_{j_{n_h}, \uparrow} \ket{\mathrm{FM}},
\end{align}
where we have used $e^{i\mathbf{j} \cdot \mathbf{k}} = e^{ik_{a}} e^{i \mathbf{j}^{(a-1)} \cdot \mathbf{k}}$ since $\mathbf{j}^{(a-1)} = (j_{1}, \cdots, j_{a-1}, j_{a}-1, j_{a+1}, \cdots, j_{n_h})$. Note that in the summation over $j'$ in $A_{a,+}$, we impose the constraints for any $i=1,\cdots,n_{\downarrow}$
\begin{align}
    \label{eq:anyholeanydown_Aa+_constraint}
    r(\mathbf{j};j',\mathbf{d})_i &\neq j_{b}, \quad b=1,\cdots, n_{h}, \\ \nonumber
    r(\mathbf{j};j',\mathbf{d})_i &\neq j_{a} - 1.  
\end{align}
Apart from the constraint $r(\mathbf{j};j',\mathbf{d})_i \neq j_{a}$, the remaining constraints are automatically satisfied due to the fermion creation operators.

The $B_{a,-}$ with $a=1,\cdots,n_{h}$ in Eq.~\eqref{eq:anyholesanydown_eigenstate_withH} is given by
\begin{align}
    \label{eq:anyholeanydownBa-}
    B_{a,-} &= - \sum_{\mathbf{j}} \sum_{j'=0}^{N-1} e^{ij_{\alpha}(\mathbf{j}; j') k_{\alpha}} e^{i \mathbf{j} \cdot \mathbf{k}}  \left( \sum_{b=1}^{n_{\downarrow}}  c^{\dagger}_{r(\mathbf{j};j',\mathbf{d})_b -1, \downarrow} c_{r(\mathbf{j};j',\mathbf{d})_b, \downarrow}  \right)  \left( \prod_{i=1}^{n_\downarrow} c^{\dagger}_{r(\mathbf{j}; j', \mathbf{d})_i, \downarrow} c_{r(\mathbf{j}; j', \mathbf{d})_i, \uparrow} \right)  \left(  \prod_{i=1}^{n_{h}} c_{j_{i}, \uparrow} \right) \ket{\mathrm{FM}} \nonumber \\ 
&= - \sum_{\mathbf{j}} \sum_{j'=0}^{N-1} e^{ij_{\alpha}(\mathbf{j}; j') k_{\alpha}} e^{i \mathbf{j} \cdot \mathbf{k}} \left(   \sum_{b=1}^{n_{\downarrow}}  c^{\dagger}_{r(\mathbf{j};j',\mathbf{d})_b -1, \downarrow} c_{r(\mathbf{j};j',\mathbf{d})_b, \uparrow} \right)  \left( \prod_{\substack{i=1 \\ i \neq b}}^{n_\downarrow} c^{\dagger}_{r(\mathbf{j}; j', \mathbf{d})_i, \downarrow} c_{r(\mathbf{j}; j', \mathbf{d})_i, \uparrow} \right)  \left(  \prod_{i=1}^{n_{h}} c_{j_{i}, \uparrow} \right) \ket{\mathrm{FM}} \nonumber \\ 
&= - \sum_{\mathbf{j}} \sum_{j'=0}^{N-1} e^{ij_{\alpha}(\mathbf{j}; j')k_{\alpha}} e^{i\mathbf{j} \cdot \mathbf{k}}  \sum_{b=1}^{n_{\downarrow}}  c^{\dagger}_{r(\mathbf{j};j',\mathbf{d})_b -1, \downarrow} c_{r(\mathbf{j};j',\mathbf{d})_b, \uparrow} \delta_{r(\mathbf{j};j',\mathbf{d})_b -1, j_{a}}   \left( \prod_{\substack{i=1 \\ i \neq b}}^{n_\downarrow} c^{\dagger}_{r(\mathbf{j}; j', \mathbf{d})_i, \downarrow} c_{r(\mathbf{j}; j', \mathbf{d})_i, \uparrow} \right)  \left(  \prod_{i=1}^{n_{h}} c_{j_{i}, \uparrow} \right) \ket{\mathrm{FM}}  \nonumber \\
&= - \sum_{\mathbf{j}} \sum_{b=1}^{n_{\downarrow}} e^{i\mathbf{j} \cdot \mathbf{k}}  e^{ij_{\alpha}(\mathbf{j}; l(\mathbf{j}; j_{a}+1, \mathbf{d})_{b}) k_{\alpha}}    c^{\dagger}_{j_{a}, \downarrow} c_{j_{a}+1, \uparrow}   \left( \prod_{\substack{i=1 \\ i \neq b } }^{n_\downarrow} c^{\dagger}_{r(\mathbf{j}; l(\mathbf{j}; j_{a}+1, \mathbf{d})_{b}, \mathbf{d})_i, \downarrow} c_{r(\mathbf{j}; l(\mathbf{j}; j_{a}+1, \mathbf{d})_{b}, \mathbf{d})_i, \uparrow} \right)  \left(  \prod_{i=1}^{n_{h}} c_{j_{i}, \uparrow} \right) \ket{\mathrm{FM}} \nonumber \\ 
& = e^{-ik_a} \sum_{\mathbf{j}} \sum_{b=1}^{n_{\downarrow}} e^{i \mathbf{j}^{(a+1)} \cdot \mathbf{k}}  e^{ij_{\alpha}(\mathbf{j}; l(\mathbf{j}; j_{a}+1, \mathbf{d})_{b}) k_{\alpha}}    c^{\dagger}_{j_{a}, \downarrow} c_{j_{a}, \uparrow}    \left( \prod_{\substack{i=1 \\ i \neq b } }^{n_\downarrow} c^{\dagger}_{r(\mathbf{j}; l(\mathbf{j}; j_{a}+1, \mathbf{d})_{b}, \mathbf{d})_i, \downarrow} c_{r(\mathbf{j}; l(\mathbf{j}; j_{a}+1, \mathbf{d})_{b}, \mathbf{d})_i, \uparrow} \right)  \nonumber \\ 
& \quad \quad \quad c_{j_1, \uparrow} \cdots c_{j_{a-1}, \uparrow} c_{j_{a}+1, \uparrow} c_{j_{a+1}, \uparrow} \cdots c_{j_{n_h}, \uparrow}   \ket{\mathrm{FM}}. 
\end{align}
Note that these terms correspond to the cases where 
$r(\mathbf{j}; j', \mathbf{d})_{i} = j_{a}$ for $i = 1, \cdots, n_{\downarrow}$, 
which are excluded in $A_{a,-}$, as shown in Eq.~\eqref{eq:anyholeanydown_Aa-_constraint}.

Similarly, by replacing $\mathbf{j}^{(a+1)}$ with $\mathbf{j}^{(a-1)}$ and $j_{a}+1$ with $j_{a}-1$, the $B_{a,+}$ with $a=1,\cdots,n_{h}$ in Eq.~\eqref{eq:anyholesanydown_eigenstate_withH} is given by
\begin{align}
    \label{eq:anyholeanydownBa+}
    B_{a,+}  & = e^{ik_a} \sum_{\mathbf{j}} \sum_{b=1}^{n_{\downarrow}} e^{i\mathbf{j}^{(a-1)} \cdot \mathbf{k}}    e^{ij_{\alpha}(\mathbf{j}; l(\mathbf{j}; j_{a}-1, \mathbf{d})_{b}) k_{\alpha}}   c^{\dagger}_{j_{a}, \downarrow} c_{j_{a}, \uparrow}    \left( \prod_{\substack{i=1 \\ i \neq b }}^{n_\downarrow} c^{\dagger}_{r(\mathbf{j}; l(\mathbf{j}; j_{a}-1, \mathbf{d})_{b} , \mathbf{d})_i, \downarrow} c_{r(\mathbf{j}; l(\mathbf{j}; j_{a}-1, \mathbf{d})_{b}, \mathbf{d})_i, \uparrow} \right) \nonumber \\ 
& c_{j_1, \uparrow} \cdots c_{j_{a-1}, \uparrow} c_{j_{a}-1, \uparrow} c_{j_{a+1}, \uparrow} \cdots c_{j_{n_h}, \uparrow}   \ket{\mathrm{FM}}. 
\end{align}
Note that these terms correspond to the cases where $r(\mathbf{j}; j', \mathbf{d})_{i} = j_{a}$ for $i=1,\cdots, n_{\downarrow}$, which have been excluded in $A_{a,+}$, as shown in Eq.~\eqref{eq:anyholeanydown_Aa+_constraint}.

Next, we first consider $A_{a,-}$ in Eq.~\eqref{eq:anyholeanydownAa-}.
\begin{align}
    \label{eq:anyholeanydownAa-_2}
    A_{a,-} & = e^{-ik_a} \sum_{\mathbf{j}} \sum_{j'=0}^{N-1} e^{ij_{\alpha}(\mathbf{j}; j')k_{\alpha}}  e^{i \mathbf{j}^{(a+1)} \cdot \mathbf{k}}  \left( \prod_{i=1}^{n_\downarrow} c^{\dagger}_{r(\mathbf{j}; j', \mathbf{d})_i, \downarrow} c_{r(\mathbf{j}; j', \mathbf{d})_i, \uparrow} \right)  c_{j_1, \uparrow} \cdots c_{j_{a-1}, \uparrow} c_{j_{a}+1, \uparrow} c_{j_{a+1}, \uparrow} \cdots c_{j_{n_h}, \uparrow} \ket{\mathrm{FM}} \nonumber \\ 
& =  e^{-ik_a} \sum_{\mathbf{j}_a} \sum_{j'=0}^{N-1} \sum_{j_a=0}^{N-2} e^{ij_{\alpha}(\mathbf{j}; j')k_{\alpha}}  e^{i \mathbf{j}^{(a+1)} \cdot \mathbf{k}}  \left( \prod_{i=1}^{n_\downarrow} c^{\dagger}_{r(\mathbf{j}; j', \mathbf{d})_i, \downarrow} c_{r(\mathbf{j}; j', \mathbf{d})_i, \uparrow} \right)  c_{j_1, \uparrow} \cdots c_{j_{a-1}, \uparrow} c_{j_{a}+1, \uparrow} c_{j_{a+1}, \uparrow} \cdots c_{j_{n_h}, \uparrow} \ket{\mathrm{FM}} \nonumber \\ 
& + e^{-ik_a} \sum_{\mathbf{j}_a} \sum_{j'=0}^{N-1}  e^{ij_{\alpha}((\mathbf{j}_a, N-1); j')k_{\alpha}}  e^{i (\mathbf{j}_a, N) \cdot \mathbf{k}}  \left( \prod_{i=1}^{n_\downarrow} c^{\dagger}_{r((\mathbf{j}_a, N-1); j', \mathbf{d})_i, \downarrow} c_{r((\mathbf{j}_a, N-1); j', \mathbf{d})_i, \uparrow} \right)  c_{j_1, \uparrow} \cdots c_{j_{a-1}, \uparrow} c_{0, \uparrow} c_{j_{a+1}, \uparrow} \cdots c_{j_{n_h}, \uparrow} \ket{\mathrm{FM}} \nonumber \\ 
& =  e^{-ik_a} \sum_{\mathbf{j}_a} \sum_{j'=0}^{N-1} \sum_{j_a=0}^{N-2} e^{ij_{\alpha}(\mathbf{j}^{(a+1)}; j')k_{\alpha}}  e^{i \mathbf{j}^{(a+1)} \cdot \mathbf{k}}  \left( \prod_{i=1}^{n_\downarrow} c^{\dagger}_{r(\mathbf{j}^{(a+1)}; j', \mathbf{d})_i, \downarrow} c_{r(\mathbf{j}^{(a+1)}; j', \mathbf{d})_i, \uparrow} \right)  c_{j_1, \uparrow} \cdots c_{j_{a-1}, \uparrow} c_{j_{a}+1, \uparrow} c_{j_{a+1}, \uparrow} \cdots c_{j_{n_h}, \uparrow} \ket{\mathrm{FM}} \nonumber \\ 
& + e^{-ik_a} \sum_{\mathbf{j}_a} \sum_{j'=0}^{N-1}  e^{ik_{\alpha}}e^{ij_{\alpha}((\mathbf{j}_a, 0); j')k_{\alpha}} e^{iNk_{a}}  e^{i (\mathbf{j}_a, 0) \cdot \mathbf{k}}  \left( \prod_{i=1}^{n_\downarrow} c^{\dagger}_{r((\mathbf{j}_a, 0); j', \mathbf{d})_i, \downarrow} c_{r((\mathbf{j}_a, 0); j', \mathbf{d})_i, \uparrow} \right)   \nonumber \\
& \quad \quad \quad c_{j_1, \uparrow} \cdots c_{j_{a-1}, \uparrow} c_{0, \uparrow} c_{j_{a+1}, \uparrow} \cdots c_{j_{n_h}, \uparrow} \ket{\mathrm{FM}},
\end{align}
where in the third equality we have used
\begin{subequations}\label{eq:jalpha_r_shift_bulk}
\begin{align}
    j_{\alpha}(\mathbf{j}; j') &= j_{\alpha}(\mathbf{j}^{(a+1)}; j'),
    \label{eq:jalpha_shift_Aa-_bulk} \\ 
    r(\mathbf{j}; j', \mathbf{d})_i &= r(\mathbf{j}^{(a+1)}; j', \mathbf{d})_i.
    \label{eq:r_shift_Aa-_bulk}  
\end{align}
\end{subequations}
and 
\begin{subequations}\label{eq:jalpha_r_shift_boundary}
\begin{align}
    j_{\alpha}((\mathbf{j}_a, N-1); j') &= j_{\alpha}((\mathbf{j}_a, 0); j') + 1,
    \label{eq:jalpha_shift_Aa-_boundary} \\ 
    r((\mathbf{j}_a, N-1); j', \mathbf{d})_i &= r((\mathbf{j}_a, 0); j', \mathbf{d})_i.
    \label{eq:r_shift_Aa-_boundary}
\end{align}
\end{subequations}
(the proof is shown in Sec.~\ref{sec:Proofofequalitiesusedintheeigenstateconstruction}).

We then consider $B_{a,-}$ in Eq.~\eqref{eq:anyholeanydownBa-}.
\begin{align}
        \label{eq:anyholeanydownBa-_2}
        B_{a,-} & = e^{-ik_a} \sum_{\mathbf{j}} \sum_{b=1}^{n_{\downarrow}} e^{i \mathbf{j}^{(a+1)} \cdot \mathbf{k}}  e^{ij_{\alpha}(\mathbf{j}; l(\mathbf{j}; j_{a}+1, \mathbf{d})_b) k_{\alpha}}    c^{\dagger}_{j_{a}, \downarrow} c_{j_{a}, \uparrow}    \left( \prod_{\substack{i=1 \\ i \neq b } }^{n_\downarrow} c^{\dagger}_{r(\mathbf{j}; l(\mathbf{j}; j_{a}+1, \mathbf{d})_{b}, \mathbf{d})_i, \downarrow} c_{r(\mathbf{j}; l(\mathbf{j}; j_{a}+1, \mathbf{d})_{b}, \mathbf{d})_i, \uparrow} \right) \nonumber \\ 
        & \quad \quad \quad c_{j_1, \uparrow} \cdots c_{j_{a-1}, \uparrow} c_{j_{a}+1, \uparrow} c_{j_{a+1}, \uparrow} \cdots c_{j_{n_h}, \uparrow}   \ket{\mathrm{FM}} \nonumber \\ 
     & = e^{-ik_a} \sum_{\mathbf{j}_a} \sum_{j_a = 0}^{N-2} \sum_{b=1}^{n_{\downarrow}}  e^{i \mathbf{j}^{(a+1)} \cdot \mathbf{k}} e^{ij_{\alpha}(\mathbf{j}; l(\mathbf{j}; j_{a}+1, \mathbf{d})_b) k_{\alpha}}    c^{\dagger}_{j_{a}, \downarrow} c_{j_{a}, \uparrow}    \left( \prod_{\substack{i=1 \\ i \neq b } }^{n_\downarrow} c^{\dagger}_{r(\mathbf{j}; l(\mathbf{j}; j_{a}+1, \mathbf{d})_{b}, \mathbf{d})_i, \downarrow} c_{r(\mathbf{j}; l(\mathbf{j}; j_{a}+1, \mathbf{d})_{b}, \mathbf{d})_i, \uparrow} \right) \nonumber \\ 
     & \quad \quad \quad c_{j_1, \uparrow} \cdots c_{j_{a-1}, \uparrow} c_{j_{a}+1, \uparrow} c_{j_{a+1}, \uparrow} \cdots c_{j_{n_h}, \uparrow}   \ket{\mathrm{FM}} \nonumber \\ 
& + e^{-ik_a} \sum_{\mathbf{j}_a} \sum_{b=1}^{n_{\downarrow}} e^{i (\mathbf{j}_a, N)\cdot \mathbf{k}}  e^{ij_{\alpha}((\mathbf{j}_a,N-1); l((\mathbf{j}_a,N-1); 0, \mathbf{d})_b) k_{\alpha}}    c^{\dagger}_{N-1, \downarrow} c_{N-1, \uparrow}  \nonumber \\ 
& \left( \prod_{\substack{i=1 \\ i \neq b } }^{n_\downarrow} c^{\dagger}_{r((\mathbf{j}_a,N-1); l((\mathbf{j}_a,N-1); 0, \mathbf{d})_{b}, \mathbf{d})_i, \downarrow} c_{r((\mathbf{j}_a,N-1); l((\mathbf{j}_a,N-1); 0, \mathbf{d})_{b}, \mathbf{d})_i, \uparrow} \right)  c_{j_1, \uparrow} \cdots c_{j_{a-1}, \uparrow} c_{0, \uparrow} c_{j_{a+1}, \uparrow} \cdots c_{j_{n_h}, \uparrow}   \ket{\mathrm{FM}} \nonumber \\ 
&=  e^{-ik_a} \sum_{\mathbf{j}_a} \sum_{j_a = 0}^{N-2} \sum_{b=1}^{n_{\downarrow}} e^{i \mathbf{j}^{(a+1)} \cdot \mathbf{k}}  e^{ij_{\alpha}(\mathbf{j}^{(a+1)}; l(\mathbf{j}^{(a+1)}; j_{a}, \mathbf{d})_b) k_{\alpha}}    c^{\dagger}_{j_{a}, \downarrow} c_{j_{a}, \uparrow}    \left( \prod_{\substack{i=1 \\ i \neq b } }^{n_\downarrow} c^{\dagger}_{r(\mathbf{j}^{(a+1)}; l(\mathbf{j}^{(a+1)}; j_{a}, \mathbf{d})_{b}, \mathbf{d})_i, \downarrow} c_{r(\mathbf{j}^{(a+1)}; l(\mathbf{j}^{(a+1)}; j_{a}, \mathbf{d})_{b}, \mathbf{d})_i, \uparrow} \right) \nonumber  \\ 
& \quad \quad \quad c_{j_1, \uparrow} \cdots c_{j_{a-1}, \uparrow} c_{j_{a}+1, \uparrow} c_{j_{a+1}, \uparrow} \cdots c_{j_{n_h}, \uparrow}   \ket{\mathrm{FM}} \nonumber \\ 
&+e^{-ik_a} \sum_{\mathbf{j}_a}\sum_{b=1}^{n_{\downarrow}} e^{iNk_a} e^{i (\mathbf{j}_a,  0 ) \cdot \mathbf{k}}  e^{-i(N-n_{h}-1)k_{\alpha}} e^{ij_{\alpha}((\mathbf{j}_a, 0); l((\mathbf{j}_a, 0); N-1, \mathbf{d})_b) k_{\alpha}}    c^{\dagger}_{N-1, \downarrow} c_{N-1, \uparrow} \nonumber  \\ 
& \left( \prod_{\substack{i=1 \\ i \neq b} }^{n_\downarrow} c^{\dagger}_{r((\mathbf{j}_a, 0); l((\mathbf{j}_a, 0); N-1, \mathbf{d})_{b}, \mathbf{d})_i, \downarrow} c_{r((\mathbf{j}_a, 0); l((\mathbf{j}_a, 0); N-1, \mathbf{d})_{b}, \mathbf{d})_i, \uparrow} \right) c_{j_1, \uparrow} \cdots c_{j_{a-1}, \uparrow} c_{0, \uparrow} c_{j_{a+1}, \uparrow} \cdots c_{j_{n_h}, \uparrow}   \ket{\mathrm{FM}},
\end{align}
where in the third equality we have used
\begin{align}
    j_{\alpha}(\mathbf{j}; l(\mathbf{j}; j_{a}+1, \mathbf{d})_b)  &= j_{\alpha}(\mathbf{j}^{(a+1)}; l(\mathbf{j}^{(a+1)}; j_{a}, \mathbf{d})_b) \label{eq:jalpha_shift_Ba-_bulk} \\ 
r(\mathbf{j}; l(\mathbf{j}; j_{a}+1, \mathbf{d})_{b}, \mathbf{d})_i &= r(\mathbf{j}^{(a+1)}; l(\mathbf{j}^{(a+1)}; j_{a}, \mathbf{d})_{b}, \mathbf{d})_i \label{eq:r_shift_Ba-_bulk} 
\end{align}
(the proof is shown in Sec.~\ref{sec:Proofofequalitiesusedintheeigenstateconstruction}) and 
\begin{align}
    j_{\alpha}((\mathbf{j}_a, N-1); l((\mathbf{j}_a, N-1); 0, \mathbf{d})_b) &=   j_{\alpha}((\mathbf{j}_a, 0); l((\mathbf{j}_a, 0); N-1, \mathbf{d})_b) - (N-n_h-1) \label{eq:jalpha_shift_Ba-_boundary} \\ 
    r((\mathbf{j}_a,N-1); l((\mathbf{j}_a,N-1); 0, \mathbf{d})_{b}, \mathbf{d})_i &= r((\mathbf{j}_a, 0); l((\mathbf{j}_a, 0); N-1, \mathbf{d})_{b}, \mathbf{d})_i \label{eq:r_shift_Ba-_boundary}
\end{align}
(the proof is shown in Sec.~\ref{sec:Proofofequalitiesusedintheeigenstateconstruction}).

Combining $A_{a,-}$ in Eq.~\eqref{eq:anyholeanydownAa-_2} and $B_{a,-}$ in Eq.~\eqref{eq:anyholeanydownBa-_2}, and imposing the conditions
$e^{i(N-n_h) k_{\alpha}} = 1$ and $e^{iN k_{a}} e^{ik_{\alpha}} = 1$ for $a=1,\cdots, n_{h}$, we obtain
\begin{align}
    \sum_{a=1}^{n_h} \left(A_{a,-} + B_{a,-}\right)
    =
    \sum_{a=1}^{n_{h}} e^{-ik_{a}}\, \ket{\mathbf{k}, k_{\alpha}, \mathbf{d}}.
\end{align}

Next, we consider $A_{a,+}$ in Eq.~\eqref{eq:anyholeanydownAa+}.
\begin{align}
    \label{eq:anyholeanydownAa+_2}
    A_{a,+} & = e^{ik_a} \sum_{\mathbf{j}} \sum_{j'=0}^{N-1} e^{ij_{\alpha}(\mathbf{j}; j')k_{\alpha}}  e^{i \mathbf{j}^{(a-1)} \cdot \mathbf{k}}  \left( \prod_{i=1}^{n_\downarrow} c^{\dagger}_{r(\mathbf{j}; j', \mathbf{d})_i, \downarrow} c_{r(\mathbf{j}; j', \mathbf{d})_i, \uparrow} \right)  c_{j_1, \uparrow} \cdots c_{j_{a-1}, \uparrow} c_{j_{a}-1, \uparrow} c_{j_{a+1}, \uparrow} \cdots c_{j_{n_h}, \uparrow} \ket{\mathrm{FM}} \nonumber \\ 
& =  e^{ik_a} \sum_{\mathbf{j}_a} \sum_{j'=0}^{N-1} \sum_{j_a=1}^{N-1} e^{ij_{\alpha}(\mathbf{j}; j')k_{\alpha}}  e^{i \mathbf{j}^{(a-1)} \cdot \mathbf{k}}  \left( \prod_{i=1}^{n_\downarrow} c^{\dagger}_{r(\mathbf{j}; j', \mathbf{d})_i, \downarrow} c_{r(\mathbf{j}; j', \mathbf{d})_i, \uparrow} \right)  c_{j_1, \uparrow} \cdots c_{j_{a-1}, \uparrow} c_{j_{a}-1, \uparrow} c_{j_{a+1}, \uparrow} \cdots c_{j_{n_h}, \uparrow} \ket{\mathrm{FM}} \nonumber \\ 
& + e^{ik_a} \sum_{\mathbf{j}_a} \sum_{j'=0}^{N-1}  e^{ij_{\alpha}((\mathbf{j}_a, 0); j')k_{\alpha}}  e^{i (\mathbf{j}_a, -1) \cdot \mathbf{k}}  \left( \prod_{i=1}^{n_\downarrow} c^{\dagger}_{r((\mathbf{j}_a, 0); j', \mathbf{d})_i, \downarrow} c_{r((\mathbf{j}_a, 0); j', \mathbf{d})_i, \uparrow} \right)  c_{j_1, \uparrow} \cdots c_{j_{a-1}, \uparrow} c_{N-1, \uparrow} c_{j_{a+1}, \uparrow} \cdots c_{j_{n_h}, \uparrow} \ket{\mathrm{FM}} \nonumber \\ 
& =  e^{ik_a} \sum_{\mathbf{j}_a} \sum_{j'=0}^{N-1} \sum_{j_a=1}^{N-1} e^{ij_{\alpha}(\mathbf{j}^{(a-1)}; j')k_{\alpha}}  e^{i \mathbf{j}^{(a-1)} \cdot \mathbf{k}}  \left( \prod_{i=1}^{n_\downarrow} c^{\dagger}_{r(\mathbf{j}^{(a-1)}; j', \mathbf{d})_i, \downarrow} c_{r(\mathbf{j}^{(a-1)}; j', \mathbf{d})_i, \uparrow} \right)  c_{j_1, \uparrow} \cdots c_{j_{a-1}, \uparrow} c_{j_{a}-1, \uparrow} c_{j_{a+1}, \uparrow} \cdots c_{j_{n_h}, \uparrow} \ket{\mathrm{FM}} \nonumber \\ 
& + e^{ik_a} \sum_{\mathbf{j}_a} \sum_{j'=0}^{N-1}  e^{-ik_{\alpha}}e^{ij_{\alpha}((\mathbf{j}_a, N-1); j')k_{\alpha}} e^{-iNk_{a}}  e^{i (\mathbf{j}_a, N-1) \cdot \mathbf{k}}  \left( \prod_{i=1}^{n_\downarrow} c^{\dagger}_{r((\mathbf{j}_a, N-1); j', \mathbf{d})_i, \downarrow} c_{r((\mathbf{j}_a, N-1); j', \mathbf{d})_i, \uparrow} \right)  \nonumber \\
& \quad \quad \quad c_{j_1, \uparrow} \cdots c_{j_{a-1}, \uparrow} c_{N-1, \uparrow} c_{j_{a+1}, \uparrow} \cdots c_{j_{n_h}, \uparrow} \ket{\mathrm{FM}},
\end{align}
where, in the third equality, we have used
\begin{align}
    j_{\alpha}(\mathbf{j}; j')  &=  j_{\alpha}(\mathbf{j}^{(a-1)}; j'), \\
    r(\mathbf{j}; j', \mathbf{d})_i  &= r(\mathbf{j}^{(a-1)}; j', \mathbf{d})_i,
\end{align}
(the proof is the same as that of 
Eqs.~\eqref{eq:jalpha_shift_Aa-_bulk} and \eqref{eq:r_shift_Aa-_bulk} by replacing $\mathbf{j}^{(a+1)}$ with $\mathbf{j}^{(a-1)}$), and
\begin{align}
    j_{\alpha}((\mathbf{j}_a, 0); j') &= j_{\alpha}((\mathbf{j}_a, N-1); j') - 1, \\
    r((\mathbf{j}_a, 0); j', \mathbf{d})_i &= r((\mathbf{j}_a, N-1); j', \mathbf{d})_i,
\end{align}
which are the same as 
Eqs.~\eqref{eq:jalpha_shift_Aa-_boundary} and \eqref{eq:r_shift_Aa-_boundary}.

Finally, we consider $B_{a,+}$ in Eq.~\eqref{eq:anyholeanydownBa+}.
\begin{align}
        \label{eq:anyholeanydownBa+_2}
        B_{a,+} & = e^{ik_a} \sum_{\mathbf{j}} \sum_{b=1}^{n_{\downarrow}} e^{i \mathbf{j}^{(a-1)} \cdot \mathbf{k}}  e^{ij_{\alpha}(\mathbf{j}; l(\mathbf{j}; j_{a}-1, \mathbf{d})_b) k_{\alpha}}    c^{\dagger}_{j_{a}, \downarrow} c_{j_{a}, \uparrow}    \left( \prod_{\substack{i=1 \\ i \neq b } }^{n_\downarrow} c^{\dagger}_{r(\mathbf{j}; l(\mathbf{j}; j_{a}-1, \mathbf{d})_{b}, \mathbf{d})_i, \downarrow} c_{r(\mathbf{j}; l(\mathbf{j}; j_{a}-1, \mathbf{d})_{b}, \mathbf{d})_i, \uparrow} \right) \nonumber \\ 
        & \quad \quad \quad c_{j_1, \uparrow} \cdots c_{j_{a-1}, \uparrow} c_{j_{a}-1, \uparrow} c_{j_{a+1}, \uparrow} \cdots c_{j_{n_h}, \uparrow}   \ket{\mathrm{FM}} \nonumber \\ 
     & = e^{ik_a} \sum_{\mathbf{j}_a} \sum_{j_a = 1}^{N-1} \sum_{b=1}^{n_{\downarrow}}  e^{i \mathbf{j}^{(a-1)} \cdot \mathbf{k}} e^{ij_{\alpha}(\mathbf{j}; l(\mathbf{j}; j_{a}-1, \mathbf{d})_b) k_{\alpha}}    c^{\dagger}_{j_{a}, \downarrow} c_{j_{a}, \uparrow}    \left( \prod_{\substack{i=1 \\ i \neq b } }^{n_\downarrow} c^{\dagger}_{r(\mathbf{j}; l(\mathbf{j}; j_{a}-1, \mathbf{d})_{b}, \mathbf{d})_i, \downarrow} c_{r(\mathbf{j}; l(\mathbf{j}; j_{a}-1, \mathbf{d})_{b}, \mathbf{d})_i, \uparrow} \right)  \nonumber \\
     & \quad \quad \quad c_{j_1, \uparrow} \cdots c_{j_{a-1}, \uparrow} c_{j_{a}-1, \uparrow} c_{j_{a+1}, \uparrow} \cdots c_{j_{n_h}, \uparrow}   \ket{\mathrm{FM}} \nonumber \\ 
& + e^{ik_a} \sum_{\mathbf{j}_a} \sum_{b=1}^{n_{\downarrow}} e^{i (\mathbf{j}_a, -1)\cdot \mathbf{k}}  e^{ij_{\alpha}((\mathbf{j}_a,0); l((\mathbf{j}_a,0); N-1, \mathbf{d})_b) k_{\alpha}}    c^{\dagger}_{0, \downarrow} c_{0, \uparrow} \nonumber  \\ 
& \left( \prod_{\substack{i=1 \\ i \neq b } }^{n_\downarrow} c^{\dagger}_{r((\mathbf{j}_a,0); l((\mathbf{j}_a,0); N-1, \mathbf{d})_{b}, \mathbf{d})_i, \downarrow} c_{r((\mathbf{j}_a,0); l((\mathbf{j}_a,0); N-1, \mathbf{d})_{b}, \mathbf{d})_i, \uparrow} \right)  c_{j_1, \uparrow} \cdots c_{j_{a-1}, \uparrow} c_{N, \uparrow} c_{j_{a+1}, \uparrow} \cdots c_{j_{n_h}, \uparrow}   \ket{\mathrm{FM}} \nonumber \\ 
&=  e^{ik_a} \sum_{\mathbf{j}_a} \sum_{j_a = 1}^{N-1} \sum_{b=1}^{n_{\downarrow}} e^{i \mathbf{j}^{(a-1)} \cdot \mathbf{k}}  e^{ij_{\alpha}(\mathbf{j}^{(a-1)}; l(\mathbf{j}^{(a-1)}; j_{a}, \mathbf{d})_b) k_{\alpha}}    c^{\dagger}_{j_{a}, \downarrow} c_{j_{a}, \uparrow}    \left( \prod_{\substack{i=1 \\ i \neq b } }^{n_\downarrow} c^{\dagger}_{r(\mathbf{j}^{(a-1)}; l(\mathbf{j}^{(a-1)}; j_{a}, \mathbf{d})_{b}, \mathbf{d})_i, \downarrow} c_{r(\mathbf{j}^{(a-1)}; l(\mathbf{j}^{(a-1)}; j_{a}, \mathbf{d})_{b}, \mathbf{d})_i, \uparrow} \right) \nonumber \\  
& \quad \quad \quad c_{j_1, \uparrow} \cdots c_{j_{a-1}, \uparrow} c_{j_{a}-1, \uparrow} c_{j_{a+1}, \uparrow} \cdots c_{j_{n_h}, \uparrow}   \ket{\mathrm{FM}} \nonumber \\ 
&+e^{ik_a} \sum_{\mathbf{j}_a}\sum_{b=1}^{n_{\downarrow}} e^{-iNk_a} e^{i (\mathbf{j}_a,  N-1 ) \cdot \mathbf{k}}  e^{i(N-n_{h}-1)k_{\alpha}} e^{ij_{\alpha}((\mathbf{j}_a, N-1); l((\mathbf{j}_a, N-1); 0, \mathbf{d})_b) k_{\alpha}}    c^{\dagger}_{0, \downarrow} c_{0, \uparrow}   \nonumber  \\ 
& \left( \prod_{\substack{i=1 \\ i \neq b} }^{n_\downarrow} c^{\dagger}_{r((\mathbf{j}_a, N-1); l((\mathbf{j}_a, N-1); 0, \mathbf{d})_{b}, \mathbf{d})_i, \downarrow} c_{r((\mathbf{j}_a, N-1); l((\mathbf{j}_a, N-1); 0, \mathbf{d})_{b}, \mathbf{d})_i, \uparrow} \right) c_{j_1, \uparrow} \cdots c_{j_{a-1}, \uparrow} c_{N-1, \uparrow} c_{j_{a+1}, \uparrow} \cdots c_{j_{n_h}, \uparrow}   \ket{\mathrm{FM}},
\end{align}
where, in the third equality, we have used
\begin{align}
j_{\alpha}(\mathbf{j}; l(\mathbf{j}; j_{a}-1, \mathbf{d})_b)
&=
j_{\alpha}(\mathbf{j}^{(a-1)}; l(\mathbf{j}^{(a-1)}; j_{a}, \mathbf{d})_b), \\
r(\mathbf{j}; l(\mathbf{j}; j_{a}-1, \mathbf{d})_{b}, \mathbf{d})_i
&=
r(\mathbf{j}^{(a-1)}; l(\mathbf{j}^{(a-1)}; j_{a}, \mathbf{d})_{b}, \mathbf{d})_i,
\end{align}
(the proof is the same as that of
Eqs.~\eqref{eq:jalpha_shift_Ba-_bulk} and \eqref{eq:r_shift_Ba-_bulk} by replacing $\mathbf{j}^{(a+1)}$ with $\mathbf{j}^{(a-1)}$),
and
\begin{align}
j_{\alpha}((\mathbf{j}_a, 0); l((\mathbf{j}_a, 0); N-1, \mathbf{d})_b)
&=
j_{\alpha}((\mathbf{j}_a, N-1); l((\mathbf{j}_a, N-1); 0, \mathbf{d})_b)
+ (N-n_h-1),  \\
r((\mathbf{j}_a, 0); l((\mathbf{j}_a, 0); N-1, \mathbf{d})_{b}, \mathbf{d})_i
&=
r((\mathbf{j}_a, N-1); l((\mathbf{j}_a, N-1); 0, \mathbf{d})_{b}, \mathbf{d})_i,
\end{align}
which are the same as those in
Eqs.~\eqref{eq:jalpha_shift_Ba-_boundary} and \eqref{eq:r_shift_Ba-_boundary}.

Combining $A_{a,+}$ in Eq.~\eqref{eq:anyholeanydownAa+_2} and $B_{a,+}$ in Eq.~\eqref{eq:anyholeanydownBa+_2}, and imposing the conditions $e^{i(N-n_h)k_{\alpha}}=1$ and $e^{iNk_a} e^{ik_{\alpha}}=1$ for $a=1,\cdots, n_{h}$, we have
\begin{align}
    \sum_{a=1}^{n_h} (A_{a,+} + B_{a,+})  = \sum_{a=1}^{n_h} e^{ik_{a}} \ket{\mathbf{k}, k_{\alpha}, \mathbf{d}}.
\end{align}

In conclusion, when
\begin{align}
    e^{i(N-n_h)k_{\alpha}} &=1, \\ 
    e^{ik_{\alpha}} e^{iNk_{a}} &= 1, \quad a = 1, \cdots, n_{h},
\end{align}
we have
\begin{align}
    \hat{H} \ket{\mathbf{k}, k_{\alpha}, \mathbf{d}} = \sum_{a=1}^{n_h} 2\cos(k_{a}) \ket{\mathbf{k}, k_{\alpha}, \mathbf{d}}.
    \label{eq:energyspectrum_anyhole_anyspindown}
\end{align}
Moreover, $k_{\alpha}$ and $k_a$ with $a=1,\cdots, n_h$ can be expressed as 
\begin{align}
    k_{\alpha} &= -\frac{2\pi m_{\downarrow}}{N-n_{h}}, \quad \mathrm{with} \quad m_{\downarrow} = 0, \cdots, N-n_h-1, \\
    k_{a} &= \frac{2\pi m_a}{N} - \frac{k_{\alpha}}{N} = \frac{2\pi m_{a}}{N} + \frac{2\pi m_{\downarrow}}{N(N-n_h)}, \quad a = 1, \cdots, n_h.
\end{align}
Since any two $k_i$ and $k_j$ ($i,j=1,\cdots, n_h$) are distinct when $i \neq j$, $0 \leq m_1 < m_2 < \cdots < m_{n_h} \leq N-1$. Physically, the condition $e^{i (N-n_h) k_{\alpha}} = 1$ arises because, after removing the $n_h$ sites occupied by the holes, the spin configurations with the same $\mathbf d$ form an orbit on the $(N-n_h)$-site chain.

Since the eigenenergy does not depend on the spin configuration $\mathbf{d}$, for each energy level 
$\sum_{a=1}^{n_h} 2\cos(k_a)$ the degeneracy is 
$\binom{N-n_h}{n_{\downarrow}}/(N-n_h)$, i.e., the number of orbits formed by the spin configurations.

Next, we prove that $\ket{\mathbf{k},k_{\alpha}, \mathbf{d}}$ is an eigenstate of current operator $\hat{J}$ with eigenvalue $\sum_{a=1}^{n_h} 2\sin(k_a)$. Under the action of current operator $\hat{J}$,
\begin{align}
    \hat{J} \ket{\mathbf{k}, k_{\alpha}, \mathbf{d}} &= i \sum_{\mathbf{j}}\sum_{j'=0}^{N-1} e^{ij_{\alpha}(\mathbf{j}; j') k_{\alpha}} e^{i \mathbf{j} \cdot \mathbf{k}}  \\ \nonumber
    & \left(\sum_{a=1}^{n_h}( -c^{\dagger}_{j_a, \uparrow} c_{j_a+1, \uparrow}  + c^{\dagger}_{j_a, \uparrow} c_{j_a-1,\uparrow}) +  \sum_{b=1}^{n_{\downarrow}} (c^{\dagger}_{r(\mathbf{j};j',\mathbf{d})_b +1, \downarrow} - c^{\dagger}_{r(\mathbf{j};j',\mathbf{d})_b -1, \downarrow})c_{r(\mathbf{j};j', \mathbf{d})_b, \downarrow}  \right) \\ \nonumber
&  \left( \prod_{i=1}^{n_\downarrow} c^{\dagger}_{r(\mathbf{j}; j', \mathbf{d})_i, \downarrow} c_{r(\mathbf{j}; j', \mathbf{d})_i, \uparrow} \right)  \left(  \prod_{i=1}^{n_{h}} c_{j_{i}, \uparrow} \right) \ket{\mathrm{FM}}. 
\end{align}
Comparing with Eq.~\eqref{eq:anyholesanydown_eigenstate_withH}, the only difference is that $-\sum_{a=1}^{n_h}c^{\dagger}_{j_a, \uparrow} c_{j_a+1, \uparrow} - \sum_{b=1}^{n_{\downarrow}} c^{\dagger}_{r(\mathbf{j};j',\mathbf{d})_b -1, \downarrow}c_{r(\mathbf{j};j', \mathbf{d})_b, \downarrow}$ in Eq.~\eqref{eq:anyholesanydown_eigenstate_withH} becomes $-i\sum_{a=1}^{n_h}c^{\dagger}_{j_a, \uparrow} c_{j_a+1, \uparrow} - i\sum_{b=1}^{n_{\downarrow}} c^{\dagger}_{r(\mathbf{j};j',\mathbf{d})_b -1, \downarrow}c_{r(\mathbf{j};j', \mathbf{d})_b, \downarrow}$ and  $-\sum_{a=1}^{n_h}c^{\dagger}_{j_a, \uparrow} c_{j_a-1, \uparrow} - \sum_{b=1}^{n_{\downarrow}} c^{\dagger}_{r(\mathbf{j};j',\mathbf{d})_b +1, \downarrow}c_{r(\mathbf{j};j', \mathbf{d})_b, \downarrow}$ in Eq.~\eqref{eq:anyholesanydown_eigenstate_withH} becomes $i\sum_{a=1}^{n_h}c^{\dagger}_{j_a, \uparrow} c_{j_a-1, \uparrow} + i\sum_{b=1}^{n_{\downarrow}} c^{\dagger}_{r(\mathbf{j};j',\mathbf{d})_b +1, \downarrow}c_{r(\mathbf{j};j', \mathbf{d})_b, \downarrow}$. From the derivation for Eq.~\eqref{eq:anyholesanydown_eigenstate_withH}, we know the former contributes $\sum_{a=1}^{n_h} e^{-ik_a} \ket{\mathbf{k}, k_{\alpha}, \mathbf{d}}$ and the latter contributes $\sum_{a=1}^{n_h} e^{ik_a} \ket{\mathbf{k}, k_{\alpha}, \mathbf{d}}$. Therefore, we have 
\begin{align}
    \label{eq:ProofOfEigenstateJWithSpinDown}
    \hat{J} \ket{\mathbf{k}, k_{\alpha}, \mathbf{d}} &= (i \sum_{a=1}^{n_h} e^{-ik_a} -i\sum_{a=1}^{n_h} e^{ik_a}) \ket{\mathbf{k}, k_{\alpha}, \mathbf{d}} \nonumber \\ 
    &= \sum_{a=1}^{n_h}2\sin(k_a) \ket{\mathbf{k}, k_{\alpha}, \mathbf{d}}.
\end{align}

\subsubsection{Proof of equalities used in the eigenstate construction}
\label{sec:Proofofequalitiesusedintheeigenstateconstruction}

In this section, we present the proofs of the equalities given in
Eqs.~\eqref{eq:jalpha_shift_Aa-_bulk}, \eqref{eq:r_shift_Aa-_bulk},
\eqref{eq:jalpha_shift_Aa-_boundary}, \eqref{eq:r_shift_Aa-_boundary},
\eqref{eq:jalpha_shift_Ba-_bulk}, \eqref{eq:r_shift_Ba-_bulk},
\eqref{eq:jalpha_shift_Ba-_boundary}, and \eqref{eq:r_shift_Ba-_boundary}.

\textbf{Proof of the identities given in Eqs.~\eqref{eq:jalpha_shift_Aa-_bulk}, \eqref{eq:r_shift_Aa-_bulk} used in $A_{a,-}$ :}
We first present the proof of equalities given in 
Eqs.~\eqref{eq:jalpha_shift_Aa-_bulk}, \eqref{eq:r_shift_Aa-_bulk}, under the constraints given in Eq.~\eqref{eq:anyholeanydown_Aa-_constraint}.

Since the two hole configurations $\mathbf j$ and $\mathbf j^{(a+1)}$ differ
only in that the removed site $j_a$ is replaced by $j_a+1$,  
by definition shown in Eq.~\eqref{eq:DefinitionOfGeneralJalpha}, we have
\begin{align}
    j_\alpha(\mathbf j;j')
    -
    j_\alpha(\mathbf j^{(a+1)};j')
    =
    -\Theta(j'-j_a)+\Theta(j'-(j_a+1)).
\end{align}
For $j'\neq j_a,j_a+1$, the right-hand side vanishes. Therefore,
\begin{align}
    j_{\alpha}(\mathbf j;j')
    =
    j_{\alpha}(\mathbf j^{(a+1)};j'),
\end{align}
which proves Eq.~\eqref{eq:jalpha_shift_Aa-_bulk}.

Next, when the first spin-down fermion is located at site $j'$, the reduced-chain coordinate of the
$i$-th spin-down fermion is given by
\begin{align}
    j_\alpha(\mathbf j;j')
    +
    \sum_{\ell=1}^{i-1}d_\ell
    \pmod{N-n_h}.
\end{align}
Since Eq.~\eqref{eq:jalpha_shift_Aa-_bulk} shows that the reduced-chain
coordinate of the first spin-down fermion is the same for $\mathbf j$ and
$\mathbf j^{(a+1)}$, the reduced-chain coordinate of the $i$-th spin-down
fermion is also the same in the two configurations. Furthermore, due to constraints in Eq.~\eqref{eq:anyholeanydown_Aa-_constraint}, no spin-down fermion is located at sites $j_a$ or $j_a+1$,
moving the hole from $j_a$ to $j_a+1$ does not change the
physical occupied site corresponding to this reduced-chain coordinate. Hence,
\begin{align}
    r(\mathbf j;j',\mathbf d)_i
    =
    r(\mathbf j^{(a+1)};j',\mathbf d)_i,
\end{align}
which proves Eq.~\eqref{eq:r_shift_Aa-_bulk}.

\textbf{Proof of the identities given in Eqs.~\eqref{eq:jalpha_shift_Aa-_boundary}, \eqref{eq:r_shift_Aa-_boundary} used in $A_{a,-}$ :}
We then present the proof of equalities given in 
Eqs.~\eqref{eq:jalpha_shift_Aa-_boundary}, \eqref{eq:r_shift_Aa-_boundary}, under the constraints given in Eq.~\eqref{eq:anyholeanydown_Aa-_constraint}. Thus, there is no spin-down fermion located at sites $0$ or $N-1$.

By definition shown in Eq.~\eqref{eq:DefinitionOfGeneralJalpha}, we obtain
\begin{align}
    j_{\alpha}((\mathbf j_a,N-1);j')
    -
    j_{\alpha}((\mathbf j_a,0);j')
    =
    \Theta(j'-0)-\Theta(j'-(N-1)).
\end{align}
For $j'\neq 0,N-1$, one has
\begin{align}
    \Theta(j')=1,
    \qquad
    \Theta(j'-(N-1))=0,
\end{align}
and therefore
\begin{align}
    j_{\alpha}((\mathbf j_a,N-1);j')
    =
    j_{\alpha}((\mathbf j_a,0);j')+1,
\end{align}
which proves Eq.~\eqref{eq:jalpha_shift_Aa-_boundary}.

Due to Eq.~\eqref{eq:jalpha_shift_Aa-_boundary}, we have
\begin{align}
    j_{\alpha}((\mathbf{j}_a, N-1); j') + \sum_{\ell=1}^{i-1} d_\ell
    =
    j_{\alpha}((\mathbf{j}_a, 0); j') + 1 + \sum_{\ell=1}^{i-1} d_\ell
    \pmod{N-n_h}.
\end{align}

Therefore, the reduced-chain coordinate of the $i$-th spin-down fermion in the
configuration $(\mathbf{j}_a,N-1)$ is
\begin{align}
    y_i^{(N-1)}
    &=
    j_\alpha((\mathbf{j}_a,N-1);j')
    +
    \sum_{\ell=1}^{i-1}d_\ell
    \pmod{N-n_h}
    \nonumber\\
    &=
    \left[
    j_\alpha((\mathbf{j}_a,0);j')+1
    \right]
    +
    \sum_{\ell=1}^{i-1}d_\ell
    \pmod{N-n_h}
    \nonumber\\
    &=
    y_i^{(0)}+1
    \pmod{N-n_h},
    \label{eq:yi_shift_explicit}
\end{align}
where $y^{(0)}_{i}$ is the reduced-chain coordinate of the $i$-th spin-down fermion in the configuration $(\mathbf{j}_a, 0)$,
\begin{align}
    y_i^{(0)}
    =
    j_\alpha((\mathbf{j}_a,0);j')
    +
    \sum_{\ell=1}^{i-1}d_\ell
    \pmod{N-n_h}.
\end{align}

Therefore, we obtain (see the definition of $r(\cdot)$ function in Eq.~\eqref{eq:DefinitionOfRfunction})
\begin{align}
    r((\mathbf{j}_a,N-1);j',\mathbf d)_i
    &=
    j_\alpha^{-1}((\mathbf{j}_a,N-1);y_i^{(N-1)})
    \nonumber\\
    &=
    j_\alpha^{-1}((\mathbf{j}_a,0);y_i^{(N-1)}-1)
    \nonumber\\
    &=
    j_\alpha^{-1}((\mathbf{j}_a,0);y_i^{(0)})
    \nonumber\\
    &=
    r((\mathbf{j}_a,0);j',\mathbf d)_i.
\end{align}
This proves Eq.~\eqref{eq:r_shift_Aa-_boundary}.

\textbf{Proof of the identities given in Eqs.~\eqref{eq:jalpha_shift_Ba-_bulk}, \eqref{eq:r_shift_Ba-_bulk} used in $B_{a,-}$ :}
We then present the proof of equalities given in 
Eqs.~\eqref{eq:jalpha_shift_Ba-_bulk}, \eqref{eq:r_shift_Ba-_bulk}.

For
\begin{align}
    l(\mathbf{j}; j_a+1, \mathbf{d})_b,
    \qquad
    l(\mathbf{j}^{(a+1)}; j_a, \mathbf{d})_b,
\end{align}
when $b=1$, we have
\begin{align}
    l(\mathbf{j}; j_a+1, \mathbf{d})_1 &= j_a+1, \\
    l(\mathbf{j}^{(a+1)}; j_a, \mathbf{d})_1 &= j_a.
\end{align}
Therefore, proving Eq.~\eqref{eq:jalpha_shift_Ba-_bulk} is equivalent to proving
\begin{align}
    j_{\alpha}(\mathbf{j}; j_a+1) - j_{\alpha}(\mathbf{j}^{(a+1)}; j_a) = 0.
\end{align}
Using the definition of $j_\alpha$ shown in Eq.~\eqref{eq:DefinitionOfGeneralJalpha}, we have
\begin{align}
    j_{\alpha}(\mathbf{j}; j_a+1) - j_{\alpha}(\mathbf{j}^{(a+1)}; j_a)
    &=
    \left[j_a+1-\Theta\!\bigl((j_a+1)-j_a\bigr)\right]
    -
    \left[j_a-\Theta\!\bigl(j_a-(j_a+1)\bigr)\right]
    \nonumber\\
    &=
    \left(j_a+1-1\right)-\left(j_a-0\right)
    \nonumber\\
    &=0.
\end{align}

When $b \neq 1$, 
let
\begin{align}
    x_0
    :=
    j_\alpha(\mathbf{j};j_a+1)
    =
    j_\alpha(\mathbf{j}^{(a+1)};j_a),
\end{align}
which is the common reduced-chain coordinate of the occupied sites
$j_a+1$ in the configuration $\mathbf j$ and $j_a$ in the configuration
$\mathbf j^{(a+1)}$. By definition, if the $b$-th spin-down fermion is fixed at reduced-chain
coordinate $x_0$, then the first spin-down fermion must be located at
\begin{align}
    x_0-\sum_{\ell=1}^{b-1}d_\ell
    \pmod{N-n_h}.
\end{align}
Hence,
\begin{align}
    j_\alpha\!\left(\mathbf{j};l(\mathbf{j};j_a+1,\mathbf{d})_b\right)
    =
    j_\alpha\!\left(\mathbf{j}^{(a+1)};l(\mathbf{j}^{(a+1)};j_a,\mathbf{d})_b\right)
    =
    x_0-\sum_{\ell=1}^{b-1}d_\ell
    \pmod{N-n_h}.
\end{align}
This proves Eq.~\eqref{eq:jalpha_shift_Ba-_bulk}.

Next, we prove Eq.~\eqref{eq:r_shift_Ba-_bulk},
\begin{align}
    r(\mathbf{j}; l(\mathbf{j}; j_{a}+1, \mathbf{d})_{b}, \mathbf{d})_i &= r(\mathbf{j}^{(a+1)}; l(\mathbf{j}^{(a+1)}; j_{a}, \mathbf{d})_{b}, \mathbf{d})_i. \nonumber
\end{align}
Note that we require $b \neq i$.
When $b=1$, 
\begin{align}
    r(\mathbf{j}; l(\mathbf{j}; j_{a}+1, \mathbf{d})_{1}, \mathbf{d})_i &= r(\mathbf{j};  j_{a}+1, \mathbf{d})_i  \\ 
    r(\mathbf{j}^{(a+1)}; l(\mathbf{j}^{(a+1)}; j_{a}, \mathbf{d})_{b}, \mathbf{d})_i & = r(\mathbf{j}^{(a+1)}; j_{a}, \mathbf{d})_i
\end{align}

Therefore, proving Eq.~\eqref{eq:r_shift_Ba-_bulk} reduces to proving
\begin{align}
    r(\mathbf j; j_a+1,\mathbf d)_i
    =
    r(\mathbf j^{(a+1)}; j_a,\mathbf d)_i,
    \qquad i\neq 1.
    \label{eq:r_shift_Ba_bulk_b1_goal}
\end{align}

The definition of $r(\cdot)$ function is shown in Eq.~\eqref{eq:DefinitionOfRfunction}.
Applying this formula to the two sides of Eq.~\eqref{eq:r_shift_Ba_bulk_b1_goal},
we obtain
\begin{align}
    r(\mathbf j; j_a+1,\mathbf d)_i
    &=
    j_\alpha^{-1}\!\left(
    \mathbf j;\,
    j_\alpha(\mathbf j;j_a+1)
    +
    \sum_{\ell=1}^{i-1}d_\ell
    \pmod{N-n_h}
    \right),
    \label{eq:r_b1_left}
    \\
    r(\mathbf j^{(a+1)}; j_a,\mathbf d)_i
    &=
    j_\alpha^{-1}\!\left(
    \mathbf j^{(a+1)};\,
    j_\alpha(\mathbf j^{(a+1)};j_a)
    +
    \sum_{\ell=1}^{i-1}d_\ell
    \pmod{N-n_h}
    \right).
    \label{eq:r_b1_right}
\end{align}

From Eq.~\eqref{eq:jalpha_shift_Ba-_bulk}, specialized to $b=1$, we already know
that
\begin{align}
    j_\alpha(\mathbf j;j_a+1)
    =
    j_\alpha(\mathbf j^{(a+1)};j_a).
    \label{eq:jalpha_b1_equal}
\end{align}
Hence the reduced-chain coordinates appearing in
Eqs.~\eqref{eq:r_b1_left} and \eqref{eq:r_b1_right} differ only through the
inverse maps $j_\alpha^{-1}(\mathbf j;\cdot)$ and
$j_\alpha^{-1}(\mathbf j^{(a+1)};\cdot)$.

Now let
\begin{align}
    x_0 := j_\alpha(\mathbf j;j_a+1) = j_\alpha(\mathbf j^{(a+1)};j_a),
\end{align}
which is precisely the reduced-chain coordinate of the occupied site $j_a+1$ in
the configuration $\mathbf j$ and of the occupied site $j_a$ in the
configuration $\mathbf j^{(a+1)}$.

For $i\neq 1$, one has
\begin{align}
    \sum_{\ell=1}^{i-1}d_\ell >0,
\end{align}
and therefore
\begin{align}
    x_0+\sum_{\ell=1}^{i-1}d_\ell
    \not\equiv x_0
    \pmod{N-n_h}.
\end{align}
That is, the reduced-chain coordinate of the $i$-th spin-down fermion is not
the special coordinate $x_0$ at which the inverse maps for $\mathbf j$ and
$\mathbf j^{(a+1)}$ differ. Therefore,
\begin{align}
    j_\alpha^{-1}\!\left(
    \mathbf j;\,
    x_0+\sum_{\ell=1}^{i-1}d_\ell
    \right)
    =
    j_\alpha^{-1}\!\left(
    \mathbf j^{(a+1)};\,
    x_0+\sum_{\ell=1}^{i-1}d_\ell
    \right),
\end{align}
which immediately implies
\begin{align}
    r(\mathbf j; j_a+1,\mathbf d)_i
    =
    r(\mathbf j^{(a+1)}; j_a,\mathbf d)_i,
    \qquad i\neq 1.
\end{align}

For $b\ge 2$, we have
\begin{align}
    l(\mathbf j;j_a+1,\mathbf d)_b
    =
    l(\mathbf j^{(a+1)};j_a,\mathbf d)_b.
\end{align}
 Moreover, Eq.~\eqref{eq:jalpha_shift_Aa-_bulk} gives
\begin{align}
    j_\alpha\!\left(\mathbf j;l(\mathbf j;j_a+1,\mathbf d)_b\right)
    =
    j_\alpha\!\left(\mathbf j^{(a+1)};l(\mathbf j^{(a+1)};j_a,\mathbf d)_b\right).
\end{align}
Therefore the reduced-chain coordinate of the $i$-th spin-down fermion is the
same in the two descriptions. Since $i\neq b$, the $i$-th spin-down fermion is
not the one occupying the special sites $j_a+1$ in $\mathbf j$ and $j_a$ in
$\mathbf j^{(a+1)}$. Hence the corresponding physical site is unchanged after
mapping back to the original chain, and we obtain
\begin{align}
    r(\mathbf{j}; l(\mathbf{j}; j_{a}+1, \mathbf{d})_{b}, \mathbf{d})_i
    =
    r(\mathbf{j}^{(a+1)}; l(\mathbf{j}^{(a+1)}; j_{a}, \mathbf{d})_{b}, \mathbf{d})_i,
    \qquad i\neq b.
\end{align}
This completes the proof of Eq.~\eqref{eq:r_shift_Ba-_bulk}.

\textbf{Proof of the identities given in Eqs.~\eqref{eq:jalpha_shift_Ba-_boundary}, \eqref{eq:r_shift_Ba-_boundary} used in $B_{a,-}$ :}
We then present the proof of equalities given in 
Eqs.~\eqref{eq:jalpha_shift_Ba-_boundary}, \eqref{eq:r_shift_Ba-_boundary}.

We now prove Eq.~\eqref{eq:jalpha_shift_Ba-_boundary}
\begin{align}
    j_{\alpha}\!\left((\mathbf{j}_a,N-1);\,l((\mathbf{j}_a,N-1);0,\mathbf{d})_b\right)
    =
    j_{\alpha}\!\left((\mathbf{j}_a,0);\,l((\mathbf{j}_a,0);N-1,\mathbf{d})_b\right)
    -(N-n_h-1). \nonumber 
\end{align}

Recall that for a given hole configuration $\mathbf j$ and reference site $j'$, the
position of the $i$-th spin-down fermion in the reduced $(N-n_h)$-site chain is
\begin{align}
    j_{\alpha}(\mathbf j;j')
    +
    \sum_{\ell=1}^{i-1} d_\ell
    \pmod{N-n_h}.
    \label{eq:reduced_coordinate_general}
\end{align}
Equivalently, if the $b$-th spin-down fermion is fixed at reduced-chain coordinate
$x$, then the first spin-down fermion must be located at
\begin{align}
    x-\sum_{\ell=1}^{b-1} d_\ell
    \pmod{N-n_h}.
    \label{eq:first_spin_down_coordinate_general}
\end{align}

We first consider the hole configuration $(\mathbf j_a,N-1)$. In this case, the
site $0$ is occupied and is the leftmost site in the reduced chain. Therefore,
its reduced-chain coordinate is
\begin{align}
    j_{\alpha}((\mathbf j_a,N-1);0)=0.
    \label{eq:left_boundary_reduced_coordinate}
\end{align}
By definition,
\begin{align}
    l((\mathbf j_a,N-1);0,\mathbf d)_b
\end{align}
is the position of the first spin-down fermion in the original $N$-site chain
such that the $b$-th spin-down fermion is located at site $0$. Since the latter
has reduced-chain coordinate $0$, Eq.~\eqref{eq:first_spin_down_coordinate_general}
gives
\begin{align}
    j_{\alpha}\!\left((\mathbf j_a,N-1);\,l((\mathbf j_a,N-1);0,\mathbf d)_b\right)
    =
    -\sum_{\ell=1}^{b-1} d_\ell
    \pmod{N-n_h}.
    \label{eq:left_boundary_first_spin_down_coordinate}
\end{align}

Next we consider the hole configuration $(\mathbf j_a,0)$. In this case, the
site $N-1$ is occupied and is the rightmost site in the reduced chain. Hence its
reduced-chain coordinate is
\begin{align}
    j_{\alpha}((\mathbf j_a,0);N-1)=N-n_h-1.
    \label{eq:right_boundary_reduced_coordinate}
\end{align}
Similarly, by definition,
\begin{align}
    l((\mathbf j_a,0);N-1,\mathbf d)_b
\end{align}
is the position of the first spin-down fermion in the original chain such that
the $b$-th spin-down fermion is located at site $N-1$. Since the latter has
reduced-chain coordinate $N-n_h-1$, Eq.~\eqref{eq:first_spin_down_coordinate_general}
implies
\begin{align}
    j_{\alpha}\!\left((\mathbf j_a,0);\,l((\mathbf j_a,0);N-1,\mathbf d)_b\right)
    =
    N-n_h-1-\sum_{\ell=1}^{b-1} d_\ell
    \pmod{N-n_h}.
    \label{eq:right_boundary_first_spin_down_coordinate}
\end{align}

Comparing Eqs.~\eqref{eq:left_boundary_first_spin_down_coordinate} and
\eqref{eq:right_boundary_first_spin_down_coordinate}, we obtain
\begin{align}
    j_{\alpha}\!\left((\mathbf j_a,N-1);\,l((\mathbf j_a,N-1);0,\mathbf d)_b\right)
    &=
    -\sum_{\ell=1}^{b-1} d_\ell
    \nonumber\\
    &=
    \left(N-n_h-1-\sum_{\ell=1}^{b-1} d_\ell\right)-(N-n_h-1)
    \nonumber\\
    &=
    j_{\alpha}\!\left((\mathbf j_a,0);\,l((\mathbf j_a,0);N-1,\mathbf d)_b\right)
    -(N-n_h-1),
\end{align}
which proves Eq.~\eqref{eq:jalpha_shift_Ba-_boundary}.

Next, we prove Eq.~\eqref{eq:r_shift_Ba-_boundary}.
Let
\begin{align}
    x_b^{(L)}
    &:=
    j_\alpha\!\left((\mathbf j_a,N-1);\,l((\mathbf j_a,N-1);0,\mathbf d)_b\right),
    \\
    x_b^{(R)}
    &:=
    j_\alpha\!\left((\mathbf j_a,0);\,l((\mathbf j_a,0);N-1,\mathbf d)_b\right).
\end{align}
From Eq.~\eqref{eq:jalpha_shift_Ba-_boundary}, we have
\begin{align}
    x_b^{(L)}=x_b^{(R)}-(N-n_h-1).
\end{align}
Hence the reduced-chain coordinates of the $i$-th spin-down fermion satisfy
\begin{align}
    x_b^{(L)}+\sum_{\ell=1}^{i-1}d_\ell
    =
    x_b^{(R)}+\sum_{\ell=1}^{i-1}d_\ell +1
    \pmod{N-n_h}.
\end{align}
Since the hole configurations $(\mathbf j_a,N-1)$ and $(\mathbf j_a,0)$ differ
only by a cyclic relabeling of the reduced chain by one site, their inverse maps
satisfy
\begin{align}
    j_\alpha^{-1}((\mathbf j_a,N-1);y)
    =
    j_\alpha^{-1}((\mathbf j_a,0);y-1).
\end{align}
Therefore,
\begin{align}
    r((\mathbf{j}_a,N-1);\,l((\mathbf{j}_a,N-1);0,\mathbf d)_b,\mathbf d)_i
    =
    r((\mathbf{j}_a,0);\,l((\mathbf{j}_a,0);N-1,\mathbf d)_b,\mathbf d)_i.
\end{align}
This proves Eq.~\eqref{eq:r_shift_Ba-_boundary}.

\section{Analytical expressions for the Drude weight and specific heat for a fixed number of holes}
\label{sec:fixed_number_hole_Drude_weight}

In this section, we derive analytical expressions for the Drude weight and specific heat for a fixed number of holes \(n_h\), i.e., with \(n_h\) fixed in the thermodynamic limit. We have also shown that \(\sigma_{\mathrm{diff}}(\omega, T) = 2\pi D(T)\,\delta(\omega)\); see the definition in Sec.~\ref{sec:Optical_conductivity_model}.

\subsection{Single-hole case}
\subsubsection{Partition function}
\label{sec:Analytical_expression_of_partition_function_1H}

Based on the exact solutions for the energy spectrum derived in Sec.~\ref{sec:Exact_solutions_with_single_hole}, we now provide the analytical expression for the partition function with a single hole.
The partition function is defined as 
\begin{eqnarray}
Z
= \Tr(e^{-\beta \hat{H}}) = 
\sum_{n} e^{-\beta E_n}.
\end{eqnarray}
Explicitly, it reads
\begin{eqnarray}
\label{eq:1HPartitionFunction}
Z
&=&
2 \sum_{m=0}^{N-1}
e^{-2\beta \cos\!\left(\frac{2\pi m}{N}\right)}
+
\sum_{n_{\downarrow}=1}^{N-2}
\frac{\binom{N-1}{n_{\downarrow}}}{N-1}
\sum_{m=0}^{N-1}
\sum_{m_{\downarrow}=0}^{N-2}
e^{-2\beta \cos\!\left(
\frac{2\pi m}{N}
+
\frac{2\pi m_{\downarrow}}{N(N-1)}
\right)} \nonumber \\ 
&=& 2 \sum_{m=0}^{N-1}
e^{-2\beta \cos\!\left(\frac{2\pi m}{N}\right)}
+
\frac{2^{\,N-1}-2}{N-1}
\sum_{m=0}^{N-1}
\sum_{m_{\downarrow}=0}^{N-2}
e^{-2\beta \cos\!\left(
\frac{2\pi m}{N}
+
\frac{2\pi m_{\downarrow}}{N(N-1)}
\right)} \nonumber \\ 
&=& 2 \sum_{m=0}^{N-1}
e^{-2\beta \cos\!\left(\frac{2\pi m}{N}\right)}
+
\frac{2^{\,N-1}-2}{N-1}
\sum_{m=0}^{N-1}
\sum_{m_{\downarrow}=0}^{N-2}
e^{-2\beta \cos\!\left(
\frac{2\pi ((N-1)m + m_{\downarrow} )}{N(N-1)}
\right)}  \nonumber \\ 
&=& 2 \sum_{m=0}^{N-1}
e^{-2\beta \cos\!\left(\frac{2\pi m}{N}\right)}
+
\frac{2^{\,N-1}-2}{N-1}
\sum_{q=0}^{N(N-1)-1}
e^{-2\beta \cos\!\left(
\frac{2\pi q}{N(N-1)}
\right)},
\end{eqnarray}
where the factor $2$ in the first term arises because it accounts for the contribution from both sectors with $n_{\downarrow}=0$ and $n_{\downarrow}=N-1$. These two sectors have the same energy spectrum, as one can be obtained from the other by flipping the spins of all fermions. In the second equality we have used the binomial identity $\sum_{n_{\downarrow}=1}^{N-2} \binom{N-1}{n_{\downarrow}}=\sum_{n_{\downarrow}=0}^{N-1}
\binom{N-1}{n_{\downarrow}} -2  = 2^{N-1}-2$.
The large degeneracy originates from the fact that, in the
$U\to\infty$ limit, the Hubbard model reduces to the $t$-$J$ model
with $J=0$, so that the antiferromagnetic superexchange vanishes and
only the charge degrees of freedom remain active. In the fourth equality, we define $q=(N-1)m+m_{\downarrow}$, so that $q=0,\cdots, N(N-1)-1$.

Using the Jacobi--Anger expansion~\cite{colton1998inverse, cuyt2008handbook} 
\begin{align}
e^{-2\beta\cos\theta}
&=
\sum_{l=-\infty}^{\infty}
(-1)^l I_l(2\beta)\, e^{i l\theta}
\nonumber\\
&=
I_0(2\beta)
+2\sum_{l=1}^{\infty}(-1)^l I_l(2\beta)\cos(l\theta),
\label{eq:Jacobi-Anger}
\end{align}
where $I_l(\cdot)$ is the $l$-th modified Bessel function of the first kind,
the partition function is rewritten as
\begin{align}
Z
&=
2\sum_{m=0}^{N-1}
\sum_{l=-\infty}^{\infty}
(-1)^l I_l(2\beta)\,
e^{i l \frac{2\pi m}{N}} +
\frac{2^{N-1}-2}{N-1}
\sum_{q=0}^{N(N-1)-1}
\sum_{l=-\infty}^{\infty}
(-1)^l I_l(2\beta)\,
e^{i l \frac{2\pi q}{N(N-1)}} \nonumber \\  
&=
2
\sum_{l=-\infty}^{\infty}
(-1)^l I_l(2\beta)\, \sum_{m=0}^{N-1}
e^{i l \frac{2\pi m}{N}} +
\frac{2^{N-1}-2}{N-1}
\sum_{l=-\infty}^{\infty}
(-1)^l I_l(2\beta)\, \sum_{q=0}^{N(N-1)-1}
e^{i l \frac{2\pi q}{N(N-1)}}
\end{align}
where in the second equality we have simply exchanged the order of summation.
Since 
\begin{align}
\sum_{m=0}^{N-1} e^{i l \frac{2\pi m}{N}}
=
\begin{cases}
N, & l \in N\mathbb{Z},\\[4pt]
0, & l \notin N\mathbb{Z},
\end{cases}
\end{align}
and 
\begin{align}
\sum_{q=0}^{N(N-1)-1} e^{i l \frac{2\pi q}{N(N-1)}}
=
\begin{cases}
N(N-1), & l \in N(N-1)\mathbb{Z},\\[4pt]
0, & l \notin N(N-1)\mathbb{Z},
\end{cases}
\end{align}
The partition function is 
\begin{align}
   Z &=
2 N
\sum_{p=-\infty}^{\infty}
(-1)^{pN} I_{pN}(2\beta) +
N(N-1)\frac{2^{N-1}-2}{N-1}
\sum_{p=-\infty}^{\infty}
(-1)^{pN(N-1)} I_{pN(N-1)}(2\beta) \nonumber \\ 
& =2 N
\sum_{p=-\infty}^{\infty}
(-1)^{pN} I_{pN}(2\beta) +
N(2^{N-1}-2)
\sum_{p=-\infty}^{\infty}
(-1)^{pN(N-1)} I_{pN(N-1)}(2\beta) \nonumber \\ 
&= 2 N I_{0}(2\beta) + 
4N\sum_{p=1}^{\infty} 
(-1)^{pN} I_{pN}(2\beta) + N(2^{N-1}-2) I_{0}(2\beta) + 2N(2^{N-1}-2) \sum_{p=1}^{\infty}
(-1)^{pN(N-1)} I_{pN(N-1)}(2\beta) \nonumber\\ 
&= N 2^{N-1} I_{0}(2\beta) + 
4N\sum_{p=1}^{\infty} 
(-1)^{pN} I_{pN}(2\beta) + 2N(2^{N-1}-2) \sum_{p=1}^{\infty}
(-1)^{pN(N-1)} I_{pN(N-1)}(2\beta),
\end{align}
where in the third equality, we have used $I_{-p}(2\beta) = I_p(2 \beta)$. In the thermodynamic limit, the asymptotic expressions for $I_{pN}(2\beta)$ and $I_{pN(N-1)}(2\beta)$ are
\begin{align}
I_{pN}(2\beta)
&\sim
\frac{1}{\sqrt{2\pi pN}}
\left(\frac{e\beta}{pN}\right)^{pN}, \label{eq:IpN}  \\ 
I_{pN(N-1)}(2\beta)
&\sim
\frac{1}{\sqrt{2\pi pN(N-1)}}
\left(\frac{e\beta}{pN(N-1)}\right)^{pN(N-1)}, \label{eq:IpNN-1}
\end{align}
which are exponentially suppressed. Therefore, the thermodynamic term, i.e., the leading term in $N$, for the partition function is given by $N 2^{N-1} I_{0}(2\beta)$. Next, we consider the subleading correction in terms of $N$. When $p=1$, using the asymptotic forms shown in Eqs.~\eqref{eq:IpN} and ~\eqref{eq:IpNN-1}, we obtain
\begin{align}
\frac{2N \bigl(2^{N-1}-2\bigr) I_{N(N-1)}(2\beta)}{4N I_N(2\beta)}
&=
\frac{\bigl(2^{N-1}-2\bigr) I_{N(N-1)}(2\beta)}{2 I_N(2\beta)} \nonumber \\ 
&\sim
\frac{2^{N-1}-2}{2\sqrt{N-1}}
\left(\frac{e\beta}{N(N-1)}\right)^{N(N-1)}
\left(\frac{N}{e\beta}\right)^N.
\end{align}
Taking the logarithm gives
\begin{align}
\ln\!\left[
\frac{\bigl(2^{N-1}-2\bigr) I_{N(N-1)}(2\beta)}{2 I_N(2\beta)}
\right]
\sim
N\ln 2
-
N(N-1)\ln\!\left(\frac{N(N-1)}{e\beta}\right)
+
N\ln\!\left(\frac{N}{e\beta}\right),
\end{align}
whose leading term scales as $-2N^2\ln N$. Therefore,
\begin{align}
\label{eq:ComparisonININN-1}
\frac{2N \bigl(2^{N-1}-2\bigr) I_{N(N-1)}(2\beta)}{4N I_N(2\beta)}
\to 0,
\qquad N\to\infty.
\end{align}
Consequently, the subleading term is $4(-1)^{N}N I_{N}(2\beta)$.

In conclusion, in the thermodynamic limit, the partition function is 
\begin{align}
    Z  \approx N\,2^{\,N-1}\, I_{0}(2\beta) + 4(-1)^{N}N I_{N}(2\beta),
\end{align}
where $4(-1)^{N}N I_{N}(2\beta)$ is the subleading term in $N$. Here, ``$\approx$'' indicates that higher-order terms in $N$ have been neglected.

\subsubsection{Drude weight}
\label{sec:Analytical_expression_for_the_Drude_weight_1H}
In this section, we derive an analytical expression for the Drude weight in the single-hole sector. As discussed in Sec.~\ref{sec:Optical_conductivity_model}, only the Drude weight contributes to the optical conductivity. 

For the present model,
due to $[\hat{J}, \hat{H}]=0$, 
the Drude weight is given by
\begin{eqnarray}
D(T) 
= \frac{1}{2N} \langle -\hat{H} \rangle 
= -\frac{1}{2N} \frac{\Tr(\hat{H} e^{-\beta \hat{H}})}{Z}.
\end{eqnarray}
We therefore evaluate the trace $
\Tr(\hat{H} e^{-\beta \hat{H}})$.
Explicitly, it reads
\begin{align}
    \Tr(\hat{H} e^{-\beta \hat{H}}) &= 2 \sum_{m=0}^{N-1}
e^{-2\beta \cos\!\left(\frac{2\pi m}{N}\right)}  2\cos\!\left(\frac{2\pi m}{N}\right)
+
\sum_{n_{\downarrow}=1}^{N-2}
\frac{\binom{N-1}{n_{\downarrow}}}{N-1}
\sum_{m=0}^{N-1}
\sum_{m_{\downarrow}=0}^{N-2}
e^{-2\beta \cos\!\left(
\frac{2\pi m}{N}
+
\frac{2\pi m_{\downarrow}}{N(N-1)}
\right)} 2\cos\!\left(
\frac{2\pi m}{N}
+
\frac{2\pi m_{\downarrow}}{N(N-1)}
\right) \nonumber \\ 
& = 2 \sum_{m=0}^{N-1}
e^{-2\beta \cos\!\left(\frac{2\pi m}{N}\right)}  2\cos\!\left(\frac{2\pi m}{N}\right)
+ \frac{2^{N-1}-2}{N-1}
\sum_{m=0}^{N-1}
\sum_{m_{\downarrow}=0}^{N-2}
e^{-2\beta \cos\!\left(
\frac{2\pi m}{N}
+
\frac{2\pi m_{\downarrow}}{N(N-1)}
\right)} 2\cos\!\left(
\frac{2\pi m}{N}
+
\frac{2\pi m_{\downarrow}}{N(N-1)}
\right)  \nonumber \\ 
& = 2 \sum_{m=0}^{N-1}
e^{-2\beta \cos\!\left(\frac{2\pi m}{N}\right)}  2\cos\!\left(\frac{2\pi m}{N}\right)
+ \frac{2^{N-1}-2}{N-1}
\sum_{m=0}^{N-1}
\sum_{m_{\downarrow}=0}^{N-2}
e^{-2\beta \cos\!\left(
\frac{2\pi ((N-1)m + m_{\downarrow})}{N(N-1)}
\right)} 2\cos\!\left(
\frac{2\pi((N-1)m +  m_{\downarrow})}{N(N-1)}
\right) \nonumber \\ 
& = 2 \sum_{m=0}^{N-1}
e^{-2\beta \cos\!\left(\frac{2\pi m}{N}\right)}  2\cos\!\left(\frac{2\pi m}{N}\right)
+ \frac{2^{N-1}-2}{N-1}
\sum_{q=0}^{N(N-1)-1}
e^{-2\beta \cos\!\left(
\frac{2\pi q}{N(N-1)}
\right)} 2\cos\!\left(
\frac{2\pi q}{N(N-1)}
\right). 
\end{align}

Applying the Jacobi--Anger expansion (see Eq.~\eqref{eq:Jacobi-Anger}),
we obtain
\begin{align}
    \Tr(\hat{H} e^{-\beta \hat{H}}) &= 2 I_{0}(2\beta) \sum_{m=0}^{N-1} 2\cos(\frac{2\pi m}{N})  +4\sum_{l=1}^{\infty}(-1)^l I_l(2\beta) \sum_{m=0}^{N-1}2 \cos(l \frac{2\pi m}{N}) \cos(\frac{2\pi m}{N}) \nonumber \\ 
    & +  \frac{2^{N-1}-2}{N-1} I_{0}(2\beta) \sum_{q=0}^{N(N-1)-1} 2 \cos(\frac{2 \pi q}{N(N-1)})  \nonumber \\
    & +  2 \frac{2^{N-1}-2}{N-1}\sum_{l=1}^{\infty}(-1)^l I_l(2\beta) \sum_{q=0}^{N(N-1)-1}2 \cos(l \frac{2\pi q}{N(N-1)}) \cos(\frac{2\pi q}{N(N-1)}) \nonumber \\ 
&= 2 I_{0}(2\beta) \sum_{m=0}^{N-1} 2\cos(\frac{2\pi m}{N})  +4\sum_{l=1}^{\infty}(-1)^l I_l(2\beta) \sum_{m=0}^{N-1} \left( \cos((l+1) \frac{2\pi m}{N}) +  \cos((l-1)\frac{2\pi m}{N})  \right) \nonumber \\ 
    & +  \frac{2^{N-1}-2}{N-1} I_{0}(2\beta) \sum_{q=0}^{N(N-1)-1}  2 \cos(\frac{2 \pi q}{N(N-1)})  \nonumber \\ 
    & +  2 \frac{2^{N-1}-2}{N-1}\sum_{l=1}^{\infty}(-1)^l I_l(2\beta) \sum_{q=0}^{N(N-1)-1} \left( \cos((l+1) \frac{2\pi q}{N(N-1)}) + \cos((l-1)\frac{2\pi q}{N(N-1)}) \right) \nonumber \\ 
&= 4\sum_{l=1}^{\infty}(-1)^l I_l(2\beta) \sum_{m=0}^{N-1} \left( \cos((l+1) \frac{2\pi m}{N}) +  \cos((l-1)\frac{2\pi m}{N})  \right) \nonumber \\ 
    & +  2 \frac{2^{N-1}-2}{N-1}\sum_{l=1}^{\infty}(-1)^l I_l(2\beta) \sum_{q=0}^{N(N-1)-1} \left( \cos((l+1) \frac{2\pi q}{N(N-1)}) + \cos((l-1)\frac{2\pi q}{N(N-1)}) \right). 
\end{align}
In the second equality, we have used
$2\cos\theta\cos(l\theta)
=
\cos((l-1)\theta)+\cos((l+1)\theta)$. In the third equality, we have used $\sum_{m=0}^{N-1} \cos(\frac{2\pi m}{N})=0$ and $\sum_{q=0}^{N(N-1)-1}  \cos(\frac{2 \pi q}{N(N-1)}) =0$. Since
\begin{align}
\sum_{m=0}^{N-1}\cos\!\left((l+1)\frac{2\pi m}{N}\right)
=
\begin{cases}
N, & l+1 \in N\mathbb{Z},\\[4pt]
0, & l+1 \notin N\mathbb{Z},
\end{cases}
\end{align}
\begin{align}
\sum_{m=0}^{N-1}\cos\!\left((l-1)\frac{2\pi m}{N}\right)
=
\begin{cases}
N, & l-1 \in N\mathbb{Z},\\[4pt]
0, & l-1 \notin N\mathbb{Z},
\end{cases}
\end{align}
\begin{align}
\sum_{q=0}^{N(N-1)-1}\cos\!\left((l+1)\frac{2\pi q}{N(N-1)}\right)
=
\begin{cases}
N(N-1), & l+1 \in N(N-1)\mathbb{Z},\\[4pt]
0, & l+1 \notin N(N-1)\mathbb{Z},
\end{cases}
\end{align}
\begin{align}
\sum_{q=0}^{N(N-1)-1}\cos\!\left((l-1)\frac{2\pi q}{N(N-1)}\right)
=
\begin{cases}
N(N-1), & l-1 \in N(N-1)\mathbb{Z},\\[4pt]
0, & l-1 \notin N(N-1)\mathbb{Z},
\end{cases}
\end{align}
we obtain 
\begin{align}
        \Tr(\hat{H} e^{-\beta \hat{H}}) 
&= 4N \sum_{p=1}^{\infty}(-1)^{pN-1} I_{pN-1}(2\beta) + 4N \sum_{p=0}^{\infty}(-1)^{pN+1} I_{pN+1}(2\beta) \nonumber \\ 
& + N (2^N-4)  \sum_{p=1}^{\infty}(-1)^{pN(N-1)-1} I_{pN(N-1)-1}(2\beta) + N (2^N-4)  \sum_{p=0}^{\infty}(-1)^{pN(N-1)+1} I_{pN(N-1)+1}(2\beta) \nonumber \\ 
&= -N2^N I_1(2\beta) +
4N \sum_{p=1}^{\infty}(-1)^{pN-1} I_{pN-1}(2\beta) + 4N \sum_{p=1}^{\infty}(-1)^{pN+1} I_{pN+1}(2\beta) \nonumber \\ 
& + N (2^N-4)  \sum_{p=1}^{\infty}(-1)^{pN(N-1)-1} I_{pN(N-1)-1}(2\beta) + N (2^N-4)  \sum_{p=1}^{\infty}(-1)^{pN(N-1)+1} I_{pN(N-1)+1}(2\beta).
\end{align}
Therefore, as discussed in Eqs.~\eqref{eq:IpN} and ~\eqref{eq:IpNN-1}, the thermodynamic term is $-N2^N I_1(2\beta)$. Next, we consider the subleading term in $N$. When $p=1$, there are four terms: $4N (-1)^{N-1} I_{N-1}(2\beta)$, $4N (-1)^{N+1} I_{N+1}(2\beta)$, $N(2^N-4) (-1)^{N(N-1)-1} I_{N(N-1)-1}(2\beta)$, $N(2^N-4) (-1)^{N(N-1)+1} I_{N(N-1)+1}(2\beta)$. As in the discussion in Eqs.~\eqref{eq:ComparisonININN-1}, the last two terms are negligible compared to the first two terms in the thermodynamic limit. We then compare the first two terms.
\begin{align}
\frac{4N I_{N+1}(2\beta)}{4N I_{N-1}(2\beta)}
&=
\frac{I_{N+1}(2\beta)}{I_{N-1}(2\beta)} \sim
\frac{\beta^2}{N(N+1)}
\to 0,
\qquad N\to\infty.
\end{align}
Therefore, the subleading term in $N$ is given by $4N (-1)^{N-1} I_{N-1}(2\beta)$.

In conclusion, 
\begin{eqnarray}
\Tr(\hat{H}e^{-\beta \hat{H}})
\approx
-\,N2^{N} I_{1}(2\beta) + 4N(-1)^{N-1}I_{N-1}(2\beta).
\end{eqnarray}
Here, ``$\approx$'' indicates that we keep only leading and subleading terms in $N$.

Consequently, the Drude weight $D(T)$ is
\begin{eqnarray}
\label{eq:1HAnalyticalExpressionOfDrude}
D(T)
&=&
\frac{1}{2N}\langle -\hat{H}\rangle
\nonumber \\ 
&\approx&\frac{1}{2N}
\frac{N2^{N} I_{1}(2\beta)}
{N2^{N-1}I_{0}(2\beta)}
\nonumber \\ 
&=&
\frac{1}{N}\frac{I_{1}(2\beta)}{I_{0}(2\beta)}
\nonumber \\ 
&=&
\frac{\beta}{N}
\left(
1-\frac{I_2(2\beta)}{I_0(2\beta)}
\right).
\end{eqnarray}
Here, ``$\approx$'' indicates that we keep only the leading term in $N$. In the last equality, we have used $\frac{I_1(2\beta)}{I_{0}(2\beta)} = \beta (1- \frac{I_{2}(2\beta)}{I_{0}(2\beta)})$.

After considering the subleading term in $N$, the Drude weight is given by
\begin{align}
    \label{eq:1HDrudeWithFiniteSiz}
    D(T) & \approx \frac{1}{2N} \frac{N2^N I_1(2\beta) + 4N(-1)^{N} I_{N-1}(2\beta)}{N2^{N-1} I_0(2\beta) + 4N(-1)^{N}  I_N(2\beta)} \nonumber \\ 
    &= \frac{1}{2N} \frac{2^N I_1(2\beta) + 4(-1)^{N} I_{N-1}(2\beta)}{2^{N-1} I_0(2\beta) + 4(-1)^{N}  I_N(2\beta)}. 
\end{align}

\subsubsection{$D_{\mathrm{diff}}(T)$}
\label{sec:Drude_weight_regular_1H}

In Eq.~\eqref{eq:1HAnalyticalExpressionOfDrude},
we provided
the analytical expression for the Drude weight in the single-hole sector. 

In this section, we present the analytical expression for $D_{\mathrm{diff}}(T)$ defined in Eq.~\eqref{eq:DefinitionOfSigmadiff}. We have proven that $D_{\mathrm{diff}}(T) = D(T)$, which indicates that the Meissner stiffness is zero. See Sec.~\ref{sec:Model_and_observables} for more details.

We denote
\begin{equation}
Z'
=
\sum_{n}
e^{-\beta E_n}
\bigl|
J_{nn}
\bigr|^2,
\end{equation}
such that $\sigma_{\mathrm{diff}}(\omega, T)$ in Eq.~\eqref{eq:DefinitionOfSigmadiff} is given by $\frac{\pi \beta}{N} \frac{Z'}{Z} \delta(\omega)$.
Using the exact expressions for $\vert J_{nn} \vert$ derived in Eqs.~\eqref{eq:1HNoDownJexpression}, ~\eqref{eq:1H1DJexpression} and ~\eqref{eq:ProofOfEigenstateJWithSpinDown}
we obtain
\begin{align}
Z'
& = 2 \sum_{m=0}^{N-1} e^{-2\beta \cos(\frac{2 \pi m}{N})} 4 \sin^2(\frac{2\pi m}{N}) +
\sum_{n_{\downarrow}=1}^{N-2}
\frac{\binom{N-1}{n_{\downarrow}}}{N-1}
\sum_{m=0}^{N-1}
\sum_{m_{\downarrow}=0}^{N-2}
e^{-2\beta \cos(\frac{2\pi m}{N} + \frac{2 \pi m_{\downarrow}}{N(N-1)})}
\,4\sin^2(\frac{2\pi m}{N} + \frac{2 \pi m_{\downarrow}}{N(N-1)})
\nonumber \\
& = 2 \sum_{m=0}^{N-1} e^{-2\beta \cos(\frac{2 \pi m}{N})} 2(1-\cos(2\frac{2\pi m}{N}))  \nonumber \\
&+
\frac{2^{N-1}-2}{N-1}
\sum_{m=0}^{N-1}
\sum_{m_{\downarrow}=0}^{N-2}
e^{-2\beta \cos(\frac{2\pi m}{N} + \frac{2 \pi m_{\downarrow}}{N(N-1)})} 2 (1-\cos(2(\frac{2\pi m}{N} + \frac{2 \pi m_{\downarrow}}{N(N-1)})))
\nonumber \\ 
& = 4 \sum_{m=0}^{N-1} e^{-2\beta \cos(\frac{2 \pi m}{N})} +
2 \frac{2^{N-1}-2}{N-1}
\sum_{m=0}^{N-1}
\sum_{m_{\downarrow}=0}^{N-2}
e^{-2\beta \cos(\frac{2\pi m}{N} + \frac{2 \pi m_{\downarrow}}{N(N-1)})} 
\nonumber \\
& - 4 \sum_{m=0}^{N-1} e^{-2\beta \cos(\frac{2 \pi m}{N})}\cos(2\frac{2\pi m}{N}) - 2 \frac{2^{N-1}-2}{N-1}
\sum_{m=0}^{N-1}
\sum_{m_{\downarrow}=0}^{N-2}
e^{-2\beta \cos(\frac{2\pi m}{N} + \frac{2 \pi m_{\downarrow}}{N(N-1)})} \cos(2(\frac{2\pi m}{N} + \frac{2 \pi m_{\downarrow}}{N(N-1)})) \nonumber \\ 
& = 2Z - S_{\mathrm{cos}}.
\end{align}
In the second equality, we have used $4\sin^2\theta = 2(1-\cos 2\theta)$. Here
\begin{align}
    S_{\mathrm{cos}} &=  4 \sum_{m=0}^{N-1} e^{-2\beta \cos(\frac{2 \pi m}{N})}\cos(2\frac{2\pi m}{N}) +  \frac{2^{N}-4}{N-1}
\sum_{m=0}^{N-1}
\sum_{m_{\downarrow}=0}^{N-2}
e^{-2\beta \cos(\frac{2\pi m}{N} + \frac{2 \pi m_{\downarrow}}{N(N-1)})} \cos(2(\frac{2\pi m}{N} + \frac{2 \pi m_{\downarrow}}{N(N-1)}))  \nonumber \\
& = 4 \sum_{m=0}^{N-1} e^{-2\beta \cos(\frac{2 \pi m}{N})}\cos(2\frac{2\pi m}{N}) +  \frac{2^{N}-4}{N-1}
\sum_{q=0}^{N(N-1)-1}
e^{-2\beta \cos(\frac{2 \pi q}{N(N-1)})} \cos(2 \frac{2 \pi q}{N(N-1)}).
\end{align}

Applying the Jacobi--Anger expansion (see Eq.~\eqref{eq:Jacobi-Anger}),
we obtain
\begin{align}
S_{\cos}
&=
4 \sum_{m=0}^{N-1} e^{-2\beta \cos\!\left(\frac{2 \pi m}{N}\right)}\cos\!\left(\frac{4\pi m}{N}\right)
+  \frac{2^{N}-4}{N-1}
\sum_{q=0}^{N(N-1)-1}
e^{-2\beta \cos\!\left(\frac{2 \pi q}{N(N-1)}\right)} \cos\!\left(\frac{4 \pi q}{N(N-1)}\right)
\nonumber\\
&=
4 I_0(2\beta)\sum_{m=0}^{N-1}\cos\!\left(\frac{4\pi m}{N}\right)
+
8\sum_{l=1}^{\infty}(-1)^l I_l(2\beta)
\sum_{m=0}^{N-1}
\cos\!\left(l\frac{2\pi m}{N}\right)\cos\!\left(\frac{4\pi m}{N}\right)
\nonumber\\
&\quad+
\frac{2^{N}-4}{N-1} I_0(2\beta)\sum_{q=0}^{N(N-1)-1}\cos\!\left(\frac{4\pi q}{N(N-1)}\right)
\nonumber\\
&\quad+
2\frac{2^{N}-4}{N-1}
\sum_{l=1}^{\infty}(-1)^l I_l(2\beta)
\sum_{q=0}^{N(N-1)-1}
\cos\!\left(l\frac{2\pi q}{N(N-1)}\right)\cos\!\left(\frac{4\pi q}{N(N-1)}\right)
\nonumber\\
&=
4 I_0(2\beta)\sum_{m=0}^{N-1}\cos\!\left(\frac{4\pi m}{N}\right)
+
4\sum_{l=1}^{\infty}(-1)^l I_l(2\beta)
\sum_{m=0}^{N-1}
\left[
\cos\!\left((l+2)\frac{2\pi m}{N}\right)
+
\cos\!\left((l-2)\frac{2\pi m}{N}\right)
\right]
\nonumber\\
&\quad+
\frac{2^{N}-4}{N-1} I_0(2\beta)\sum_{q=0}^{N(N-1)-1}\cos\!\left(\frac{4\pi q}{N(N-1)}\right)
\nonumber\\
&\quad+
\frac{2^{N}-4}{N-1}
\sum_{l=1}^{\infty}(-1)^l I_l(2\beta)
\sum_{q=0}^{N(N-1)-1}
\left[
\cos\!\left((l+2)\frac{2\pi q}{N(N-1)}\right)
+
\cos\!\left((l-2)\frac{2\pi q}{N(N-1)}\right)
\right] \nonumber \\ 
&=
4\sum_{l=1}^{\infty}(-1)^l I_l(2\beta)
\sum_{m=0}^{N-1}
\left[
\cos\!\left((l+2)\frac{2\pi m}{N}\right)
+
\cos\!\left((l-2)\frac{2\pi m}{N}\right)
\right]
\nonumber\\
&\quad+
\frac{2^{N}-4}{N-1}
\sum_{l=1}^{\infty}(-1)^l I_l(2\beta)
\sum_{q=0}^{N(N-1)-1}
\left[
\cos\!\left((l+2)\frac{2\pi q}{N(N-1)}\right)
+
\cos\!\left((l-2)\frac{2\pi q}{N(N-1)}\right)
\right] \nonumber \\ 
&=4 N\sum_{p=1}^{\infty}(-1)^{pN-2} I_{pN-2}(2\beta) + 4 N\sum_{p=0}^{\infty}(-1)^{pN+2} I_{pN+2}(2\beta) \nonumber  \\ 
&\quad+
N(2^{N}-4) \sum_{p=1}^{\infty} (-1)^{pN(N-1)-2} I_{pN(N-1)-2}(2\beta) +
N(2^{N}-4) \sum_{p=0}^{\infty} (-1)^{pN(N-1)+2} I_{pN(N-1)+2}(2\beta) \nonumber  \\  
& = N2^N I_{2}(2\beta) \nonumber \\ 
& \quad+ 4 N\sum_{p=1}^{\infty}(-1)^{pN-2} I_{pN-2}(2\beta) + 4 N\sum_{p=1}^{\infty}(-1)^{pN+2} I_{pN+2}(2\beta)  \nonumber \\ 
&\quad+
N(2^{N}-4) \sum_{p=1}^{\infty} (-1)^{pN(N-1)-2} I_{pN(N-1)-2}(2\beta) +
N(2^{N}-4) \sum_{p=0}^{\infty} (-1)^{pN(N-1)+2} I_{pN(N-1)+2}(2\beta). 
\end{align}
In the fourth equality, we have used $\sum_{m=0}^{N-1}\cos\!\left(\frac{4\pi m}{N}\right)=0$ and $\sum_{q=0}^{N(N-1)-1}\cos\!\left(\frac{4\pi q}{N(N-1)}\right)=0$. Therefore, the thermodynamic term (the leading term in $N$) is $S_{\mathrm{cos}} = N2^N I_2(2\beta)$. And the subleading term in $N$ is $4N(-1)^{N-2} I_{N-2}(2\beta)$.

Therefore, in the thermodynamic limit, 
\begin{align}
    \sigma_{\mathrm{diff}}(\omega, T) &= \frac{\pi \beta}{N} \frac{Z'}{Z} \delta(\omega) \nonumber \\ 
    &= \frac{\pi \beta}{N} \frac{2Z - S_{\mathrm{cos}}}{Z} \delta(\omega) \nonumber \\ 
    & \approx \frac{\pi \beta}{N} (2 - \frac{N2^N I_2(2\beta)}{N2^{N-1} I_0(2\beta)} )  \delta(\omega) \nonumber \\
    & = \frac{2\pi \beta}{N} (1- \frac{I_2(2\beta)}{I_0(2\beta)})  \delta(\omega).
\end{align}
Here, ``$\approx$'' means that we keep only the leading term in the thermodynamic limit. Therefore,
\begin{align}
    D_{\mathrm{diff}}(T) = \frac{\beta}{N} (1- \frac{I_2(2\beta)}{I_0(2\beta)}).
\end{align}
Comparing with the analytical expression for the Drude weight in Eq.~\eqref{eq:1HAnalyticalExpressionOfDrude}, we have proved in the thermodynamic limit
\begin{align}
    D_{\mathrm{diff}}(T) = D(T),
\end{align}
which indicates that the Meissner stiffness is zero.

Next, we consider the subleading term.
\begin{align}
    \sigma_{\mathrm{diff}}(\omega, T) &= \frac{\pi \beta}{N} (2 - \frac{N2^N I_2(2\beta) + 4 (-1)^N  N I_{N-2}(2\beta)}{N2^{N-1} I_0(2\beta) + 4(-1)^N N I_{N}(2\beta)})  \delta(\omega)\nonumber \\ 
    &= \frac{\pi \beta}{N} (2 - \frac{2^N I_2(2\beta) + 4 (-1)^N   I_{N-2}(2\beta)}{2^{N-1} I_0(2\beta) + 4(-1)^N  I_{N}(2\beta)}) \delta(\omega).
\end{align}
Therefore, comparing with the Drude weight shown in Eq.~\eqref{eq:1HDrudeWithFiniteSiz}, after considering the subleading term in $N$, $2\pi D(T) \delta(\omega)$ and $\sigma_{\mathrm{diff}}(\omega, T) = 2 \pi D_{\mathrm{diff}}(T) \delta(\omega)$ are different.

\subsubsection{Specific heat}
\label{sec:Analytical_expression_for_the_specific_heat_1H}
In this section, we also provide an analytical expression for the specific heat (see definition in Eq.~\eqref{eq:specificheat}). Since we have obtained the analytical expression for $\langle \hat{H} \rangle$, we then calculate $\langle \hat{H}^2\rangle$ in the following.

For the present model with a single hole, we have
\begin{align}
    \Tr(\hat{H}^2 e^{-\beta \hat{H}}) &= \Tr((4-\hat{J}^2) e^{-\beta \hat{H}}) \nonumber \\ 
    & = 4 \Tr(e^{-\beta \hat{H}}) - \Tr(\hat{J}^2 e^{-\beta \hat{H}}) \nonumber \\ 
    & = 4Z - (2Z-S_{\mathrm{cos}}) \nonumber \\ 
    & = 2Z + S_{\mathrm{cos}} \nonumber \\ 
    & \approx N2^N I_{0}(2\beta) + N2^N I_{2}(2\beta).
\end{align}
Here, ``$\approx$'' means that we keep only the leading term in $N$.

Therefore, the specific heat in the thermodynamic limit is  
\begin{align}
    C_V(T) &= \beta^2 ( \langle \hat{H}^2 \rangle - \langle \hat{H} \rangle^2) \nonumber \\ 
    &\approx \beta^2 (\frac{N2^N I_{0}(2\beta) + N2^N I_2(2\beta)}{N2^{N-1}I_{0}(2\beta)} - (\frac{N2^N I_1(2\beta)}{N2^{N-1}I_0(2\beta)})^2) \nonumber \\ 
    &=  \beta^2 (2 + 2\frac{I_2(2\beta)}{I_0(2\beta)} - 4(\frac{ I_1(2\beta)}{I_0(2\beta)})^2), 
\end{align}
where ``$\approx$'' indicates that we keep only the leading order term in $N$.

In the low-temperature limit ($\beta\to\infty$), this expression simplifies to
\begin{eqnarray}
C_V(T)
\approx
\frac{1}{2}
+
\frac{1}{8\beta}.
\end{eqnarray}

The origin of the finite zero-temperature limit $C_V\to 1/2$ can be understood from the low-energy dispersion. For the single-hole case, 
the band dispersion is
\begin{eqnarray}
\varepsilon(k)
=
2\cos k,
\end{eqnarray}
where $k = \frac{2\pi m}{N} + \frac{2 \pi m_{\downarrow}}{N(N-1)} = \frac{2 \pi q}{N(N-1)}$ with $q= (N-1) m + m_{\downarrow}$.
At low temperatures, the dominant contributions arise from excitations near the band minimum at $k=\pi$. Expanding around $k=\pi$ yields a quadratic dispersion,
\begin{eqnarray}
\varepsilon(k)
\approx
-2
+
(k-\pi)^2.
\end{eqnarray}
Treating $k$ as a continuous variable and shifting $k\to k+\pi$, the partition function reduces to a Gaussian integral,
\begin{eqnarray}
Z
\sim
\int dk \, e^{-\beta k^2}.
\end{eqnarray}
This corresponds to a single quadratic degree of freedom, which contributes a constant heat capacity $C_V = 1/2$ in the low-temperature limit, in agreement with the analytical result.

\subsection{Two-hole case}
\subsubsection{Partition function}
In this section, we derive the analytical expression for the partition
function $Z$ in the two-hole sector. To simplify the calculation and
obtain exact results, we assume that $N-2$ is a prime number, so that
only the normal orbits of length $N-2$ appear.

The partition function with system size $N$ is given by 
\begin{eqnarray}
Z
&=&
2
\sum_{0 \le m_1 < m_2 \le N-1}
e^{-2\beta
\left(
\cos\!\left(\frac{2\pi m_1}{N}\right)
+
\cos\!\left(\frac{2\pi m_2}{N}\right)
\right)}
\nonumber\\
&&
+
\sum_{n_{\downarrow}=1}^{N-3}
\frac{\binom{N-2}{n_{\downarrow}}}{N-2}
\sum_{0 \le m_1 < m_2 \le N-1}
\sum_{m_{\downarrow}=0}^{N-3}
e^{-2\beta
\Bigl(
\cos\!\left(\frac{2\pi m_1}{N}
+ \frac{2\pi m_{\downarrow}}{N(N-2)}\right)
+
\cos\!\left(\frac{2\pi m_2}{N}
+ \frac{2\pi m_{\downarrow}}{N(N-2)}\right)
\Bigr)}
\nonumber\\
& =& 2
\sum_{0 \le m_1 < m_2 \le N-1}
e^{-2\beta
\left(
\cos\!\left(\frac{2\pi m_1}{N}\right)
+
\cos\!\left(\frac{2\pi m_2}{N}\right)
\right)} \nonumber \\ 
&& +
\frac{2^{N-2}-2}{N-2}
\sum_{0 \le m_1 < m_2 \le N-1}
\sum_{m_{\downarrow}=0}^{N-3} e^{-2\beta
\Bigl(
\cos\!\left(\frac{2\pi m_1}{N}
+ \frac{2\pi m_{\downarrow}}{N(N-2)}\right)
+
\cos\!\left(\frac{2\pi m_2}{N}
+ \frac{2\pi m_{\downarrow}}{N(N-2)}\right)
\Bigr)},
\end{eqnarray}
where in the last equality we have used $\sum_{n_{\downarrow}=1}^{N-3} \binom{N-2}{n_{\downarrow}} = \sum_{n_{\downarrow}=0}^{N-2} \binom{N-2}{n_{\downarrow}} - 2 = 2^{N-2}-2$.

For convenience, we introduce the notations
\begin{eqnarray}
\alpha_m = \frac{2\pi m}{N},
\qquad
\phi_{m_{\downarrow}} = \frac{2\pi m_{\downarrow}}{N(N-2)}.
\end{eqnarray}

The partition function can then be written as
\begin{eqnarray}
\label{eq:2HPartitionZ}
Z
&=& 2\sum_{0 \leq m_1 < m_2 \leq N-1} e^{-2\beta \left[\cos(\alpha_{m_1}) + \cos(\alpha_{m_2}) \right] } + 
\frac{2^{N-2}-2}{N-2}
\sum_{0 \le m_1 < m_2 \le N-1}
\sum_{m_{\downarrow}=0}^{N-3}
e^{-2\beta\left[
\cos(\alpha_{m_1}+\phi_{m_{\downarrow}})
+
\cos(\alpha_{m_2}+\phi_{m_{\downarrow}})
\right]}
\nonumber\\
&=& \left( \sum_{0 \leq m_1, m_2 \leq N-1} e^{-2\beta \left[ \cos(\alpha_{m_1}) + \cos(\alpha_{m_2}) \right]}  - \sum_{0 \leq m \leq N-1}  e^{-4\beta \cos(\alpha_{m})} \right)  \nonumber \\
&& + \frac{2^{N-3}-1}{N-2} 
\Biggl(
\sum_{0 \le m_1,m_2 \le N-1}
\sum_{m_{\downarrow}=0}^{N-3}
e^{-2\beta\left[
\cos(\alpha_{m_1}+\phi_{m_{\downarrow}})
+
\cos(\alpha_{m_2}+\phi_{m_{\downarrow}})
\right]}
-
\sum_{0 \le m_1 \le N-1}
\sum_{m_{\downarrow}=0}^{N-3}
e^{-4\beta \cos(\alpha_{m_1}+\phi_{m_{\downarrow}})}
\Biggr) \nonumber \\ 
&=& \left( (\sum_{0 \leq m \leq N-1} e^{-2\beta  \cos(\alpha_{m})})^2  - \sum_{0 \leq m \leq N-1}  e^{-4\beta \cos(\alpha_{m})} \right) \nonumber \\ 
&& + \frac{2^{N-3}-1}{N-2} 
\Biggl(
\sum_{0 \le m_1,m_2 \le N-1}
\sum_{m_{\downarrow}=0}^{N-3}
e^{-2\beta\left[
\cos(\alpha_{m_1}+\phi_{m_{\downarrow}})
+
\cos(\alpha_{m_2}+\phi_{m_{\downarrow}})
\right]}
-
\sum_{0 \le m_1 \le N-1}
\sum_{m_{\downarrow}=0}^{N-3}
e^{-4\beta \cos(\alpha_{m_1}+\phi_{m_{\downarrow}})}
\Biggr) \nonumber \\ 
&=& S_1 + \frac{2^{N-3}-1}{N-2} S_2.
\end{eqnarray}
In the following, we calculate $S_1$ and $S_2$ separately.

Using the Jacobi--Anger expansion (see Eq.~\eqref{eq:Jacobi-Anger}),
we obtain
\begin{align}
\sum_{m=0}^{N-1} e^{-2\beta\cos(\alpha_m)}
&=
N I_0(2\beta)
+
2\sum_{l=1}^{\infty}(-1)^l I_l(2\beta)
\sum_{m=0}^{N-1}\cos(l\alpha_m),
\\
\sum_{m=0}^{N-1} e^{-4\beta\cos(\alpha_m)}
&=
N I_0(4\beta)
+
2\sum_{l=1}^{\infty}(-1)^l I_l(4\beta)
\sum_{m=0}^{N-1}\cos(l\alpha_m).
\end{align}
Therefore, $S_1$ in Eq.~\eqref{eq:2HPartitionZ} is 
\begin{align}
\label{eq:2HS1}
S_1 &=
\left(
N I_0(2\beta)
+
2\sum_{l=1}^{\infty}(-1)^l I_l(2\beta)
\sum_{m=0}^{N-1}\cos(l\alpha_m)
\right)^2-
\left(
N I_0(4\beta)
+
2\sum_{l=1}^{\infty}(-1)^l I_l(4\beta)
\sum_{m=0}^{N-1}\cos(l\alpha_m)
\right) \nonumber \\ 
&=
\left(
N I_0(2\beta)
+
2N \sum_{p=1}^{\infty}(-1)^{pN} I_{pN}(2\beta)
\right)^2 -
\left(
N I_0(4\beta)
+
2N \sum_{p=1}^{\infty}(-1)^{pN} I_{pN}(4\beta)
\right) \nonumber \\ 
& \approx N^2 I_0(2\beta)^2 - NI_{0}(4\beta),
\end{align}
where in the second equality we used that 
$\sum_{m=0}^{N-1}\cos(l\alpha_m)=N$ for $l\in N\mathbb{Z}$ and vanishes otherwise. Here, ``$\approx$'' indicates that we keep only the leading and subleading terms in $N$.

For $S_2$ in Eq.~\eqref{eq:2HPartitionZ}, with $\theta_{m,m_\downarrow}=\alpha_m+\phi_{m_\downarrow}$, we have
\begin{align}
&e^{-2\beta\left[
\cos(\alpha_{m_1}+\phi_{m_{\downarrow}})
+
\cos(\alpha_{m_2}+\phi_{m_{\downarrow}})
\right]} \nonumber \\ 
&=
e^{-2\beta\cos(\theta_{m_1,m_\downarrow})}
e^{-2\beta\cos(\theta_{m_2,m_\downarrow})}
\nonumber\\
&=
\left[
I_0(2\beta)+2\sum_{l_1=1}^{\infty}(-1)^{l_1}I_{l_1}(2\beta)
\cos\!\bigl(l_1\theta_{m_1,m_\downarrow}\bigr)
\right]
\left[
I_0(2\beta)+2\sum_{l_2=1}^{\infty}(-1)^{l_2}I_{l_2}(2\beta)
\cos\!\bigl(l_2\theta_{m_2,m_\downarrow}\bigr)
\right],
\end{align}
and
\begin{align}
e^{-4\beta\cos(\alpha_{m_1}+\phi_{m_\downarrow})}
=
I_0(4\beta)
+
2\sum_{l=1}^{\infty}(-1)^l I_l(4\beta)
\cos\!\bigl(l\theta_{m_1,m_\downarrow}\bigr).
\end{align}
Hence
\begin{align}
\label{eq:2HS2}
S_2 &=
\sum_{m_\downarrow=0}^{N-3}
\Biggl[
\sum_{m=0}^{N-1}
e^{-2\beta\cos(\alpha_m+\phi_{m_\downarrow})}
\Biggr]^2 - \sum_{m_1=0}^{N-1}
\sum_{m_{\downarrow}=0}^{N-3}
e^{-4\beta \cos(\alpha_{m_1}+\phi_{m_{\downarrow}})}
\nonumber\\
&=
\sum_{m_\downarrow=0}^{N-3}
\Biggl[
N I_0(2\beta)
+
2\sum_{l=1}^{\infty}(-1)^l I_l(2\beta)
\sum_{m=0}^{N-1}
\cos\!\bigl(l(\alpha_m+\phi_{m_\downarrow})\bigr)
\Biggr]^2 \nonumber \\ 
&- \sum_{m_\downarrow=0}^{N-3}
\Biggl[
N I_0(4\beta)
+
2\sum_{l=1}^{\infty}(-1)^l I_l(4\beta)
\sum_{m_1=0}^{N-1}
\cos\!\bigl(l(\alpha_{m_1}+\phi_{m_\downarrow})\bigr)
\Biggr] \nonumber \\ 
&=
\sum_{m_\downarrow=0}^{N-3}
\Biggl[
N I_0(2\beta)
+
2N \sum_{p=1}^{\infty}(-1)^{pN} I_{pN}(2\beta)
\cos\!\bigl(pN\phi_{m_\downarrow}\bigr)
\Biggr]^2 \nonumber \\ 
&- \sum_{m_\downarrow=0}^{N-3}
\Biggl[
N I_0(4\beta)
+
2N\sum_{p=1}^{\infty}(-1)^{pN} I_{pN}(4\beta)
\cos\!\bigl(l\phi_{m_\downarrow}\bigr)
\Biggr] \nonumber \\ 
&=
\sum_{m_\downarrow=0}^{N-3}
\Biggl[
N^2 (I_0(2\beta))^2 + 4N^2 I_0(2\beta) \sum_{p=1}^{\infty}(-1)^{pN} I_{pN}(2\beta)
\cos\!\bigl(pN\phi_{m_\downarrow}\bigr)
+
(2N \sum_{p=1}^{\infty}(-1)^{pN} I_{pN}(2\beta)
\cos\!\bigl(pN\phi_{m_\downarrow}\bigr))^2
\Biggr]\nonumber \\ 
&- 
\Biggl[
N(N-2) I_0(4\beta)
+
2N(N-2)\sum_{p'=1}^{\infty}(-1)^{p'N(N-2)} I_{p'N(N-2)}(4\beta)
\Biggr]  \nonumber \\
&=
\Biggl[
N^2 (N-2) (I_0(2\beta))^2 + 4N^2 (N-2) I_0(2\beta) \sum_{p'=1}^{\infty}(-1)^{p'N(N-2)} I_{p'N(N-2)}(2\beta)
+
\sum_{m_\downarrow=0}^{N-3}(2N \sum_{p=1}^{\infty}(-1)^{pN} I_{pN}(2\beta)
\cos\!\bigl(pN\phi_{m_\downarrow}\bigr))^2
\Biggr] \nonumber \\ 
&- 
\Biggl[
N(N-2) I_0(4\beta)
+
2N(N-2)\sum_{p'=1}^{\infty}(-1)^{p'N(N-2)} I_{p'N(N-2)}(4\beta)
\Biggr] \nonumber \\ 
& \approx  (N-2)N^2 I_0(2\beta)^2 - (N-2) N I_{0}(4\beta),
\end{align}
where we keep only the leading and subleading terms in $N$.

Combining $S_1$ in Eq.~\eqref{eq:2HS1} with $S_{2}$ in Eq.~\eqref{eq:2HS2}, we obtain
\begin{align}
    Z &= S_1 + \frac{2^{N-3}-1}{N-2} S_2 \nonumber \\ 
      & \approx (N^2 I_{0}(2\beta)^2 - N I_{0}(4\beta)) +  \frac{2^{N-3}-1}{N-2} (N^2(N-2) I_0(2\beta)^2 - N(N-2) I_{0}(4\beta)) \nonumber \\ 
      & = N^2 2^{N-3} I_0(2\beta)^2 - N2^{N-3} I_0(4\beta),
\end{align}
where $N^2 2^{N-3} I_0(2\beta)^2$ is the leading term in $N$ and $- N2^{N-3} I_0(4\beta)$ is the subleading term in $N$.

\subsubsection{Drude weight}
In this section, we derive an analytical expression for the Drude weight $D(T)$ in the two-hole sector.

Recall that the Drude weight is given by
\begin{eqnarray}
    D(T) = \frac{1}{2N} \langle - \hat{H} \rangle = -\frac{1}{2N} \frac{\Tr(\hat{H} e^{-\beta \hat{H}})}{Z}.
\end{eqnarray}
We therefore evaluate $\Tr(\hat{H} e^{-\beta \hat{H}})$. 

We start from
\begin{align}
\label{eq:trace_start_sym}
\Tr(\hat{H} e^{-\beta \hat{H}})
&= 2 \sum_{0 \leq m_1 < m_2 \leq N-1}  \exp\!\Bigl[
-2\beta\bigl(
\cos\alpha_{m_1}+\cos\alpha_{m_2}
\bigr)
\Bigr] (2\cos\alpha_{m_1}+2\cos\alpha_{m_2}) \nonumber \\ 
&+ \sum_{n_{\downarrow}=1}^{N-3}
\frac{\binom{N-2}{n_{\downarrow}}}{N-2}
\sum_{0 \le m_1 < m_2 \le N-1}
\sum_{m_{\downarrow}=0}^{N-3}
\exp\!\Bigl[
-2\beta\bigl(
\cos\theta_{m_1,m_{\downarrow}}+\cos\theta_{m_2,m_{\downarrow}}
\bigr)
\Bigr]\,
\bigl(2\cos\theta_{m_1,m_{\downarrow}}+2\cos\theta_{m_2,m_{\downarrow}}\bigr) \nonumber \\ 
&= 2 \sum_{0 \leq m_1 < m_2 \leq N-1}  \exp\!\Bigl[
-2\beta\bigl(
\cos\alpha_{m_1}+\cos\alpha_{m_2}
\bigr)
\Bigr] (2\cos\alpha_{m_1}+2\cos\alpha_{m_2}) \nonumber \\ 
&+ \frac{2^{N-2}-2}{N-2}
\sum_{0 \le m_1 < m_2 \le N-1}
\sum_{m_{\downarrow}=0}^{N-3}
\exp\!\Bigl[
-2\beta\bigl(
\cos\theta_{m_1,m_{\downarrow}}+\cos\theta_{m_2,m_{\downarrow}}
\bigr)
\Bigr]\,
\bigl(2\cos\theta_{m_1,m_{\downarrow}}+2\cos\theta_{m_2,m_{\downarrow}}\bigr)  \nonumber \\
&= \left( \sum_{0 \leq m_1, m_2 \leq N-1}  \exp\!\Bigl[
-2\beta\bigl(
\cos\alpha_{m_1}+\cos\alpha_{m_2}
\bigr)
\Bigr] (2\cos\alpha_{m_1}+2\cos\alpha_{m_2}) \right. \nonumber \\  
&\left.  -  \sum_{0 \leq m \leq N-1}  \exp\!\Bigl[
-4\beta
\cos\alpha_{m}
\Bigr] (4\cos\alpha_{m})  \right) \nonumber \\ 
&+ \frac{2^{N-3}-1}{N-2}
\left(\sum_{0 \le m_1, m_2 \le N-1}
\sum_{m_{\downarrow}=0}^{N-3}
\exp\!\Bigl[
-2\beta\bigl(
\cos\theta_{m_1,m_{\downarrow}}+\cos\theta_{m_2,m_{\downarrow}}
\bigr)
\Bigr]\,
\bigl(2\cos\theta_{m_1,m_{\downarrow}}+2\cos\theta_{m_2,m_{\downarrow}}\bigr) \right. \nonumber \\ 
& \left. - \sum_{0 \le m_1 \le N-1}
\sum_{m_{\downarrow}=0}^{N-3}
\exp\!\Bigl[
-4\beta\bigl(
\cos\theta_{m_1,m_{\downarrow}}
\bigr)
\Bigr]\,
4\cos\theta_{m_1,m_{\downarrow}} \right) \nonumber \\ 
& = S_1 + \frac{2^{N-3}-1}{N-2} S_2.
\end{align}
Next, we calculate $S_1$ and $S_2$ separately.

$S_1$ in Eq.~\eqref{eq:trace_start_sym} can be written as 
\begin{align}
S_1 &= 2\left(\sum_{m=0}^{N-1} e^{-2\beta \cos\alpha_m}\right)
\left(\sum_{m=0}^{N-1} e^{-2\beta \cos\alpha_m}\, 2\cos\alpha_m\right)
-\sum_{m=0}^{N-1} e^{-4\beta \cos\alpha_m}\left(4\cos\alpha_m\right) \nonumber \\ 
&=
2\left[
\sum_{m=0}^{N-1}
\left(
I_0(2\beta)+2\sum_{l=1}^{\infty}(-1)^l I_l(2\beta)\cos(l\alpha_m)
\right)
\right]
\left[
\sum_{m=0}^{N-1}
\left(
I_0(2\beta)+2\sum_{l=1}^{\infty}(-1)^l I_l(2\beta)\cos(l\alpha_m)
\right)
2\cos\alpha_m
\right]
\nonumber \\ 
&\quad-
\sum_{m=0}^{N-1}
\left(
I_0(4\beta)+2\sum_{l=1}^{\infty}(-1)^l I_l(4\beta)\cos(l\alpha_m)
\right)
(4\cos\alpha_m) \nonumber \\ 
&=
2\left[
\left(
N I_0(2\beta)+2N \sum_{p=1}^{\infty}(-1)^{pN} I_{pN}(2\beta)
\right)
\right]
\left[
\sum_{m=0}^{N-1}
\left(2\sum_{l=1}^{\infty}(-1)^l I_l(2\beta)(\cos((l+1)\alpha_m) + \cos((l-1)\alpha_m))
\right)
\right]
\nonumber \\ 
&\quad-
\sum_{m=0}^{N-1}
\left(4\sum_{l=1}^{\infty}(-1)^l I_l(4\beta) (\cos((l+1)\alpha_m) + \cos((l-1)\alpha_m) ) 
\right) \nonumber \\ 
&=
2\left[
\left(
N I_0(2\beta)+2N \sum_{p=1}^{\infty}(-1)^{pN} I_{pN}(2\beta)
\right)
\right]
\left[
\left(2N \sum_{p=1}^{\infty}(-1)^{pN-1} I_{pN-1}(2\beta) 
\right) + \left(2N \sum_{p=0}^{\infty}(-1)^{pN+1} I_{pN+1}(2\beta) 
\right)
\right]
\nonumber \\ 
&\quad-
\left(4N\sum_{p=1}^{\infty}(-1)^{pN-1} I_{pN-1}(4\beta)  + 4N\sum_{p=0}^{\infty}(-1)^{pN+1} I_{pN+1}(4\beta) 
\right)  \nonumber \\ 
& \approx -4N^2 I_0(2\beta) I_1(2\beta) + 4NI_1(4\beta),
\end{align}
where we keep only the leading and subleading terms in $N$.

$S_2$ in Eq.~\eqref{eq:trace_start_sym} can be written as 
\begin{align}
S_2 = &\sum_{m_1,m_2=0}^{N-1}
\sum_{m_{\downarrow}=0}^{N-3}
\exp\!\Bigl[
-2\beta\bigl(
\cos\theta_{m_1,m_{\downarrow}}+\cos\theta_{m_2,m_{\downarrow}}
\bigr)
\Bigr]\,
\bigl(2\cos\theta_{m_1,m_{\downarrow}}+2\cos\theta_{m_2,m_{\downarrow}}\bigr)
\nonumber\\
&\quad
-
\sum_{m_1=0}^{N-1}
\sum_{m_{\downarrow}=0}^{N-3}
\exp\!\Bigl[
-4\beta\cos\theta_{m_1,m_{\downarrow}}
\Bigr]\,
4\cos\theta_{m_1,m_{\downarrow}}
\nonumber\\
&=
\sum_{m_{\downarrow}=0}^{N-3}
\Biggl[
2
\left(\sum_{m=0}^{N-1} e^{-2\beta \cos\theta_{m,m_{\downarrow}}}\right)
\left(\sum_{m=0}^{N-1} e^{-2\beta \cos\theta_{m,m_{\downarrow}}}\,2\cos\theta_{m,m_{\downarrow}}\right)
-
\sum_{m=0}^{N-1} e^{-4\beta \cos\theta_{m,m_{\downarrow}}}\,4\cos\theta_{m,m_{\downarrow}}
\Biggr] \nonumber \\ 
&=
\sum_{m_{\downarrow}=0}^{N-3}
\Biggl\{
2
\left[
\sum_{m=0}^{N-1}
\left(
I_0(2\beta)+2\sum_{l=1}^{\infty}(-1)^l I_l(2\beta)\cos(l\theta_{m,m_{\downarrow}})
\right)
\right]
\left[
\sum_{m=0}^{N-1}
\left(
I_0(2\beta)+2\sum_{l=1}^{\infty}(-1)^l I_l(2\beta)\cos(l\theta_{m,m_{\downarrow}})
\right)
2\cos\theta_{m,m_{\downarrow}}
\right]
\nonumber\\
&\hspace{2.6cm}
-
\sum_{m=0}^{N-1}
\left(
I_0(4\beta)+2\sum_{l=1}^{\infty}(-1)^l I_l(4\beta)\cos(l\theta_{m,m_{\downarrow}})
\right)
(4\cos\theta_{m,m_{\downarrow}})
\Biggr\}\nonumber \\ 
&=
\sum_{m_{\downarrow}=0}^{N-3}
\Biggl\{
2
\left(
N I_0(2\beta)+2N\sum_{p=1}^{\infty}(-1)^{pN} I_{pN}(2\beta)\cos(pN\phi_{m_{\downarrow}})
\right) \nonumber \\ 
&\quad \quad \quad \left(
2N\sum_{p=1}^{\infty}(-1)^{pN-1} I_{pN-1}(2\beta)\cos(pN\phi_{m_{\downarrow}}) + 2N\sum_{p=0}^{\infty}(-1)^{pN+1} I_{pN+1}(2\beta)\cos(pN\phi_{m_{\downarrow}})
\right)
\nonumber\\
&\hspace{1.6cm}
-
\left(
4N\sum_{p=1}^{\infty}(-1)^{pN-1} I_{pN-1}(4\beta)\cos(pN\phi_{m_{\downarrow}}) + 4N\sum_{p=0}^{\infty}(-1)^{pN+1} I_{pN+1}(4\beta)\cos(pN\phi_{m_{\downarrow}})
\right)
\Biggr\} \nonumber \\ 
& \approx -4N^2(N-2) I_{0}(2\beta) I_1(2\beta) + 4N(N-2) I_{1}(4\beta),
\end{align}
where we keep only the leading and subleading terms in $N$.

Therefore,
\begin{align}
    \Tr(\hat{H} e^{-\beta \hat{H}}) &= S_1 + \frac{2^{N-3}-1}{N-2} S_{2} \nonumber \\ 
    & \approx -4 N^2 I_0(2\beta) I_1(2\beta) + 4NI_1(4\beta) + (2^{N-3}-1) (-4 N^2 I_0(2\beta) I_1(2\beta) + 4N I_1(4\beta)) \nonumber \\ 
    &= - 2^{N-1} N^2 I_0(2\beta) I_1(2\beta) + 2^{N-1} N I_1(4 \beta).
\end{align}
Here, ``$\approx$'' indicates that we keep only the leading and subleading terms in $N$ and neglect all higher-order terms in $N$.

Therefore, the Drude weight is 
\begin{align}
     \label{eq:2HDrudeFinal}
     D(T) &= -\frac{1}{2N} \frac{\Tr(\hat{H} e^{-\beta \hat{H}})}{Z} \nonumber \\ 
     & \approx \frac{1}{2N} \frac{2^{N-1} N^2 I_{0}(2\beta) I_1(2\beta) - 2^{N-1} NI_{1}(4\beta)}{N^2 2^{N-3} I_{0}(2\beta)^2 - N 2^{N-3} I_{0}(4\beta)} \nonumber \\ 
     & \approx \frac{2}{N} \frac{I_{1}(2\beta)}{I_{0}(2\beta)} \nonumber \\ 
     & = \frac{2\beta}{N}(1 - \frac{I_2(2\beta)}{I_0(2\beta)}),
\end{align}
where the second ``$\approx$'' indicates that we have neglected all finite-size corrections.

Consequently, in the thermodynamic limit, the Drude weight in the two-hole sector is twice that in the one-hole sector.

\subsubsection{$D_{\mathrm{diff}}(T)$}

Here, we present the analytical expression for $D_{\mathrm{diff}}(T)$ defined in Eq.~\eqref{eq:DefinitionOfSigmadiff} in the two-hole sector. We have proved that $D_{\mathrm{diff}}(T) = D(T)$ in the thermodynamic limit as shown below.

We define
\(
Z'=\sum_{n} e^{-\beta E_n}\, \bigl|J_{nn}\bigr|^2,
\)
namely,
\begin{align}
\label{eq:2HZprime}
Z'
&=
\sum_{n} e^{-\beta E_n}\, \bigl|J_{nn}\bigr|^2
\nonumber \\
&=  2 \sum_{0 \leq m_1 < m_2 \leq N-1} e^{-2\beta (\cos(\frac{2\pi m_1}{N}) + \cos(\frac{2 \pi m_2}{N}))} \times 4(\sin(\frac{2\pi m_1}{N}) + \sin(\frac{2\pi m_2}{N}))^2 \nonumber \\ 
&+ \frac{2^{N-2}-2}{N-2}
\sum_{0 \le m_1 < m_2 \le N-1}
\sum_{m_{\downarrow}=0}^{N-3}
e^{-2\beta\Bigl[
\cos\!\Bigl(\frac{2\pi m_1}{N}+\frac{2\pi m_{\downarrow}}{N(N-2)}\Bigr)
+
\cos\!\Bigl(\frac{2\pi m_2}{N}+\frac{2\pi m_{\downarrow}}{N(N-2)}\Bigr)
\Bigr]}
\nonumber \\
&\qquad\qquad\qquad\qquad\times
4\Biggl[
\sin\!\Bigl(\frac{2\pi m_1}{N}+\frac{2\pi m_{\downarrow}}{N(N-2)}\Bigr)
+
\sin\!\Bigl(\frac{2\pi m_2}{N}+\frac{2\pi m_{\downarrow}}{N(N-2)}\Bigr)
\Biggr]^2
\nonumber \\ 
&=  2 \sum_{0 \leq m_1 < m_2 \leq N-1} e^{-2\beta (\cos(\alpha_{m_1}) + \cos(\alpha_{m_2}))} \times 4(\sin(\alpha_{m_1}) + \sin(\alpha_{m_2}))^2 \nonumber \\ 
& + \frac{2^{N-2}-2}{N-2}
\sum_{0 \le m_1 < m_2 \le N-1}
\sum_{m_{\downarrow}=0}^{N-3}
e^{-2\beta\left[\cos(\alpha_{m_1}+\phi_{m_{\downarrow}})+\cos(\alpha_{m_2}+\phi_{m_{\downarrow}})\right]}
\,4\left[\sin(\alpha_{m_1}+\phi_{m_{\downarrow}})+\sin(\alpha_{m_2}+\phi_{m_{\downarrow}})\right]^2
\nonumber \\ 
&=  \left( \sum_{0 \leq m_1, m_2 \leq N-1} e^{-2\beta (\cos(\alpha_{m_1}) + \cos(\alpha_{m_2}))} \times 4(\sin(\alpha_{m_1}) + \sin(\alpha_{m_2}))^2 - \sum_{0 \leq m \leq N-1} e^{-4\beta \cos(\alpha_{m})} \times 16(\sin(\alpha_{m}))^2  \right)\nonumber\\ 
& + \frac{2^{N-3}-1}{N-2}
\Biggl(
\sum_{0 \le m_1,m_2 \le N-1}
\sum_{m_{\downarrow}=0}^{N-3}
e^{-2\beta\left[\cos(\alpha_{m_1}+\phi_{m_{\downarrow}})+\cos(\alpha_{m_2}+\phi_{m_{\downarrow}})\right]}
\,4\left[\sin(\alpha_{m_1}+\phi_{m_{\downarrow}})+\sin(\alpha_{m_2}+\phi_{m_{\downarrow}})\right]^2
\nonumber \\ 
&\qquad\qquad
-
\sum_{0 \le m \le N-1}
\sum_{m_{\downarrow}=0}^{N-3}
e^{-4\beta\cos(\alpha_m+\phi_{m_{\downarrow}})}\,
16\sin^2(\alpha_m+\phi_{m_{\downarrow}})
\Biggr) \nonumber \\ 
& = S_1 + \frac{2^{N-3}-1}{N-2} S_2,
\end{align}
where $\alpha_m = \frac{2\pi m}{N}$ and $\phi_{m_{\downarrow}} = \frac{2\pi m_{\downarrow}}{N(N-2)}$.

Next, we calculate $S_1$ and $S_2$ in Eq.~\eqref{eq:2HZprime} separately.

The $S_1$ in Eq.~\eqref{eq:2HZprime} is given by
\begin{align}
S_1& = \sum_{m_1,m_2=0}^{N-1} e^{-2\beta (\cos\alpha_{m_1} + \cos\alpha_{m_2})}
\,4(\sin\alpha_{m_1} + \sin\alpha_{m_2})^2
-\sum_{m=0}^{N-1} e^{-4\beta \cos\alpha_{m}}\,16\sin^2\alpha_{m}
\nonumber\\
&=
8
\left(\sum_{m=0}^{N-1} e^{-2\beta\cos\alpha_m}\sin^2\alpha_m\right)
\left(\sum_{m=0}^{N-1} e^{-2\beta\cos\alpha_m}\right)
+
8
\left(\sum_{m=0}^{N-1} e^{-2\beta\cos\alpha_m}\sin\alpha_m\right)^2
\nonumber\\
&\quad-
16\sum_{m=0}^{N-1} e^{-4\beta\cos\alpha_m}\sin^2\alpha_m \nonumber \\ 
&=
8
\left(\sum_{m=0}^{N-1} e^{-2\beta\cos\alpha_m}\sin^2\alpha_m\right)
\left(\sum_{m=0}^{N-1} e^{-2\beta\cos\alpha_m}\right)
-
16\sum_{m=0}^{N-1} e^{-4\beta\cos\alpha_m}\sin^2\alpha_m \nonumber \\ 
&=
4\left(\sum_{m=0}^{N-1} e^{-2\beta\cos\alpha_m}\right)^2
-
4\left(\sum_{m=0}^{N-1} e^{-2\beta\cos\alpha_m}\right)
\left(\sum_{m=0}^{N-1} e^{-2\beta\cos\alpha_m}\cos 2\alpha_m\right)
\nonumber\\
&\quad-
8\sum_{m=0}^{N-1} e^{-4\beta\cos\alpha_m}
+
8\sum_{m=0}^{N-1} e^{-4\beta\cos\alpha_m}\cos 2\alpha_m.
\end{align}
In the third equality, we have used $\sum_{m=0}^{N-1} e^{-2\beta\cos\alpha_m}\sin\alpha_m=0$. In the last equality, we have used $\sin^2(\alpha) = \frac{1-\cos(2\alpha)}{2}$.

Using the Jacobi--Anger expansion (see Eq.~\eqref{eq:Jacobi-Anger})
we have

\begin{align}
S_1&=
4\left[
\sum_{m=0}^{N-1}
\left(
I_0(2\beta)+2\sum_{l=1}^{\infty}(-1)^l I_l(2\beta)\cos(l\alpha_m)
\right)
\right]^2
\nonumber \\ 
&\quad-
4\left[
\sum_{m=0}^{N-1}
\left(
I_0(2\beta)+2\sum_{l=1}^{\infty}(-1)^l I_l(2\beta)\cos(l\alpha_m)
\right)
\right]
\left[
\sum_{m=0}^{N-1}
\left(
I_0(2\beta)+2\sum_{l=1}^{\infty}(-1)^l I_l(2\beta)\cos(l\alpha_m)
\right)\cos 2\alpha_m
\right]
\nonumber \\ 
&\quad-
8\sum_{m=0}^{N-1}
\left(
I_0(4\beta)+2\sum_{l=1}^{\infty}(-1)^l I_l(4\beta)\cos(l\alpha_m)
\right)
\nonumber \\ 
&\quad+
8\sum_{m=0}^{N-1}
\left(
I_0(4\beta)+2\sum_{l=1}^{\infty}(-1)^l I_l(4\beta)\cos(l\alpha_m)
\right)\cos 2\alpha_m \nonumber \\ 
&=
4
\left(
N I_0(2\beta)+2N\sum_{p=1}^{\infty}(-1)^{pN} I_{pN}(2\beta)
\right)^2
\nonumber \\ 
&\quad-
4
\left(
N I_0(2\beta)+2N \sum_{p=1}^{\infty}(-1)^{pN} I_{pN}(2\beta)
\right)
\left(
N \sum_{p=1}^{\infty}(-1)^{pN-2} I_{pN-2}(2\beta) + N \sum_{p=0}^{\infty}(-1)^{pN+2} I_{pN+2}(2\beta)
\right)
\nonumber \\ 
&\quad-
8
\left(
N I_0(4\beta)+2N \sum_{p=1}^{\infty}(-1)^{pN} I_{pN}(4\beta)
\right)
\nonumber \\ 
&\quad+
8
\left(N \sum_{p=1}^{\infty}(-1)^{pN-2} I_{pN-2}(4\beta)  + N \sum_{p=0}^{\infty}(-1)^{pN+2} I_{pN+2}(4\beta) 
\right)   \nonumber \\
& \approx 4N^2 I_{0}(2\beta)^2 - 4N^2 I_{0}(2\beta) I_{2}(2\beta) - 8NI_{0}(4\beta) + 8N I_{2}(4\beta),
\end{align}
where we keep only the leading and subleading terms in $N$.

Next, we consider $S_2$ in Eq.~\eqref{eq:2HZprime}. We have
\begin{eqnarray}
\label{eq:2HJ2S2Part1}
&&
\sum_{0 \le m_1,m_2 \le N-1}
\sum_{m_{\downarrow}=0}^{N-3}
e^{-2\beta\left(\cos(\alpha_{m_1}+\phi_{m_{\downarrow}})+\cos(\alpha_{m_2}+\phi_{m_{\downarrow}})\right)}
\,4\left(\sin(\alpha_{m_1}+\phi_{m_{\downarrow}})+\sin(\alpha_{m_2}+\phi_{m_{\downarrow}})\right)^2
\nonumber\\
&=&
4\sum_{0 \le m_1,m_2 \le N-1}
\sum_{m_{\downarrow}=0}^{N-3}
\sum_{l_1,l_2\in\mathbb{Z}}
I_{l_1}(-2\beta)\,I_{l_2}(-2\beta)\,
e^{il_1(\alpha_{m_1}+\phi_{m_{\downarrow}})}\,
e^{il_2(\alpha_{m_2}+\phi_{m_{\downarrow}})}
\nonumber\\
&&\qquad\qquad\times
\Bigl(
\sin^2(\alpha_{m_1}+\phi_{m_{\downarrow}})
+2\sin(\alpha_{m_1}+\phi_{m_{\downarrow}})\sin(\alpha_{m_2}+\phi_{m_{\downarrow}})
+\sin^2(\alpha_{m_2}+\phi_{m_{\downarrow}})
\Bigr)
\nonumber\\
&=&
8\sum_{0 \le m_1,m_2 \le N-1}
\sum_{m_{\downarrow}=0}^{N-3}
\sum_{l_1,l_2\in\mathbb{Z}}
I_{l_1}(-2\beta)\,I_{l_2}(-2\beta)\,
e^{il_1(\alpha_{m_1}+\phi_{m_{\downarrow}})}\,
e^{il_2(\alpha_{m_2}+\phi_{m_{\downarrow}})}
\nonumber\\
&&\qquad\qquad\times
\Bigl(
\sin^2(\alpha_{m_1}+\phi_{m_{\downarrow}})
+\sin(\alpha_{m_1}+\phi_{m_{\downarrow}})\sin(\alpha_{m_2}+\phi_{m_{\downarrow}})
\Bigr)
\nonumber\\
&=&
8\sum_{0 \le m_1,m_2 \le N-1}
\sum_{m_{\downarrow}=0}^{N-3}
\sum_{l_1,l_2\in\mathbb{Z}}
I_{l_1}(-2\beta)\,I_{l_2}(-2\beta)\,
e^{il_1(\alpha_{m_1}+\phi_{m_{\downarrow}})}\,
e^{il_2(\alpha_{m_2}+\phi_{m_{\downarrow}})}\,
\sin^2(\alpha_{m_1}+\phi_{m_{\downarrow}})
\nonumber\\
&=&
-2\sum_{0 \le m_1,m_2 \le N-1}
\sum_{m_{\downarrow}=0}^{N-3}
\sum_{l_1,l_2\in\mathbb{Z}}
I_{l_1}(-2\beta)\,I_{l_2}(-2\beta)\,
e^{il_1(\alpha_{m_1}+\phi_{m_{\downarrow}})}\,
e^{il_2(\alpha_{m_2}+\phi_{m_{\downarrow}})}
\Bigl(e^{i(\alpha_{m_1}+\phi_{m_{\downarrow}})}-e^{-i(\alpha_{m_1}+\phi_{m_{\downarrow}})}\Bigr)^2
\nonumber\\
&=&
-2\sum_{0 \le m_1,m_2 \le N-1}
\sum_{m_{\downarrow}=0}^{N-3}
\sum_{l_1,l_2\in\mathbb{Z}}
I_{l_1}(-2\beta)\,I_{l_2}(-2\beta)\,
e^{il_1(\alpha_{m_1}+\phi_{m_{\downarrow}})}\,
e^{il_2(\alpha_{m_2}+\phi_{m_{\downarrow}})}
\Bigl(e^{2i(\alpha_{m_1}+\phi_{m_{\downarrow}})}-2+e^{-2i(\alpha_{m_1}+\phi_{m_{\downarrow}})}\Bigr)
\nonumber\\
&=&
-2\sum_{0 \le m_1,m_2 \le N-1}
\sum_{m_{\downarrow}=0}^{N-3}
\sum_{l_1,l_2\in\mathbb{Z}}
I_{l_1}(-2\beta)\,I_{l_2}(-2\beta)\,
e^{il_2\alpha_{m_2}}
\nonumber\\
&&\qquad\qquad\times
\Bigl(
e^{i(l_1+l_2+2)\phi_{m_{\downarrow}}}\,e^{i(l_1+2)\alpha_{m_1}}
-2e^{i(l_1+l_2)\phi_{m_{\downarrow}}}\,e^{il_1\alpha_{m_1}}
+e^{i(l_1+l_2-2)\phi_{m_{\downarrow}}}\,e^{i(l_1-2)\alpha_{m_1}}
\Bigr)
\nonumber\\
&=&
-2N^2
\sum_{m_{\downarrow}=0}^{N-3}
\sum_{n_1,n_2\in\mathbb{Z}}
e^{i(n_1N+n_2N)\phi_{m_{\downarrow}}}
\Bigl(
I_{n_1N-2}(-2\beta)\,I_{n_2N}(-2\beta)
-2I_{n_1N}(-2\beta)\,I_{n_2N}(-2\beta)
\nonumber\\
&&\qquad\qquad\qquad\qquad\qquad\qquad
+I_{n_1N+2}(-2\beta)\,I_{n_2N}(-2\beta)
\Bigr)
\nonumber\\
&=&
-2N^2(N-2)
\sum_{\substack{n_1,n_2\in\mathbb{Z}\\ n_1+n_2\equiv 0\;(\mathrm{mod}\;N-2)}}
\Bigl(
I_{n_1N-2}(-2\beta)\,I_{n_2N}(-2\beta)
-2I_{n_1N}(-2\beta)\,I_{n_2N}(-2\beta) \nonumber \\
&&\qquad\qquad\qquad\qquad\qquad\qquad
+I_{n_1N+2}(-2\beta)\,I_{n_2N}(-2\beta)
\Bigr),
\end{eqnarray}
and
\begin{eqnarray}
\label{eq:2HJ2S2Part2}
&&
\sum_{0 \le m_1 \le N-1}
\sum_{m_{\downarrow}=0}^{N-3}
e^{-4\beta\cos(\alpha_{m_1}+\phi_{m_{\downarrow}})}\,
16\sin^2(\alpha_{m_1}+\phi_{m_{\downarrow}})
\nonumber\\
&=&
-4\sum_{0 \le m_1 \le N-1}
\sum_{m_{\downarrow}=0}^{N-3}
\sum_{\ell\in\mathbb{Z}}
I_{\ell}(-4\beta)\,
e^{i\ell(\alpha_{m_1}+\phi_{m_{\downarrow}})}
\Bigl(e^{i(\alpha_{m_1}+\phi_{m_{\downarrow}})}-e^{-i(\alpha_{m_1}+\phi_{m_{\downarrow}})}\Bigr)^2
\nonumber\\
&=&
-4\sum_{0 \le m_1 \le N-1}
\sum_{m_{\downarrow}=0}^{N-3}
\sum_{\ell\in\mathbb{Z}}
I_{\ell}(-4\beta)\,
e^{i\ell(\alpha_{m_1}+\phi_{m_{\downarrow}})}
\Bigl(e^{2i(\alpha_{m_1}+\phi_{m_{\downarrow}})}-2+e^{-2i(\alpha_{m_1}+\phi_{m_{\downarrow}})}\Bigr)
\nonumber\\
&=&
-4N(N-2)
\sum_{\substack{n\in\mathbb{Z}\\ n\equiv 0\;(\mathrm{mod}\;N-2)}}
\Bigl(
I_{nN-2}(-4\beta)
-2I_{nN}(-4\beta)
+I_{nN+2}(-4\beta)
\Bigr).
\end{eqnarray}
Combining Eq.~\eqref{eq:2HJ2S2Part1} and Eq.~\eqref{eq:2HJ2S2Part2}, we obtain
\begin{align}
    S_{2} = -4N^2(N-2) (I_2(2\beta) I_{0}(2\beta) - I_{0}(2\beta)^2) + 8N(N-2) (I_2(4\beta) - I_{0}(4\beta)),
\end{align}
where we keep only the leading and subleading terms in $N$.

Therefore, 
\begin{eqnarray}
Z'
&=&
\sum_{n} e^{-\beta E_n}\, \bigl|J_{nn}\bigr|^2
\nonumber\\
&\approx& 4N^2 I_{0}(2\beta)^2 - 4N^2 I_{0}(2\beta) I_{2}(2\beta) - 8NI_{0}(4\beta) + 8N I_{2}(4\beta) \nonumber \\ 
&+ & \frac{2^{N-3}-1}{N-2}
\Bigl(
-4N^2(N-2)\bigl(I_2(2\beta)I_0(2\beta)-I_0(2\beta)^2\bigr)
+8N(N-2)\bigl(I_2(4\beta)-I_0(4\beta)\bigr)
\Bigr)
\nonumber\\
&\approx&
2^{N-1}
\Bigl(
N^2\bigl(I_0(2\beta)^2-I_0(2\beta)I_2(2\beta)\bigr)
+2N\bigl(I_2(4\beta)-I_0(4\beta)\bigr)
\Bigr),
\end{eqnarray}
where in the second line we keep both leading and subleading terms in $N$ while in the third line we keep only the leading term.

Therefore, $\sigma_{\mathrm{diff}}(\omega, T)$ is 
\begin{align}
    \sigma_{\mathrm{diff}}(\omega, T) &= \frac{\pi \beta}{N} \frac{Z'}{Z} \delta(\omega)  \nonumber \\
    & \approx \frac{\pi \beta}{N } \frac{2^{N-1}
\Bigl(
N^2\bigl(I_0(2\beta)^2-I_0(2\beta)I_2(2\beta)\bigr)
+2N\bigl(I_2(4\beta)-I_0(4\beta)\bigr)
\Bigr)}{N^2 2^{N-3} I_0(2\beta)^2 - N2^{N-3} I_0(4\beta)}\delta(\omega) \nonumber \\ 
& \approx \frac{\pi \beta}{N } \frac{2^{N-1}
N^2\bigl(I_0(2\beta)^2-I_0(2\beta)I_2(2\beta)\bigr)
}{N^2 2^{N-3} I_0(2\beta)^2} \delta(\omega)  \nonumber \\
& = \frac{4 \pi \beta}{N} (1 - \frac{I_{2}(2\beta)}{I_0(2\beta)}) \delta(\omega).
\end{align}
Therefore, $D_{\mathrm{diff}}(T) = \frac{2 \beta}{N} (1 - \frac{I_{2}(2\beta)}{I_0(2\beta)})$ in the thermodynamic limit and thus $D_{\mathrm{diff}}(T) = D(T)$.

\subsubsection{Specific heat}
\label{sec:Analytical_expression_for_the_specific_heat_2H}
In this section, we present the exact solution for the specific heat in the two-hole sector in the thermodynamic limit.
As shown in Eq.~\eqref{eq:2HDrudeFinal}, the expectation value of the energy is 
\begin{align}
    \langle \hat{H} \rangle = -\frac{4 I_1(2\beta)}{I_0(2\beta)}.
\end{align}

Then we consider $\langle \hat{H}^2 \rangle$. We have
\begin{align}
\Tr(\hat{H}^2  e^{-\beta \hat{H}})
&=  2 \sum_{0 \leq m_1 < m_2 \leq N-1} e^{-2\beta (\cos(\frac{2\pi m_1}{N}) + \cos(\frac{2 \pi m_2}{N}))} \times 4(\cos(\frac{2\pi m_1}{N}) + \cos(\frac{2\pi m_2}{N}))^2  \nonumber \\
&+ \frac{2^{N-2}-2}{N-2}
\sum_{0 \le m_1 < m_2 \le N-1}
\sum_{m_{\downarrow}=0}^{N-3}
e^{-2\beta\Bigl[
\cos\!\Bigl(\frac{2\pi m_1}{N}+\frac{2\pi m_{\downarrow}}{N(N-2)}\Bigr)
+
\cos\!\Bigl(\frac{2\pi m_2}{N}+\frac{2\pi m_{\downarrow}}{N(N-2)}\Bigr)
\Bigr]}
\nonumber \\ 
&\qquad\qquad\qquad\qquad\times
4\Biggl[
\cos\!\Bigl(\frac{2\pi m_1}{N}+\frac{2\pi m_{\downarrow}}{N(N-2)}\Bigr)
+
\cos\!\Bigl(\frac{2\pi m_2}{N}+\frac{2\pi m_{\downarrow}}{N(N-2)}\Bigr)
\Biggr]^2
\nonumber \\ 
&=  2 \sum_{0 \leq m_1 < m_2 \leq N-1} e^{-2\beta (\cos(\alpha_{m_1}) + \cos(\alpha_{m_2}))} \times 4(\cos(\alpha_{m_1}) + \cos(\alpha_{m_2}))^2 \nonumber \\  
& + \frac{2^{N-2}-2}{N-2}
\sum_{0 \le m_1 < m_2 \le N-1}
\sum_{m_{\downarrow}=0}^{N-3}
e^{-2\beta\left[\cos(\alpha_{m_1}+\phi_{m_{\downarrow}})+\cos(\alpha_{m_2}+\phi_{m_{\downarrow}})\right]}
\,4\left[\cos(\alpha_{m_1}+\phi_{m_{\downarrow}})+\cos(\alpha_{m_2}+\phi_{m_{\downarrow}})\right]^2
\nonumber \\ 
&=  \left( \sum_{0 \leq m_1, m_2 \leq N-1} e^{-2\beta (\cos(\alpha_{m_1}) + \cos(\alpha_{m_2}))} \times 4(\cos(\alpha_{m_1}) + \cos(\alpha_{m_2}))^2 - \sum_{0 \leq m \leq N-1} e^{-4\beta \cos(\alpha_{m})} \times 16(\cos(\alpha_{m}))^2  \right) \nonumber\\ 
& + \frac{2^{N-3}-1}{N-2}
\Biggl(
\sum_{0 \le m_1,m_2 \le N-1}
\sum_{m_{\downarrow}=0}^{N-3}
e^{-2\beta\left[\cos(\alpha_{m_1}+\phi_{m_{\downarrow}})+\cos(\alpha_{m_2}+\phi_{m_{\downarrow}})\right]}
\,4\left[\cos(\alpha_{m_1}+\phi_{m_{\downarrow}})+\cos(\alpha_{m_2}+\phi_{m_{\downarrow}})\right]^2
\nonumber \\ 
&\qquad\qquad
-
\sum_{0 \le m \le N-1}
\sum_{m_{\downarrow}=0}^{N-3}
e^{-4\beta\cos(\alpha_m+\phi_{m_{\downarrow}})}\,
16\cos^2(\alpha_m+\phi_{m_{\downarrow}})
\Biggr) \nonumber \\ 
& = S_1 + \frac{2^{N-3}-1}{N-2} S_2,
\end{align}
where
\begin{align}
S_1
&=
\sum_{m_1,m_2=0}^{N-1} e^{-2\beta (\cos\alpha_{m_1} + \cos\alpha_{m_2})} \,4(\cos\alpha_{m_1} + \cos\alpha_{m_2})^2
-\sum_{m=0}^{N-1} e^{-4\beta \cos\alpha_m}\,16\cos^2\alpha_m
\nonumber\\
&=
8\left(\sum_{m=0}^{N-1} e^{-2\beta\cos\alpha_m}\cos^2\alpha_m\right)
\left(\sum_{m=0}^{N-1} e^{-2\beta\cos\alpha_m}\right)
+
8\left(\sum_{m=0}^{N-1} e^{-2\beta\cos\alpha_m}\cos\alpha_m\right)^2
\nonumber\\
&\quad-
16\sum_{m=0}^{N-1} e^{-4\beta\cos\alpha_m}\cos^2\alpha_m
\nonumber\\
&=
4\left(\sum_{m=0}^{N-1} e^{-2\beta\cos\alpha_m}(\cos(2\alpha_m) + 1)\right) 
\left(\sum_{m=0}^{N-1} e^{-2\beta\cos\alpha_m}\right)
+
8\left(\sum_{m=0}^{N-1} e^{-2\beta\cos\alpha_m}\cos\alpha_m\right)^2
\nonumber\\
&\quad-
8\sum_{m=0}^{N-1} e^{-4\beta\cos\alpha_m} (\cos(2\alpha_m) + 1)
\nonumber\\
& = 4\left(
N I_0(2\beta)
+
2N\sum_{p=1}^{\infty}(-1)^{pN}I_{pN}(2\beta)
\right)^2
\nonumber\\
&\quad+
4\left(
N I_0(2\beta)
+
2N\sum_{p=1}^{\infty}(-1)^{pN}I_{pN}(2\beta)
\right)
\left(
N\sum_{p=1}^{\infty}(-1)^{pN-2}I_{pN-2}(2\beta)
+
N\sum_{p=0}^{\infty}(-1)^{pN+2}I_{pN+2}(2\beta)
\right)
\nonumber\\
&\quad+
8\left(
N\sum_{p=1}^{\infty}(-1)^{pN-1}I_{pN-1}(2\beta)
+
N\sum_{p=0}^{\infty}(-1)^{pN+1}I_{pN+1}(2\beta)
\right)^2
\nonumber\\
&\quad-
8\left(
N I_0(4\beta)
+
2N\sum_{p=1}^{\infty}(-1)^{pN}I_{pN}(4\beta)
\right)
\nonumber\\
&\quad-
8\left(
N\sum_{p=1}^{\infty}(-1)^{pN-2}I_{pN-2}(4\beta)
+
N\sum_{p=0}^{\infty}(-1)^{pN+2}I_{pN+2}(4\beta)
\right).
\end{align}
When we keep only the leading and subleading terms in $N$, we obtain
\begin{align}
    S_{1} \approx 4N^2 (I_{0}(2\beta)^2  + I_{0}(2\beta) I_2(2\beta) + 2 I_{1}(2\beta)^2) - 8N(I_{0}(4\beta) + I_{2}(4\beta)).
\end{align}

For $S_{2}$, we have 
\begin{align}
    &\sum_{0 \le m_1,m_2 \le N-1}
\sum_{m_{\downarrow}=0}^{N-3}
e^{-2\beta\left[\cos(\alpha_{m_1}+\phi_{m_{\downarrow}})+\cos(\alpha_{m_2}+\phi_{m_{\downarrow}})\right]}
\,4\left[\cos(\alpha_{m_1}+\phi_{m_{\downarrow}})+\cos(\alpha_{m_2}+\phi_{m_{\downarrow}})\right]^2 \nonumber\\ 
& =8\sum_{m_{\downarrow}=0}^{N-3}
\left(\sum_{m=0}^{N-1} e^{-2\beta\cos(\alpha_m+\phi_{m_{\downarrow}})}\cos^2(\alpha_m+\phi_{m_{\downarrow}})\right)
\left(\sum_{m=0}^{N-1} e^{-2\beta\cos(\alpha_m+\phi_{m_{\downarrow}})}\right)
\nonumber\\
&\quad+
8\sum_{m_{\downarrow}=0}^{N-3}
\left(\sum_{m=0}^{N-1} e^{-2\beta\cos(\alpha_m+\phi_{m_{\downarrow}})}\cos(\alpha_m+\phi_{m_{\downarrow}})\right)^2
\nonumber\\
&=
4\sum_{m_{\downarrow}=0}^{N-3}
\left(\sum_{m=0}^{N-1} e^{-2\beta\cos(\alpha_m+\phi_{m_{\downarrow}})}
\bigl[1+\cos\!\bigl(2(\alpha_m+\phi_{m_{\downarrow}})\bigr)\bigr]\right)
\left(\sum_{m=0}^{N-1} e^{-2\beta\cos(\alpha_m+\phi_{m_{\downarrow}})}\right)
\nonumber\\
&\quad+
8\sum_{m_{\downarrow}=0}^{N-3}
\left(\sum_{m=0}^{N-1} e^{-2\beta\cos(\alpha_m+\phi_{m_{\downarrow}})}\cos(\alpha_m+\phi_{m_{\downarrow}})\right)^2
\nonumber\\
&=
4\sum_{m_{\downarrow}=0}^{N-3}
\left(\sum_{m=0}^{N-1} e^{-2\beta\cos(\alpha_m+\phi_{m_{\downarrow}})}\right)^2
\nonumber\\
&\quad+
4\sum_{m_{\downarrow}=0}^{N-3}
\left(\sum_{m=0}^{N-1} e^{-2\beta\cos(\alpha_m+\phi_{m_{\downarrow}})}
\cos\!\bigl(2(\alpha_m+\phi_{m_{\downarrow}})\bigr)\right)
\left(\sum_{m=0}^{N-1} e^{-2\beta\cos(\alpha_m+\phi_{m_{\downarrow}})}\right)
\nonumber\\
&\quad+
8\sum_{m_{\downarrow}=0}^{N-3}
\left(\sum_{m=0}^{N-1} e^{-2\beta\cos(\alpha_m+\phi_{m_{\downarrow}})}\cos(\alpha_m+\phi_{m_{\downarrow}})\right)^2 \nonumber \\
&=
4\sum_{m_{\downarrow}=0}^{N-3}
\Biggl[
\sum_{m=0}^{N-1}
\left(
I_0(2\beta)+2\sum_{l=1}^{\infty}(-1)^l I_l(2\beta)\cos\!\bigl(l(\alpha_m+\phi_{m_{\downarrow}})\bigr)
\right)
\Biggr]^2
\nonumber\\
&\quad+
4\sum_{m_{\downarrow}=0}^{N-3}
\Biggl[
\sum_{m=0}^{N-1}
\left(
I_0(2\beta)+2\sum_{l=1}^{\infty}(-1)^l I_l(2\beta)\cos\!\bigl(l(\alpha_m+\phi_{m_{\downarrow}})\bigr)
\right)
\cos\!\bigl(2(\alpha_m+\phi_{m_{\downarrow}})\bigr)
\Biggr]
\nonumber\\
&\qquad\qquad\times
\Biggl[
\sum_{m=0}^{N-1}
\left(
I_0(2\beta)+2\sum_{l=1}^{\infty}(-1)^l I_l(2\beta)\cos\!\bigl(l(\alpha_m+\phi_{m_{\downarrow}})\bigr)
\right)
\Biggr]
\nonumber\\
&\quad+
8\sum_{m_{\downarrow}=0}^{N-3}
\Biggl[
\sum_{m=0}^{N-1}
\left(
I_0(2\beta)+2\sum_{l=1}^{\infty}(-1)^l I_l(2\beta)\cos\!\bigl(l(\alpha_m+\phi_{m_{\downarrow}})\bigr)
\right)
\cos(\alpha_m+\phi_{m_{\downarrow}})
\Biggr]^2  \nonumber \\
&=
4\sum_{m_{\downarrow}=0}^{N-3}
\Biggl[
N I_0(2\beta)
+
2N\sum_{p=1}^{\infty}(-1)^{pN} I_{pN}(2\beta)\cos\!\bigl(pN\phi_{m_{\downarrow}}\bigr)
\Biggr]^2
\nonumber\\
&\quad+
4\sum_{m_{\downarrow}=0}^{N-3}
\Biggl[
N\sum_{p=1}^{\infty}(-1)^{pN-2} I_{pN-2}(2\beta)\cos\!\bigl(pN\phi_{m_{\downarrow}}\bigr)
+
N\sum_{p=0}^{\infty}(-1)^{pN+2} I_{pN+2}(2\beta)\cos\!\bigl(pN\phi_{m_{\downarrow}}\bigr)
\Biggr]
\nonumber\\
&\qquad\qquad\times
\Biggl[
N I_0(2\beta)
+
2N\sum_{p=1}^{\infty}(-1)^{pN} I_{pN}(2\beta)\cos\!\bigl(pN\phi_{m_{\downarrow}}\bigr)
\Biggr]
\nonumber\\
&\quad+
8\sum_{m_{\downarrow}=0}^{N-3}
\Biggl[
N\sum_{p=1}^{\infty}(-1)^{pN-1} I_{pN-1}(2\beta)\cos\!\bigl(pN\phi_{m_{\downarrow}}\bigr)
+
N\sum_{p=0}^{\infty}(-1)^{pN+1} I_{pN+1}(2\beta)\cos\!\bigl(pN\phi_{m_{\downarrow}}\bigr)
\Biggr]^2 \nonumber \\
& \approx 4(N-2)N^2 I_0(2\beta)^2 + 4(N-2)N^2 I_0(2\beta) I_{2}(2\beta) + 8(N-2)N^2 I_{1}(2\beta)^2,
\end{align}
and 
\begin{align}
&\sum_{m=0}^{N-1}
\sum_{m_{\downarrow}=0}^{N-3}
e^{-4\beta\cos(\alpha_m+\phi_{m_{\downarrow}})}\,
16\cos^2(\alpha_m+\phi_{m_{\downarrow}})
\nonumber\\
&=
8
\sum_{m=0}^{N-1}
\sum_{m_{\downarrow}=0}^{N-3}
e^{-4\beta\cos(\alpha_m+\phi_{m_{\downarrow}})}
\Bigl[
1+\cos\!\bigl(2(\alpha_m+\phi_{m_{\downarrow}})\bigr)
\Bigr]
\nonumber\\
&=
8
\sum_{m=0}^{N-1}
\sum_{m_{\downarrow}=0}^{N-3}
e^{-4\beta\cos(\alpha_m+\phi_{m_{\downarrow}})}
+
8
\sum_{m=0}^{N-1}
\sum_{m_{\downarrow}=0}^{N-3}
e^{-4\beta\cos(\alpha_m+\phi_{m_{\downarrow}})}
\cos\!\bigl(2(\alpha_m+\phi_{m_{\downarrow}})\bigr) \nonumber \\
&=
8\sum_{m_{\downarrow}=0}^{N-3}
\Biggl[
N I_0(4\beta)
+
2N\sum_{p=1}^{\infty}(-1)^{pN} I_{pN}(4\beta)\cos\!\bigl(pN\phi_{m_{\downarrow}}\bigr)
\Biggr]
\nonumber\\
&\quad+
8\sum_{m_{\downarrow}=0}^{N-3}
\Biggl[
N\sum_{p=1}^{\infty}(-1)^{pN-2} I_{pN-2}(4\beta)\cos\!\bigl(pN\phi_{m_{\downarrow}}\bigr)
+
N\sum_{p=0}^{\infty}(-1)^{pN+2} I_{pN+2}(4\beta)\cos\!\bigl(pN\phi_{m_{\downarrow}}\bigr)
\Biggr] \nonumber \\
& \approx 8N(N-2) I_0(4\beta) + 8N(N-2) I_{2}(4\beta).
\end{align}
Therefore, we have
\begin{align}
    S_2 \approx 4(N-2)N^2 I_0(2\beta)^2 + 4(N-2)N^2 I_0(2\beta) I_{2}(2\beta) + 8(N-2)N^2 I_{1}(2\beta)^2 - (8N(N-2) I_0(4\beta) + 8N(N-2) I_{2}(4\beta)).
\end{align}
Consequently,
\begin{align}
    \Tr(\hat{H}^2 e^{-\beta \hat{H}}) &= S_1 + \frac{2^{N-3}-1}{N-2} S_{2} \nonumber \\ 
    & \approx 4N^2 (I_{0}(2\beta)^2  + I_{0}(2\beta) I_2(2\beta) + 2 I_{1}(2\beta)^2) - 8N(I_{0}(4\beta) + I_{2}(4\beta)) \nonumber \\ 
    & + \frac{2^{N-3}-1}{N-2} ( 4(N-2)N^2 (I_0(2\beta)^2 +  I_0(2\beta) I_{2}(2\beta) +  2 I_{1}(2\beta)^2) - (8N(N-2) (I_0(4\beta) +I_{2}(4\beta)))) \nonumber \\ 
    & = N^2 2^{N-1} (I_{0}(2\beta)^2  + I_{0}(2\beta) I_2(2\beta) + 2 I_{1}(2\beta)^2) - N2^{N}(I_{0}(4\beta) + I_{2}(4\beta)). 
\end{align}
Therefore, when we keep only the leading term in $N$, 
the specific heat is
\begin{align}
    C_V^{2h}(T) &= \beta^2 (\frac{N^2 2^{N-1} (I_{0}(2\beta)^2  + I_{0}(2\beta) I_2(2\beta) + 2 I_{1}(2\beta)^2)}{N^2 2^{N-3} I_{0}(2\beta)^2} - (\frac{4I_{1}(2\beta)}{I_{0}(2\beta)})^2 ) \nonumber\\ 
    &= \beta^2 (4+ 4 \frac{I_{2}(2\beta)}{I_0(2\beta)}-8(\frac{I_1(2\beta)}{I_0(2\beta)})^2)  \nonumber \\
    & = 2C_V^{1h}(T).
\end{align}

\subsection{Any number of holes}
\label{sec:Thedistinctionbetweenfixednumberofholesandfixeddensityofholes}

In this section, we first show that when the number of holes is fixed, the system can be regarded as consisting of \(n_h\) free particles. Consequently, at leading order in \(N\), \(n_h\) contributes only an overall factor to the Drude weight, \(\sigma_{\mathrm{diff}}(\omega, T)\), and the specific heat, as summarized in Table~\ref{tab:fixed-holes-results-Drude-Specificheat}. We then compare the thermodynamic limits corresponding to a fixed number of holes and a fixed hole density.

Specifically, for an eigenstate $\ket{\mathbf{k},k_{\alpha},\mathbf{d}}$ with
$\mathbf{k}=(k_1,\cdots,k_{n_h})$, we require $k_i\neq k_j$ for any
$i\neq j$ with $i,j=1,\cdots,n_h$; otherwise the state
$\ket{\mathbf{k},k_{\alpha},\mathbf{d}}$ vanishes. Therefore, in the
calculation of the Drude weight or other observable, the summation of states appears in the form
\[
\sum_{0\le m_1<m_2<\cdots<m_{n_h}\le N-1},
\]
which enforces $m_i\neq m_j$ for any $i\neq j$. The number of allowed
choices of $(m_1,m_2,\cdots,m_{n_h})$ is $\binom{N}{n_h}$.

We may replace
$\sum_{0\le m_1<m_2<\cdots<m_{n_h}\le N-1}$ by
$\frac{1}{n_h!}\sum_{m_1,m_2,\cdots,m_{n_h}=0}^{N-1}$ and then subtract the
contributions from the disallowed combinations. The number of choices of $(m_1, m_2, \cdots, m_{n_h})$ in $\frac{1}{n_h!}\sum_{m_1,m_2,\cdots,m_{n_h}=0}^{N-1}$ is $\frac{N^{n_h}}{n_h!}$.

We define
\begin{align}
R(N,n_h)
&\equiv
\frac{\frac{N^{n_h}}{n_h!}-\binom{N}{n_h}}{\frac{N^{n_h}}{n_h!}}
=
1-\frac{n_h!\binom{N}{n_h}}{N^{n_h}}
\nonumber\\
&=
1-\frac{N(N-1)\cdots(N-n_h+1)}{N^{n_h}}\nonumber \\ 
&=
1-\prod_{r=0}^{n_h-1}\left(1-\frac{r}{N}\right).
\end{align}

When we fix the number of holes, we have 
\begin{align}
  1 - \prod_{r=0}^{n_h-1}(1-\frac{r}{N}) \to 0, \quad N \to \infty.
\end{align}
Thus, the system can be regarded as consisting of $n_h$ free particles. Therefore, in the sector with $n_h$ holes, $n_h$ merely contributes an overall factor at leading order in $N$, as shown in Table~\ref{tab:fixed-holes-results-Drude-Specificheat}.

When we fix the density of holes $\delta = \frac{n_h}{N}$, we define
\begin{align}
P_N
\equiv
\prod_{r=0}^{\delta N-1}\left(1-\frac{r}{N}\right).
\end{align}
Then
\begin{align}
R(N,\delta N)=1-P_N.
\end{align}
Taking the logarithm, we obtain
\begin{align}
\ln P_N
&=
\sum_{r=0}^{\delta N-1}\ln\left(1-\frac{r}{N}\right)
\nonumber\\
&=
N\left[
\frac{1}{N}\sum_{r=0}^{\delta N-1}\ln\left(1-\frac{r}{N}\right)
\right].
\end{align}
In the thermodynamic limit $N\to\infty$, the sum becomes a Riemann sum,
\begin{align}
\frac{1}{N}\sum_{r=0}^{\delta N-1}\ln\left(1-\frac{r}{N}\right)
\to
\int_0^\delta \ln(1-x)\,dx.
\end{align}
The integral is
\begin{align}
\int_0^\delta \ln(1-x)\,dx
=
-\delta-(1-\delta)\ln(1-\delta),
\end{align}
and therefore
\begin{align}
\ln P_N
=
-N\Bigl[\delta+(1-\delta)\ln(1-\delta)\Bigr]+o(N).
\end{align}
Hence
\begin{align}
P_N
=
\exp\!\left(
-N\Bigl[\delta+(1-\delta)\ln(1-\delta)\Bigr]+o(N)
\right),
\end{align}
so that
\begin{align}
R(N,\delta N)
=
1-
\exp\!\left(
-N\Bigl[\delta+(1-\delta)\ln(1-\delta)\Bigr]+o(N)
\right).
\end{align}
Since
\begin{align}
\delta+(1-\delta)\ln(1-\delta)>0,
\qquad 0<\delta<1,
\end{align}
we conclude that
\begin{align}
R(N,\delta N)\to 1,
\qquad N\to\infty.
\end{align}
Thus, for fixed hole density, replacing
\begin{align}
\sum_{0\le m_1<\cdots<m_{n_h}\le N-1}
\end{align}
by
\begin{align}
\frac{1}{n_h!}\sum_{m_1,\cdots,m_{n_h}=0}^{N-1}
\end{align}
is not a controlled approximation in the thermodynamic limit.

Therefore, at fixed hole density, the holes form a finite-density Fermi sea in the thermodynamic limit, so the low-temperature physics is that of a degenerate Fermi gas. Consequently, the low-energy excitations are governed by the linearized dispersion near the Fermi points and the Sommerfeld expansion yields the standard $T^{2}$ correction to the Drude weight. In Sec.~\ref{sec:finitedoping}, we have provided the analytical expression for the Drude weight with fixed $\delta$ in both low- and high-temperature limits.

By contrast, when the number of holes is kept fixed as $N \to \infty$, the hole density vanishes, $\delta \to 0$, and the system remains in the dilute, nondegenerate regime rather than entering a degenerate Fermi-gas regime. For example, in the single-hole sector, the low-energy excitations are governed by the quadratic dispersion near the band minimum,
\begin{align}
    \varepsilon(k)
    =
    2\cos k
    =
    2\cos(\pi+q)
    =
    -2+q^2+O(q^4).
\end{align}
Accordingly, at low temperatures the thermal expectation value of the energy is
\begin{align}
    \langle \hat{H} \rangle
    \approx
    -2+
    \frac{\int_{-\infty}^{\infty} dq\, q^2 e^{-\beta q^2}}
    {\int_{-\infty}^{\infty} dq\, e^{-\beta q^2}}
    =
    -2+\frac{T}{2}+O(T^2).
\end{align}
It then follows that the Drude weight acquires a linear-in-$T$ correction at low temperatures.

\section{Analytical expression for the Drude weight at fixed hole density}
\label{sec:finitedoping}

In this section, we derive the analytical expression for the Drude weight of the one-dimensional infinite-$U$ Hubbard model at finite doping, i.e., with $\delta = \frac{n_h}{N}$  fixed in the thermodynamic limit. We analyze both the high-temperature and low-temperature limits.

We denote by $n_{\downarrow}$ the number of spin-down fermions. Thus, the number of spin-up fermions is $n_{\uparrow} = N- n_h - n_{\downarrow}$. As shown in Eq.~\eqref{eq:energyspectrum_anyhole_anyspindown}, when $1 \leq n_{\downarrow} \leq N-n_h-1$, 
the energy spectrum is given by
\begin{align}
    E(m_1, m_2, \cdots, m_{n_{h}}; m_{\downarrow}) = \sum_{i=1}^{n_{h}}  2\cos(\frac{2 \pi m_i}{N} + \frac{2 \pi m_{\downarrow}}{N(N-n_h)}) = \sum_{i=1}^{n_{h}} E(m_{i}; m_{\downarrow}),
\end{align}
where $0 \leq m_{1} < m_{2} < \cdots < m_{n_{h}} \leq N-1$ and $0 \leq m_{\downarrow} \leq N-n_{h}-1$ (see the explanations below Eq.~\eqref{eq:energyspectrum_anyhole_anyspindown}). The degeneracy of each energy level is 
\begin{align}
   \frac{\binom{N-n_h}{n_{\downarrow}}}{N-n_{h}}.
   \label{eq:finite_doping_degeneracy}
\end{align}

We focus on the sector with $n_{\downarrow}=1$, since the energy spectrum remains unchanged in sectors with other values of $1 \leq n_{\downarrow} \leq N-n_h-1$ (see Eq.~\eqref{eq:energyspectrum_anyhole_anyspindown}). Including those sectors would only contribute an overall factor, which cancels between the numerator and denominator in the calculation of the Drude weight. Moreover, the degeneracy shown in Eq.~\eqref{eq:finite_doping_degeneracy} also cancels between the numerator and denominator in the calculation of the Drude weight. Note that we have neglected the contributions from the $n_{\downarrow}=0$ and $n_{\downarrow}=N-n_h$ sectors, since the numbers of states in these two sectors are negligible compared with those in the other $n_{\downarrow}$ sectors.
Therefore, in the following, we directly calculate the thermal expectation value of the energy at temperature $T$, which is given by
\begin{align}
    \langle \hat{H} \rangle &= \frac{\sum_{0 \leq m_1 < m_2 < \cdots < m_{n_h} \leq N-1} \sum_{m_{\downarrow}=0}^{N-n_h-1} e^{-\beta E(m_1, m_2, \cdots, m_{n_h}; m_{\downarrow})} E(m_1, m_2, \cdots, m_{n_h}; m_{\downarrow})}{\sum_{0 \leq m_1 < m_2 < \cdots < m_{n_h} \leq N-1} \sum_{m_{\downarrow}=0}^{N-n_h-1} e^{-\beta E(m_1, m_2, \cdots, m_{n_h}; m_{\downarrow})}} \nonumber \\ 
    &=  \frac{\sum_{\mathbf m} \sum_{m_{\downarrow}=0}^{N-n_h-1} e^{-\beta E(\mathbf{m}; m_{\downarrow})}E(\mathbf{m}; m_{\downarrow})}{\sum_{\mathbf m} \sum_{m_{\downarrow}=0}^{N-n_h-1} e^{-\beta E(\mathbf{m}; m_{\downarrow})}},
\label{eq:finite_doping_thermal_energy}
\end{align}
where for notational simplicity, we have used $\sum_{\mathbf{m}}$ with 
$\mathbf{m} = (m_1, m_2, \cdots, m_{n_h})$ as the shorthand for $\sum_{0 \leq m_1 < m_2 < \cdots < m_{n_h} \leq N-1}$.

\subsection{High temperature limit}
\label{sec:highT_energy_fixed_N_nh}

We first derive the high-temperature expansion of the thermal energy (Eq.~\eqref{eq:finite_doping_thermal_energy}) as well as the Drude weight directly from the exact energy spectrum. Note that the high-temperature derivation applies both to the case of a fixed number of holes and to the case of a fixed hole density, as will become clear in the following derivation (see also the explanation at the end of this section).

In the high-temperature limit, \(\beta=1/T\to 0\), we expand
\begin{align}
e^{-\beta E}
=
1-\beta E+O(\beta^2).
\end{align}
In the following, we keep only the leading order in $\beta$.
Substituting this into Eq.~\eqref{eq:finite_doping_thermal_energy}, we obtain
\begin{align}
\sum_{\mathbf m} \sum_{m_{\downarrow}=0}^{N-n_h-1} E(\mathbf m; m_{\downarrow}) e^{-\beta E(\mathbf{m}; m_{\downarrow})}
&=
\sum_{\mathbf m} \sum_{m_{\downarrow}=0}^{N-n_h-1} E(\mathbf m; m_{\downarrow})
-\beta \sum_{\mathbf m} \sum_{m_{\downarrow}=0}^{N-n_h-1} E(\mathbf m; m_{\downarrow})^2
+O(\beta^2) \nonumber \\
& =
-\beta \sum_{\mathbf m} \sum_{m_{\downarrow}=0}^{N-n_h-1} E(\mathbf m; m_{\downarrow})^2
+O(\beta^2),
\end{align}
where we have used $\sum_{\mathbf m} \sum_{m_{\downarrow}=0}^{N-n_h-1} E(\mathbf m; m_{\downarrow})=0$ and similarly
\begin{align}
\sum_{\mathbf m} \sum_{m_{\downarrow}=0}^{N-n_h-1} e^{-\beta E(\mathbf{m}; m_{\downarrow})}
&=
\sum_{\mathbf m} \sum_{m_{\downarrow}=0}^{N-n_h-1} 1
-\beta \sum_{\mathbf m} \sum_{m_{\downarrow}=0}^{N-n_h-1} E(\mathbf m; m_{\downarrow}) + O(\beta^2)  \nonumber \\
& = \sum_{\mathbf m} \sum_{m_{\downarrow}=0}^{N-n_h-1} 1 + O(\beta^2) \nonumber \\
& = (N-n_h) \binom{N}{n_h} + O(\beta^2).
\end{align}
Therefore, in the high temperature limit, the expectation value of energy is given by
\begin{align}
    \langle \hat{H} \rangle = -\beta \frac{\sum_{\mathbf m} \sum_{m_{\downarrow}=0}^{N-n_h-1} E(\mathbf m; m_{\downarrow})^2} {(N-n_h) \binom{N}{n_h}} + O(\beta^2)
    \label{eq:finite_doping_thermalE_highT}.
\end{align}
Physically, because the infinite-temperature mean energy vanishes, the leading term in $\langle\hat H\rangle$ is proportional to $-\beta$ times the infinite-temperature energy variance.

With fixed $\mathbf m$ and $m_{\downarrow}$, we have 
\begin{align}
    E(\mathbf m; m_{\downarrow})^2 &= (\sum_{i=1}^{n_h} E(m_i; m_{\downarrow}))^2 \nonumber \\
    & = \sum_{i=1}^{n_h} E(m_i; m_{\downarrow})^2 + \sum_{\substack{i,j=1\\ i\neq j}}^{n_h} E(m_i; m_{\downarrow}) E(m_j; m_{\downarrow}).
\end{align}
Therefore, with fixed $m_{\downarrow}$, we have 
\begin{align}
    \sum_{\mathbf m}E(\mathbf m; m_{\downarrow})^2 &= \sum_{\mathbf m} \left( \sum_{i=1}^{n_h} E(m_i; m_{\downarrow})^ 2 + \sum_{\substack{i,j=1\\ i\neq j}}^{n_h} E(m_i; m_{\downarrow}) E(m_j; m_{\downarrow}) \right) \nonumber \\ 
    & = \frac{n_h}{N} \binom{N}{n_h} \sum_{j=0}^{N-1} E(j; m_{\downarrow})^2 + \frac{\binom{n_h}{2} \binom{N}{n_h}}{\binom{N}{2}} \sum_{\substack{j,j'=0\\ j \neq j'}}^{N-1} E(j; m_{\downarrow}) E(j'; m_{\downarrow}) \nonumber \\
    & = \frac{n_h}{N} \binom{N}{n_h} \sum_{j=0}^{N-1} E(j; m_{\downarrow})^2 + \frac{\binom{n_h}{2} \binom{N}{n_h}}{\binom{N}{2}} \left(    (\sum_{j=0}^{N-1} E(j; m_{\downarrow}))^2 - \sum_{j=0}^{N-1} E(j; m_{\downarrow})^2  \right) \nonumber \\ 
    & = \left( \frac{n_h}{N} \binom{N}{n_h} - \frac{\binom{n_h}{2} \binom{N}{n_h}}{\binom{N}{2}} \right) \sum_{j=0}^{N-1}  E(j; m_{\downarrow})^2 + \frac{\binom{n_h}{2} \binom{N}{n_h}}{\binom{N}{2}}   (\sum_{j=0}^{N-1} E(j; m_{\downarrow}))^2  \nonumber \\
    &= \left( \frac{n_h}{N} \binom{N}{n_h} - \frac{\binom{n_h}{2} \binom{N}{n_h}}{\binom{N}{2}} \right) 2N.
    \label{eq:finite_doping_E2_highT}
\end{align}
In the second equality, we have used the fact that $m_i=0,\cdots,N-1$. Although, for a given $\mathbf m$, one has $m_i \neq m_{i'}$ when $i \neq i'$, after summing over all possible $\mathbf m$, each energy level $j=0,\cdots,N-1$ appears the same number of times. In the last equality we have used $\sum_{j=0}^{N-1} E(j; m_{\downarrow}) = 0$ and
\begin{align}
    \sum_{j=0}^{N-1}  E(j; m_{\downarrow})^2  &= 4\sum_{j=0}^{N-1} \cos^2(\frac{2\pi j}{N} + \frac{ 2 \pi m_{\downarrow}}{N(N-n_h)}) \nonumber \\
    & = 2\sum_{j=0}^{N-1} (1+ \cos(\frac{4\pi j}{N} + \frac{4\pi m_{\downarrow}}{N(N-n_h)}) )\nonumber \\
    & = 2N.
\end{align}
Substituting Eq.~\eqref{eq:finite_doping_E2_highT} into Eq.~\eqref{eq:finite_doping_thermalE_highT}, we have 
\begin{align}
    \langle \hat{H} \rangle &= -\beta \frac{\sum_{m_{\downarrow}=0}^{N-n_h-1} \left( \frac{n_h}{N} \binom{N}{n_h} - \frac{\binom{n_h}{2} \binom{N}{n_h}}{\binom{N}{2}} \right) 2N}{(N-n_h) \binom{N}{n_h}} + O(\beta^2) \nonumber \\
    & = -\beta (\frac{n_h}{N} - \frac{n_h(n_h-1)}{N(N-1)}) 2N + O(\beta^2).
\end{align}

Therefore, the Drude weight for the present model in the high temperature limit is 
\begin{align}
    D(T) &= -\frac{1}{2N} \langle E \rangle \nonumber \\
    &= \beta (\frac{n_h}{N} - \frac{n_h(n_h-1)}{N(N-1)}) \nonumber \\
    & \approx \frac{\delta(1-\delta)}{T},
    \label{eq:appendix_Drude_highT}
\end{align}
where we have used the approximation $\frac{n_h-1}{N-1} \approx \frac{n_h}{N} = \delta$. We note that the above derivation applies to both fixed-hole-number and fixed-hole-density cases. Taking $\delta=1/N$, corresponding to a single doped hole, reproduces the result shown in the main text.

We now explain why, in the high-temperature limit, the derivation of Drude weight (inverse-temperature behavior, $D \sim 1/T$) does not depend on whether one keeps $n_h$ fixed or $\delta$ fixed.
Regardless of whether the number of holes or the hole density is fixed, the eigenenergy of a given many-body eigenstate is always given by the sum of $n_h$ distinct single-particle mode energies.
In the high-temperature limit, to leading order in $\beta$, the Drude weight is proportional to the variance of the single-particle energy spectrum [see Eq.~\eqref{eq:finite_doping_E2_highT}], which is the same in both cases and $n_h$ just contributes an overall factor.
Therefore, the above derivation applies to both fixed $n_h$ and fixed $\delta$.

\subsection{Zero temperature}
\label{sec:ZeroT_energy_fixed_density}
Before deriving the analytical expression for the Drude weight with fixed density of holes in the low-temperature limit, we first consider the case $T=0$.

Note that the thermal expectation value of energy in Eq.~\eqref{eq:finite_doping_thermal_energy} can be expressed as 
\begin{align}
    \langle \hat{H} \rangle = - \frac{\partial \ln Z}{\partial \beta},
\end{align}
where 
\begin{align}
    Z = \sum_{m_{\downarrow}=0}^{N-n_h-1} \sum_{\mathbf m}  e^{-\beta E(\mathbf m; m_{\downarrow})} = \sum_{m_{\downarrow}=0}^{N-n_h-1} Z_{m_{\downarrow}}.
\end{align}
Here, $Z$ is the partition function and 
$Z_{m_{\downarrow}} = \sum_{\mathbf m}  e^{-\beta E(\mathbf m; m_{\downarrow})}$ is the partition function with fixed $m_{\downarrow}$.

We define
\begin{align}
\theta_m^{m_\downarrow}
=
\frac{2\pi m}{N}
+
\frac{2\pi m_\downarrow}{N(N-n_h)},
\label{eq:theta_def_lowT_fixed_density}
\end{align}
so that 
$E(\mathbf m;m_\downarrow)
=
\sum_{i=1}^{n_h} 2\cos\!\bigl(\theta_{m_i}^{m_\downarrow}\bigr)
$.
For brevity, we also write
\begin{align}
\phi_{m_\downarrow}
=
\frac{2\pi m_\downarrow}{N(N-n_h)},
\end{align}
so that $\theta_m^{m_\downarrow}
=
\frac{2\pi m}{N}+\phi_{m_\downarrow}$.

We first determine the ground state energy with fixed \(m_\downarrow\), i.e., the ground state energy corresponding to $Z_{m_{\downarrow}}$. In the derivation below, it will become clear that the ground-state energy, to leading order in system size $N$, does not depend on $m_{\downarrow}$.

Since the single-particle energy
\begin{align}
\varepsilon(\theta)=2\cos\theta
\end{align}
is minimized at \(\theta=\pi\), the ground state energy is obtained
by choosing those \(n_h\) distinct integers \(m\) whose corresponding angles
\(\theta_m^{m_\downarrow}\) lie closest to \(\pi\). Because the allowed values of \(m\) differ by integer steps, the corresponding
angles \(\theta_m^{m_\downarrow}\) are equally spaced with spacing
\begin{align}
\theta_{m+1}^{m_\downarrow}-\theta_m^{m_\downarrow}
=
\frac{2\pi}{N}.
\end{align}
Hence, the \(n_h\) angles closest to \(\pi\) necessarily correspond to a block
of \(n_h\) consecutive integers. Therefore, with fixed \(m_\downarrow\),
the $\mathbf m_0$ of the ground state is of the form
\begin{align}
\mathbf m_0=\{M,M+1,\dots,M+n_h-1\},
\label{eq:ground_state_block_fixed_density}
\end{align}
for some integer \(M\) chosen such that the corresponding block of angles is centered
as closely as possible around \(\theta=\pi\). In the following, we refer to $\mathbf m_0$ in Eq.~\eqref{eq:ground_state_block_fixed_density} as the ground-state block.
The left and right endpoints of this block are
\begin{align}
\theta_L^{(m_\downarrow)}
&=
\theta_M^{m_\downarrow}
=
\frac{2\pi M}{N}+\phi_{m_\downarrow},
\label{eq:thetaL_exact}
\\
\theta_R^{(m_\downarrow)}
&=
\theta_{M+n_h-1}^{m_\downarrow}
=
\frac{2\pi(M+n_h-1)}{N}+\phi_{m_\downarrow}.
\label{eq:thetaR_exact}
\end{align}
Their difference is
\begin{align}
\theta_R^{(m_\downarrow)}-\theta_L^{(m_\downarrow)}
=
\frac{2\pi(n_h-1)}{N}.
\label{eq:width_block_exact}
\end{align}
Since \(\delta=n_h/N\) is fixed, Eq.~\eqref{eq:width_block_exact} gives
\begin{align}
\theta_R^{(m_\downarrow)}-\theta_L^{(m_\downarrow)}
=
2\pi\delta+O(N^{-1}).
\label{eq:width_block_asymptotic}
\end{align}

Next, consider the center of the block (Eq.~\eqref{eq:ground_state_block_fixed_density})
\begin{align}
\theta_c^{(m_\downarrow)}
=
\frac{\theta_L^{(m_\downarrow)}+\theta_R^{(m_\downarrow)}}{2}
=
\frac{2\pi}{N}\left(M+\frac{n_h-1}{2}\right)+\phi_{m_\downarrow}.
\label{eq:center_block_exact}
\end{align}
By construction, the integer \(M\) is chosen such that \(\theta_c^{(m_\downarrow)}\)
is as close as possible to \(\pi\). Since adjacent possible values of \(\theta_c^{(m_\downarrow)}\)
differ by \(2\pi/N\), this implies
\begin{align}
\theta_c^{(m_\downarrow)}=\pi+O(N^{-1}).
\label{eq:center_block_asymptotic}
\end{align}
Combining Eqs.~\eqref{eq:width_block_asymptotic} and \eqref{eq:center_block_asymptotic},
we obtain
\begin{align}
\theta_L^{(m_\downarrow)}
=
\theta_c^{(m_\downarrow)}
-\frac12\bigl(\theta_R^{(m_\downarrow)}-\theta_L^{(m_\downarrow)}\bigr)
=
\pi-\pi\delta+O(N^{-1}),
\label{eq:thetaL_asymptotic}
\\
\theta_R^{(m_\downarrow)}
=
\theta_c^{(m_\downarrow)}
+\frac12\bigl(\theta_R^{(m_\downarrow)}-\theta_L^{(m_\downarrow)}\bigr)
=
\pi+\pi\delta+O(N^{-1}).
\label{eq:thetaR_asymptotic}
\end{align}
Thus, the leading-order positions of $\theta_L^{(m_\downarrow)}$ and $\theta_R^{(m_\downarrow)}$ depend only on $\delta$ and do not depend on $m_{\downarrow}$.

We now derive the ground state energy which is given by
\begin{align}
E_0(m_\downarrow)
=
\sum_{j=0}^{n_h-1}
2\cos\!\left(
\frac{2\pi(M+j)}{N}
+
\phi_{m_\downarrow}
\right),
\qquad
\phi_{m_\downarrow}
=
\frac{2\pi m_\downarrow}{N(N-n_h)}.
\label{eq:E0_before_centering}
\end{align}
According to Eq.~\eqref{eq:center_block_asymptotic}, the ground state energy can be expressed as 
\begin{align}
E_0(m_\downarrow)
&=
\sum_{j=0}^{n_h-1}
2\cos\!\left[
\pi+ O(N^{-1})
+\frac{2\pi}{N}\left(j-\frac{n_h-1}{2}\right)
\right]
\nonumber\\
&=
-2\sum_{j=0}^{n_h-1}
\cos\!\left[
O(N^{-1})
+\frac{2\pi}{N}\left(j-\frac{n_h-1}{2}\right)
\right] \nonumber \\
&\approx
-2\sum_{j=0}^{n_h-1}
\cos\!\left[\frac{2\pi}{N}\left(j-\frac{n_h-1}{2}\right)
\right],
\label{eq:E0_with_delta}
\end{align}
because when $n_h=\delta N$, the quantity $\frac{2\pi}{N}\left(j-\frac{n_h-1}{2}\right)$ is of order unity, and the $O(N^{-1})$ correction can therefore be neglected.

Using the standard identity
\begin{align}
\sum_{j=0}^{n-1}\cos(a+jd)
=
\frac{\sin(nd/2)}{\sin(d/2)}
\cos\!\left(a+\frac{(n-1)d}{2}\right),
\label{eq:cos_sum_identity}
\end{align}
and choosing
\begin{align}
n=n_h,\qquad
a=-\frac{\pi(n_h-1)}{N},\qquad
d=\frac{2\pi}{N},
\end{align}
we obtain
\begin{align}
\sum_{j=0}^{n_h-1}
\cos\!\left[
\frac{2\pi}{N}\left(j-\frac{n_h-1}{2}\right)
\right]
=
\frac{\sin\!\left(\frac{\pi n_h}{N}\right)}
{\sin\!\left(\frac{\pi}{N}\right)}.
\end{align}

Therefore, the ground state energy is 
\begin{align}
E_0(N,n_h)
=
-2\,
\frac{\sin\!\left(\frac{\pi n_h}{N}\right)}
{\sin\!\left(\frac{\pi}{N}\right)}.
\label{eq:E0_exact_final}
\end{align}
In the thermodynamic limit with fixed hole density \(\delta=n_h/N\), one has
\begin{align}
\sin\!\left(\frac{\pi n_h}{N}\right)=\sin(\pi\delta),
\qquad
\sin\!\left(\frac{\pi}{N}\right)\sim \frac{\pi}{N},
\end{align}
and therefore
\begin{align}
E_0(N,n_h)
=
-\frac{2N}{\pi}\sin(\pi\delta).
\label{eq:E0_final}
\end{align}
Therefore, the Drude weight at zero temperature is 
\begin{align}
    D(T=0) = \frac{\sin(\pi \delta)}{\pi}.
\end{align}

\subsection{Low temperature limit}
\label{sec:LowT_energy_fixed_density}
With a fixed hole density, the low-energy states have a Fermi-sea-like structure. For the energy density at low temperatures, the Sommerfeld expansion shows that the linear-in-$T$ correction vanishes, so that the leading correction is of order $T^2$, which is the physical origin of the $T^2$ correction to the Drude weight. In the following, we derive the analytical expressions for the thermal expectation value of energy (see Eq.~\eqref{eq:finite_doping_thermal_energy}) as well as the Drude weight with fixed hole density $\delta = \frac{n_h}{N}$
in the low temperature limit directly from the explicit energy spectrum.

We first characterize the low-energy excitations above the ground state with \(\mathbf m_0\) and fixed $m_{\downarrow}$ (see Eq.~\eqref{eq:ground_state_block_fixed_density}). The crucial point is that low-temperature thermodynamics is governed only by
states whose excitation energies above the ground state vanish in the thermodynamic
limit. Excitations characterized by $\mathbf m$, where $\mathbf m$ is obtained by removing an occupied integer deep inside the block $\mathbf m_0$ (see Eq.~\eqref{eq:ground_state_block_fixed_density})
and reinserting it far away generally cost an energy of order \(O(1)\), and are
therefore exponentially suppressed as \(T\to 0\). Consequently, the only relevant
low-energy excitations are small deformations of the two edges of the consecutive
ground-state block. For example, one may consider excitations with $\mathbf m = \{M-1,M+1,\dots,M+n_h-1\}$ or $\mathbf m = \{M,M+1,\dots,M+n_h\}$. In the former case, only the left edge of the ground-state block $\mathbf m_0$ is modified, with the integer $M$ replaced by $M-1$. In the latter case, only the right edge is modified, with the integer $M+n_h-1$ replaced by $M+n_h$.
In these two examples, we have only illustrated excitations in which a single integer (i.e., a single-particle mode) near the edge of the ground state block is changed by one. In general, however, the magnitude of the change and the number of integers that change can both be greater than one.

To parameterize such excitations, we consider outward deformations of the right
and left edges separately. For the right edge, write the excited occupied integers near the right boundary as
\begin{align}
m_{n_h-j+1}
=
M+n_h-j+\lambda_j,
\qquad
j=1,2,\dots,
\label{eq:right_edge_partition_param}
\end{align}
i.e., the occupied single-particle mode changes from $M+n_h-j$ to $M+n_h-j+\lambda_j$,
where $\lambda_j \ge 0$ denotes the shift of the $j$-th occupied single-particle mode, counted from the right edge, toward larger momentum.
When all $\lambda_j=0$, this corresponds to the ground-state block $\mathbf m_0$.
The ordering condition
\begin{align}
m_{n_h-j+1}>m_{n_h-j}
\end{align}
implies
\begin{align}
\lambda_1\ge \lambda_2\ge \lambda_3\ge \cdots \ge 0.
\label{eq:lambda_partition_condition}
\end{align}
Thus, the right-edge deformation is specified by an integer partition
\(\lambda=(\lambda_1,\lambda_2,\dots)\). We note that when the number of changed integers becomes too large, or when $\lambda_{j}$ for some $j$ are too large, the associated energy cost can be substantial. Here, however, we keep all such excitations and then perform the low-temperature expansion below in which the excitations with large energy cost will be suppressed.

Similarly, for the left edge we write
\begin{align}
m_j
=
M+j-1-\mu_j,
\qquad
j=1,2,\dots,
\label{eq:left_edge_partition_param}
\end{align}
where \(\mu_j\ge 0\) measures the outward displacement of the \(j\)-th occupied
integer counted from the left edge.
The ordering condition
\begin{align}
m_j<m_{j+1}
\end{align}
implies
\begin{align}
\mu_1\ge \mu_2\ge \mu_3\ge \cdots \ge 0.
\label{eq:mu_partition_condition}
\end{align}
Hence, the left-edge deformation is specified by another integer partition
\(\mu=(\mu_1,\mu_2,\dots)\).

We define
\begin{align}
|\lambda|
=
\sum_{j=1}^{\infty}\lambda_j,
\qquad
|\mu|
=
\sum_{j=1}^{\infty}\mu_j.
\label{eq:partition_sizes_def}
\end{align}
Therefore, every low-energy excited state is labeled by a pair of integer partitions
\((\lambda,\mu)\).

We next compute the leading energy cost of moving an edge level outward by \(r\). For example, consider the excitation in which one occupied single-particle mode is shifted from \(M+n_h-1\) to \(M+n_h-1+r\).

Consider first the right edge. Since
$\varepsilon(\theta)=2\cos\theta$,
if a right-edge occupied level is shifted outward by \(r\), then
its angle changes by
\begin{align}
\Delta\theta=\frac{2\pi r}{N}.
\label{eq:delta_theta_single_shift}
\end{align}
For the low-energy excitations that contribute to the thermodynamic low-temperature
asymptotics, \(r=O(1)\), hence \(\Delta\theta=O(N^{-1})\).
We may therefore perform a Taylor expansion around the right endpoint
\begin{align}
\varepsilon\!\left(\theta_R^{(m_\downarrow)}+\Delta\theta\right)
=
\varepsilon\!\left(\theta_R^{(m_\downarrow)}\right)
+
\varepsilon'\!\left(\theta_R^{(m_\downarrow)}\right)\Delta\theta
+
\frac12\varepsilon''\!\left(\theta_R^{(m_\downarrow)}\right)(\Delta\theta)^2
+\cdots.
\label{eq:taylor_right_edge}
\end{align}
Since
\begin{align}
\varepsilon'(\theta)=-2\sin\theta,
\qquad
\varepsilon''(\theta)=-2\cos\theta,
\end{align}
and according to Eq.~\eqref{eq:thetaR_asymptotic}, 
we obtain
\begin{align}
\varepsilon'\!\left(\theta_R^{(m_\downarrow)}\right)
&=
-2\sin\!\left(\pi+\pi\delta+O(N^{-1})\right)
=
2\sin(\pi\delta)+O(N^{-1}),
\\
\varepsilon''\!\left(\theta_R^{(m_\downarrow)}\right)
&=
-2\cos\!\left(\pi+\pi\delta+O(N^{-1})\right)
=
2\cos(\pi\delta)+O(N^{-1}).
\end{align}
Substituting into Eq.~\eqref{eq:taylor_right_edge} and using
\(\Delta\theta=2\pi r/N\), we find
\begin{align}
\Delta\varepsilon_R(r)
=
\varepsilon\!\left(\theta_R^{(m_\downarrow)}+\Delta\theta\right)
-
\varepsilon\!\left(\theta_R^{(m_\downarrow)}\right)
=
\frac{4\pi\sin(\pi\delta)}{N}\,r
+
O(N^{-2}).
\label{eq:right_shift_energy}
\end{align}

The same analysis applies to the left edge. There, moving an occupied level outward
means shifting it to the left by \(r\), i.e. by angle
\(-2\pi r/N\). Expanding around \(\theta_L^{(m_\downarrow)}\), we obtain
\begin{align}
\Delta\varepsilon_L(r)
=
-\varepsilon'\!\left(\theta_L^{(m_\downarrow)}\right)\frac{2\pi r}{N}
+
O(N^{-2}).
\label{eq:left_shift_energy_intermediate}
\end{align}
According to Eq.~\eqref{eq:thetaL_asymptotic}, 
we have
\begin{align}
-\varepsilon'\!\left(\theta_L^{(m_\downarrow)}\right)
=
2\sin\!\left(\pi-\pi\delta+O(N^{-1})\right)
=
2\sin(\pi\delta)+O(N^{-1}),
\end{align}
and therefore
\begin{align}
\Delta\varepsilon_L(r)
=
\frac{4\pi\sin(\pi\delta)}{N}\,r
+
O(N^{-2}).
\label{eq:left_shift_energy}
\end{align}

Thus, for either edge, the leading energy cost of an outward displacement by \(r\) is
\begin{align}
\Delta\varepsilon(r)
=
\frac{4\pi\sin(\pi\delta)}{N}\,r
+
O(N^{-2}).
\label{eq:single_shift_energy_final}
\end{align}

We now sum the contributions of all displaced edge levels. For a right-edge partition \(\lambda\), the total right-edge displacement is $\vert \lambda \vert$ (see Eq.~\eqref{eq:partition_sizes_def}). Similarly for the left edge, the total left-edge displacement is $\vert \mu \vert$ (see Eq.~\eqref{eq:partition_sizes_def}).
Since the leading energy cost in Eq.~\eqref{eq:single_shift_energy_final}
is linear in the displacement \(r\), the total excitation energy of the pair
\((\lambda,\mu)\) is simply the sum of all individual contributions
\begin{align}
\Delta E(\lambda,\mu;m_\downarrow)
&=
\frac{4\pi\sin(\pi\delta)}{N}
\left(
\sum_{j=1}^{\infty}\lambda_j
+
\sum_{j=1}^{\infty}\mu_j
\right)
+
O(N^{-2})
\nonumber\\
&=
\frac{4\pi\sin(\pi\delta)}{N}\bigl(|\lambda|+|\mu|\bigr)
+
O(N^{-2}).
\label{eq:DeltaE_partitions_final}
\end{align}
Equation~\eqref{eq:DeltaE_partitions_final} shows that, to leading order,
the edge excitation spectrum is independent of \(m_\downarrow\).

Let \(E_0(m_\downarrow)\) denote the ground-state energy with fixed
\(m_\downarrow\) (see Eq.~\eqref{eq:E0_final} ). Then, keeping only the low-energy edge excitations,
the partition function $Z_{m_{\downarrow}}$ is
\begin{align}
Z_{\mathrm{low}}^{(m_\downarrow)}
=
e^{-\beta E_0(m_\downarrow)}
\sum_{\lambda,\mu}
\exp\!\left[
-\beta\frac{4\pi\sin(\pi\delta)}{N}\bigl(|\lambda|+|\mu|\bigr)
\right]
\left[1+o(1)\right].
\label{eq:Z_low_sector_before_q}
\end{align}
Define
\begin{align}
\alpha=\frac{4\pi\sin(\pi\delta)}{N},
\qquad
q=e^{-\beta\alpha}.
\label{eq:alpha_q_def}
\end{align}
Then Eq.~\eqref{eq:Z_low_sector_before_q} becomes
\begin{align}
Z_{\mathrm{low}}^{(m_\downarrow)}
=
e^{-\beta E_0(m_\downarrow)}
\sum_{\lambda,\mu} q^{|\lambda|+|\mu|}
\left[1+o(1)\right].
\end{align}
Because the left-edge and right-edge deformations are independent, the sum factorizes
\begin{align}
\sum_{\lambda,\mu} q^{|\lambda|+|\mu|}
=
\left(\sum_{\lambda} q^{|\lambda|}\right)
\left(\sum_{\mu} q^{|\mu|}\right)
=
\left(\sum_{\lambda} q^{|\lambda|}\right)^2.
\label{eq:partition_sum_factorization}
\end{align}

Now we use the standard generating function for integer partitions~\cite{andrews1998theory}:
\begin{align}
\sum_{\lambda} q^{|\lambda|}
=
\prod_{\ell=1}^{\infty}(1-q^\ell)^{-1},
\qquad |q|<1.
\label{eq:partition_generating_function}
\end{align}
Substituting Eq.~\eqref{eq:partition_generating_function} into
Eq.~\eqref{eq:partition_sum_factorization}, we obtain
\begin{align}
Z_{\mathrm{low}}^{(m_\downarrow)}
=
e^{-\beta E_0(m_\downarrow)}
\prod_{\ell=1}^{\infty}
\left(1-q^\ell\right)^{-2}
\left[1+o(1)\right].
\end{align}
Using the definition \(q=e^{-\beta\alpha}\), this can be written as
\begin{align}
Z_{\mathrm{low}}^{(m_\downarrow)}
=
e^{-\beta E_0(m_\downarrow)}
\prod_{\ell=1}^{\infty}
\left(
1-e^{-\beta\alpha\ell}
\right)^{-2}
\left[1+o(1)\right],
\qquad
\alpha=\frac{4\pi\sin(\pi\delta)}{N}.
\label{eq:Zlow_sector_Euler_product}
\end{align}

The full partition function $Z$ is obtained by summing $Z_{\mathrm{low}}^{(m_\downarrow)}$ in Eq.~\eqref{eq:Zlow_sector_Euler_product}
over all allowed \(m_\downarrow\).
The important point is that the Euler-product factor in
Eq.~\eqref{eq:Zlow_sector_Euler_product} is the same for all \(m_\downarrow\) sectors
at leading order, because the coefficient \(\alpha=4\pi\sin(\pi\delta)/N\) is
independent of \(m_\downarrow\).

Therefore, summing over \(m_\downarrow\) only changes an overall prefactor, which does not affect the leading \(T^2\) correction to the extensive thermal energy.
Consequently, for the purpose of extracting the leading low-temperature behavior of the
energy density, it is sufficient to analyze the Euler-product factor in
Eq.~\eqref{eq:Zlow_sector_Euler_product}.

Taking the logarithm of Eq.~\eqref{eq:Zlow_sector_Euler_product}, we obtain
\begin{align}
\ln Z_{\mathrm{low}}^{(m_\downarrow)}
=
-\beta E_0(m_\downarrow)
-2\sum_{\ell=1}^{\infty}
\ln\!\left(1-e^{-\beta\alpha\ell}\right)
+
o(1).
\label{eq:logZ_low_sector}
\end{align}
Introduce
\begin{align}
x=\beta\alpha.
\end{align}
In the thermodynamic limit, \(\alpha\sim N^{-1}\), and hence \(x\to 0^+\)
for fixed low temperature \(T\).
We now use the standard small-\(x\) asymptotics of the Euler product~\cite{apostol2012modular}:
\begin{align}
\sum_{\ell=1}^{\infty}
\ln\!\left(1-e^{-x\ell}\right)^{-1}
=
\frac{\pi^2}{6x}
+
O(\ln x),
\qquad
x\to 0^+.
\label{eq:Euler_small_x_asymptotic}
\end{align}
Substituting Eq.~\eqref{eq:Euler_small_x_asymptotic} into
Eq.~\eqref{eq:logZ_low_sector}, we find
\begin{align}
\ln Z_{\mathrm{low}}^{(m_\downarrow)}
=
-\beta E_0(m_\downarrow)
+
\frac{\pi^2}{3x}
+
O(\ln x).
\end{align}
Using \(x=\beta\alpha\), this becomes
\begin{align}
\ln Z_{\mathrm{low}}^{(m_\downarrow)}
=
-\beta E_0(m_\downarrow)
+
\frac{\pi^2}{3\beta\alpha}
+
O(\ln(\beta\alpha)).
\label{eq:logZ_low_sector_final}
\end{align}

The thermal energy is
\begin{align}
\langle \hat{H}\rangle
=
-\partial_\beta \ln Z.
\end{align}
Hence, the Euler-product contribution gives
\begin{align}
\langle \hat{H} \rangle
=
E_0
+
\frac{\pi^2}{3\alpha\beta^2}
+
\text{subleading terms},
\label{eq:E_low_before_alpha_sub}
\end{align}
where \(E_0\) denotes the ground-state energy (see Eq.~\eqref{eq:E0_final}).

Substituting
\begin{align}
\alpha=\frac{4\pi\sin(\pi\delta)}{N},
\end{align}
we obtain
\begin{align}
\frac{\pi^2}{3\alpha\beta^2}
=
\frac{\pi^2}{3\beta^2}
\frac{N}{4\pi\sin(\pi\delta)}
=
\frac{N\pi}{12\sin(\pi\delta)}\,T^2.
\end{align}
Therefore,
\begin{align}
\langle \hat{H} \rangle
=
E_0
+
\frac{N\pi}{12\sin(\pi\delta)}\,T^2
+
o(NT^2).
\label{eq:E_low_total_final}
\end{align}

Consequently, the Drude weight is 
\begin{align}
D(T) &= -\frac{1}{2N} \langle \hat{H} \rangle \nonumber \\ 
& = -\frac{E_0}{2N} -
\frac{\pi}{24\,\sin(\pi\delta)}\,T^2 +
o(T^2) \nonumber \\
&=
\frac{\sin(\pi\delta)}{\pi}
-
\frac{\pi}{24\,\sin(\pi\delta)}\,T^2
+
o(T^2).
\label{eq:final_lowT_energy_density_fixed_delta}
\end{align}
In the low-temperature limit, the leading correction to the Drude weight is of order $T^2$ when the hole density is fixed.

\section{Analytical Expression for the Effective Resistivity}

As summarized in Table~I in the main text and detailed in Secs.~\ref{sec:fixed_number_hole_Drude_weight} and \ref{sec:finitedoping}, we have obtained analytical expressions for the Drude weight at arbitrary temperatures in the single-hole and fixed-hole-number cases, as well as in the low- and high-temperature limits at fixed hole density. In this section, we derive an analytical expression for the effective resistivity by introducing a finite broadening of the Drude peak (see Sec.~\ref{sec:effective_resistivity}). More importantly, we show that the effective resistivity scales linearly with $T$ at fixed hole number but quadratically with $T$ at fixed hole density (see Sec.~\ref{sec:Temperature_dependence_of_effective_resistivity}). Although the present one-dimensional model is not intended to directly describe the strange-metal behavior observed in cuprates~\cite{doi:10.1126/science.abh4273,marelQuantumCriticalBehaviour2003, PhysRevLett.81.4720, PhysRevLett.92.167001, jinLinkSpinFluctuations2011, PhysRevB.41.846, doi:10.1126/science.1165015, daouLinearTemperatureDependence2009, PhysRevB.95.224517, doiron-leyraudPseudogapPhaseCuprate2017,legrosUniversalTlinearResistivity2019, doi:10.1126/science.aat4134} or the numerical results obtained for two-dimensional models~\cite{PhysRevLett.110.086401,PhysRevB.91.075124,PhysRevB.95.041110,PhysRevLett.123.036601,doi:10.1126/science.aau7063,fratini2025strangemetaltransportcoupling}, this qualitative distinction between the fixed-hole-number and fixed-hole-density limits in a controlled analytical setting calls for caution when extrapolating conclusions from finite-size two-dimensional simulations, for which the thermodynamic limit at fixed density remains far from accessible (see further discussion in Sec.~\ref{sec:Temperature_dependence_of_effective_resistivity}).

\subsection{Definition of the Effective Resistivity}
\label{sec:effective_resistivity}

For the present model, the optical conductivity consists solely of a Drude contribution, $\sigma(\omega,T)=2\pi D(T)\delta(\omega)$.
Because the inverse of a Dirac delta function is not well defined, a finite DC resistivity cannot be directly extracted from this singular conductivity. We therefore regularize the delta function using a Lorentzian,
\begin{align}
    \label{eq:main_Lorentzian}
    \delta_{\eta}^{\mathrm{L}}(\omega)
    =
    \frac{1}{\pi}
    \frac{\eta}{\omega^2+\eta^2},
\end{align}
where the Dirac $\delta$-function is recovered in the limit $\eta\to0$: $
    \delta(\omega)
    =
    \lim_{\eta\to0}
    \delta_{\eta}^{\mathrm{L}}(\omega)$.
Keeping $\eta$ finite, the regularized DC conductivity is
\begin{align}
    \sigma_{\mathrm{DC}}(T)
    =\left.\frac{2D(T)\eta}{\eta^2+\omega^2}\right|_{\omega=0}
    =\frac{2D(T)}{\eta}.
\end{align}
We thus define the effective resistivity extracted from the broadened Drude peak as
\begin{align}
    \label{eq:main_effective_rho}
    \rho(T)
    \equiv \frac{1}{\sigma_{\mathrm{DC}}(T)}
    =\frac{\eta}{2D(T)}.
\end{align}

This regularization and the resulting definition of the effective resistivity are motivated by the procedures commonly employed in numerical calculations of DC transport. For a generic model, irrespective of dimensionality, the optical conductivity generally contains contributions from both the Drude peak and a regular part [see Eq.~\eqref{eq:appendix_dcconductivity}]. Even in chaotic systems, for which the Drude weight is expected to vanish in the thermodynamic limit, a nonzero Drude contribution can remain in finite-size systems. Moreover, because the energy spectrum of a finite system is discrete, the delta functions appearing in both the Drude and regular contributions must be broadened to obtain a smooth conductivity. Lorentzian broadening, as introduced in Eq.~\eqref{eq:main_Lorentzian}, is a standard choice, with $\eta$ typically chosen to be of the order of the mean many-body level spacing~\cite{PhysRevB.73.035113,PhysRevB.105.L201108}. The Lorentzian broadening also admits a time-domain interpretation:
\begin{align}
    \delta_{\eta}^{\mathrm{L}}(\omega)
    =
    \frac{1}{2\pi}
    \int_{-\infty}^{\infty}
    dt\,e^{i\omega t}e^{-\eta|t|}.
\end{align}
Thus, mathematically, introducing a Lorentzian width $\eta$ is equivalent to imposing an exponential time-domain cutoff. Numerically, this cutoff represents a finite frequency resolution. Phenomenologically, $\eta$ may be interpreted as an effective damping rate if an additional physical relaxation mechanism causes the current or current--current correlation function to decay approximately as $e^{-\eta t}$. In the isolated integrable model considered here, however, $\eta$ is an artificial broadening parameter rather than an intrinsic relaxation rate generated by the Hamiltonian. The nondecaying Drude response is recovered in the limit $\eta\to0$. Therefore, although the conductivity of the present model contains only a Drude contribution, it remains meaningful to examine the temperature dependence of the regularized effective resistivity defined in Eq.~\eqref{eq:main_effective_rho}.

\subsection{Temperature Dependence of the Effective Resistivity}
\label{sec:Temperature_dependence_of_effective_resistivity}

\begin{figure}[t]
    \centering
\includegraphics[width=0.95\linewidth]{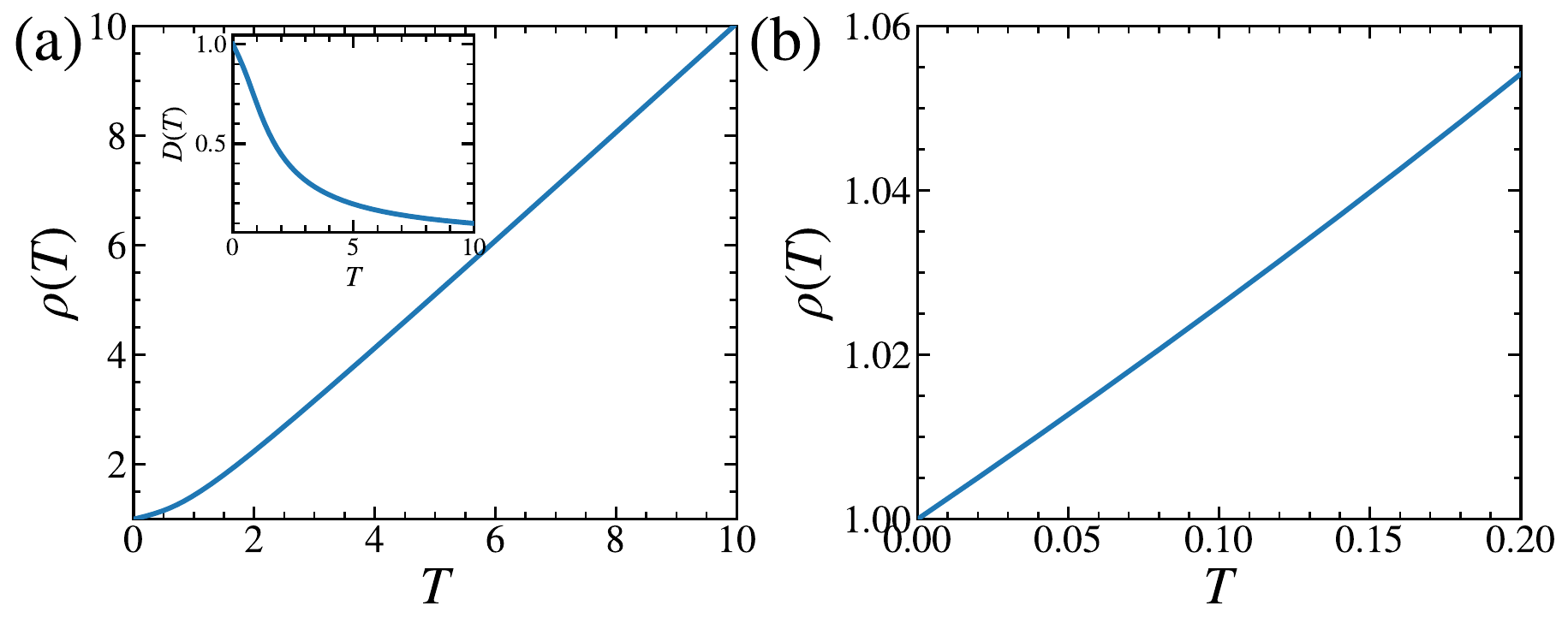}
    \caption{Rescaled effective resistivity $2\rho(T)/(N\eta)$ as a function of temperature $T$. Left panel: $T \in [0,10]$. In the high-temperature regime, the slope approaches $1$.
Right panel: $T \in [0,0.2]$. In the low-temperature regime, the slope approaches $1/4$.
}
    \label{fig:InverseDrudeweight}
\end{figure}

We first focus on the single-hole case. The analysis can be readily extended to a fixed number $n_h$ of doped holes or to a finite hole density $\delta$ using Eq.~\eqref{eq:main_effective_rho}. Figure~\ref{fig:InverseDrudeweight} shows the effective resistivity defined in Eq.~\eqref{eq:main_effective_rho}, together with the corresponding Drude weight, over the temperature range $T\in[0,10]$. Here and throughout this discussion, we set the hopping amplitude to $t=1$, so that temperature is measured in units of $t$.

We identify two regimes in which the effective resistivity exhibits a linear dependence on temperature. In the high-temperature limit, the result summarized in Table~I gives
\begin{align}
    \rho(T)
    \approx \frac{N\eta}{2}\,T.
\end{align}
Thus, the effective resistivity is linear in $T$ at high temperatures. As shown by the relation $2\pi D(T)\delta(\omega)
    =\sigma_{\mathrm{diff}}(\omega, T)$,
together with Eq.~\eqref{eq:DefinitionOfSigmadiff}, the expectation value of $\hat{J}^{2}$ becomes temperature independent in this limit. The remaining factor of $\beta=1/T$ in the conductivity therefore produces the linear-in-$T$ effective resistivity. Previous studies~\cite{PhysRevB.111.035123,y7fg-lhsm} have related high-temperature linear-in-$T$ resistivity to the atomic limit of the Hubbard model, suggesting that this behavior is a general high-temperature feature that is insensitive to dimensionality, interaction strength, and filling.

More importantly, a distinct linear-in-$T$ regime also emerges at low temperatures. According to Table~I, the effective resistivity behaves as
\begin{align}
    \rho(T)
    \approx \frac{N\eta}{2}
    \left(
        1+\frac{T}{4}
    \right).
\end{align}
The leading temperature-dependent correction is therefore again linear in $T$, but its slope differs from that in the high-temperature regime, as shown in Fig.~\ref{fig:InverseDrudeweight}(b).

For a fixed hole density, the high-temperature result summarized in Table~I likewise yields a linear-in-$T$ effective resistivity, consistent with Refs.~\cite{PhysRevB.111.035123,y7fg-lhsm}. In the low-temperature regime, however, the leading temperature-dependent correction to the effective resistivity is quadratic in $T$.

Strange-metal behavior remains a long-standing central problem in the study of strongly correlated systems~\cite{doi:10.1126/science.abh4273}. In particular, a recent numerical study of the two-dimensional infinite-$U$ Hubbard model on an 18-site rotated square lattice with a single doped hole reported a linear-in-$T$ resistivity persisting down to temperatures as low as $T=0.02t$~\cite{fratini2025strangemetaltransportcoupling}. By following the same extrapolation procedure, we have independently reproduced this result in our forthcoming work~\cite{ourupcomingpaperLinearT2}. The authors of Ref.~\cite{fratini2025strangemetaltransportcoupling} suggested that microscopic processes relevant to strange metallicity may already emerge at the single-hole level. 

Although the present one-dimensional Hubbard model is not intended to describe strange-metal behavior directly, it provides a controlled analytical example that highlights two issues requiring particular care in finite-size transport studies. First, the Drude and regular contributions to the conductivity must be identified separately. Second, the fixed-hole-number and fixed-hole-density limits need not exhibit the same temperature dependence: in the present model, the leading low-temperature correction to the effective resistivity is linear in $T$ at fixed hole number but quadratic in $T$ at fixed hole density. In exact-diagonalization studies of two-dimensional systems, the accessible system sizes are severely limited, making it difficult to distinguish these two limits reliably. These results therefore call for caution when extrapolating conclusions from finite-size numerical calculations.

Other numerical approaches to charge transport, including quantum Monte Carlo (QMC) and density matrix renormalization group (DMRG), can work directly at fixed hole density and therefore avoid the ambiguity between the fixed-hole-number and fixed-hole-density limits that may arise in small-cluster exact-diagonalization calculations. Nevertheless, obtaining a fully controlled low-temperature DC response in the two-dimensional thermodynamic limit remains challenging because of method-specific numerical limitations:

\begin{itemize}
    \item \textbf{QMC.---}
    Although QMC simulations can be performed directly at fixed density, extracting a controlled low-temperature DC response involves two additional challenges. First, away from special sign-problem-free regimes, the fermion sign problem becomes increasingly severe as the temperature is lowered and the system size is increased, thereby restricting the accessible range of parameters. Second, QMC directly evaluates the imaginary-time current-current correlation function $\Lambda(\tau)$ rather than the real-frequency conductivity. Determining $\sigma(\omega)$ therefore requires inversion of the corresponding integral relation between $\Lambda(\tau)$ and $\sigma(\omega)$. This analytic-continuation procedure is numerically ill-conditioned: small statistical uncertainties in $\Lambda(\tau)$ can lead to substantial variations in the reconstructed low-frequency spectrum, particularly in the DC limit. Finite-size and analytic-continuation uncertainties become particularly important near the lowest accessible temperatures~\cite{doi:10.1126/science.aau7063}.

    \item \textbf{DMRG.---}
    DMRG is subject to a different set of limitations. Although it can accurately treat long systems at fixed density, two-dimensional calculations are typically performed on quasi-one-dimensional cylinders of finite width. Recovering the genuine two-dimensional limit therefore requires a controlled extrapolation in the cylinder width. Such an extrapolation remains computationally demanding, especially for finite-temperature dynamical transport, where both entanglement growth and the required real-time or frequency resolution further restrict the accessible length, time, and temperature scales.
\end{itemize}

\end{document}